\documentclass[12pt,a4paper]{article}
\pdfoutput=1
\usepackage{jcappub}
\usepackage[utf8]{inputenc}
\usepackage[T1]{fontenc}
\usepackage{float}
\usepackage{amsmath}
\usepackage{epsf}
\usepackage{epsfig}
\usepackage{amssymb}
\usepackage{graphicx}
\usepackage{latexsym}
\usepackage{hyperref}
\usepackage{rotating}

\def\e{{\epsilon}}

\newcommand{\ba}{\begin{array}}
\newcommand{\ea}{\end{array}}
\newcommand{\be}{\begin{equation}}
\newcommand{\ee}{\end{equation}}
\newcommand{\bea}{\begin{eqnarray}}
\newcommand{\eea}{\end{eqnarray}}
\newcommand{\bg}{\begin{gather}}
\newcommand{\eg}{\end{gather}}
\newcommand{\bseq}{\begin{subequations}}
\newcommand{\eseq}{\end{subequations}}

\makeatletter
\def\gsim{\compoundrel>\over\sim}
\def\lsim{\compoundrel<\over\sim}
\def\compoundrel#1\over#2{\mathpalette\compoundreL{{#1}\over{#2}}}
\def\compoundreL#1#2{\compoundREL#1#2}
\def\compoundREL#1#2\over#3{\mathrel
         {\vcenter{\hbox{$\m@th\buildrel{#1#2}\over{#1#3}$}}}}
\makeatother

\numberwithin{equation}{section}

\title{Non-standard interactions and neutrinos from dark matter
  annihilation in the Sun} 

\author[a,b]{S. V. Demidov}
\affiliation[a]{Institute for Nuclear Research of the Russian Academy of
  Sciences,\\ 60th October Anniversary prospect 7a, Moscow
  117312, Russia}
\affiliation[b]{Moscow Institute of Physics and Technology,
  \\ Institutsky per. 9, Dolgoprudny 141700, Russia} 

\emailAdd{demidov@ms2.inr.ac.ru}

\date{\today}

\abstract{We perform an analysis of the influence of non-standard
neutrino interactions (NSI) on neutrino signal from dark matter
annihilations in the Sun. Taking experimentally allowed benchmark
values for the matter NSI parameters we show that the evolution of
such neutrinos with energies at GeV scale can be considerably modified. 
We simulate propagation of neutrinos 
from the Sun to the Earth for realistic dark matter annihilation
channels and find that the matter NSI can result in at most 30\%
correction to the signal rate of muon track events at neutrino
telescopes. Still present experimental bounds on dark matter from
  these searches are robust in the presence of NSI within considerable
part of their allowed parameter space. At the same time electron neutrino
flux from dark matter annihilation in the Sun can be changed by a
factor of few.} 

\begin{document}
\maketitle
\flushbottom

\section{Introduction}
\label{sec:1}
Dark matter is one of the most intriguing mysteries of the modern
particle physics. Existence of new particles is the dominant
hypothesis for explanation of this
phenomena~\cite{Bertone:2004pz}. Looking for 
neutrinos resulting from dark matter annihilation in the Sun is one of
the possible indirect way of searching 
for the signal from dark matter particles~\cite{Silk:1985ax}. The idea
is that these particles can be gravitationally trapped and accumulated inside the
Sun~\cite{Gould:1987ir} during its evolution in such amount that they start to annihilate. If
dark matter particles annihilate into SM particles than among the final
products of these annihilations can be high energy neutrinos which can
reach the surface of the Sun, traverse to the Earth and  be
observed at  neutrino telescopes such as IceCube~\cite{Aartsen:2016zhm},
Super-Kamiokande~\cite{Choi:2015ara},
ANTARES~\cite{Adrian-Martinez:2016gti} as well as
neutrino telescopes at the Baksan~\cite{Boliev:2013ai} and
Baikal~\cite{Avrorin:2014swy}.   

A lot of physical processes are involved in this scenario. The capture
process crucially depends on the mass of dark matter particles $m_{DM}$
and the size of cross section of nonrelativistic elastic scattering of dark
matter  with nucleons. In particular, effective capture is
possible only for dark matter particles heavier than about 3--5~GeV.
Otherwise, evaporation of dark
matter from the Sun is important~\cite{Busoni:2013kaa}. Dark
matter particles can annihilate over different annihilation channels,
which 
is very model-dependent. Instead of considering a particular model it
is common to work with a chosen set of annihilation channels which are
believed 
to capture the main features of the general picture. Standard
benchmark annihilation channels are $b\bar{b}$ channel with very soft
spectrum of neutrinos, $W^+W^-$ and $\tau^+\tau^-$ channels with more
energetic spectra as well as monochromatic neutrino channels
$\nu\bar{\nu}$. Each annihilation channels provides not only with unique
neutrino energy spectrum but also with a particular flavor content. Thus, different
neutrino signals can be expected from different annihilation
channels.

Produced neutrinos propagate in the Sun, from the Sun to the Earth
and in the Earth to neutrino telescope.
In most experimental searches of this type muon neutrinos are
the most important and their expected flux at the detector level  
is given by the following expression
\be
\Phi_{\nu_{\mu}} = \frac{\Gamma_{A}}{4\pi R^2}\times\sum_{\nu_\alpha, \bar{\nu_\alpha}}
\int_{E_{th}}^{m_{DM}}dE_{\nu_{\alpha}}P_{\alpha\mu}(E_{\nu_\alpha},E_{th})\frac{dN^{\rm prod}_{\nu_{\alpha}}}{dE_{\nu_{\alpha}}}\;,
\ee
where $\Gamma_A$ is dark matter annihilation rate,  $R$ is the
distance from the Sun to the Earth, $E_{th}$ is neutrino threshold energy,
$\frac{dN^{\rm prod}_{\nu_{\alpha}}}{dE_{\nu_{\alpha}}}$ is neutrino energy
spectrum at production and the function
$P_{\alpha\mu}(E_{\nu_\alpha},E_{th})$ is the probability of  
obtaining muon (anti)neutrino in the detector from (anti)neutrino of
given flavor $\alpha$ at production which encodes all effects of neutrino
propagation. This last can be quite nontrivial and it depends on
neutrino oscillations, charge current (CC) and neutral current (NC)
interactions of neutrino interactions with matter.

\begin{figure}
  \begin{center}
    \includegraphics[angle=-90,width=0.54\textwidth]{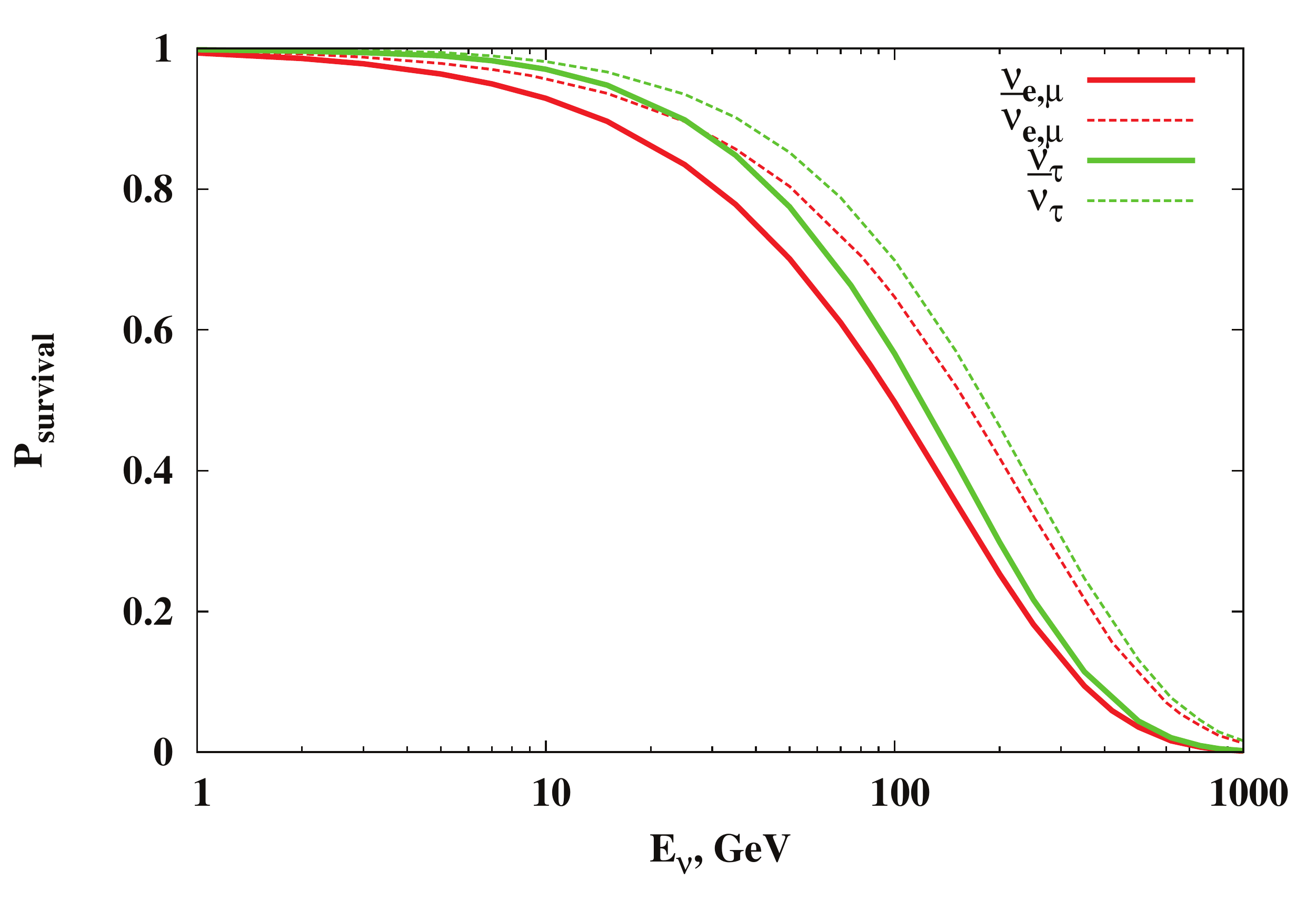}
  \end{center}
\caption{\label{fig1} Survival probability of neutrino and
  antineutrino of different flavors produced in the center of the Sun.}
\end{figure}

On Figure~\ref{fig1} we show the probability
for (anti)neutrino 
of a particular flavor to escape the Sun at the same
energy, calculated without oscillations, i.e. survival probability;
see Ref.~\cite{Cirelli:2005gh} for similar picture. One can see that
the absorption effect is small for neutrino 
of energies less than 10~GeV and very crucial for neutrino with energy
larger than about 100~GeV. Oscillations also produce dramatic effect.  
Comparing integrated final muon neutrino fluxes at the Earth level
calculated with and without oscillation effects (see e.g. right panel
of Fig.~10 in Ref.~\cite{Boliev:2013ai}) one observes that the
effect of neutrino oscillations varies from 10--40\% for $b\bar{b}$
and $W^+W^-$ annihilation channel to factor of 3.5 for dark matter
annihilations into $\tau^+\tau^-$. This indicates that  new
physics which influences neutrino interactions and oscillations could also
affect propagation of neutrinos from dark matter annihilations in the
Sun and
change corresponding neutrino signal. Studies in this direction was
performed recently in Refs.~\cite{Esmaili:2012ut,Arguelles:2012cf}
where effect of light sterile neutrino on the signal from dark
matter annihilation in the Sun was discussed.

In this paper we study influence of the non-standard interactions (NSI) of
neutrino with matter~\cite{Wolfenstein:1977ue} on neutrino signal from
dark matter annihilations in the Sun. 
Such interactions appear in different types of models of new
physics, see e.g.~\cite{Antusch:2008tz}.  Neutrino NSI  attract
recently much attention and several studies were performed
scrutinizing their different aspects, see
Refs.~\cite{Ohlsson:2012kf,Miranda:2015dra,Farzan:2017xzy} for
reviews. NSI can 
potentially influence propagation of solar
neutrinos~\cite{Friedland:2004pp,Miranda:2004nb,Palazzo:2009rb,Ghosh:2017lim}, 
neutrinos from
supernova~\cite{EstebanPretel:2007yu,Stapleford:2016jgz}, atmospheric 
neutrinos~\cite{Friedland:2004ah,Friedland:2005vy,Ohlsson:2013epa,
  Esmaili:2013fva,Chatterjee:2014gxa, Mocioiu:2014gua,
  Choubey:2014iia, Fukasawa:2015jaa, Choubey:2015xha,
  Fukasawa:2016nwn, Salvado:2016uqu}, neutrinos from
artificial sources~\cite{Kopp:2007ne,Escrihuela:2009up,Kopp:2010qt,
  Oki:2010uc, GonzalezGarcia:2011my, Adhikari:2012vc, Li:2014mlo,
  deGouvea:2015ndi, Masud:2016gcl, Blennow:2016etl, Ohlsson:2013nna,
  Fukasawa:2016lew,Deepthi:2016erc} as well as reveal themselves in
different production and decay
processes~\cite{Berkov:1987pz,Berkov:1988sd,Belotsky:2001fb}. In
particular,  matter NSI can produce 
``missing energy'' signature at collider experiments and in particular
at the LHC~\cite{Friedland:2011za,Franzosi:2015wha} although
interpretation of these results may be quite model dependent.

Here we study effect of the matter NSI on propagation of high energy
neutrinos from dark matter annihilation in the Sun.
We find that for realistic annihilation channels and
phenomenologically allowed values of NSI parameters, deviations of 
neutrino flux at the Earth level from the standard (no-NSI) case can
be considerable. The rest of the 
paper is organized as follows. 
In Section~2 we briefly remind the main facts about the matter NSI which
are relevant for the present analysis, i.e. parameters, influence on
neutrino oscillations and current experimental bounds. In Section~3 we
consider evolution of monochromatic neutrinos from dark matter
annihilations in the Sun. To single out effect of the matter NSI,
in this Section we neglect all neutrino interactions except for
the forward neutrino scattering. Turning on the matter NSI we
numerically 
simulate evolution of neutrinos from the Sun to the Earth and supply
this analysis with simplified analytical study. In Section~4 we
perform full Monte-Carlo simulations of neutrino propagation which
includes 
neutrino interactions. We consider $b\bar{b}, W^+W^-$ and
$\tau^+\tau^-$ annihilation channels and estimate the effect of
different NSI parameters on the muon track event rate at neutrino
telescopes. Section~5 contains our conclusions.

\section{NSI and modification of neutrino propagation}
At sufficiently low transverse momenta the matter NSI of neutrino can
be described by the following effective lagrangian 
\be
\label{eq:1:2}
   {\cal L}^{NC}_{NSI} =
   -\sum_{f, P=P_L,P_R} \epsilon_{\alpha\beta}^{fP} 2\sqrt{2}G_F
   (\bar{\nu}_{\alpha}\gamma^{\mu}P_L\nu_\beta)  
   (\bar{f}\gamma_{\mu}Pf).
\ee
Here $P_{L,R}$ are chirality projectors, $\epsilon_{\alpha\beta}^{fP}$
are the NSI parameters and sum is implied over all SM fermions $f$. Let us
note that apart from the neutral current type of NSI in~\eqref{eq:1:2}
in general one 
can introduce NSI of charge current type. Their main impact would be to
affect neutrino CC interaction cross sections. Present experimental
bounds on the 
parameters of the charged current type of 
NSIs are quite severe~\cite{Ohlsson:2012kf}; we expect that
their effect on high-energy neutrino propagation in the Sun and the
Earth 
would be small and focus on the influence of the matter NSIs.
One of the main consequences of the interactions~\eqref{eq:1:2} is
modification of neutrino propagation through matter. Evolution of
relativistic neutrino in media can be described by the following
Hamiltonian 
\be
\label{eq:1:3}
H = \frac{1}{2E_\nu}U{\rm diag}(m_1^2,m_2^2,m_3^2)U^{\dagger}
+ V_e\epsilon^{m},
\ee
where $m_i^2, i=1,2,3$ are neutrino masses squared, $U$ is the vacuum
Pontecorvo-Maki-Nakagawa-Sakata (PMNS) matrix and $E_\nu$ is neutrino
energy. We use standard parametrization of the PMNS matrix (omitting
Majorana phases) 
\be
\label{eq:1:4}
U = R_{23}(\theta_{23})U_\delta^\dagger R_{13}(\theta_{13})U_\delta R_{12},
\ee
where $R_{ij}$ is rotation matrix in $ij$-plane and $U_\delta$ is
phase matrix containing CP-violating parameter
\be
\label{eq:1:5}
U_\delta =
\left(
\begin{array}{ccc}
  e^{i\delta/2} & 0 & 0 \\
  0 & 1 & 0 \\
  0 & 0 & e^{-i\delta/2}
\end{array}
\right).
\ee
Further for numerical calculations we use the values of
oscillation parameters presented in Table~\ref{tab:1}
\begin{table}[!htb]
  \begin{center}
    \begin{tabular}{|c|c|c|c|c|c|c|}
      \hline
     & $\Delta m_{21}^2$, eV$^2$ & $|\Delta m_{31}^{2}|$, eV$^2$ &
      $\sin^2{\theta_{12}}$ & $\sin^2{\theta_{23}}$ &
      $\sin^2{\theta_{13}}$ & $\delta_{CP}$ \\\hline
      NH & $7.50\cdot 10^{-5}$ & $2.457\cdot 10^{-3}$ & 0.304 & 0.452 & 0.0218 & 0 \\\hline
      IH & $7.50\cdot 10^{-5}$ & $2.449\cdot 10^{-3}$ & 0.304 & 0.579 & 0.0219 & 0
      \\\hline
    \end{tabular}
    \caption{\label{tab:1} Neutrino oscillation parameters used for
      the present analysis~\cite{Gonzalez-Garcia:2015qrr}.}
  \end{center}
\end{table}
for normal (NH) and inverted (IH) hierarchy. These parameters
lie within $3\sigma $ range of their experimentally allowed
values~\cite{Gonzalez-Garcia:2015qrr}. For simplicity we assume  
CP-violating phase to be zero. The matter term in~\eqref{eq:1:3}
contains a factor $V_e\equiv \sqrt{2}G_FN_e$ for neutrino and
$V_e\equiv -\sqrt{2}G_FN_e$ for antineutrino. This term is
proportional to the electron number density $N_e$ and the matrix  
\be
\label{eq:1:6}
\e^m = 
\left(
\begin{array}{ccc}
1+\e_{ee} & \e_{e\mu} & \e_{e\tau} \\
\e_{e\mu}^*       & \e_{\mu\mu} & \e_{\mu\tau} \\
\e_{e\tau}^*       &  \e_{\mu\tau}^*          &  \e_{\tau\tau}
\end{array}
\right),
\ee
describing influence of the matter NSI as well as the Standard Model 
contribution. 
The NSI parameters $\e_{\alpha\beta}$ depend on coupling constants in
the lagrangian~\eqref{eq:1:2} and on the matter content as follows
\be
\label{eq:1:7}
\e_{\alpha\beta} = \sum_{f=e,u,d}(\e_{\alpha\beta}^{f P_L} +
\e_{\alpha\beta}^{f P_R})\frac{N_f}{N_e},
\ee
where $N_f$ is the number density of the fermion $f$.
Model independent experimental
bounds on parameters $\e_{\alpha\beta}$ are presented below 
\be
\label{eq:1:8}
|\e_{\alpha\beta}^{\rm Earth}|<
\left(
\begin{array}{ccc}
4.2 & 0.33 & 3.0 \\
... & 0.068 & 0.33 \\
... & ... & 21.0
\end{array}
\right), \;\;\;
|\e_{\alpha\beta}^{\rm Sun}|<
\left(
\begin{array}{ccc}
2.5 & 0.21 & 1.7 \\ 
... & 0.046 & 0.21 \\
... & ... & 9.0
\end{array}
\right)
\ee
for the Earth-like and Sun-like matter, see
Refs.~\cite{Ohlsson:2012kf,Biggio:2009nt}.
There are more restrictive bounds on some of the Earth NSI parameters
coming from results of the Super-Kamiokande~\cite{Mitsuka:2011ty} experiment
\be
\label{eq:1:9}
|\e^{\rm Earth}_{\mu\tau}|<0.033,\;\;\;
|\e^{\rm Earth}_{\tau\tau}-\e^{\rm Earth}_{\mu\mu}|<0.147.
\ee
Even stronger bounds on the values of $|\e^{\rm Earth}_{\mu\tau}|$  (up to
0.02) were obtained~\cite{Esmaili:2013fva,Salvado:2016uqu, Aartsen:2017xtt} from the
results of IceCube\footnote{We note that authors of
Refs.~\cite{Mitsuka:2011ty,Esmaili:2013fva,Salvado:2016uqu, 
Aartsen:2017xtt} define the NSI parameters $\e_{\alpha\beta}$ by
normalization to the density of $d$-quarks. This differs from our
definition~\eqref{eq:1:7} by a factor $r=\frac{N_d}{N_e}$ which
is about 3 for the Earth.}.
Experimental bounds on the NSI parameters
$\e_{\alpha\beta}^{fP}$ from neutrino scattering experiments was
discussed in Refs.~\cite{Ohlsson:2012kf,Miranda:2015dra,Farzan:2017xzy} in
details. Effects of the matter NSI for existing and upcoming neutrino
experiments was discussed in many papers, we refer here to a review
paper~\cite{Farzan:2017xzy}.

Let us note that not only matter NSI parameters for the
Earth can be different from those for the Sun, but also
$\e_{\alpha\beta}^{Sun}$  are position dependent because the matter
content in the Sun changes from the center to the surface. Namely, the
proton-to-neutron 
ratio $\frac{N_p}{N_n}$ varies from about 2 at the center to about 6
near the surface. 
Analysis of the most general case lies beyond the scope of
the present study in which we are going to illustrate the main effects
of the matter NSI on propagation of neutrinos in the Sun and the Earth. In what follows for simplicity we
limit ourselves by the case of position independent values of
$\e^{Sun}_{\alpha\beta}$ and  consider the simplifying situation with
$\e_{\alpha\beta}^{Earth}=\e_{\alpha\beta}^{Sun}\equiv\e_{\alpha\beta}$.

In general the NSI~\eqref{eq:1:2} could modify the NC  
neutrino-nucleon interaction cross section which could affect neutrino
propagation in the Sun and the Earth. In what follows we
neglect this effect because its influence on the final neutrino flux
will be subleading to the standard NC and CC neutrino interactions for
chosen values of the matter NSI parameters. We leave 
the detailed analysis of this effect for future study.

\section{Evolution of monochromatic WIMP neutrinos in the Sun and Earth}
Propagation of high energy neutrinos from dark matter
annihilations in the Sun has been studies both
analytically~\cite{Lehnert:2007fv,Lehnert:2010vb,Esmaili:2009ks} and   
numerically~\cite{Cirelli:2005gh,Blennow:2007tw,Baratella:2013fya,Suvorova:2013uta}.
In this Section we discuss effect of the matter NSI on propagation of
monochromatic neutrino in the Sun and the Earth. We analyze effect of
the NSI
neglecting all neutrino interactions except for the forward neutrino
scattering which directly affects neutrino oscillations. 
Bearing in mind muon track signature at neutrino telescopes in what follows we consider
neutrino energy range from 1 GeV to 1 TeV\footnote{Another strategy to
search for the signal from dark matter annihilations in the Sun
utilizes neutrinos in the MeV energy
range~\cite{Rott:2012qb,Bernal:2012qh,Rott:2015nma,Rott:2016mzs}. In
the present study we will not consider this possibility. }.
The lowest bound is determined mainly by muon energy thresholds for
such searches in neutrino experiments (about 1.5 GeV for Super-Kamiokande and
about 1~GeV for Baksan Underground Neutrino Telescope). At the same
time, neutrinos with energies larger than 1~TeV have very small
probability to escape the Sun, see Fig.~\ref{fig1}.
We numerically solve Schrodinger equation with 
the Hamiltonian~\eqref{eq:1:3} for realistic electron density profiles in
the Sun~\cite{Bahcall:2004pz} and the Earth~\cite{PREM}. Varying density
of the Sun has been 
taken into account in the following way: we divide neutrino path
in sufficiently small pieces in which electron density can be
considered as a constant and then evolve neutrino wave function with
exact evolution operator.  We use the algorithm described in
Refs.~\cite{Ohlsson:1999xb,Ohlsson:2001et} to simulate neutrino 
oscillations in $3\times 3$ scheme.
We assume that neutrinos are produced near the center of the Sun in a
flavor state $\nu_\alpha$. Production region of the neutrinos in the
Sun follows expected dark matter distribution~\cite{Griest:1986yu}
(see also~\cite{Widmark:2017yvd} for recent study)    
\be
\label{eq:2:1}
n(r) = n_0\,e^{-r^2/R_{DM}^2},\;\;{\rm with}\;\; R_{DM}\sim
0.1R_{Sun}\sqrt{\frac{1~{\rm GeV}}{m_{DM}}},
\ee
which depends on the mass of dark matter particle. Here $R_{Sun}$ is
the radius of the Sun. Numerically, size
of the DM core in the Sun varies from about $4\cdot 10^4$~km for
$m_{DM}\sim 3$~GeV to $2.2\cdot 10^3$~km for $m_{DM}\sim 1$~TeV. 
For the case of monochromatic neutrino annihilation channels we
simulate the production point according to the
distribution~\eqref{eq:2:1} with $m_{DM}=E_\nu$.
Produced neutrino evolves according to the Schrodinger equation with
the Hamiltonian~\eqref{eq:1:3}.
In the analysis we take into
account small ellipticity of the Earth orbit. Namely, we randomly
choose time of each neutrino event (i.e. fraction of the year) and
average final probabilities over positions of dark matter
annihilation and the Earth. 

\begin{figure}[t]
\begin{picture}(300,220)(0,20)
\put(210,130){\includegraphics[angle=-90,width=0.40\textwidth]{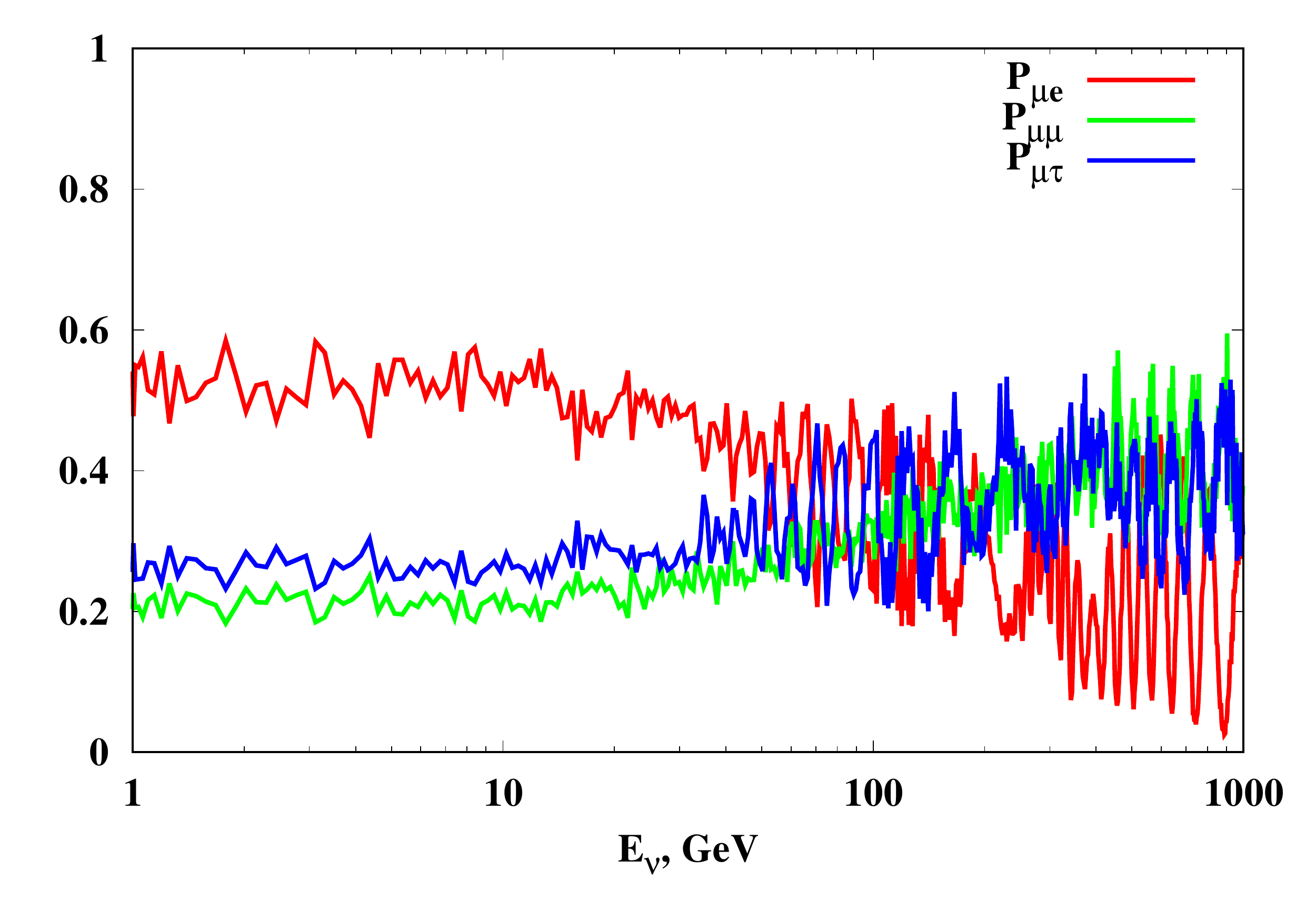}}
\put(210,250){\includegraphics[angle=-90,width=0.40\textwidth]{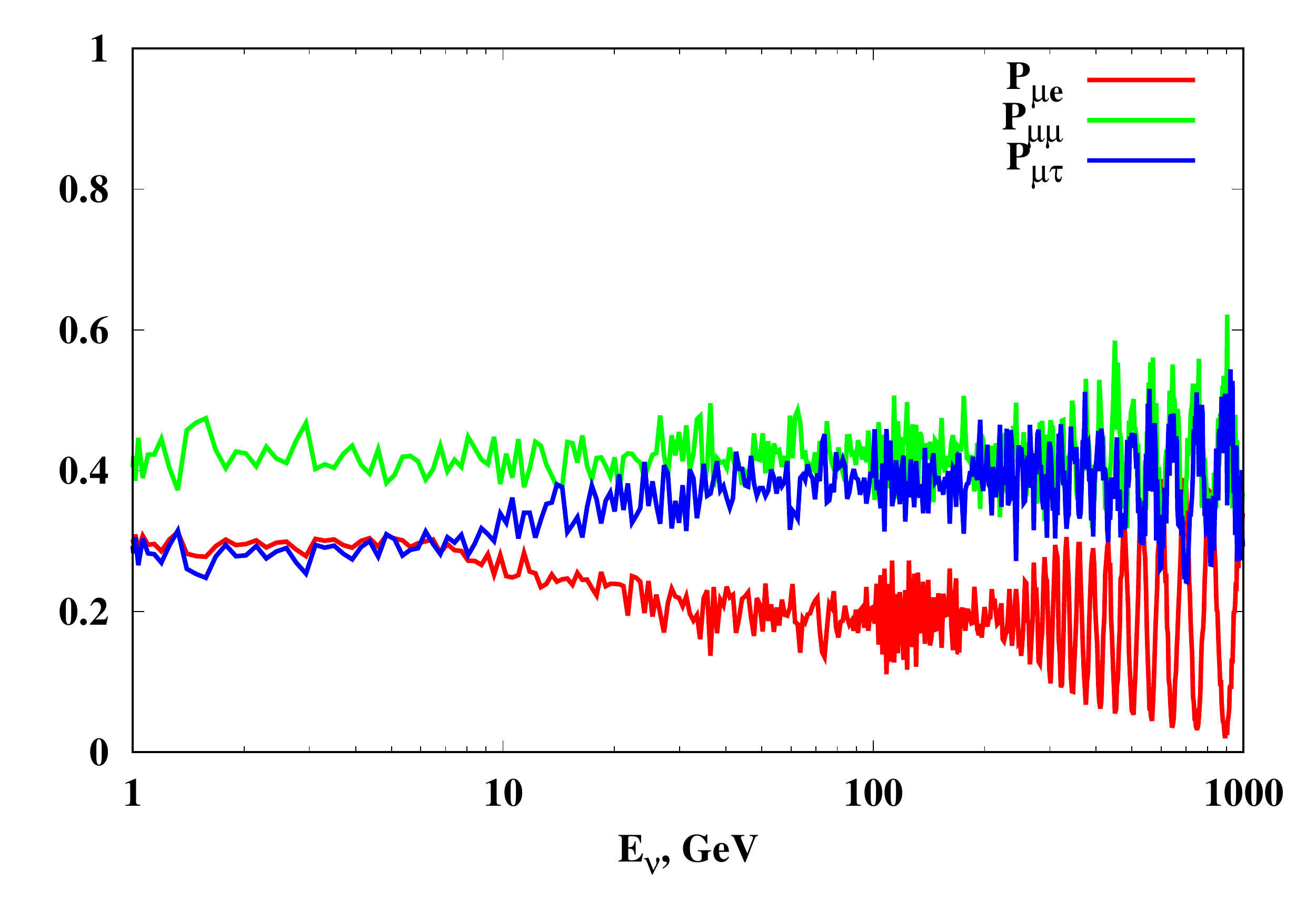}}
\put(30,130){\includegraphics[angle=-90,width=0.40\textwidth]{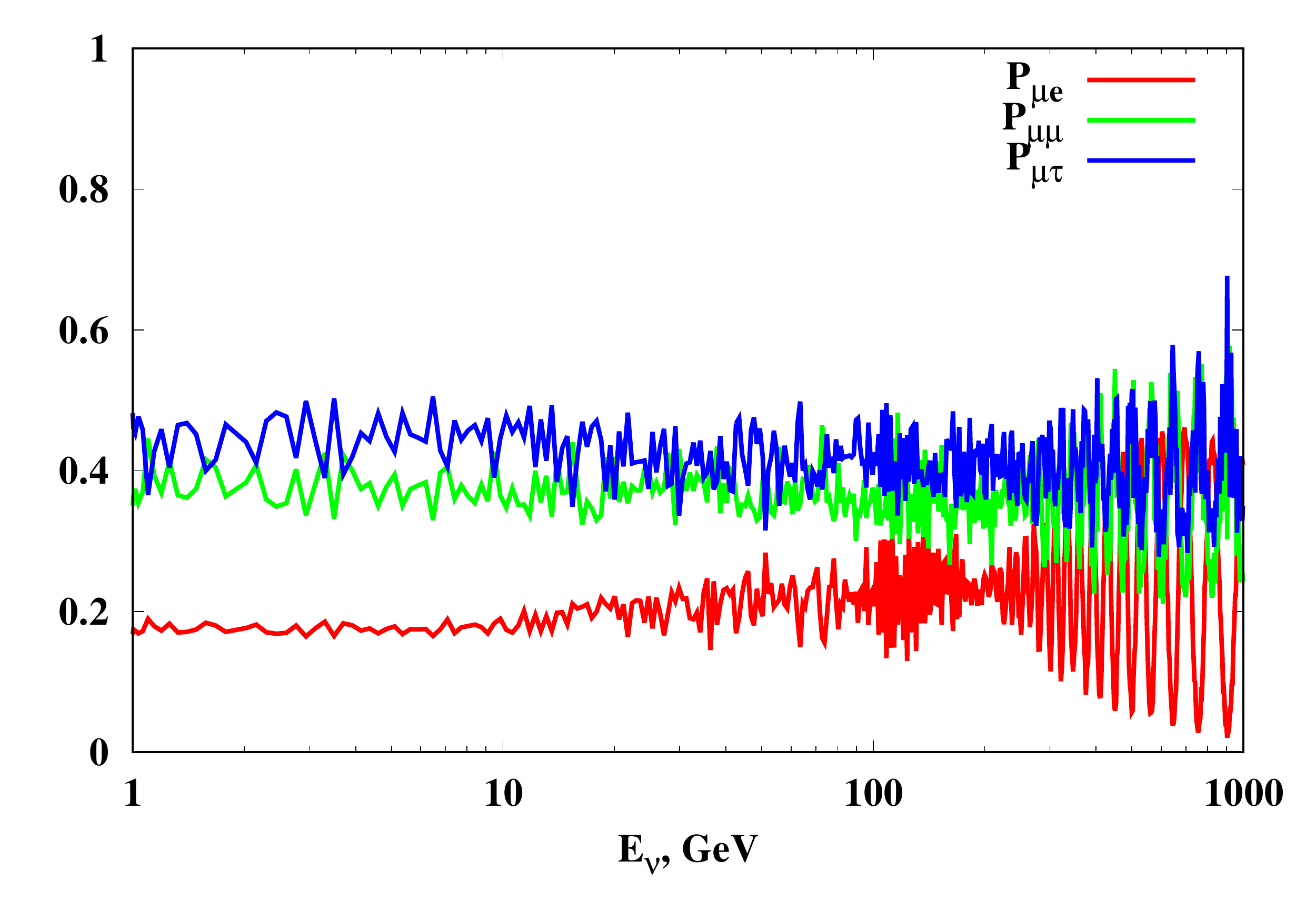}}
\put(30,250){\includegraphics[angle=-90,width=0.40\textwidth]{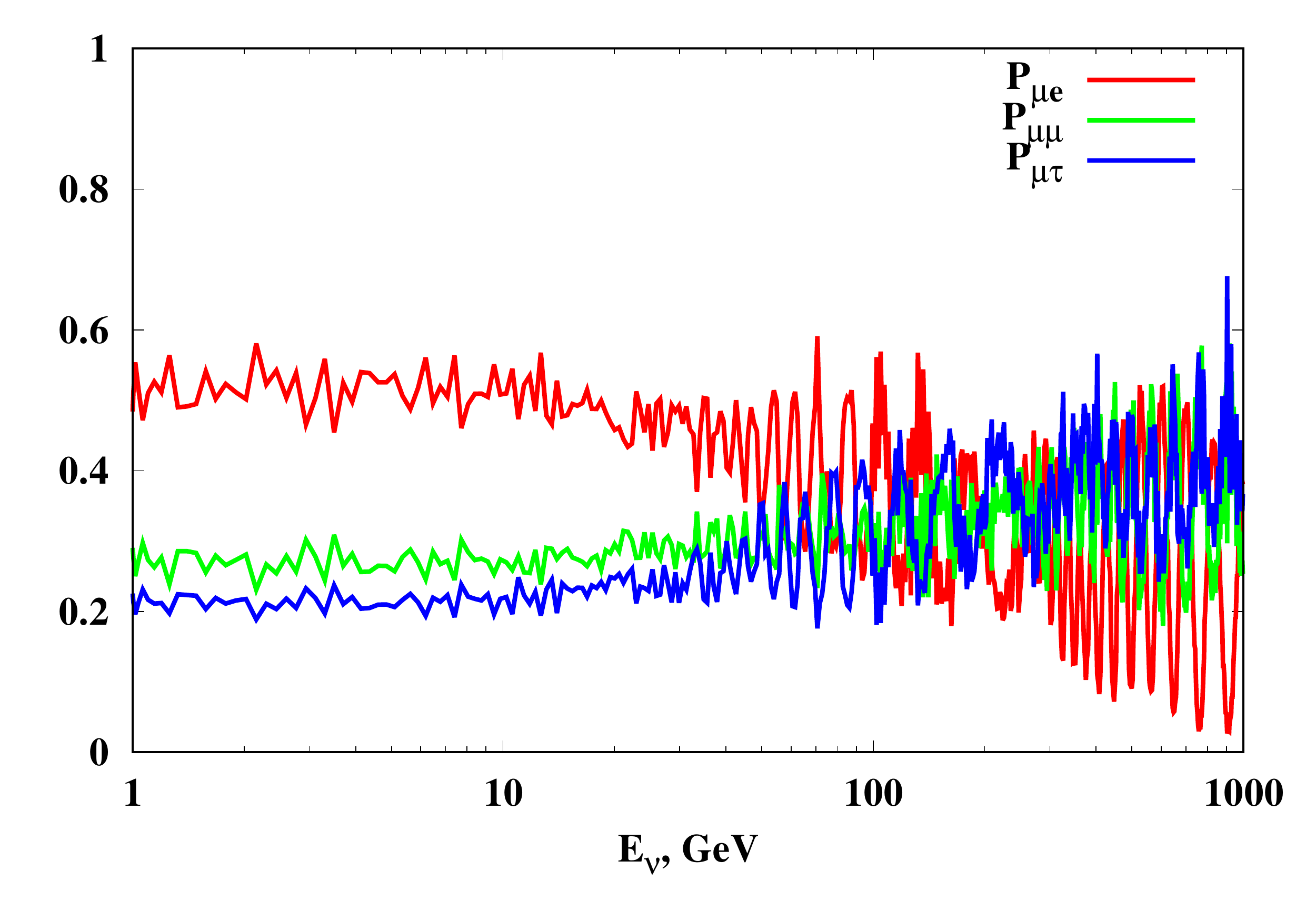}}
\end{picture}
\caption{\label{numu_sm} Probabilities $P_{\mu\alpha}$ to obtain neutrino of different
flavors $\nu_\alpha$ at the Earth orbit from $\nu_\mu$($\bar{\nu}_\mu$) in the
Sun for normal (left panels) and inverted (right panels) hierarchy for
$\e_{\alpha\beta}=0$. Plots in the upper (lower) panels correspond to
neutrino (antineutrino) case.}
\end{figure}

Let us start our analysis with the standard case without NSI. On
Fig.~\ref{numu_sm} 
\begin{figure}[hb]
\begin{picture}(300,220)(0,20)
\put(210,130){\includegraphics[angle=-90,width=0.40\textwidth]{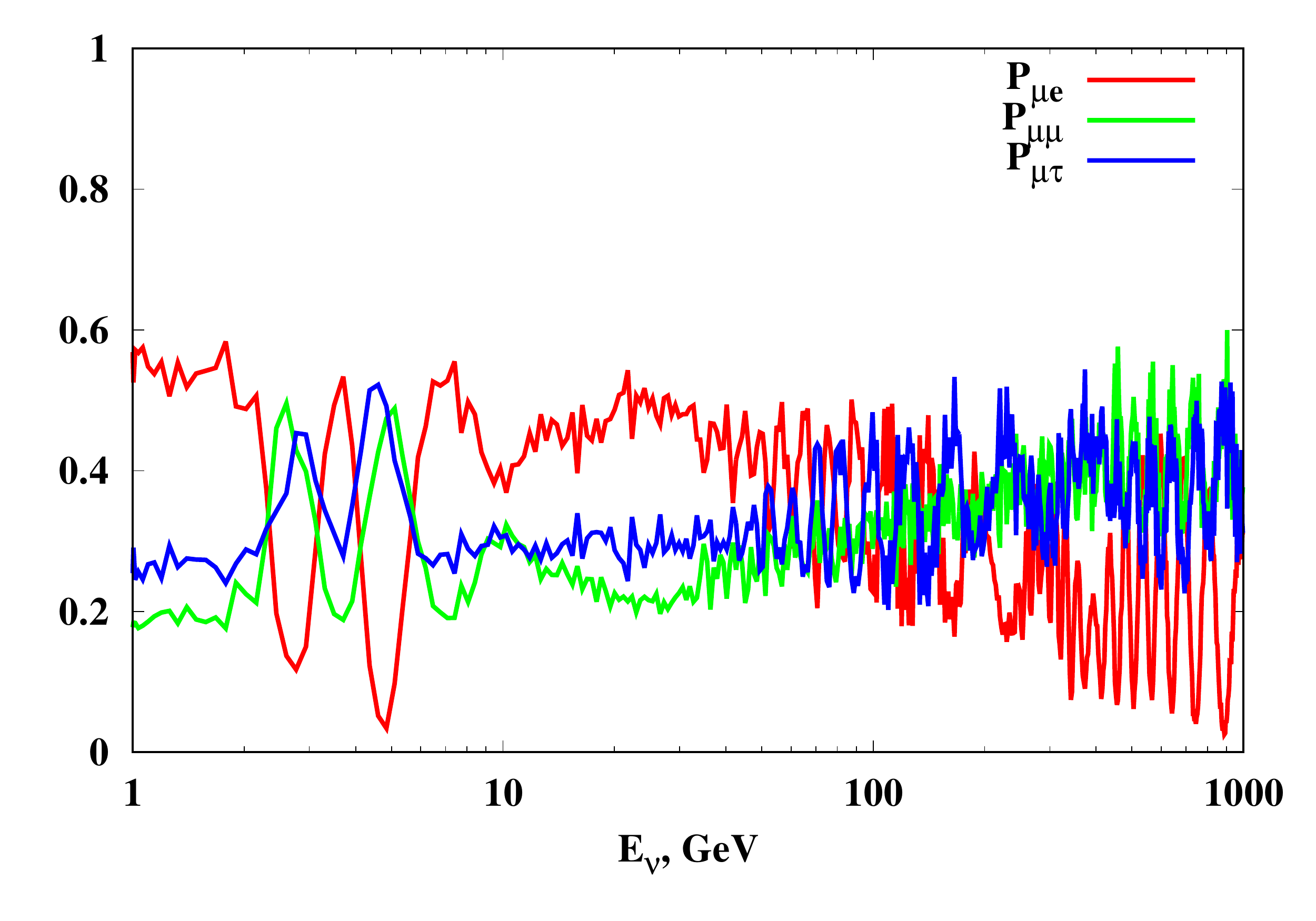}}
\put(210,250){\includegraphics[angle=-90,width=0.40\textwidth]{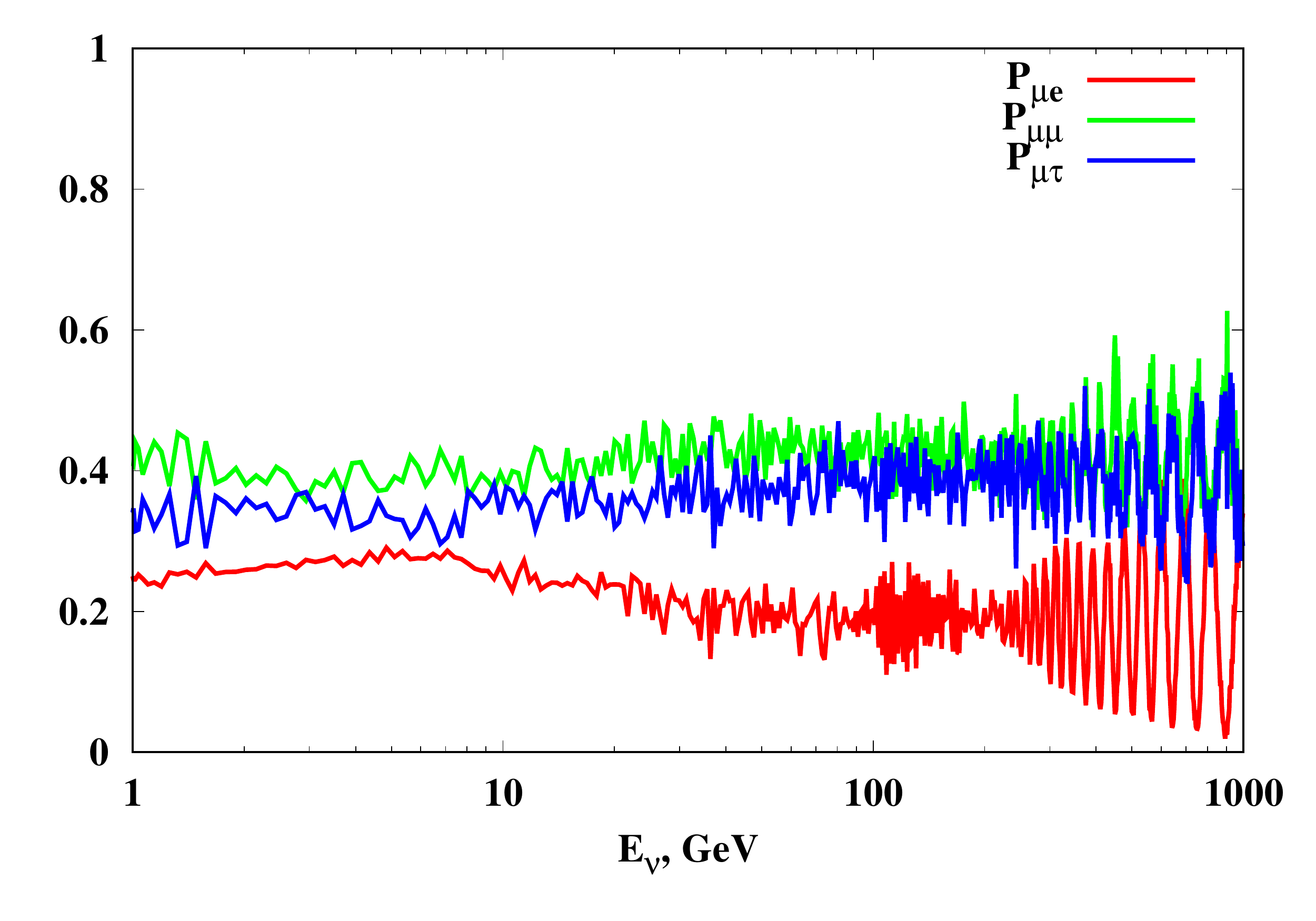}}
\put(30,130){\includegraphics[angle=-90,width=0.40\textwidth]{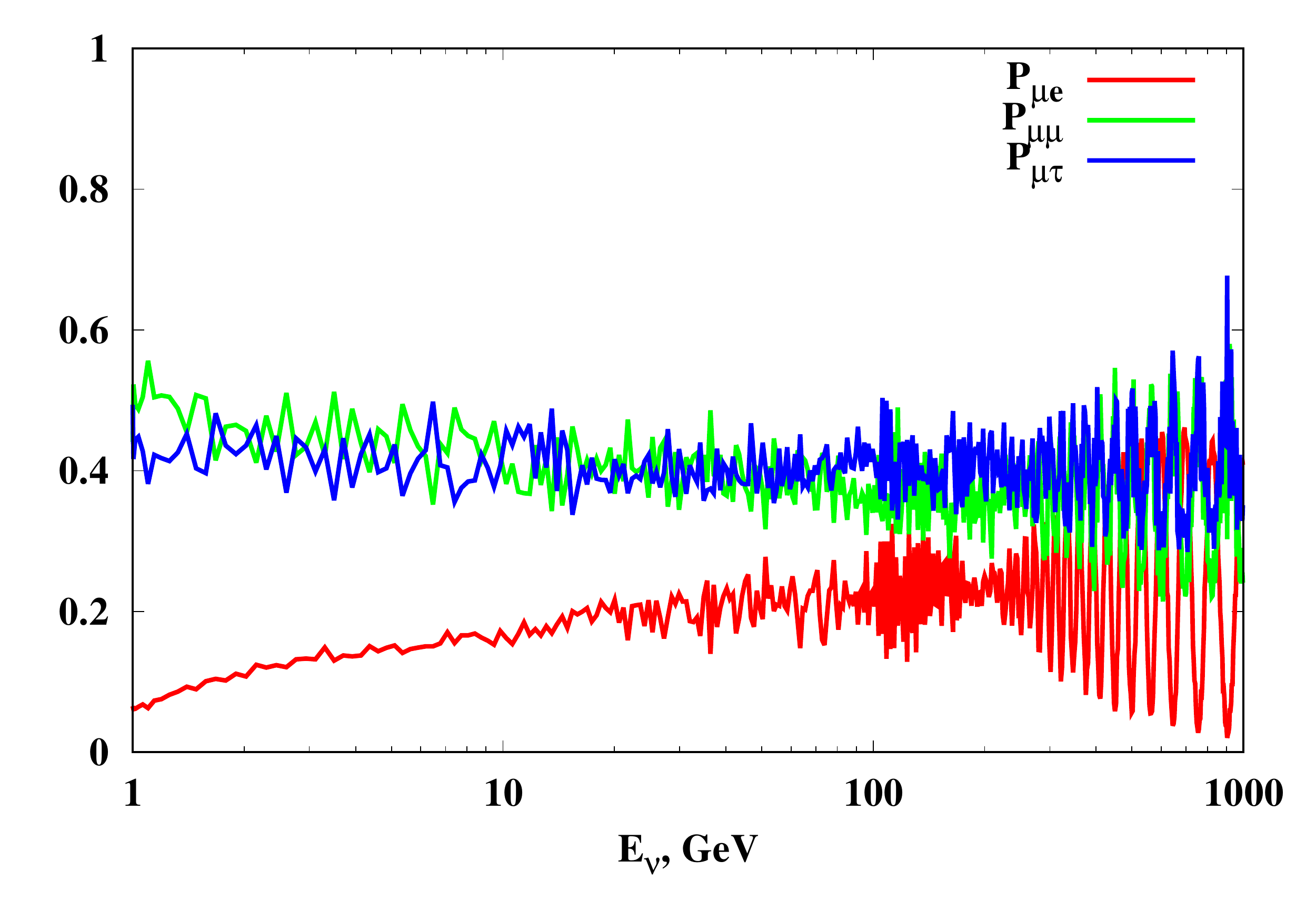}}
\put(30,250){\includegraphics[angle=-90,width=0.40\textwidth]{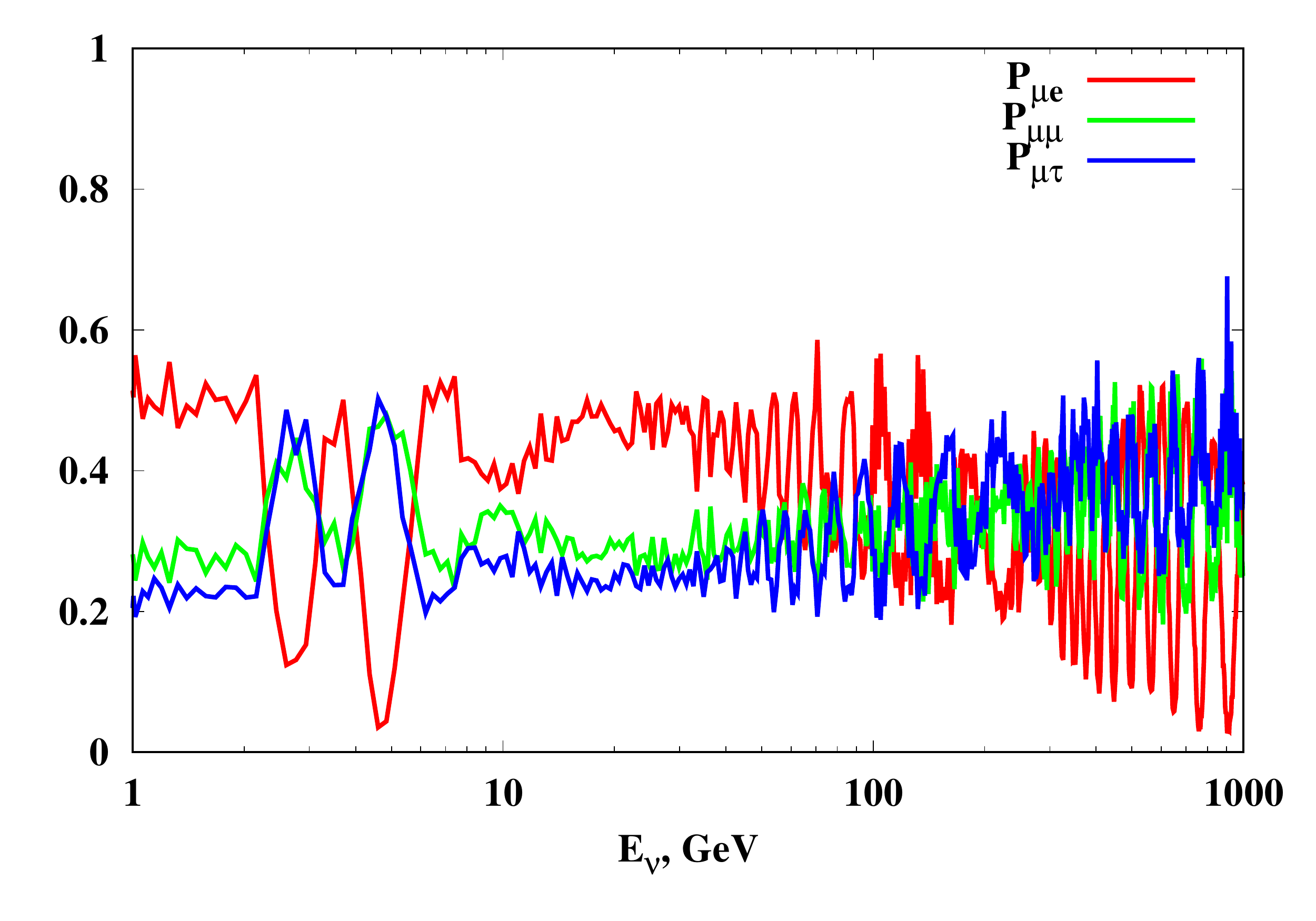}}
\end{picture}
\caption{\label{numu_sm_earth} Probabilities to obtain neutrino of different
flavors after propagation in the Earth from
$\nu_\mu$($\bar{\nu}_\mu$) produced in the 
Sun for normal (left panels) and inverted (right panels) hierarchy for
$\e_{\alpha\beta}=0$. Plots in the upper (lower) panels correspond to
neutrino (antineutrino) case. }
\end{figure}
we plot probabilities to obtain neutrinos of different
flavours at the Earth orbit (i.e. no propagation in the Earth). As an
illustrative example we consider here and in the subsequent Figures in
this Section the case of muon (anti)neutrino at production. Left and right panels correspond to
the cases of normal and inverted neutrino mass hierarchy, respectively,
with the oscillation parameters taken from Table~\ref{tab:1}. The case of neutrino is
shown in the upper panels, while the lower panels are reserved for
antineutrinos. 
To average over
varying neutrino baseline we simulate $N_{av}=100$ neutrino events for
each value of neutrino energy. In Fig.~\ref{numu_sm_earth} we present
the same probabilities after propagation in the Earth. For
illustration purposes in the Section we consider propagation though
the center of the Earth only. In the next Section we relax this
assumption when making full-fledged Monte-Carlo simulation of
neutrino propagation.  

To explain the behavior of the probabilities as functions of neutrino
energy let us introduce apart 
from neutrino flavor states $|\nu_\alpha\rangle, \alpha=e,\mu,\tau$ 
and vacuum eigenstates $|i\rangle, i=1,2,3$ also the eigenstates of
the instantaneous matter Hamiltonian as   
\be
\label{eq:2:2}
|\alpha\rangle = U^m_{\alpha j}(r)|j_m, r\rangle,
\ee
where $U^m(r)$ is the mixing matrix diagonalizing neutrino Hamiltonian in
matter: $H_m(r)=U^{m\dagger}(r)H(r)U^m(r)$.  In
the absence of the matter 
NSI for high energy neutrinos corresponding eigenstates of the matter
Hamiltonian~\eqref{eq:1:3} in the center
of the Sun are approximately $|1_m,0\rangle=|\nu_e\rangle$ and 
\be
\label{eq:2:3}
|2_m,0\rangle=c_{23}|\nu_\mu\rangle -
s_{23}|\nu_\tau\rangle,\;\;\;
|3_m,0\rangle=s_{23}|\nu_\mu\rangle +
c_{23}|\nu_\tau\rangle,
\ee
where we use the standard notations $c_{ij}=\cos{\theta_{ij}}$
and $s_{ij}=\sin{\theta_{ij}}$. The state $|\nu_e\rangle$ is decoupled
in the center due to large matter contribution and produced 
$\nu_\mu$ oscillates to $\nu_\tau$ with the mixing angle $\theta_{23}$.
If electron density changes much 
slower than neutrino oscillates, i.e. in the adiabatic regime, the
instantaneous eigenstates are approximately evolution eigenstates and
in this limit no transitions between these eigenstates occur.
If the oscillation length $L_{23}=\frac{4\pi E_\nu}{|\Delta
  m_{23}^2|}$ is small as compared to $2R_{DM}$ then $\mu-\tau$ oscillation phase
averages to zero and one obtains outside the production region a 
mixed state $c^2_{23}|2_m,r\rangle\langle 2_m,r| +
s^2_{23}|3_m,r\rangle\langle 3_m,r|$.
In the adiabatic regime the matter Hamiltonian eigenstates $|2_m\rangle$ and
$|3_m\rangle$ evolve at the solar surface into some vacuum eigenstates
$|i_2\rangle$ and $|i_3\rangle$, respectively.
For normal mass hierarchy one obtains
\begin{eqnarray}
  |2_m\rangle\to |1\rangle,\;\;\; &
  |3_m\rangle\to |2\rangle\;\; {\rm for}\; \nu\,, \\
  |2_m\rangle\to |2\rangle,\;\;\; &
  |3_m\rangle\to |3\rangle\;\; {\rm for}\; \bar{\nu}\,. \label{trans1}  
\end{eqnarray}
For the case of inverted hierarchy
\begin{eqnarray}
  |2_m\rangle\to |1\rangle,\;\;\; &
  |3_m\rangle\to |3\rangle\;\; {\rm for}\; \nu\,, \\
  |2_m\rangle\to |2\rangle,\;\;\; &
  |3_m\rangle\to |1\rangle\;\; {\rm for}\; \bar{\nu}\,.  
\end{eqnarray}
Finally one gets a mixed state describing by the following density matrix 
$c^2_{23}|i_2\rangle\langle i_2| +
s^2_{23}|i_3\rangle\langle i_3|$. Probability to detect 
neutrino $\nu_\beta$ at the Earth orbit looks as follows
\be
\label{eq:2:-1}
P(\nu_\mu\to\nu_\beta) = c^2_{23}|U_{\beta\, i_2}|^2 +
s^2_{23}|U_{\beta\, i_3}|^2.
\ee
In a more generic case for neutrino $\nu_\alpha$ at the solar center,
the neutrino state at the Earth orbit can be approximately
described~\cite{Lehnert:2007fv} by the density matrix 
\be
\label{eq:2:00}
\rho = \sum_{i,j}\left|U_{\alpha i}^m(r=0)\right|^2
P_{ij} |j\rangle\langle j|\;.
\ee
Here $P_{ij}$ are probabilities of transition between different
eigenstates during evolution in the Sun. For the adiabatic evolution,
$P_{ij}=\delta_{ij}$, taking convention for ordering of eigenvalues 
such that they smoothly approach vacuum eigenstates.
Formula~\eqref{eq:2:00} assumes that statistical  
averaging of oscillation phases to zero takes place, which results from
uncertainties in neutrino production place, small ellipticity of the
Earth orbit as well as from finite energy resolution of the detector
(see Ref.~\cite{Lehnert:2007fv} for details).
The probability to find neutrino in a
flavor state $\nu_\beta$ at the Earth orbit from neutrino state
$\nu_\alpha$ at production can be approximately
described~\cite{Lehnert:2007fv} by the following expression
\be
\label{eq:2:0}
P(\nu_\alpha\to\nu_\beta) = \sum_{i,j}\left|U_{\alpha i}^m(r=0)\right|^2
P_{ij} \left|U_{\beta j}\right|^2\;.
\ee
Adiabatic approximation may break down near
Mikheyev-Smirnov-Wolfenstein (MSW) 
resonances~\cite{Wolfenstein:1977ue,Mikheev:1986gs,Mikheev:1986wj}. For
NH there are two MSW resonances corresponding to 1--3 and 1--2
transitions in neutrino sector and none for antineutrino. For 
IH one resonance 1--2 transition happens in neutrino mode while
another 1--3 transition occurs for antineutrinos.
Corresponding resonance conditions look~~\cite{Kuo:1989qe,Lehnert:2007fv,Blennow:2013rca} as
\begin{eqnarray}
E_\nu  =  & \frac{\Delta
  m_{21}^2}{2V_e(r)c^2_{13}}\cos{2\theta_{12}}\;\;\;  &
\text{for 1--2 resonance,} \\
E_\nu  =  & \frac{\Delta
  m_{32}^2 + \Delta
  m_{21}^2c^2{12}}{2V_e(r)}\cos{2\theta_{13}}\;\;\; &
\text{for 1--3 resonance}.
\end{eqnarray}
The resonance energies in the solar core lie below GeV energy
scale. Adiabaticity for neutrino transitions through these resonances
is violated for energies $E\gsim E_{NA}$, where the onset energy for
nonadiabatic effects $E_{NA}$ can be
estimated~~\cite{Lehnert:2007fv,Kuo:1989qe} as follows
\begin{eqnarray}
  E_{NA} = & \frac{1}{3}\pi\bigg|\frac{V_e}{V^{\prime}_e}\bigg|_{r_R}\Delta
m_{21}^2s^2_{12}\;\;\; &\text{for 1--2 resonance,}\\
  E_{NA} = & \frac{1}{3}\pi\bigg|\frac{V_e}{V^{\prime}_e}\bigg|_{r_R}(\Delta
m_{32}^2 + \Delta
m_{21}^2c^2{12})s^2_{13}\;\;\; & \text{for 1--3
  resonance,} 
\end{eqnarray}
where $r_R$ is space position of the resonance.
Numerically one can find that $E_{NA}$ is about 9~GeV for 1--2
resonance and about $20$~GeV for 1--3 resonance. 
One can check that Eq.~\eqref{eq:2:-1} reproduces energy dependence of
the probabilities in
Fig.~\ref{numu_sm} in the low energy region. At energies larger than
10--20~GeV the evolution becomes more complicated. 
It happens not only due to nonadiabatic transitions through the MSW
resonance regions but also because averaging of oscillation phases
over production region may no longer take place. Moreover, at energies
larger than 100--200~GeV oscillation lengths become comparable and
even larger than ellipticity of the Earth orbit.
This can result in annual modulation of the neutrino signal discussed
in~\cite{Esmaili:2009ks,Esmaili:2010wa}.
The most important influence of subsequent propagation in the
Earth (see Fig.~\ref{numu_sm_earth}) is visible for neutrino mode (NH)
and antineutrino mode (IH) in the low energy region.
This behavior is related mainly to 2--3 and 1--3 mixings,
see~\cite{Akhmedov:2006hb, Akhmedov:2008qt}.

\subsection{Flavor conserving NSI }
\begin{figure}[t]
\begin{picture}(300,220)(0,20)
\put(210,130){\includegraphics[angle=-90,width=0.40\textwidth]{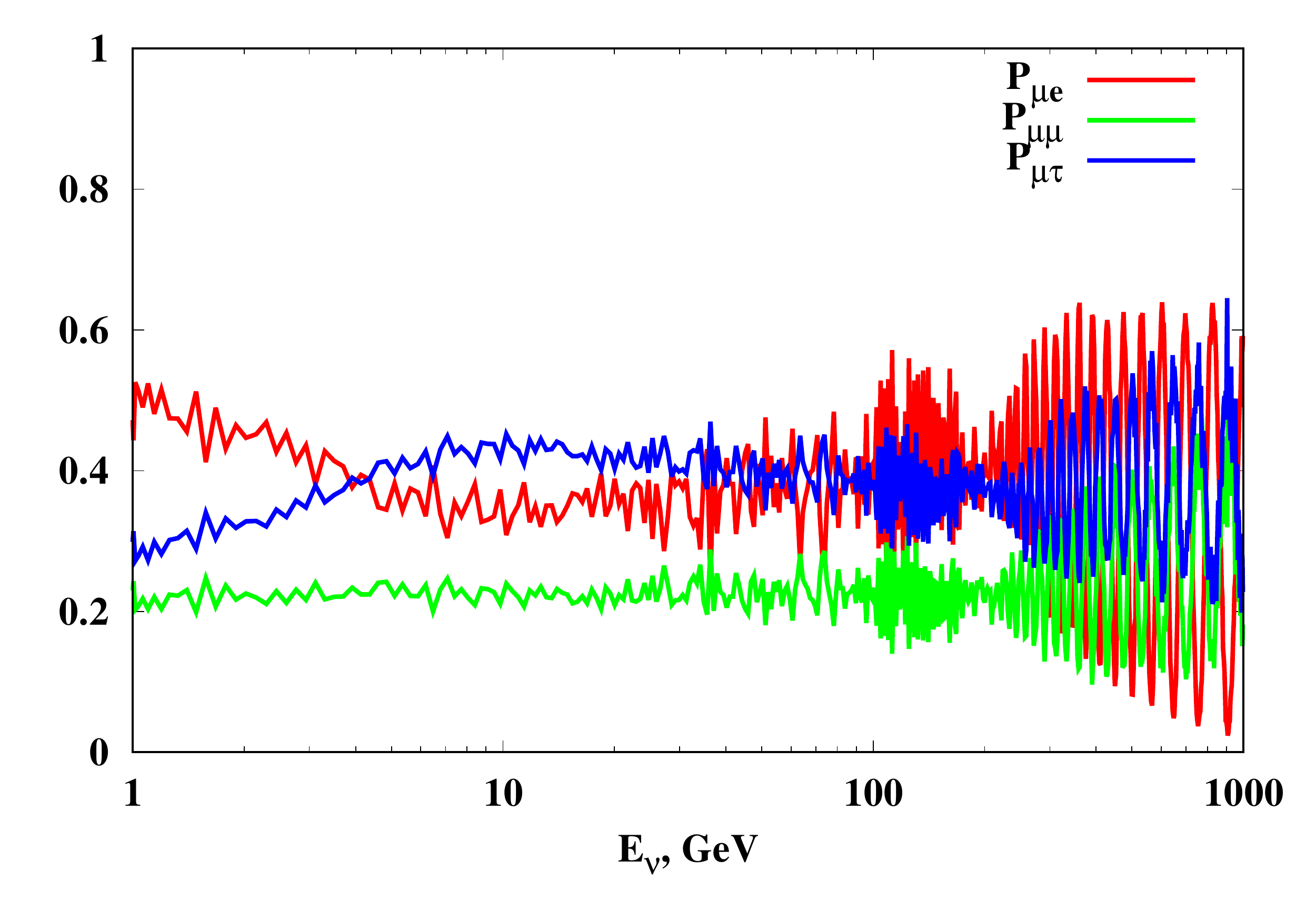}}
\put(210,250){\includegraphics[angle=-90,width=0.40\textwidth]{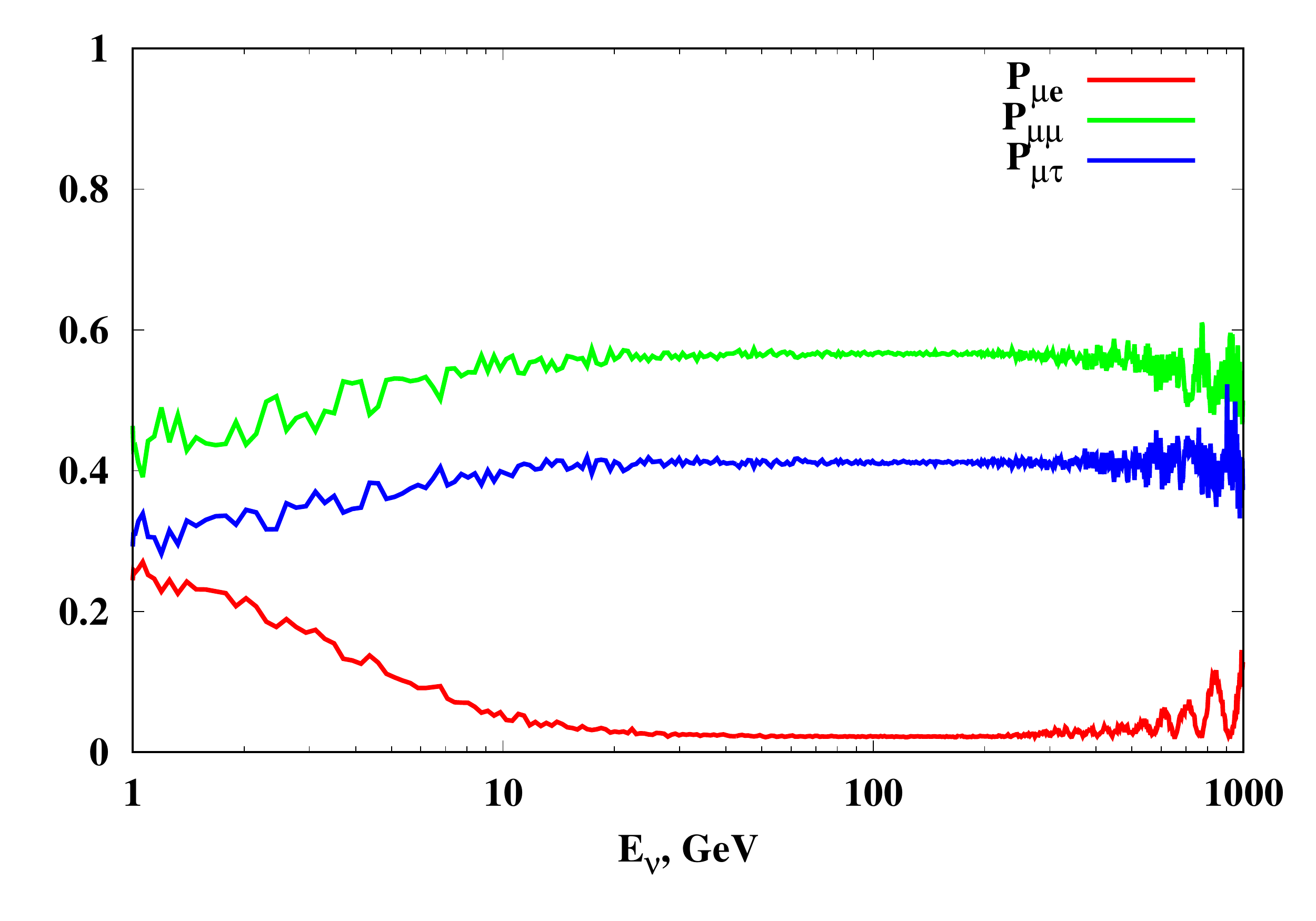}}
\put(30,130){\includegraphics[angle=-90,width=0.40\textwidth]{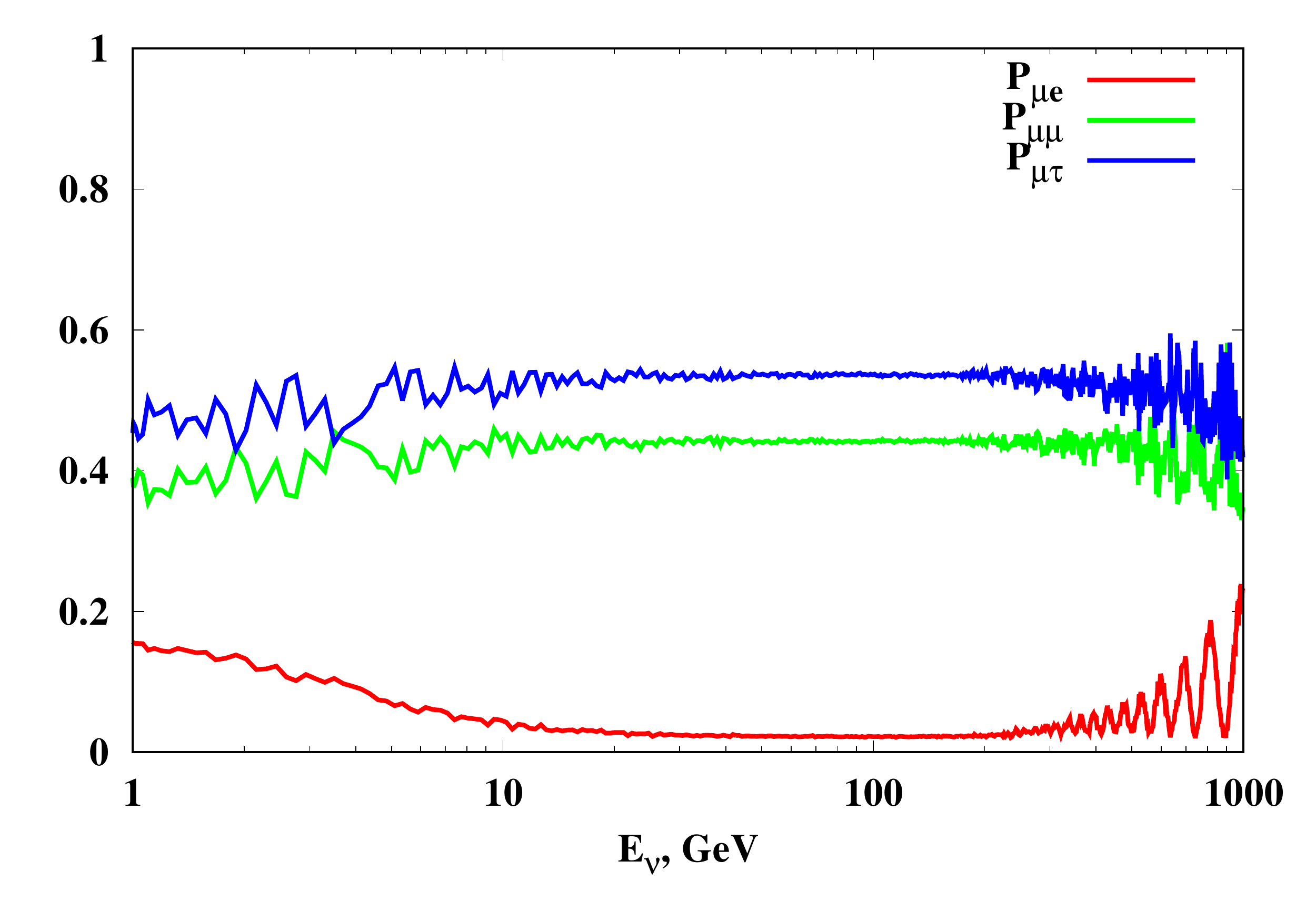}}
\put(30,250){\includegraphics[angle=-90,width=0.40\textwidth]{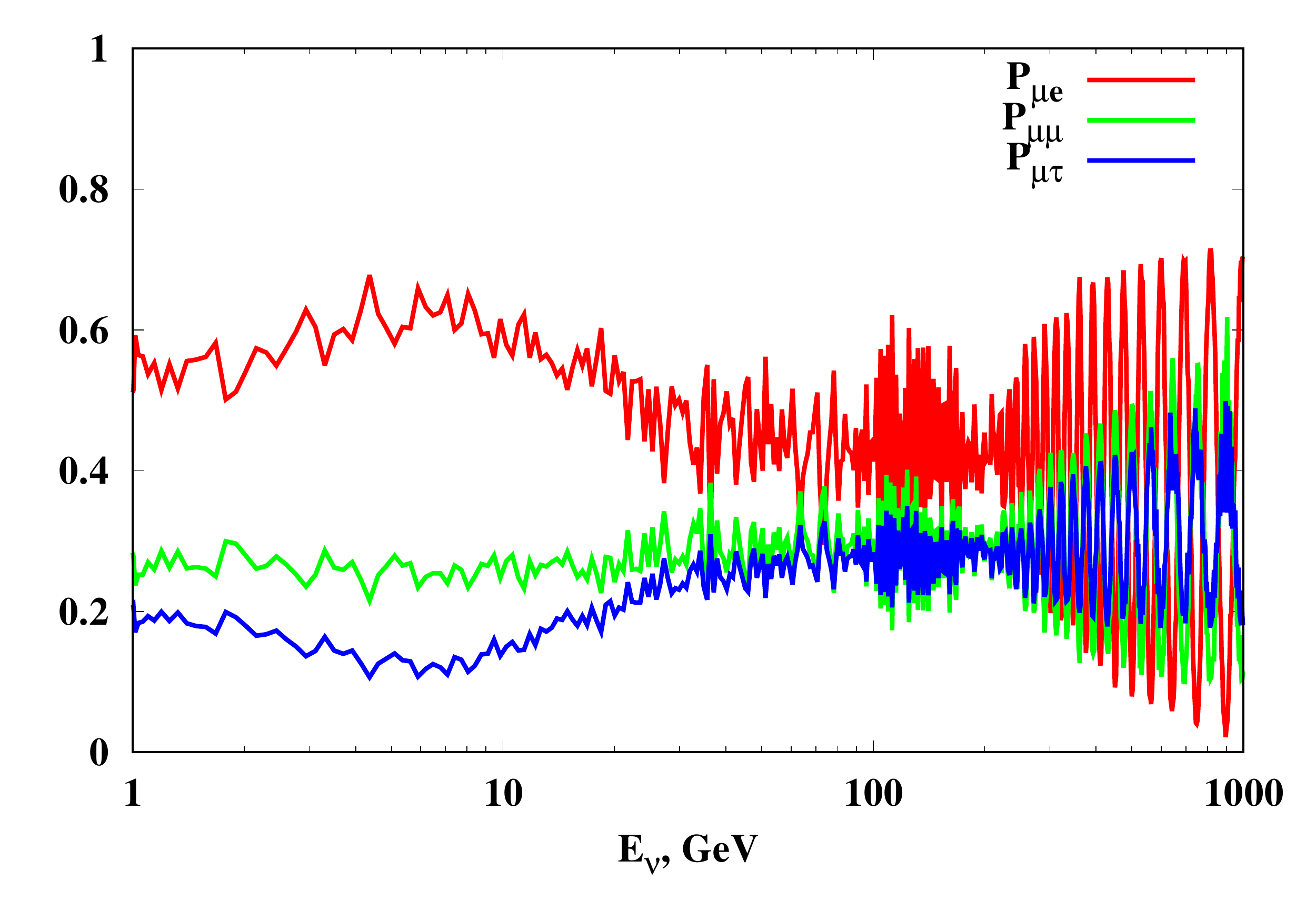}}
\end{picture}
\caption{\label{numu_tautau} The same as in Fig.~\ref{numu_sm} but for
$\e_{\tau\tau}=0.03$.}
\end{figure}

\begin{figure}[hb]
\begin{picture}(300,250)(0,20)
\put(210,130){\includegraphics[angle=-90,width=0.40\textwidth]{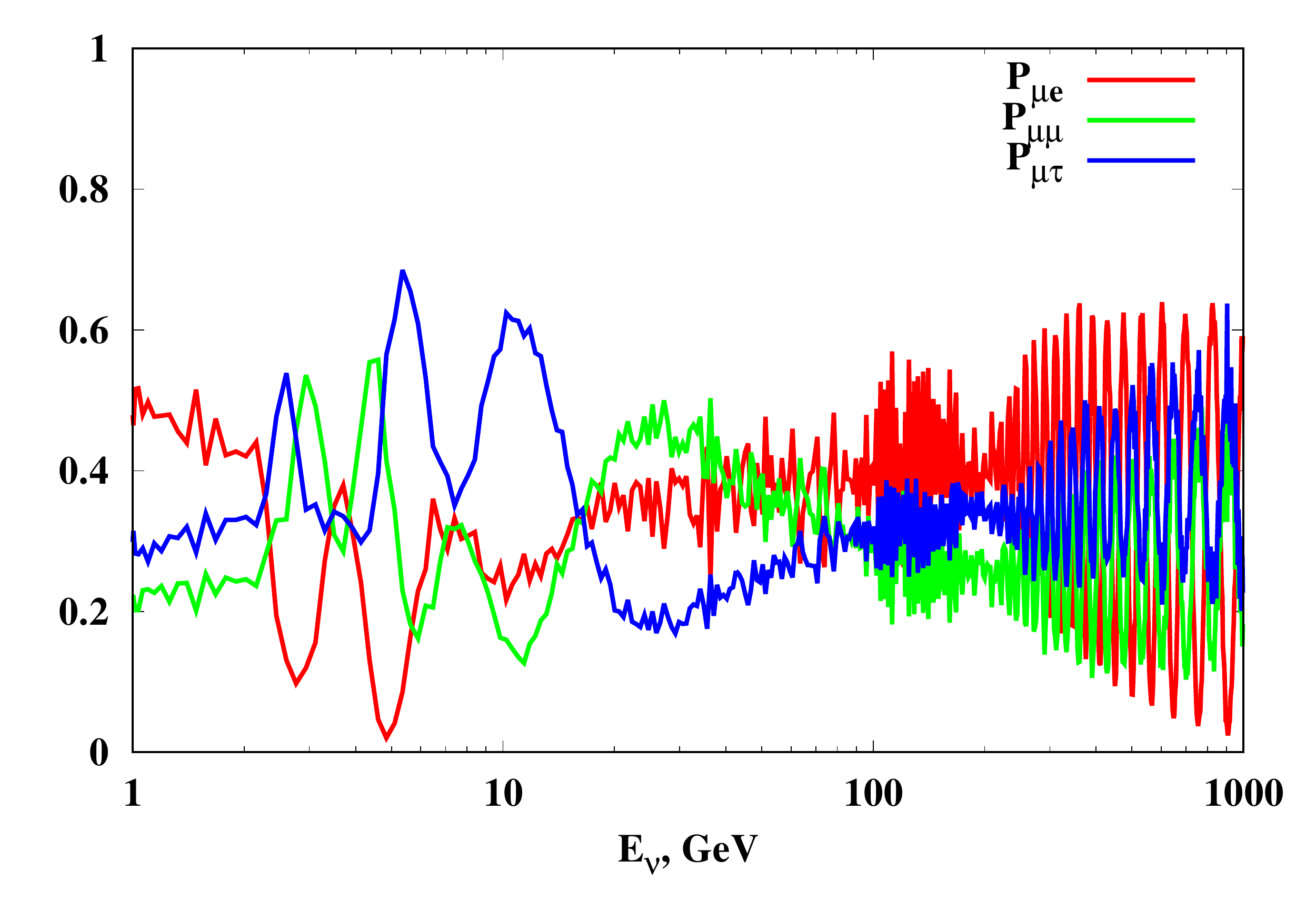}}
\put(210,250){\includegraphics[angle=-90,width=0.40\textwidth]{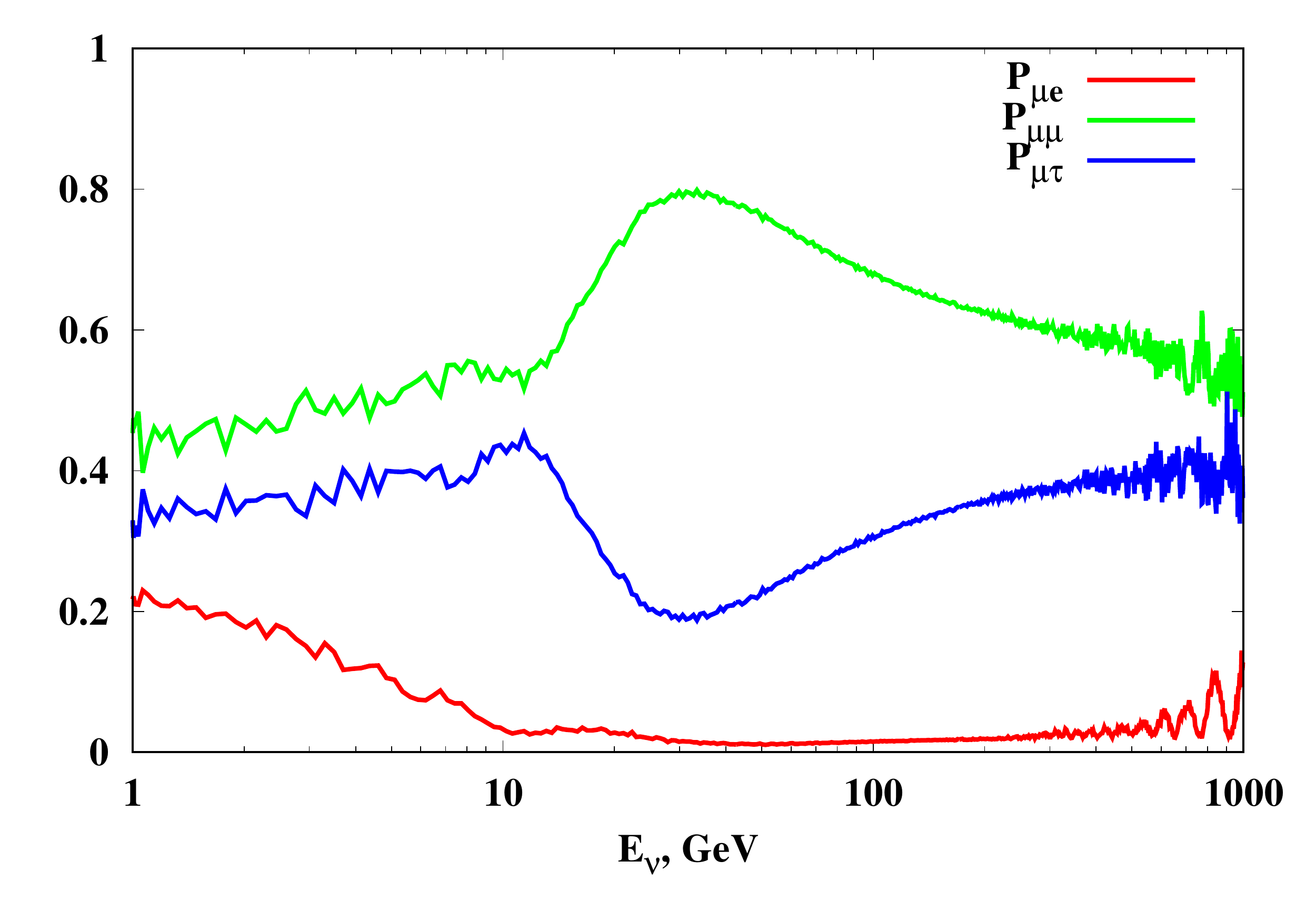}}
\put(30,130){\includegraphics[angle=-90,width=0.40\textwidth]{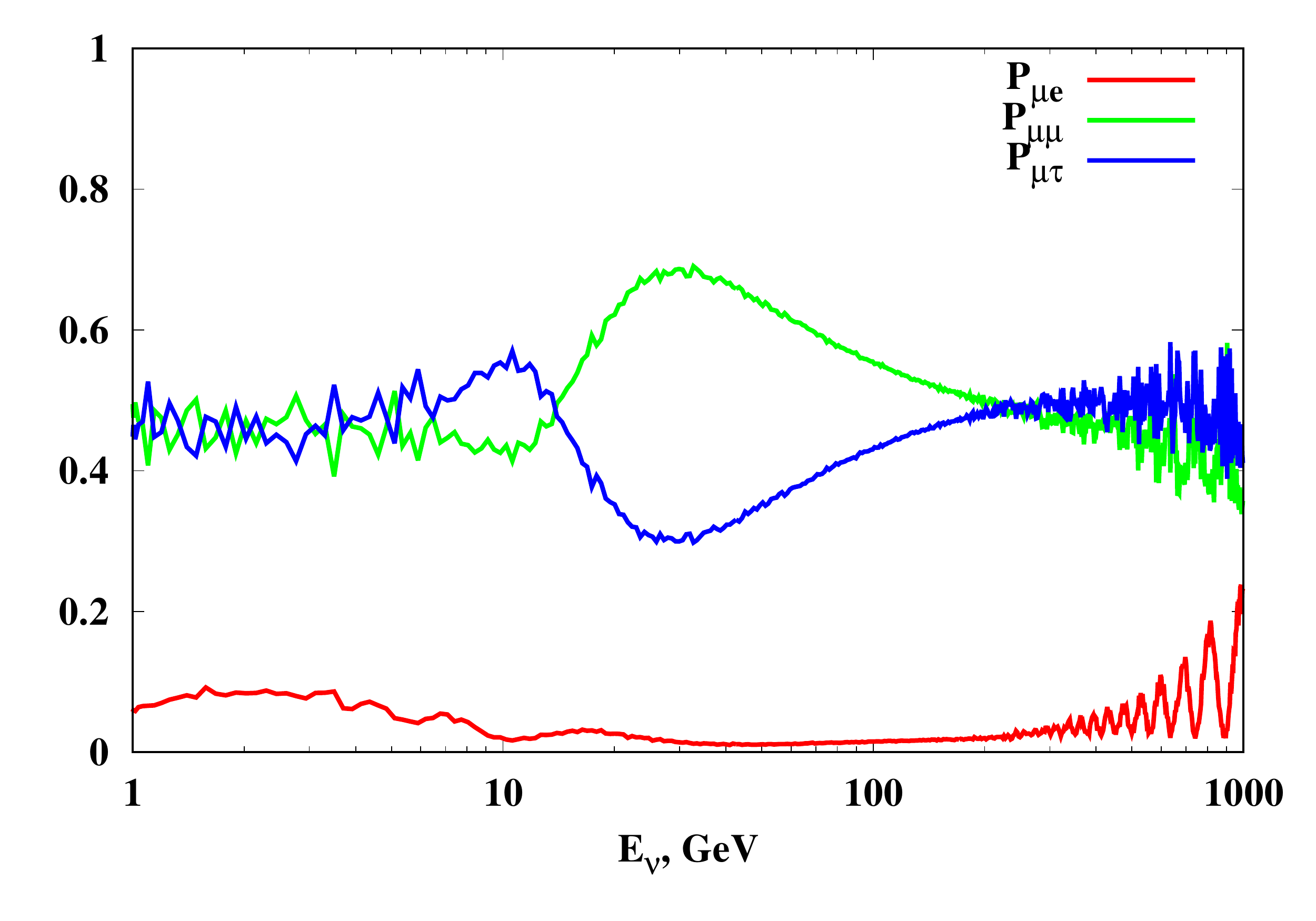}}
\put(30,250){\includegraphics[angle=-90,width=0.40\textwidth]{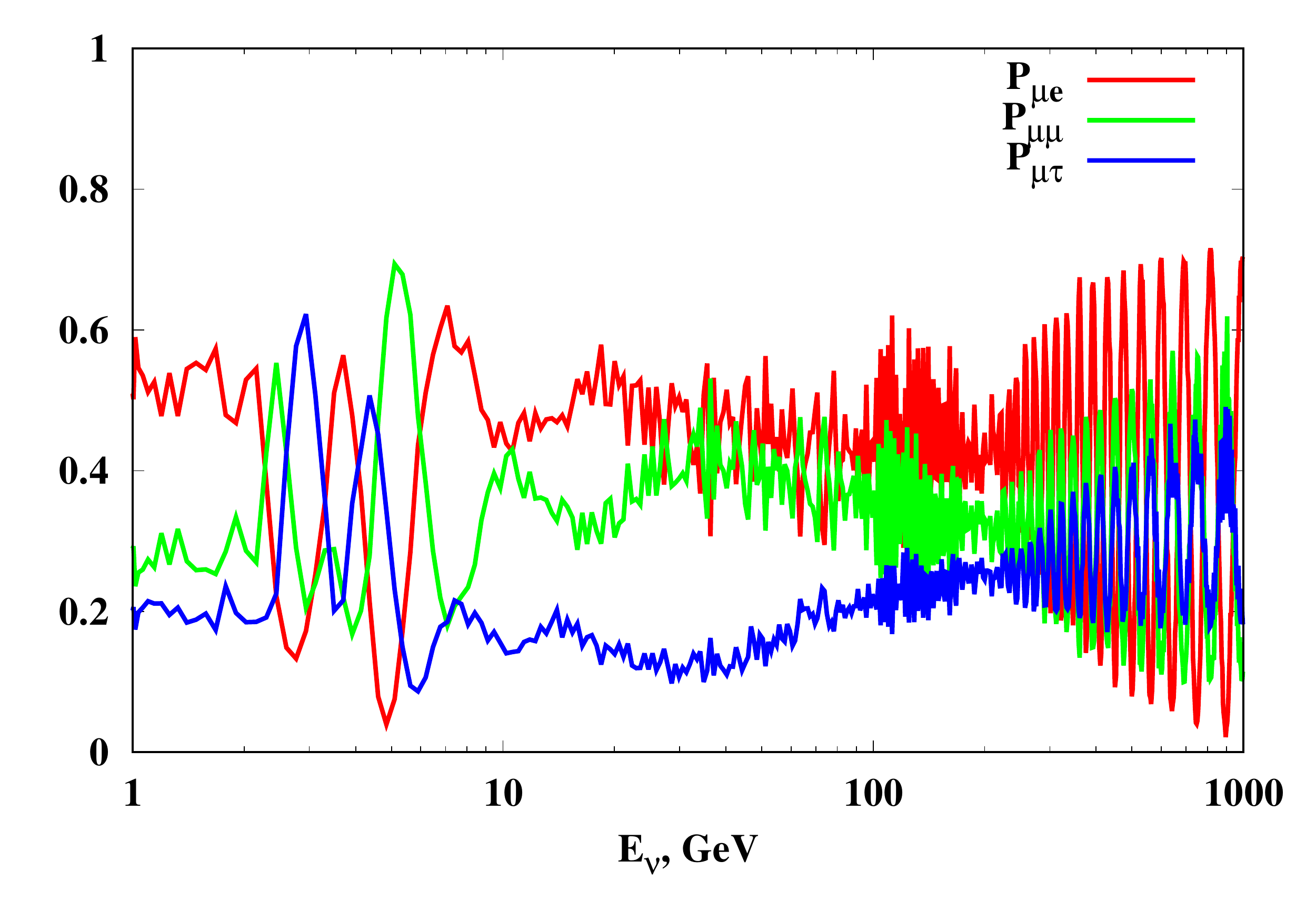}}
\end{picture}
\caption{\label{numu_tautau_earth} The same as in
  Fig.~\ref{numu_sm_earth} but for $\e_{\tau\tau}=0.03$.}
\end{figure}

In what follows we turn on the matter NSI parameters
$\e_{\alpha\beta}$ taking single non-zero parameter at a time.  
In this Section we study effect of the flavor diagonal NSI parameters.
Due to smallness of the resonance energy for 1--2 transition effect of
non-zero $\e_{ee}$ within experimentally allowed region on the
propagation of high energy neutrino is negligible. As a consequence
effect due to small non-zero $\e_{\mu\mu}$ will be the same as that
of due to $-\e_{\tau\tau}$. In what follows we consider non-zero value of
$\epsilon_{\tau\tau}$ and assume that $\epsilon_{\tau\tau}\ll 1$
in analytical expressions which is consistent with phenomenological
bounds~\eqref{eq:1:8} and~\eqref{eq:1:9}. As a benchmark point we take
$\e_{\tau\tau}=\pm 0.03$. In Figs.~\ref{numu_tautau} 
and~\ref{numu_tautau_earth} we plot probabilities to obtain neutrino
of different flavors  at the Earth orbit and after passing the Earth,
respectively, from  $\nu_\mu$ produced at the center of the Sun. These
probabilities are calculated with $\e_{\tau\tau}=0.03$. 
One can see that oscillations of neutrino in the matter of the Sun and
the Earth deviates considerably from no-NSI case. Let us discuss the
reasons for these deviations. In the center of the Sun the eigenstate
$|1_m,0\rangle=|\nu_e\rangle$ is decoupled from the others for
GeV-scale neutrinos and the
Hamiltonian for the rest 2--3 subsystem has the following form
\be
\label{eq:3.1:1}
H_{23} = \frac{\Delta m^2_{32}}{2E_\nu}R_{23}(\theta_{23})
\left(
\begin{array}{cc}
  0 & 0 \\
  0 & 1
\end{array}
\right) R^\dagger_{23}(\theta_{23}) + \e_{\tau\tau}V_e\left(
\begin{array}{cc}
  0 & 0 \\
  0 & 1
\end{array}
\right).
\ee
It can be diagonalized by corresponding rotation
$\nu=R_{23}(\tilde{\theta}_{23})\tilde{\nu}$, where
$\tilde{\theta}_{23}$ is determined by 
\be
\label{eq:3.1:2}
\tan{2\tilde{\theta}_{23}} =
\frac{\sin{2\theta_{23}}}{\cos{2\theta_{23}} + A},
\ee
where $A = \frac{2\e_{\tau\tau}V_eE_\nu}{\Delta m^2_{32}}$. When
denominator in~\eqref{eq:3.1:2} goes to zero new 2--3 resonance
occurs. The resonance energy in the center of the Sun is 
\be
\label{eq:3.1:3}
E_R^{23} = -\frac{\Delta m_{32}^2}{2\e_{\tau\tau}V_e(0)}\cos{2\theta_{23}}.
\ee
For normal mass hierarchy 2--3 resonance takes place for neutrino if ${\rm
  sign}(\e_{\tau\tau}\cos{2\theta_{23}}) = -1$ and for 
antineutrino if ${\rm sign}(\e_{\tau\tau}\cos{2\theta_{23}}) = +1$. For
inverted mass hierarchy the resonance conditions are opposite.
Taking numerical
values from Table~\ref{tab:1} one obtains\footnote{We note that for
  values 
of oscillation parameters in Table~\ref{tab:1} one finds
$\cos{2\theta_{23}}>0$ for NH and $\cos{2\theta_{23}}<0$ for
IH. However, experimentally any sign of $\cos{2\theta_{23}}$ is
allowed at $3\sigma$ level.} 
$E_R^{23} \approx 0.015/\e_{\tau\tau}$~GeV (NH) and $E_{R}^{23}\approx
0.025/\e_{\tau\tau}$~GeV (IH) and the resonance appears for
antineutrinos in both cases. We note in passing that subsequent
propagation of electron (anti)neutrino $\nu_e$  produced in the center
of Sun is not affected by non-zero $\e_{\tau\tau}$ 
($\e_{\mu\mu}$) in the considered energy range. For relatively small
values of $\e_{\tau\tau}$ the position of 2--3 resonance is well
separated in space from the other possible resonances. In this case
they can be treated separately. The resonance conversion can take
place only for neutrinos with $E>E_{R}^{23}$. For such neutrinos
flavor states are approximately Hamiltonian eigenstates and in
particular $|\nu_\mu\rangle$ is almost coincide with the second energy
level of the matter Hamiltonian for $\e_{\tau\tau}>0$.  Adiabatic evolution
through 2--3 resonance region implies
\begin{eqnarray}
  |\tilde{2}_m\rangle\to|2_m\rangle,\;\;\; &
  |\tilde{3}_m\rangle\to|3_m\rangle\;\; {\rm for}\; \nu\,,\\
  |\tilde{2}_m\rangle\to|3_m\rangle,\;\;\; &
  |\tilde{3}_m\rangle\to|2_m\rangle\;\; {\rm for}\; \bar{\nu}
\end{eqnarray}
in the case of NH and
\begin{eqnarray}
  |\tilde{2}_m\rangle\to|3_m\rangle,\;\;\; &
  |\tilde{3}_m\rangle\to|2_m\rangle\;\; {\rm for}\; \nu\,,\\
  |\tilde{2}_m\rangle\to|2_m\rangle,\;\;\; &
  |\tilde{3}_m\rangle\to|3_m\rangle\;\; {\rm for}\; \bar{\nu}\,
\end{eqnarray}
for IH. 
Let
us note that onset energy for 
non-adiabatic effects for 2--3 transition can be estimated as follows
follows~\cite{Lehnert:2007fv,Kuo:1989qe}
\be
\label{eq:3.1:5}
E_{NA} = \frac{1}{3}\pi\bigg|\frac{V_e}{V^{\prime}_e}\bigg|\Delta
m_{32}^2\sin^2{\theta_{23}} \sim 0.5~{\rm TeV},
\ee
which indicates that adiabaticity for this transition is valid almost entirely in the chosen neutrino energy
range. Effect of non-zero $\e_{\tau\tau}=0.03$ is most dramatic in
Fig.~\ref{numu_tautau} for antineutrino (NH) and neutrino (IH). In the
former case the muon antineutrino $\nu_\mu$ produced in the 
center of the Sun undergoes resonance 2--3 transition and evolves into
$|3_m\rangle$. This state meets no level crossing in subsequent
evolution to the solar surface where it becomes pure vacuum eigenstate
$|3\rangle$, see~\eqref{trans1}. This is completely different from the
evolution with no-NSI where the final state is incoherent mixture of
$|2\rangle$ and $|3\rangle$, see Fig.~\ref{numu_sm}. In the low energy
region one should take into account an admixture of
$|\tilde{3}_m\rangle$ state. Namely neglecting contribution of
$|\tilde{1}_m\rangle\equiv|\nu_e\rangle$ one finds that 
\be
\label{eq:3.1:4}
|\nu_\mu\rangle = \cos{\tilde{\theta}^0_{23}} |\tilde{2}_m,0\rangle -
\sin{\tilde{\theta}^0_{23}}|\tilde{3}_m,0\rangle,
\ee
where $\tilde{\theta}_{23}^0$ is determined by Eq.~\eqref{eq:3.1:2}
for the center of the Sun. This admixture results in deviation from
simple $P_{\mu\alpha} = |U_{\alpha 3}|^2$ law. Similar explanation can
be given for the case of neutrino (IH). For neutrino (NH) and antineutrino (IH), 
see Fig~\eqref{numu_tautau} (upper left and lower right panels),
$\nu_\mu$ state evolve into $|2_m,r\rangle$.
For neutrino (NH) in the adiabatic regime $\nu_\mu$
evolves into $|1\rangle$ at the solar surface, while the evolution at
higher energies is more complicated due to non-adiabatic effects. In
the limit of maximal adiabaticity violation the level crossing
probability approaches $c_{12}^2$, which corresponds to
incoherent vacuum oscillations and one obtains
\be
\nu_\mu\approx |\tilde{2}_m\rangle \to|2_m\rangle\to
c^2_{12}|2\rangle\langle 2| +
s^2_{12}|1\rangle \langle 1|.
\ee
For antineutrino (IH) $\nu_\mu$ evolves again into $|2_m\rangle$ state
when passing 2--3 resonance and emerges as $|2\rangle$ at the Earth
orbit in the adiabatic regime. At energies $E_\nu\lsim 10$~GeV the
main difference with no-NSI case comes from the fact that $\nu_\mu$ at
production is only an approximate eigenstate of the matter Hamiltonian.

Let us turn to subsequent evolution in the Earth, see
Fig.~\ref{numu_tautau_earth}.  The most prominent effect of non-zero
$\e_{\tau\tau}$ appears again for neutrino mode with inverted
hierarchy and for antineutrino with normal hierarchy and reaches its
culmination for $E_\nu\sim 30$~GeV. In this case almost pure
$|3\rangle$ vacuum state reaches the Earth surface, see lower left and
upper right panels on Fig.~\ref{numu_tautau}. Let us neglect
here for simplicity small 
contribution of electron neutrino $\nu_e$ related to non-zero
$\theta_{13}$. 
In the matter of the Earth this state is no longer Hamiltonian
eigenstate and thus nontrivial evolution takes place.  Eigenstates of
2--3 subsystem describing by the Hamiltonian~\eqref{eq:3.1:1} can be found again by
rotation $\nu=R_{23}(\tilde{\theta}_{23})\tilde{\nu}$ with
$\tilde{\theta}_{23}$ given by Eq.~\eqref{eq:3.1:2} with neutrino
matter potential in the Earth. To qualitatively understand the
behaviour of the probabilities in Fig.~\ref{numu_tautau_earth} let us
find the evolution for matter with constant density. In this case the
evolution of the state $|3\rangle$ after traversing the distance L can
be described as  
\be
\label{eq:3.1:6}
|3,L\rangle\approx
\sin{(\theta_{23}-\tilde{\theta}_{23})}e^{-i\tilde{E}_2L}|\tilde{2}\rangle +
\cos{(\theta_{23}-\tilde{\theta}_{23})}e^{-i\tilde{E}_3L}|\tilde{3}\rangle ,
\ee
where $\tilde{E}_{2,3}$ are the Hamiltonian eigenvalues. The probability
to find muon neutrino is then given by
\be
\label{eq:3.1:7}
P_{\mu\mu} = \sin^2{\theta_{23}} -
\sin{2\tilde{\theta}_{23}}\sin{2(\theta_{23} -
  \tilde{\theta}_{23})}\sin^2{\frac{(\tilde{E}_3-\tilde{E}_2)L}{2}}. 
\ee
Similar expression for $P_{\mu\tau}$ has the form
\be
\label{eq:3.1:8}
P_{\mu\tau} = \cos^2{\theta_{23}} +
\sin{2\tilde{\theta}_{23}}\sin{2(\theta_{23} -
  \tilde{\theta}_{23})}\sin^2{\frac{(\tilde{E}_3-\tilde{E}_2)L}{2}}. 
\ee
Numerically for $E_\nu=30$~GeV and $\e_{\tau\tau}=0.03$ one finds that
$\tilde{E}_3-\tilde{E}_2\approx  \frac{\Delta m^2_{32}}{2E_\nu}$ and
the oscillation amplitude in Eq.\eqref{eq:3.1:7} for antineutrino and
NH varies from $0.11$ for the mantle of the Earth to about $0.28$ 
for its core\footnote{For the sake of argument we consider the Earth
  as consisting of mantle with $N_e\approx 2.2 N_A$~cm$^{-3}$ and core
  with $N_e\approx 5.4 N_A$~cm$^{-3}$~\cite{Patrignani:2016xqp}.}. The
same numbers for neutrino and IH are $0.06$ and $0.23$, respectively.
One can find from~\eqref{eq:3.1:2} that
$\tilde{\theta}_{23}>\theta_{23}$ for antineutrino (NH) and neutrino
(IH) cases for $\e_{\tau\tau}>0$.  
Maximum for $P_{\mu\mu}$ and minimum for $P_{\mu\tau}$ on the lower left and upper right panels in
Fig.~\ref{numu_tautau_earth} corresponds approximately to the first
oscillation maximum for neutrino coming through the center of the
Earth. For the cases of neutrino (NH) and antineutrino (IH) (see upper
left and lower right panels in Fig.~\ref{numu_tautau_earth}) one can
observe similar bumps at $E_\nu\sim 30$~GeV although their amplitudes
are smaller due to considerable admixture of electron neutrino. Much
more complicated picture is observed at smaller energies which can be
attributed to interference of the effects of 2--3 and 1-3 mixings with
non-zero $\e_{\tau\tau}$.

\begin{figure}[t]
\begin{picture}(300,220)(0,20)
\put(210,130){\includegraphics[angle=-90,width=0.40\textwidth]{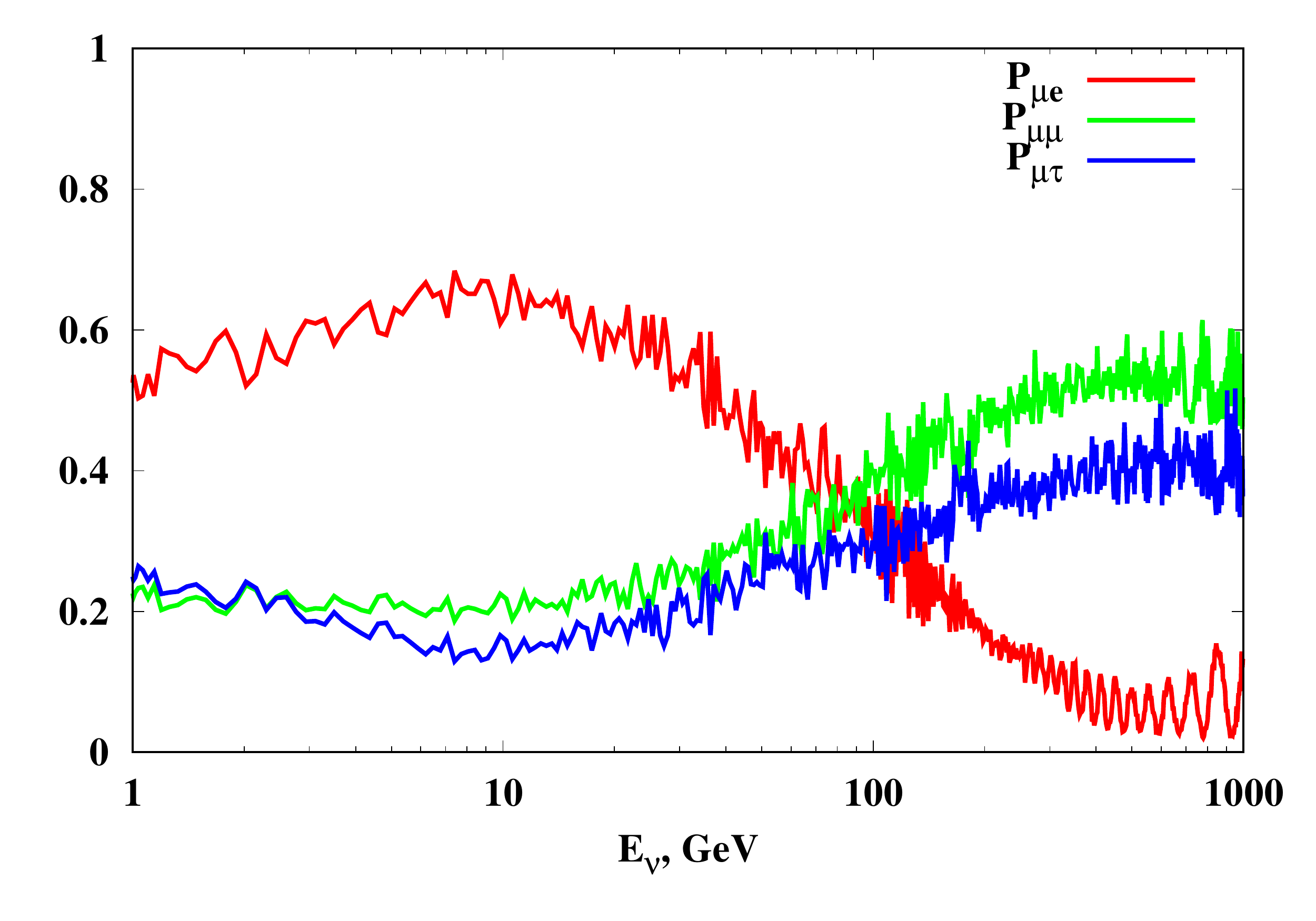}}
\put(210,250){\includegraphics[angle=-90,width=0.40\textwidth]{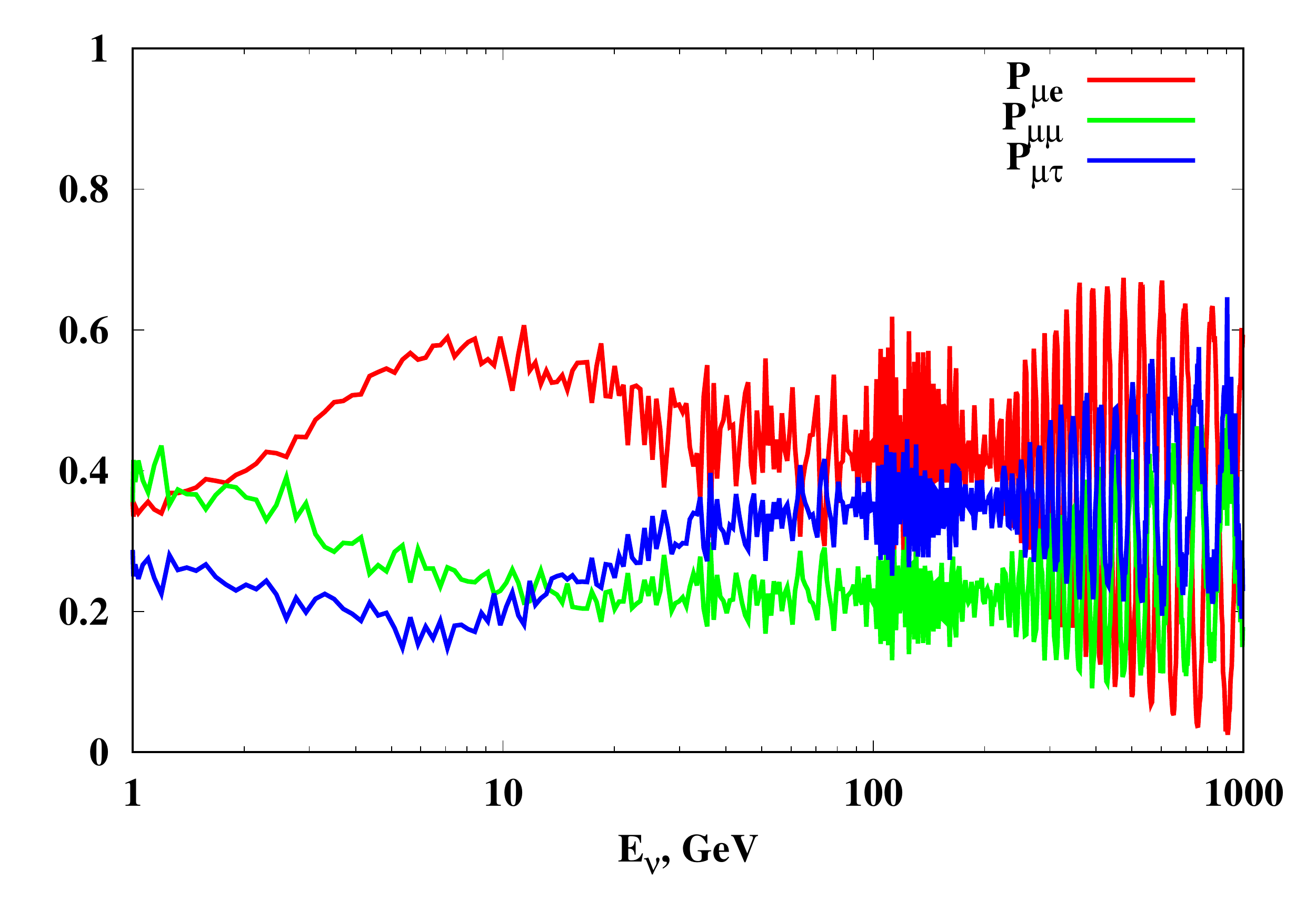}}
\put(30,130){\includegraphics[angle=-90,width=0.40\textwidth]{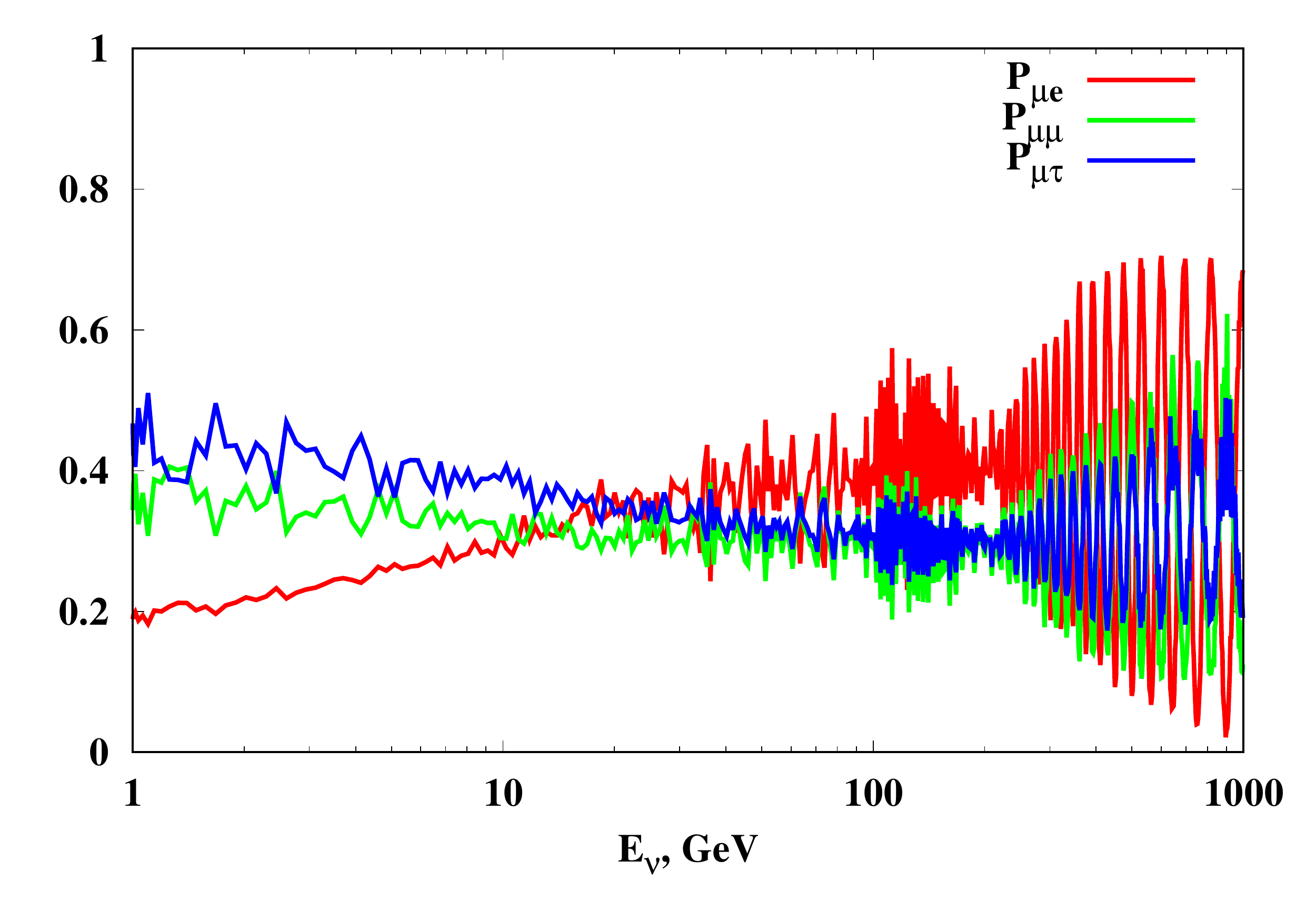}}
\put(30,250){\includegraphics[angle=-90,width=0.40\textwidth]{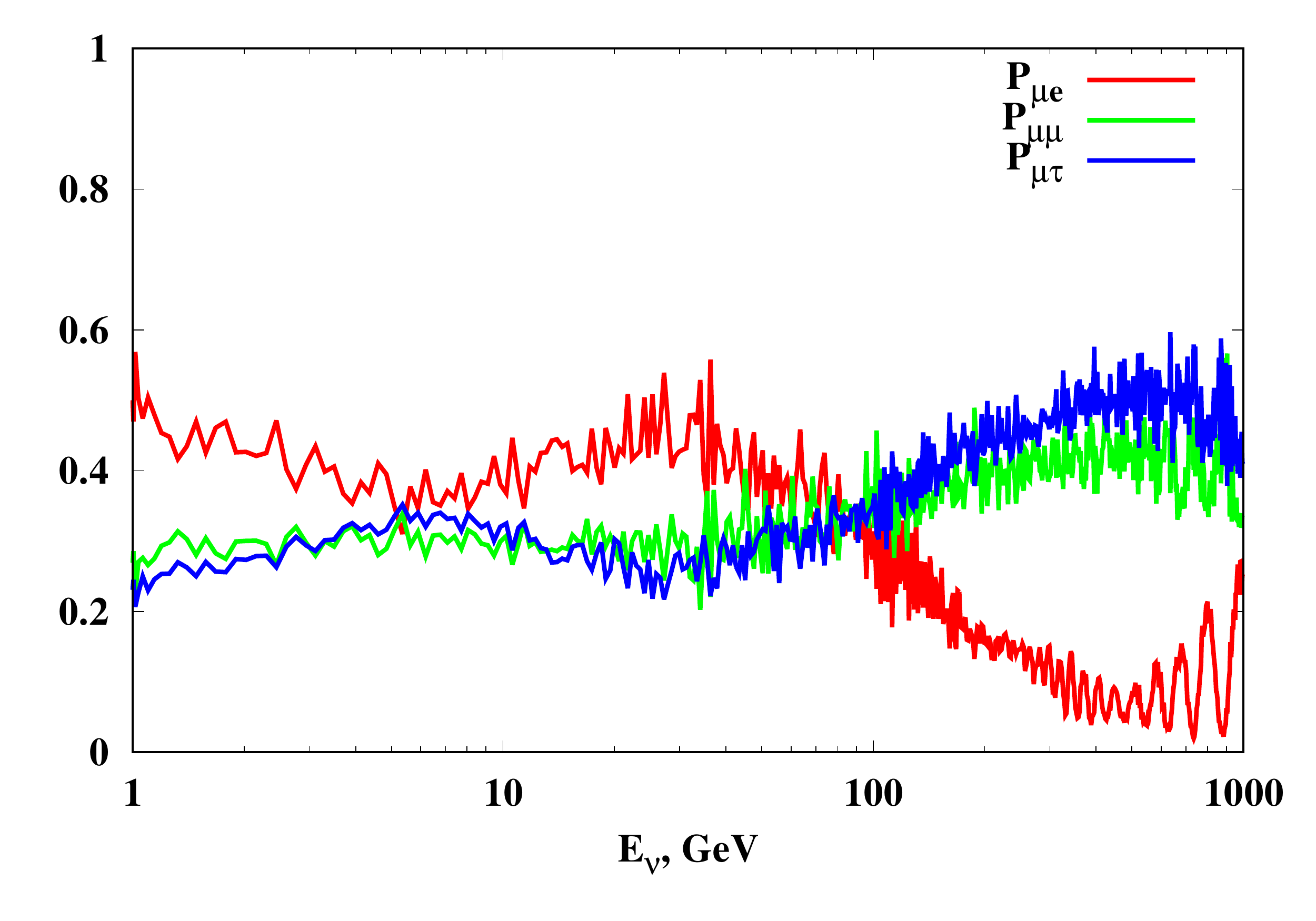}}
\end{picture}
\caption{\label{numu_tautau_sign} The same as in Fig.~\ref{numu_sm} but for
  $\e_{\tau\tau}=-0.03$.}
\end{figure}

\begin{figure}[hb]
\begin{picture}(300,220)(0,20)
\put(210,130){\includegraphics[angle=-90,width=0.40\textwidth]{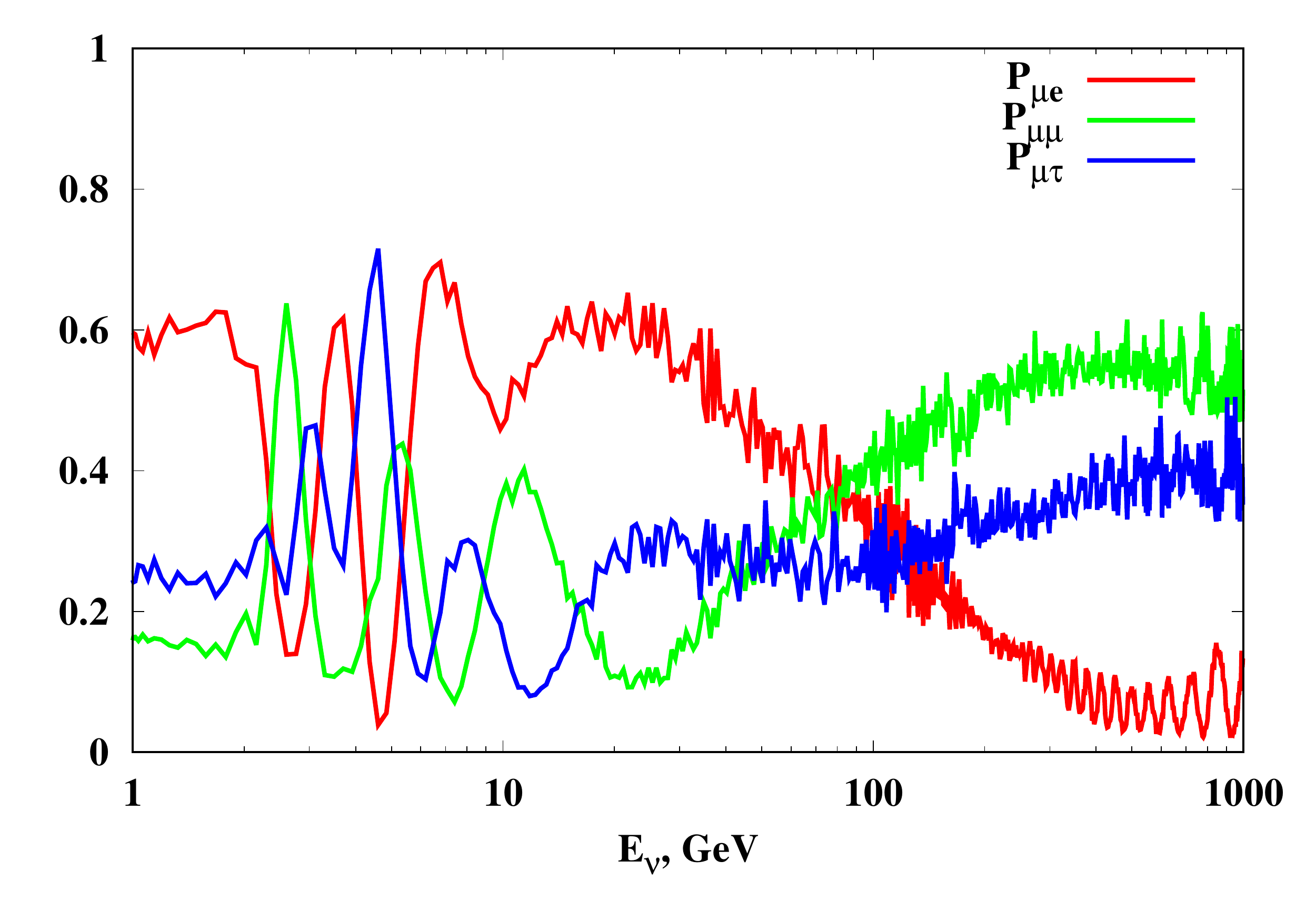}}
\put(210,250){\includegraphics[angle=-90,width=0.40\textwidth]{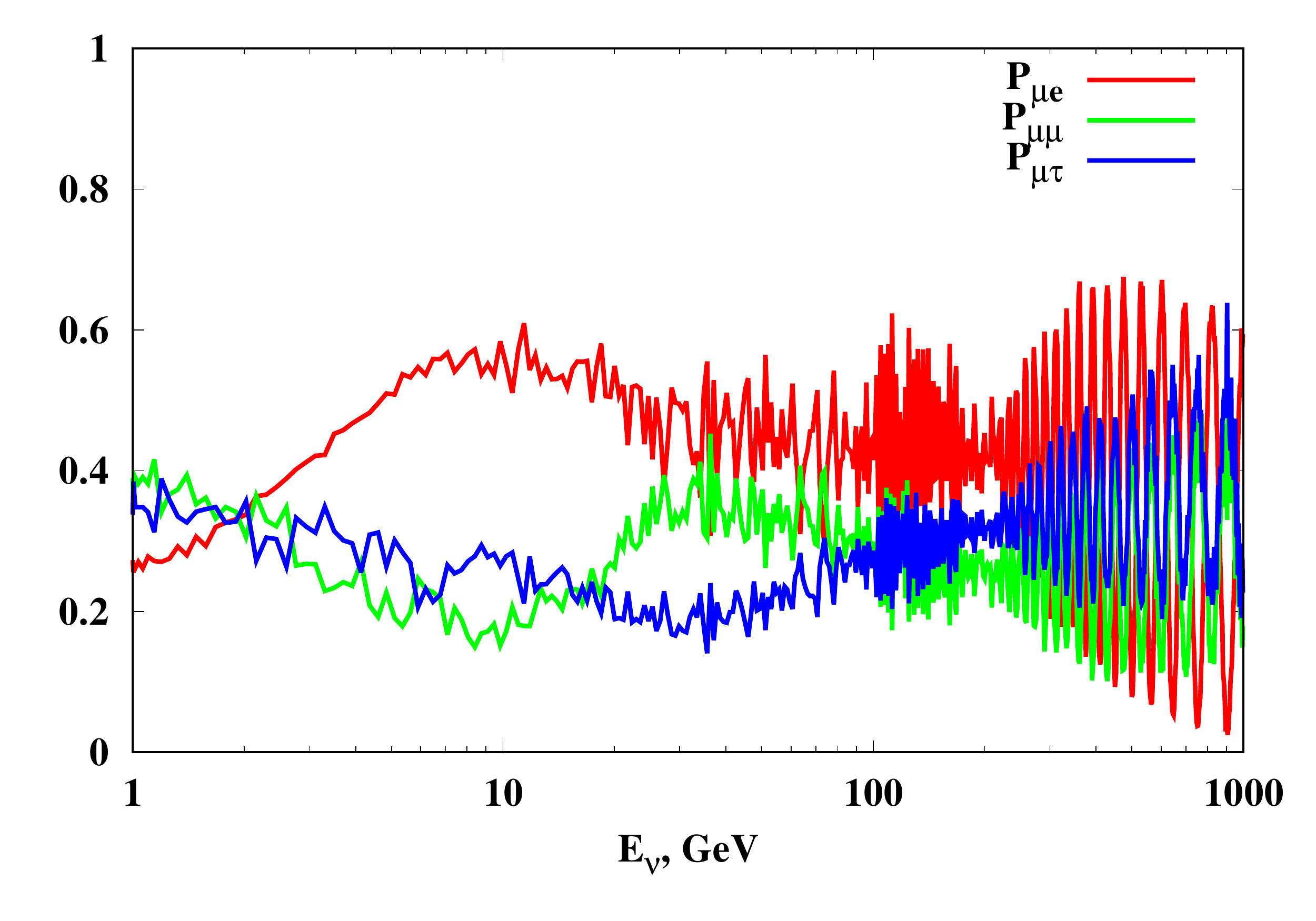}}
\put(30,130){\includegraphics[angle=-90,width=0.40\textwidth]{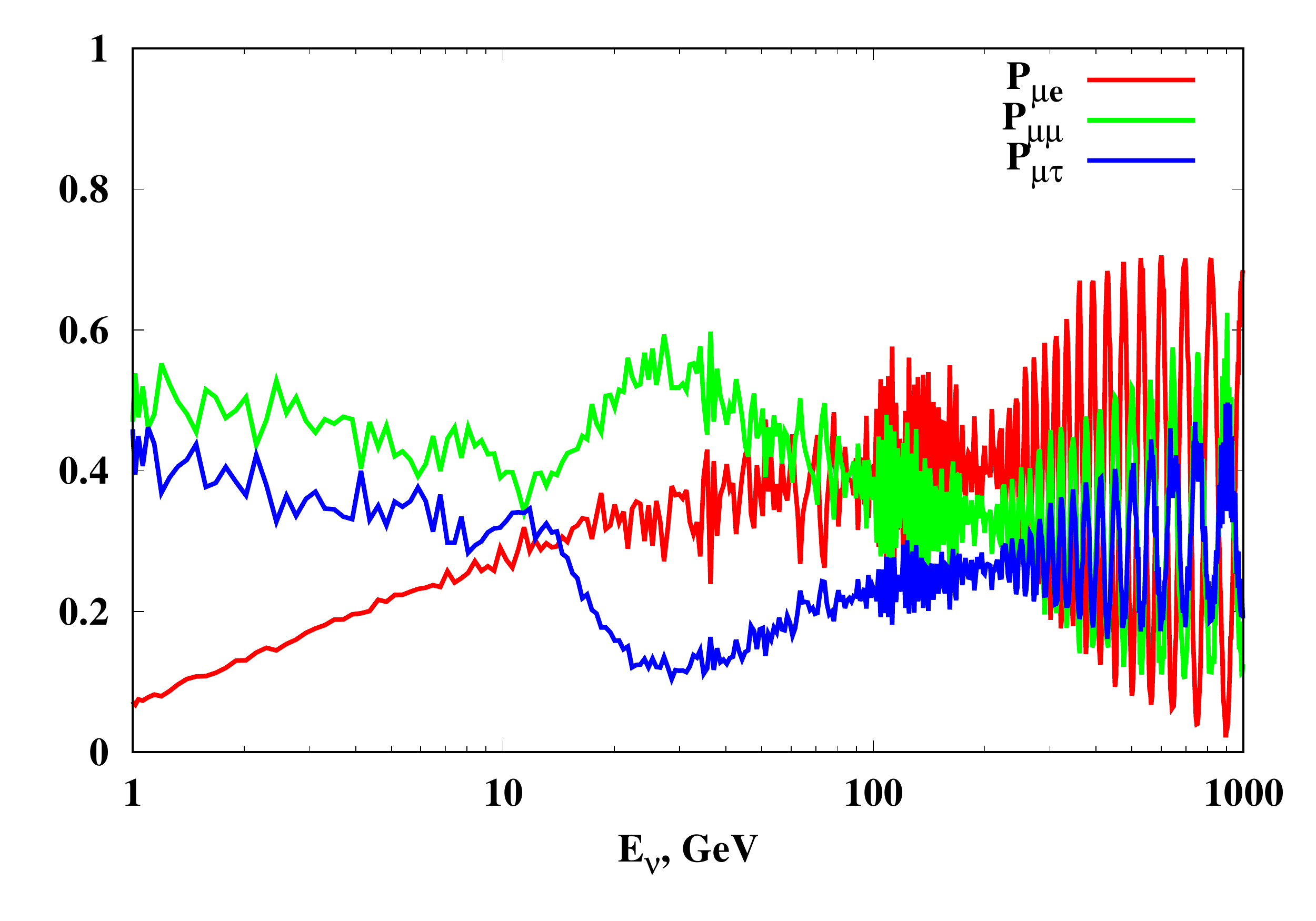}}
\put(30,250){\includegraphics[angle=-90,width=0.40\textwidth]{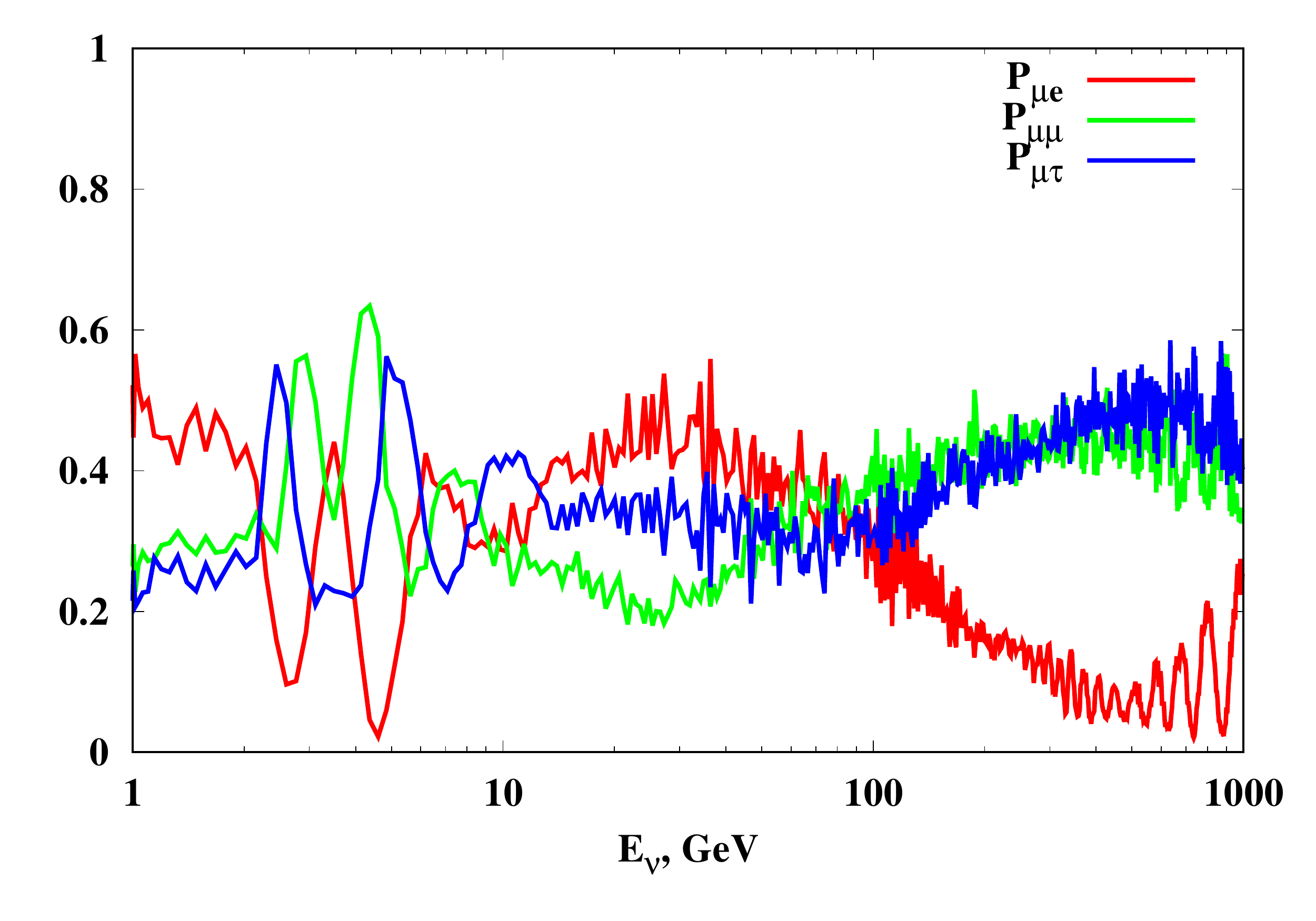}}
\end{picture}
\caption{\label{numu_tautau_sign_earth}  The same as in Fig.~\ref{numu_sm_earth}
  but for $\e_{\tau\tau}=-0.03$.}
\end{figure}

In Figs.~\ref{numu_tautau_sign} and~\ref{numu_tautau_sign_earth}
we plot the same probabilities for negative $\e_{\tau\tau}=-0.03$.
In this case 2--3 resonance takes place in neutrino
mode and muon neutrino is almost coincide with the third (first)
energy level of the matter Hamiltonian for neutrino (antineutrino) mode.
At $E_\nu\lsim 10$~GeV adiabaticity for all possible level
crossings is valid and one can apply Eq.~\eqref{eq:2:0} with
$P_{ij}=\delta_{ij}$ to verify the probabilities in
Fig.~\ref{numu_tautau_sign}. 
At higher energies nonadiabatic effects turn on. The behaviour
of the probabilities for neutrino (NH) and antineutrino (IH) again
indicates that at $E_\nu\gsim 100$~GeV neutrino escapes the Sun almost
as the vacuum eigenstate $|3\rangle$. Indeed, in this case the
muon neutrino $\nu_\mu$ at production coincides with the eigenstate
$|\tilde{2}_m\rangle$ which becomes $|3_m\rangle$ after adiabatic
2--3 transition. The same happens in antineutrino
mode for inverted mass hierarchy. At very high energies adiabaticity
for transition through 1--3 level crossing is maximally violated and
due to smallness of $s^2_{13}$ one obtains almost pure
eigenstate $|3\rangle$ outside the Sun. On
Fig.~\ref{numu_tautau_sign_earth} evolution of neutrino through the
Earth is taken into account and we 
see again the bump-like shapes for $P_{\mu\mu}$ and $P_{\mu\tau}$ at
energies around 30~GeV related to the matter effect in the Earth
discussed above.

\subsection{Flavor changing NSI}
\begin{figure}[!htb]
\begin{picture}(300,220)(0,20)
\put(210,130){\includegraphics[angle=-90,width=0.40\textwidth]{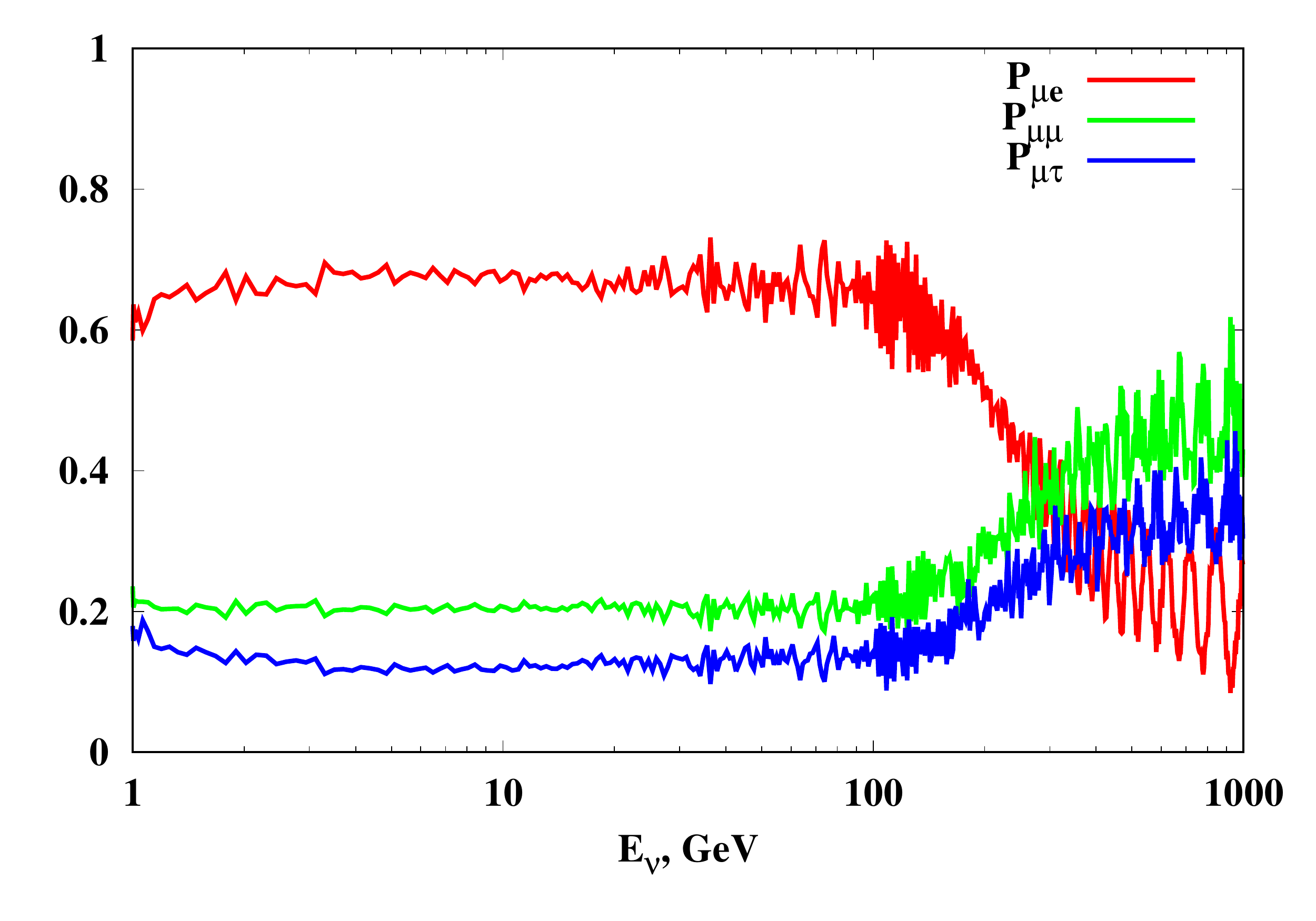}}
\put(210,250){\includegraphics[angle=-90,width=0.40\textwidth]{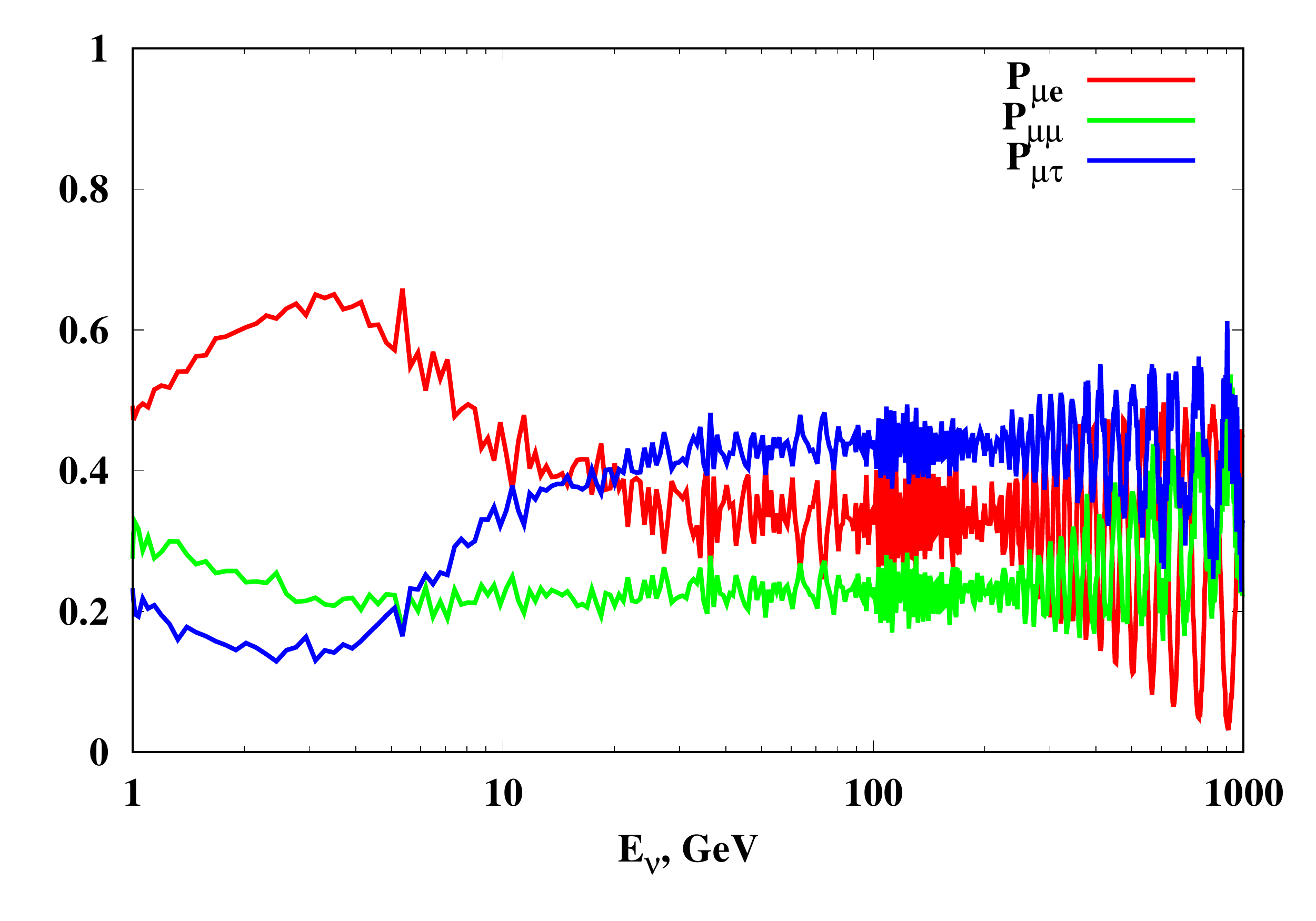}}
\put(30,130){\includegraphics[angle=-90,width=0.40\textwidth]{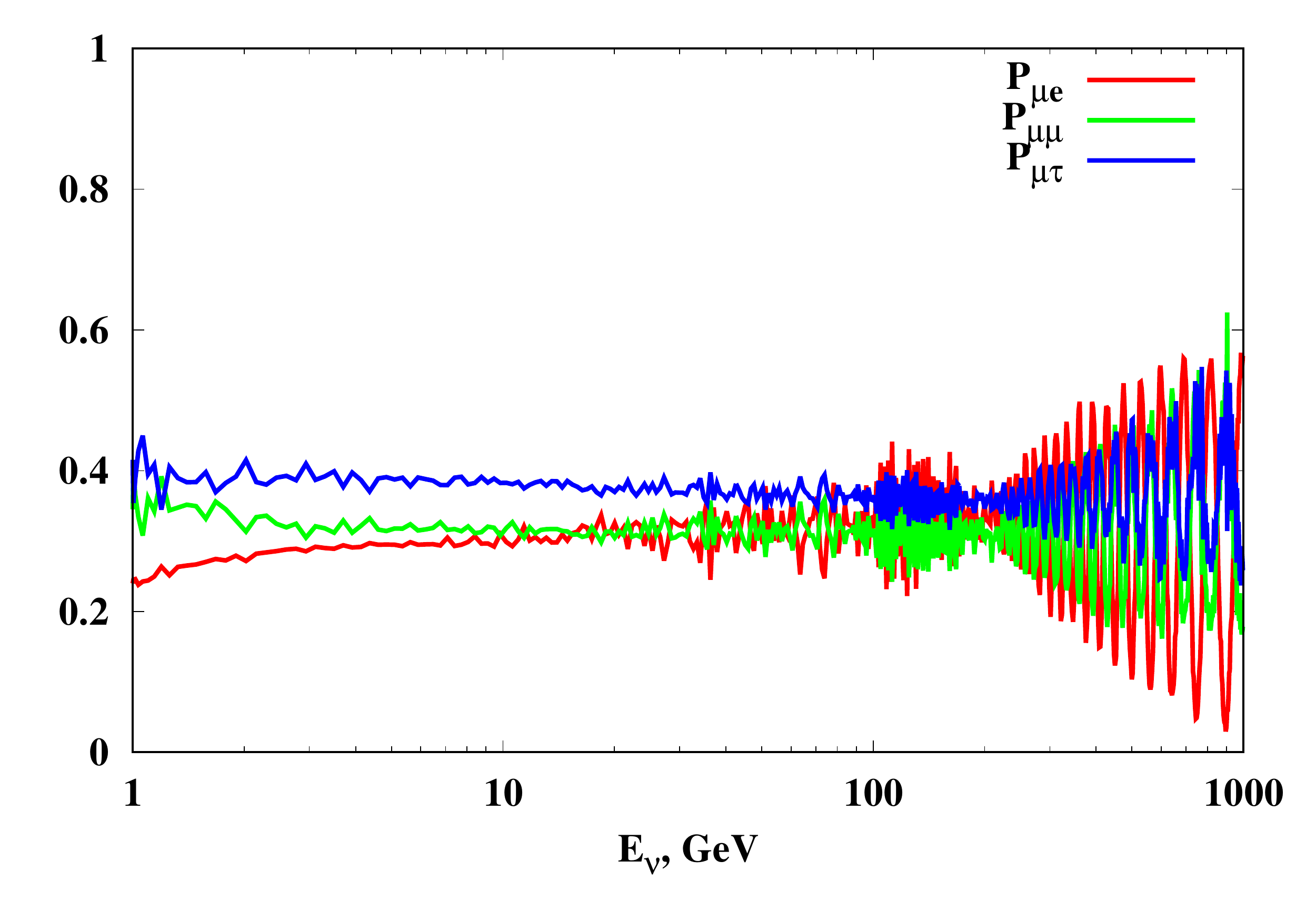}}
\put(30,250){\includegraphics[angle=-90,width=0.40\textwidth]{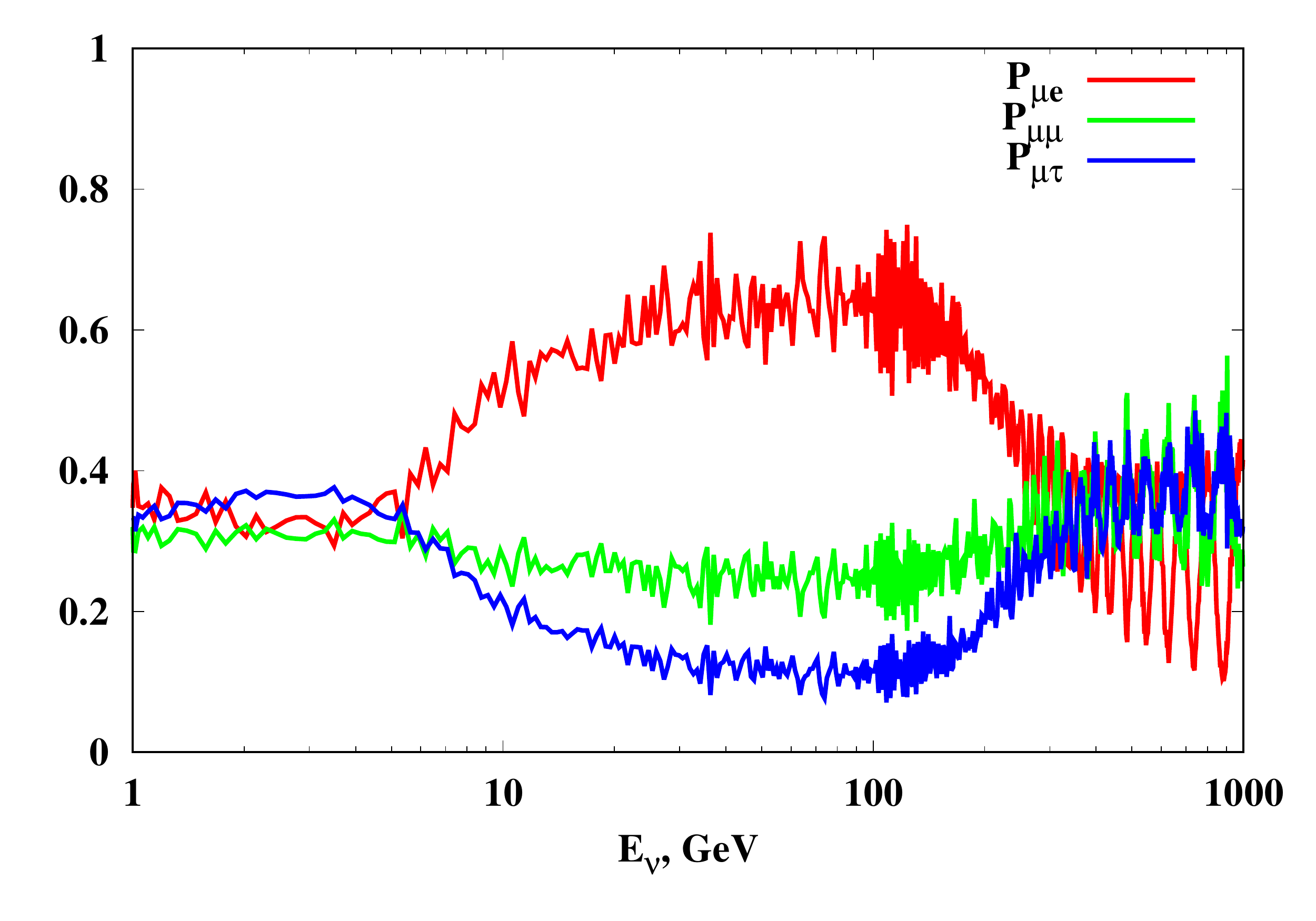}}
\end{picture}
\caption{\label{numu_etau} The same as in Fig.~\ref{numu_sm} but for $\e_{e\tau}=0.4$.}
\end{figure}

\begin{figure}[!htb]
\begin{picture}(300,220)(0,20)
\put(210,130){\includegraphics[angle=-90,width=0.40\textwidth]{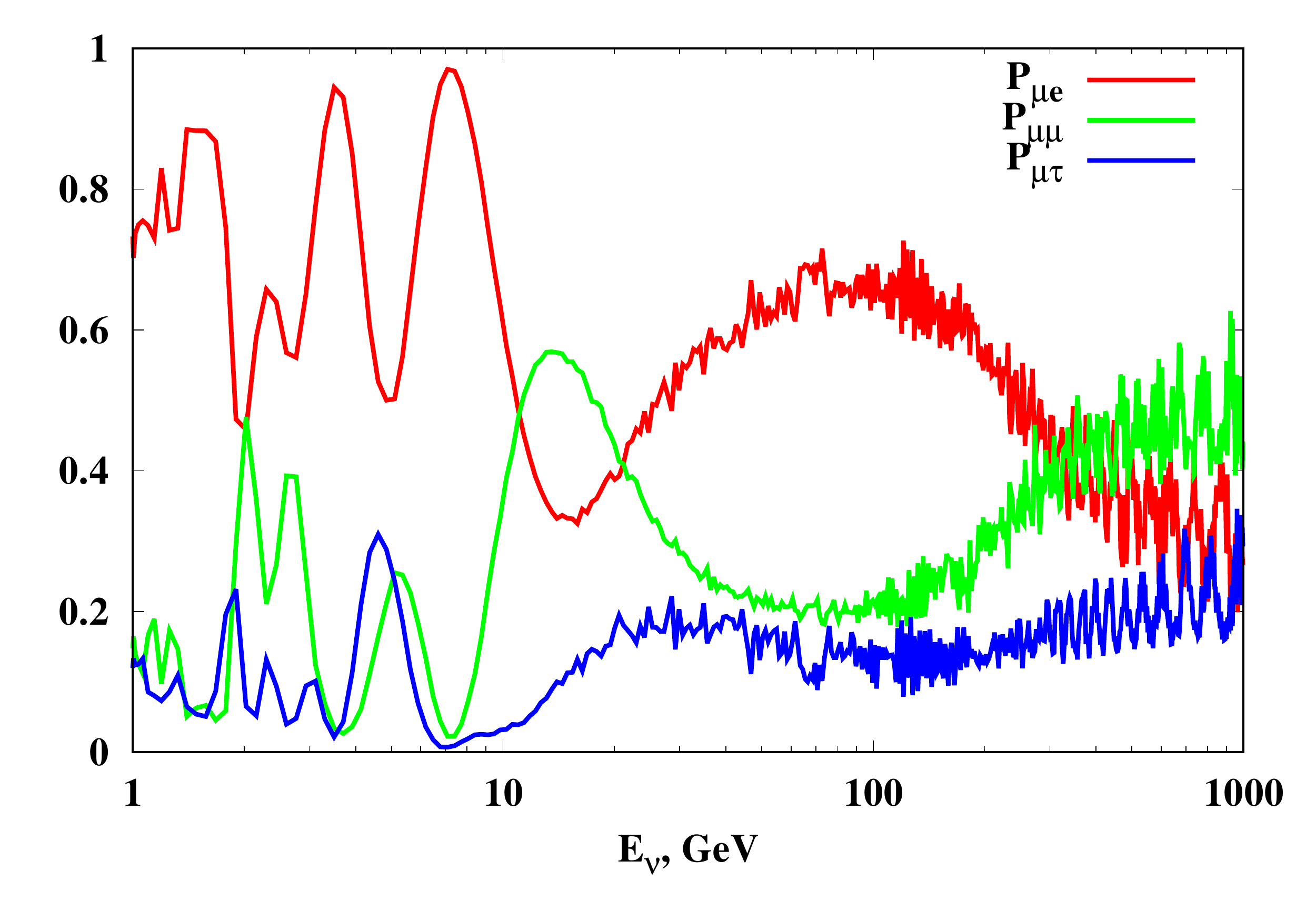}}
\put(210,250){\includegraphics[angle=-90,width=0.40\textwidth]{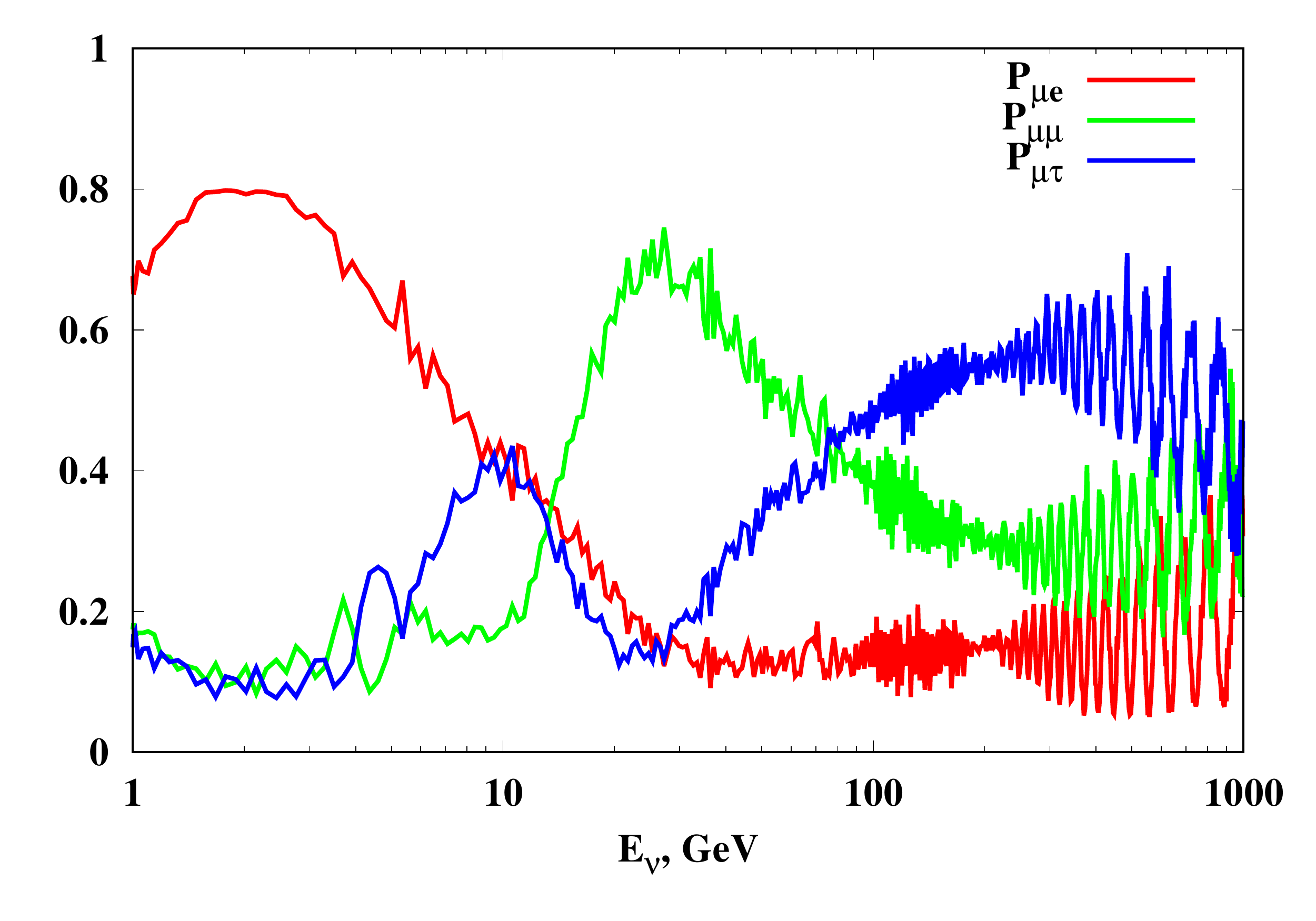}}
\put(30,130){\includegraphics[angle=-90,width=0.40\textwidth]{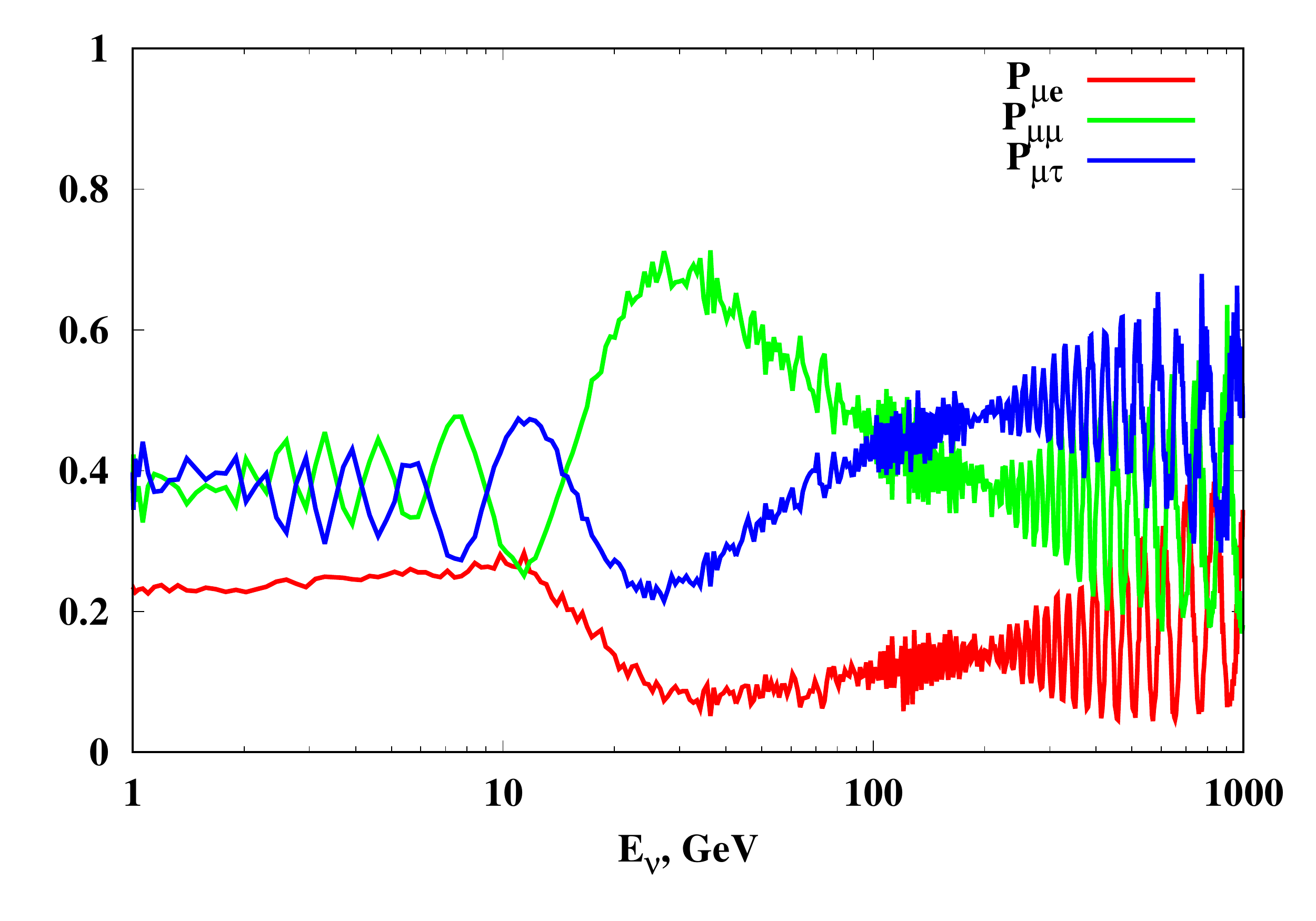}}
\put(30,250){\includegraphics[angle=-90,width=0.40\textwidth]{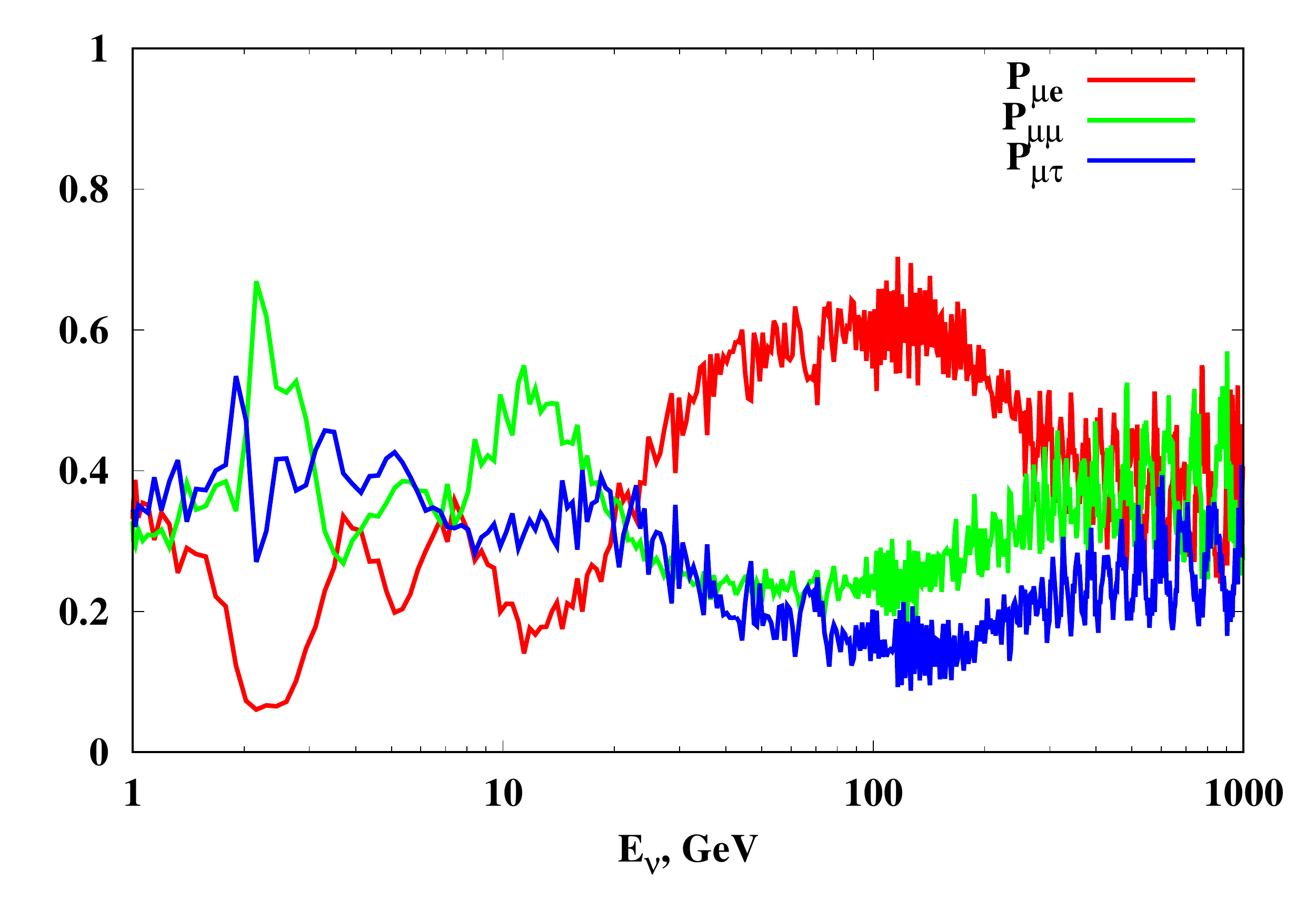}}
\end{picture}
\caption{\label{numu_etau_earth} The same as in Fig.~\ref{numu_sm_earth} but for $\e_{e\tau}=0.4$.}
\end{figure}

\begin{figure}[!htb]
\begin{picture}(300,250)(0,20)
\put(210,130){\includegraphics[angle=-90,width=0.40\textwidth]{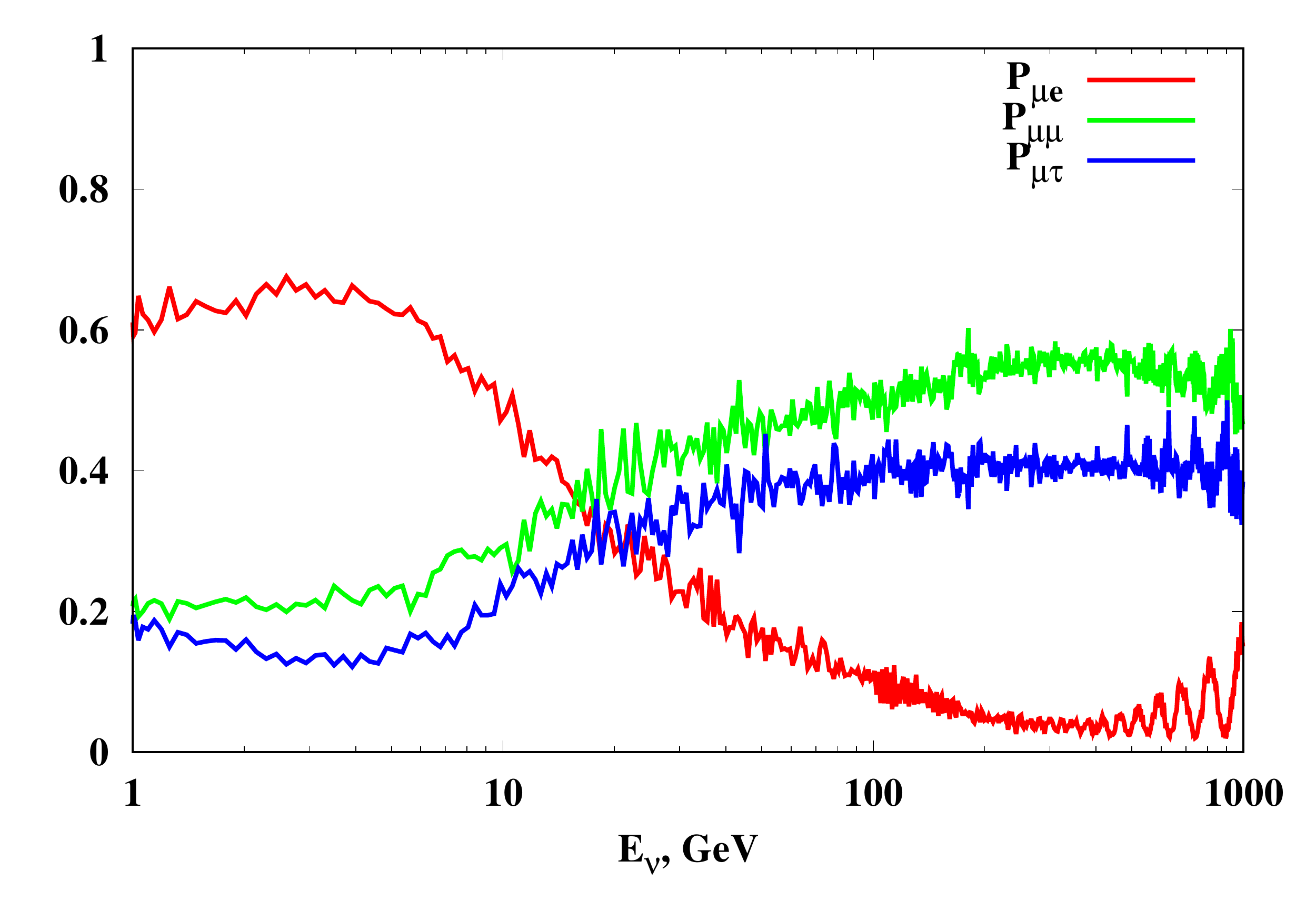}}
\put(210,250){\includegraphics[angle=-90,width=0.40\textwidth]{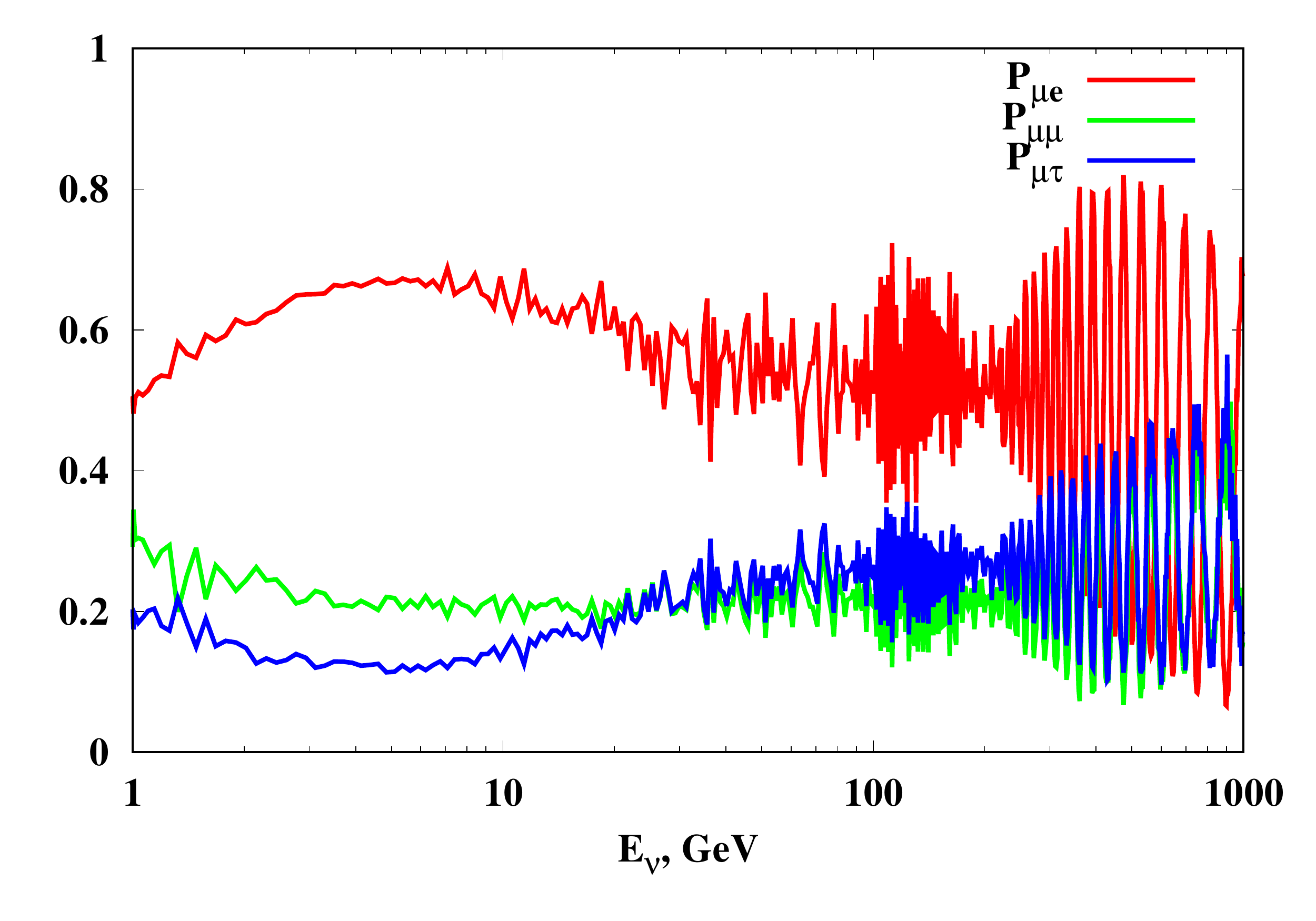}}
\put(30,130){\includegraphics[angle=-90,width=0.40\textwidth]{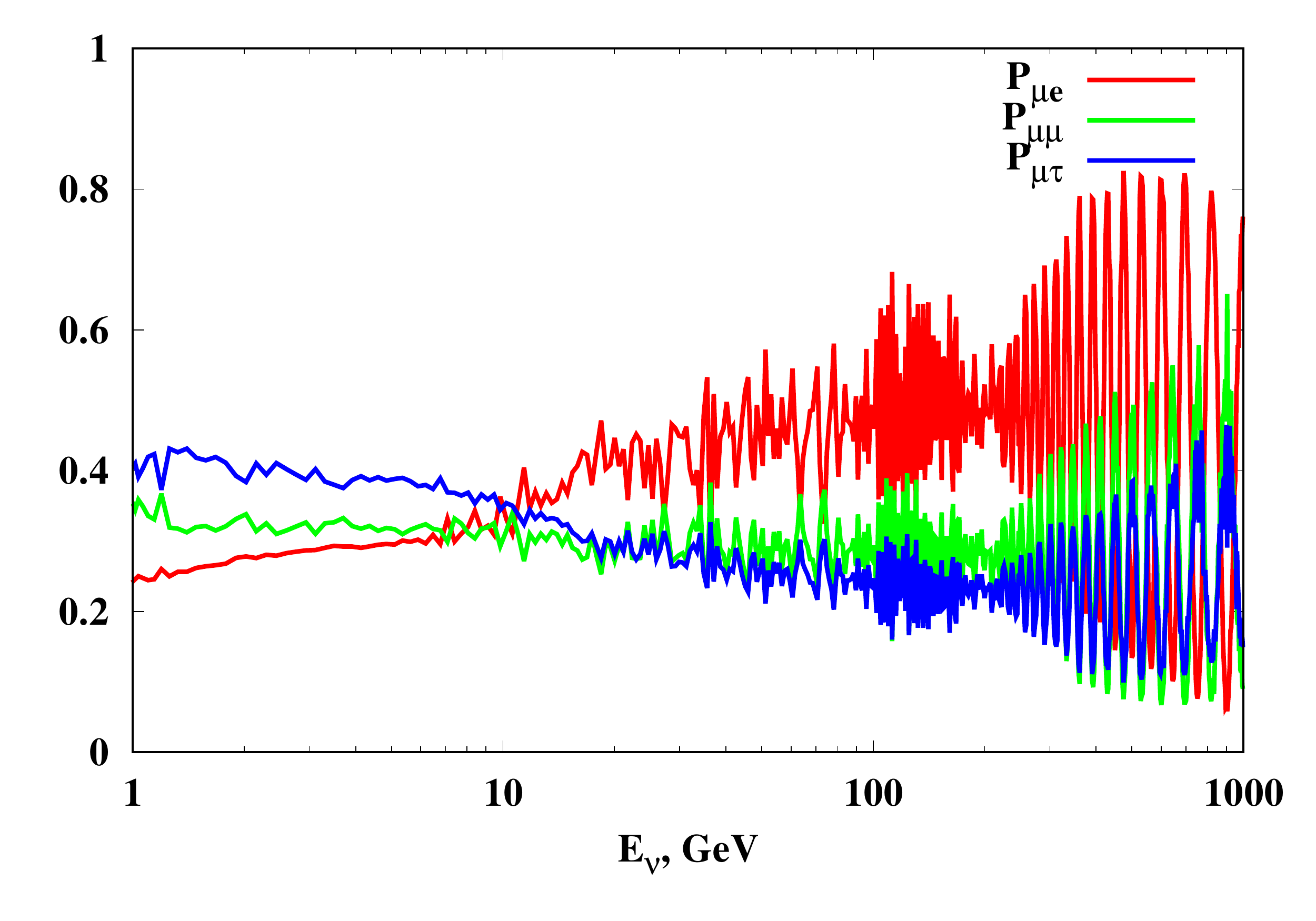}}
\put(30,250){\includegraphics[angle=-90,width=0.40\textwidth]{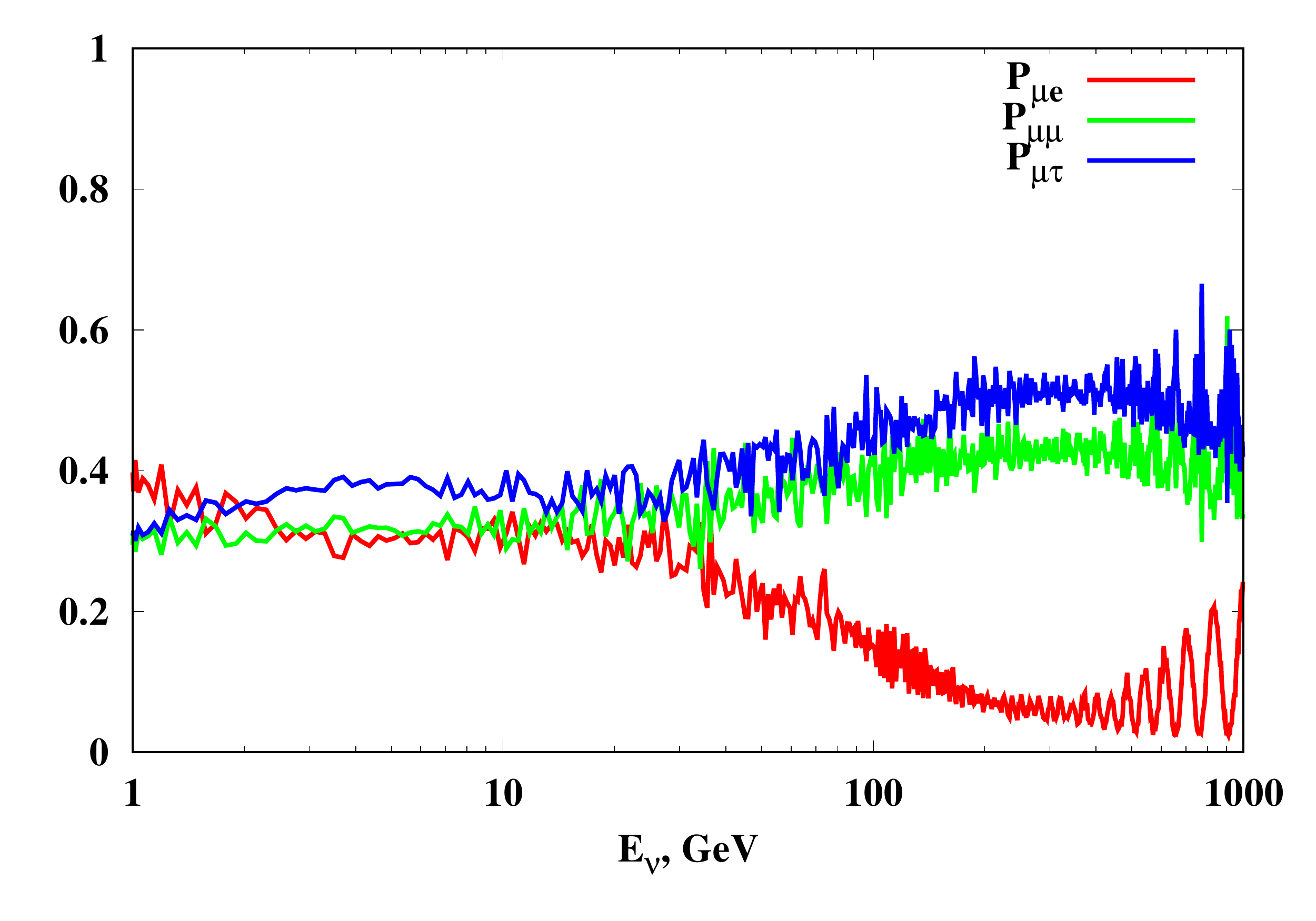}}
\end{picture}
\caption{\label{numu_etau_sign} The same as in Fig.~\ref{numu_sm} but for $\e_{e\tau}=-0.4$.}
\end{figure}

\begin{figure}[!htb]
\begin{picture}(300,250)(0,20)
\put(210,130){\includegraphics[angle=-90,width=0.40\textwidth]{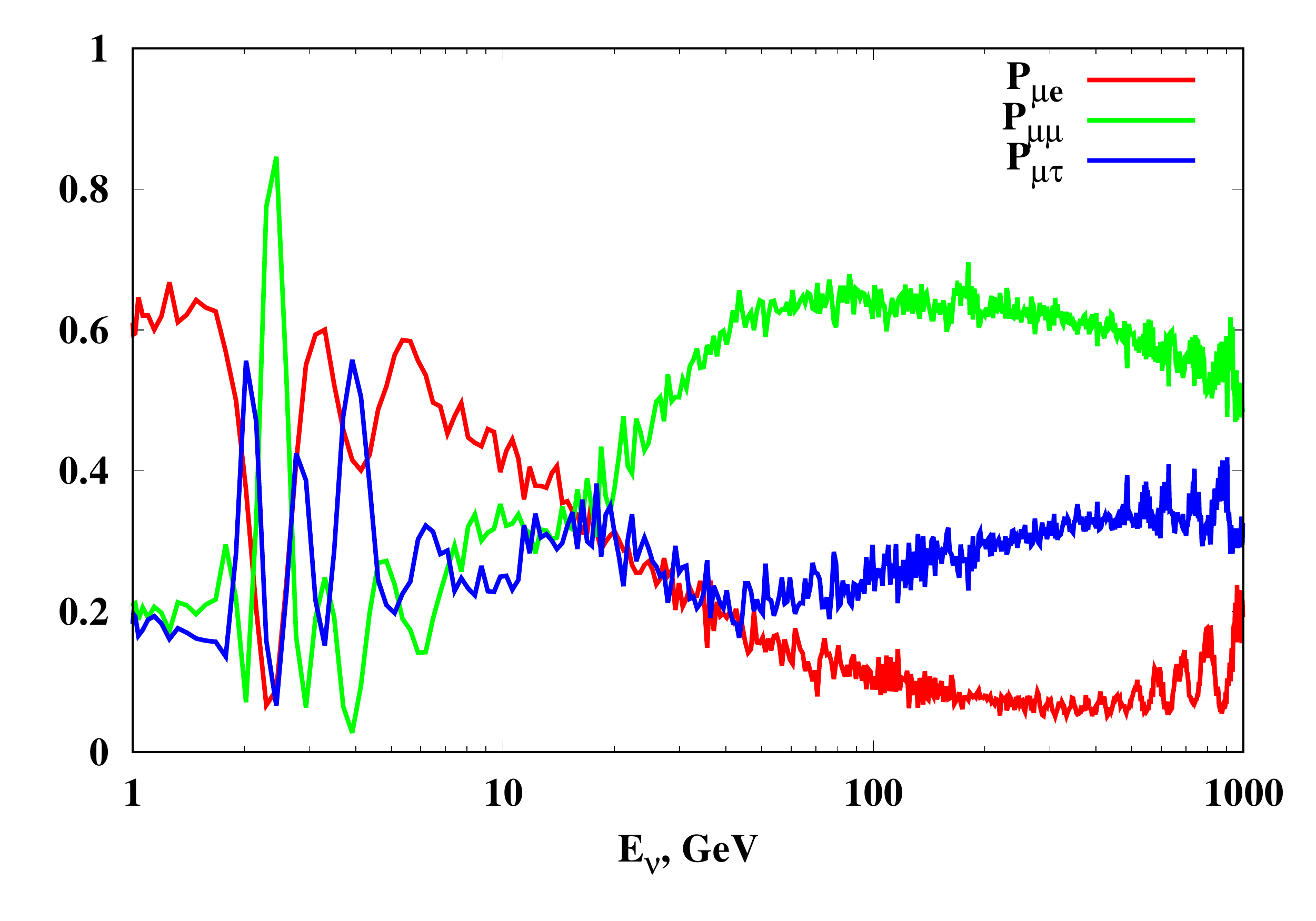}}
\put(210,250){\includegraphics[angle=-90,width=0.40\textwidth]{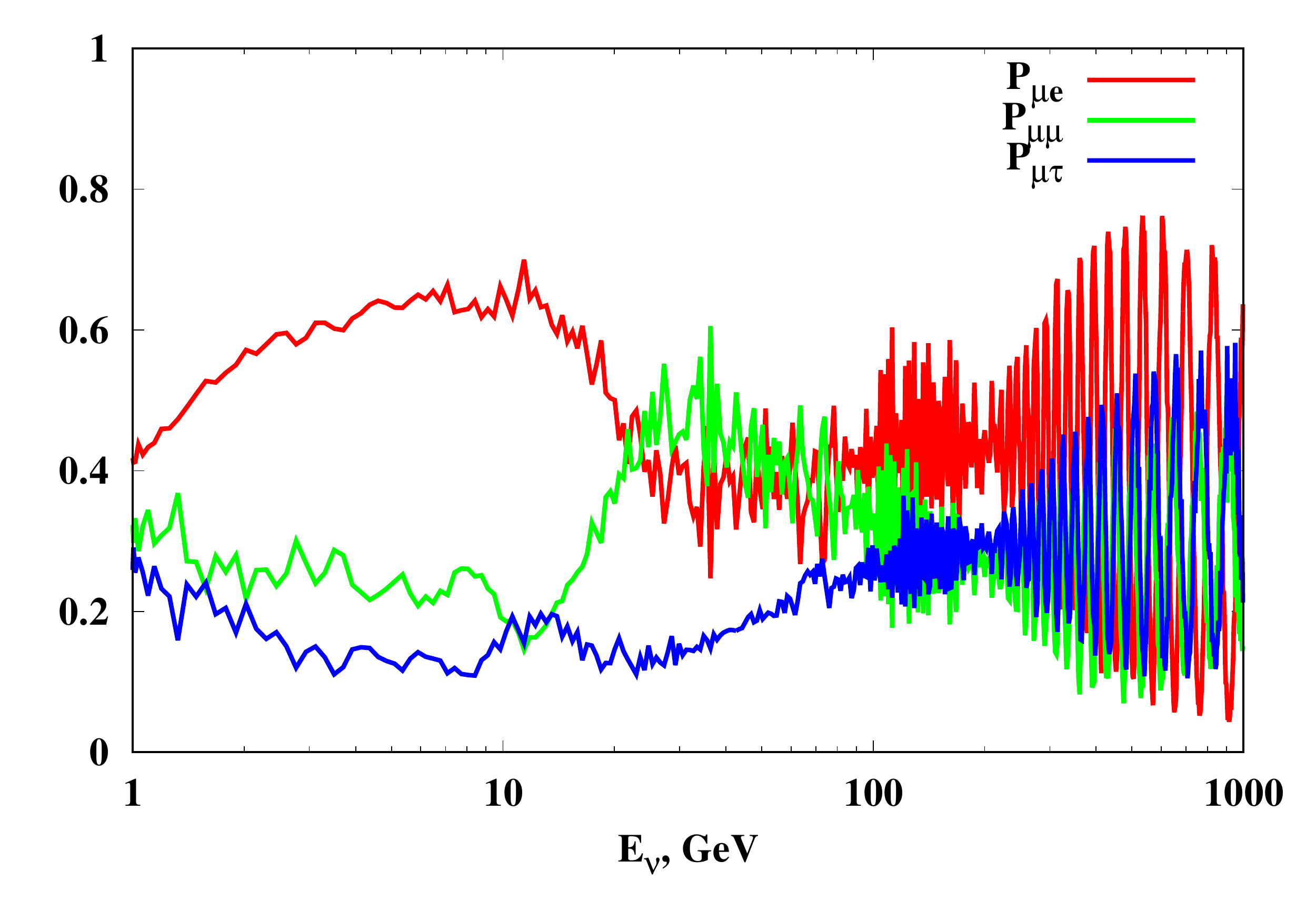}}
\put(30,130){\includegraphics[angle=-90,width=0.40\textwidth]{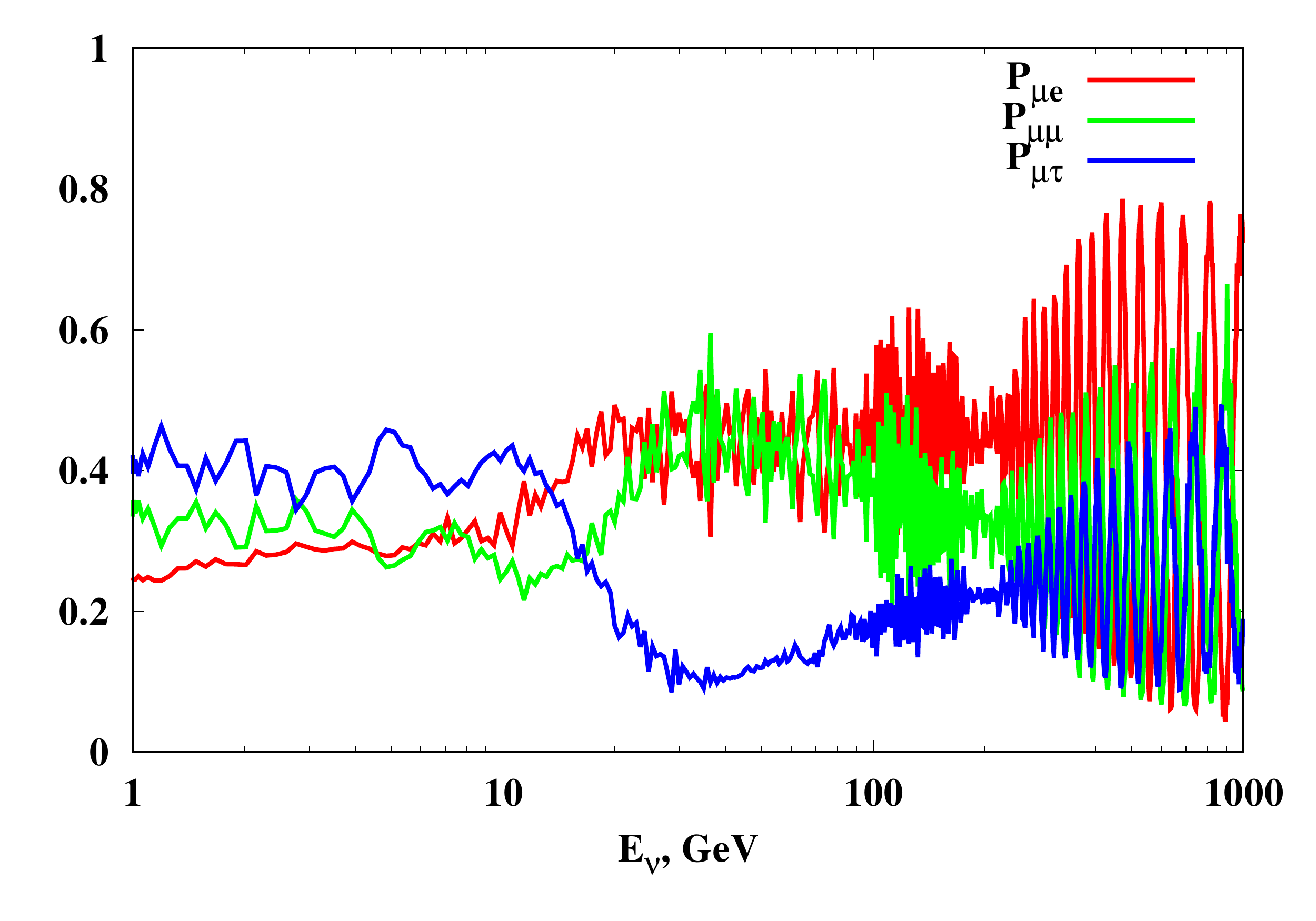}}
\put(30,250){\includegraphics[angle=-90,width=0.40\textwidth]{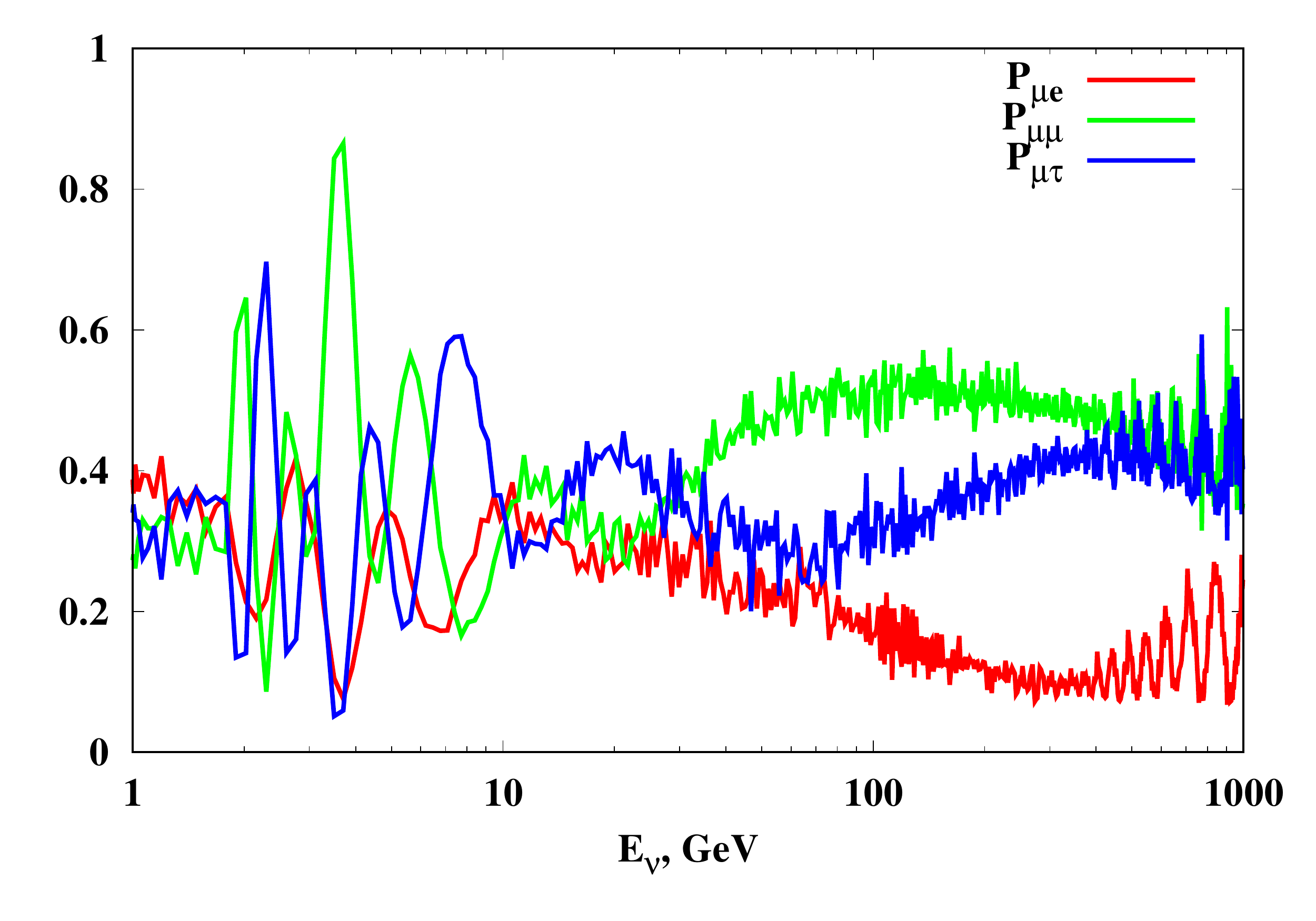}}
\end{picture}
\caption{\label{numu_etau_sign_earth} The same as in Fig.~\ref{numu_sm_earth} but for $\e_{e\tau}=-0.4$.}
\end{figure}

Now let us turn to discussion of impact of the flavor changing NSI 
parameters. We start with non-zero $\e_{e\tau}$.
Corresponding probabilities $P_{\mu\alpha}, \alpha=e,\mu,\tau$ for
$\e_{e\tau}=0.4$ chosen as a benchmark value are shown in Fig.~\ref{numu_etau} and
Fig.~\ref{numu_etau_earth} before and after propagation through the
Earth, respectively. 
Again here we assume the case of muon (anti)neutrino produced near the
center of the Sun.
Similar probabilities for $\e_{e\tau}=-0.4$ are shown in
Figs.~\ref{numu_etau_sign} and~\ref{numu_etau_sign_earth}.
Comparing plots in Figs.~\ref{numu_etau} and~\ref{numu_etau_sign} with
those in Fig.~\ref{numu_tautau_sign} for $\e_{\tau\tau}<0$ we find
similarities in the behaviour of $P_{\mu\alpha}$ as a function of
neutrino energy. The main difference as we will see shortly comes from
changes of effective mixing angles governing resonance transitions and
as a consequence from shifts of the onset energies for non-adiabatic
effects. In the following analytic expressions we limit ourselves to
small $\e_{e\tau}$ for simplicity which allows us to grasp the main
impact of the flavor changing NSI. Taking non-zero $\e_{e\tau}$
in~\eqref{eq:1:6} let us make rotation to a basis in which the matter
term in the Hamiltonian~\eqref{eq:1:3} is diagonal,
i.e. $\nu=R_{13}(\hat{\theta}_{13})\nu^\prime$, where  
\be
\label{eq:3.2:6}
\tan{2\hat{\theta}_{13}} = -2\e_{e\tau}.
\ee
The Hamiltonian in the basis has the form
\be
\label{eq:3.2:7}
H^\prime = \frac{1}{2E_\nu}U^\prime{\rm
  diag}(m_1^2,m_2^2,m_3^2)U^{\prime\dagger} + V_e
\left(
\begin{array}{ccc}
  \hat{c}_{13}^2 - 2\e_{e\tau}\hat{s}_{13}\hat{c}_{13} & 0 & 0\\
  0 & 0 & 0\\
  0 & 0 & \hat{s}_{13}^2 + 2\e_{e\tau}\hat{s}_{13}\hat{c}_{13}\\
\end{array}
\right),
\ee
where $U^\prime = R_{13}^\dagger(\hat{\theta}_{13})U$.
At small $\e_{e\tau}$ the matter term in~\eqref{eq:3.2:7} looks as
$V_e{\rm   diag}(1+\e_{e\tau}^2,0,-\e_{e\tau}^2)$. 
 Similar to the previous
case  the state $\nu^\prime_1$ decouples in the center of the 
Sun. The rest 2--3 subsystem is described by the following Hamiltonian
\be
\label{eq:3.2.8}
H_{23}^\prime = \frac{\Delta
  m_{32}^2}{2E_\nu}R_{23}(\theta^{e\tau}_{23})
\left(
\begin{array}{cc}
  0 & 0 \\
  0 & 1
\end{array}
\right)
R_{23}^\dagger(\theta_{23}^{e\tau})
- \e_{e\tau}^2V_e
\left(
\begin{array}{cc}
  0 & 0 \\
  0 & 1
\end{array}
\right),
\ee
where the mixing angle $\theta^{e\tau}_{23}$ which governs 2--3
transition can be found from
\be
\tan{\theta^{e\tau}_{23}} =
\frac{s_{23}c_{13}}{c_{23}c_{13}\hat{c}_{13} + s_{13}\hat{s}_{13}}
\ee
and for small $\e_{e\tau}$ the mixing in this sector is still close to
maximal. Here and below $\hat{c}_{ij}=\cos{\hat{\theta}_{ij}}$,
$\hat{s}_{ij}=\sin{\hat{\theta}_{ij}}$. 

Considering the vicinity of 1--3 resonance let us make rotation $\nu =
R_{23}(\theta_{23})\nu^\prime$ of the original flavor basic. The
Hamiltonian takes the form
\be
\label{eq:3.2.9}
H^\prime = \frac{1}{2E_\nu}R_{13}(\theta_{13})R_{12}(\theta_{12})
\left(
\begin{array}{ccc}
  -\Delta m_{21}^2 & & \\
  & 0 & \\
  & & \Delta m_{31}^2
\end{array}
\right)
R^\dagger_{12}(\theta_{12})R^\dagger_{13}(\theta_{13})
+ H_m^\prime,
\ee
where
\be
H_{m}^\prime = V_e\left(
\begin{array}{ccc}
  1 & -s_{23}\e_{e\tau} & c_{23}\e_{e\tau}\\
  -s_{23}\e_{e\tau} & 0 & 0 \\
  c_{23}\e_{e\tau} & 0 & 0
\end{array}
\right).
\ee
In the lowest $\e_{e\tau}$ approximation one can neglect non-zero 1--2 
element in Eq.~\eqref{eq:3.2.9} and the reduced Hamiltonian for 1--3
subsystem looks as
\be
\label{eq:3.2.10}
H_{13}^\prime = \frac{1}{2E_\nu}
R_{13}(\theta_{13})
\left(
\begin{array}{cc}
  0 & 0 \\
  0 & \Delta m_{31}^2 + \Delta m_{21}^2c_{12}^2
\end{array}
\right)
R^\dagger_{13}(\theta_{13})
+ V_e\left(
\begin{array}{cc}
  1 & c_{23}\e_{e\tau}\\
  c_{23}\e_{e\tau} & 0
\end{array}
\right).
\ee
The matter term in~\eqref{eq:3.2.10} can be diagonalized by
transformation 
$\nu^\prime = R_{13}(\hat{\theta}_{13})\tilde{\nu}$, where
$\tan{2\hat{\theta}_{13}} = -2c_{23}\e_{e\tau}$, and one obtains
\be
\tilde{H}_{13} =
\frac{1}{2E_\nu}R_{13}(\theta_{13}^{e\tau})
\left(
\begin{array}{cc}
  0 & 0 \\
  0 & \Delta m_{31}^2 + \Delta m_{21}^2c_{12}^2
\end{array}
\right)R_{13}^\dagger(\theta_{13}^{e\tau}) + V_e\left(
\begin{array}{cc}
  1+c_{23}^2\e_{e\tau}^2 & 0\\
  0 & -c_{23}^2\e_{e\tau}^2
\end{array}
\right).
\ee
Here the mixing angle for 1--3 transition is
$\theta_{13}^{e\tau} = \theta_{13}-\hat{\theta}_{13}\approx
\theta_{13} + c_{23}\e_{e\tau}$, where last equality is valid for
small $\e_{e\tau}$. 

Finally going in the vicinity of 1--2 resonance we make rotation
$\nu=R_{13}^\dagger R_{23}^\dagger\nu^{\prime}$ and obtain
\be
\label{eq:3.2.11}
H^{\prime} = \frac{1}{2E_\nu}R_{12}(\theta_{12})
\left(
\begin{array}{ccc}
  0 & & \\
  & \Delta m_{21}^2 & \\
  & & \Delta m_{31}^2
\end{array}
\right)R^\dagger_{12}(\theta_{12})
+ H_m^{\prime},
\ee
where
\be
H_m^{\prime} = V_e\left(
\begin{array}{ccc}
  c_{13}^2 - 2\e_{e\tau}s_{13}c_{13}c_{23} & ... & ... \\
  -\e_{e\tau}s_{23}c_{13} & 0 & ... \\
  s_{13}c_{13}+(c_{13}^2-s_{13}^2)c_{23}\e_{e\tau} & 0 & s_{13}^2 +
  2\e_{e\tau}s_{13}c_{13}c_{23}
\end{array}\right).
\ee
Taking the limit $\bigg|\frac{\Delta m_{31}^2}{2E_\nu}\bigg|\gg V_e$
and neglecting 1--3 element we obtain (cf. Eqs.~(2.12)--(2.14) in
Ref.~\cite{Gonzalez-Garcia:2013usa}) 
\be
\label{eq:3.2.12}
H_{12}^{\prime} = \frac{1}{2E_\nu}R_{12}(\theta_{12})
\left(
\begin{array}{cc}
  0 & \\
  & \Delta m_{21}^2
\end{array}
\right)R^\dagger_{12} + V_e\left(
\begin{array}{cc}
  c_{13}^2 - 2\e_{e\tau} s_{13}c_{13}c_{23} & -\e_{e\tau}c_{13}s_{23}\\
  -\e_{e\tau}c_{13}s_{23} & 0
\end{array}
\right).
\ee
The matter term can be diagonalized by rotation $\nu^{\prime} =
R_{12}(\hat{\theta}_{12})\tilde{\nu}$, where
$\tan{2\hat{\theta}_{12}} = \frac{2\e_{e\tau}c_{13}s_{23}}{c_{13}^2
  -2\e_{e\tau}s_{13}c_{13}c_{23}}$. The modified mixing angle governing
1--2 transition can be found as $\theta_{12}^{e\tau} =
\theta_{12}-\hat{\theta}_{12} \approx \theta_{12} -
s_{23}\e_{e\tau}$, where last equality we neglect small contribution
from non-zero $\theta_{13}$.

Let us consider the case $\e_{e\tau}=0.4$. With $\theta^{e\tau}_{23}$
close to $\frac{\pi}{4}$ the 2--3 transition remains adiabatic almost
entirely for
 1--1000~GeV neutrino energy range. The resonance
energy in the solar center is determined by Eq.~\eqref{eq:3.1:3} with
replacement $\e_{\tau\tau}\to-\e_{e\tau}^2$. 
From
Eq.~\eqref{eq:3.2:7} we see that if the matter term dominates (i.e. in the
center of the Sun) the muon neutrino is the second energy level of the
matter Hamiltonian. For normal mass ordering muon (anti)neutrino evolves into
$|2\rangle$ state in the adiabatic regime, see left panels in Fig.~\ref{numu_etau}. For neutrino mode the
answer gets modified at $E_\nu\gsim 10$~GeV where transition trough
1--2 resonance becomes non-adiabatic. In the limit of maximal
adiabaticity violation we obtain the following mixed state
$\cos^2{\theta^{e\tau}_{12}}|1\rangle\langle 1| +
\sin^2{\theta^{e\tau}_{12}}| 2\rangle\langle 2|$. For our choice
of parameters we have $\sin^2{\theta^{e\tau}_{12}}\approx 0.08-0.1$
and thus almost pure $|1\rangle$ state emerge from the Sun. In the
case of inverted mass hierarchy muon (anti)neutrino evolves in the
adiabatic regime into
$|1\rangle$ at the Earth orbit. For antineutrino the evolution
includes transition through 1--3 resonance which is adiabatic up-to
about 100~GeV due to an increase of $\theta^{e\tau}_{13}$ as compared
to $\theta_{13}$; namely $\sin^2{\theta^{e\tau}_{13}}\approx 0.14-0.16$. For neutrino mode
(IH) with $E_\nu\gsim 10$~GeV transition through 1--2 resonance
becomes non-adiabatic and at very high energies the resulting state is
described as $\cos^2{\theta^{e\tau}_{12}}|2\rangle\langle 2| +
\sin^2{\theta^{e\tau}_{12}}| 1\rangle\langle 1|$. 

Turning to the case of negative $\e_{e\tau}=-0.4$ we see the same
adiabatic evolution for $E_\nu\lsim 10$~GeV in
Fig.~\ref{numu_etau_sign} as in the case $\e_{e\tau}=0.4$. However, at
higher energies the results are quite different. In particular, 
one finds that the onset energy for non-adiabatic
effects for 1--3 transition is around 10~GeV due to smallness of
$\sin^2{\theta^{e\tau}_{13}}\approx 0.01$. In
the limit of maximal adiabaticity violation the evolution of neutrino
(NH) and antineutrino (IH)
through 1--3 resonance  results in formation of the mixed state
$\sin^2{\theta^{e\tau}_{13}}|1\rangle\langle 1| +
\cos^2{\theta^{e\tau}_{13}} | 3\rangle\langle 3|\approx
|3\rangle\langle 3|$. In the case of neutrino (IH) at very high
energies one finds again the mixed state
$\cos{\theta^{e\tau}_{12}}|2\rangle\langle 2| +
\sin^2{\theta^{e\tau}_{12}}| 1\rangle\langle 1|$. But contrary
to the  case of positive $\e_{e\tau}$ here
$\sin^2{\theta^{e\tau}_{12}}\approx 0.6$ is considerably larger. Subsequent evolution through
the Earth is shown in 
Figs.~\ref{numu_tautau_earth} and~\ref{numu_tautau_sign_earth}. One
can see that the propagation through the Earth produces the largest
effect for neutrino of low 
($E_\nu\lsim 10$~GeV) and intermediate ($10$~GeV$\lsim E_\nu\lsim
100$~GeV) energies. The resulting probabilities $P_{\mu\alpha}$ have
quite complicated energy dependence in the low energy region. At
higher energies the evolution is more smooth and can be traced
qualitatively similar to the case discussed in the previous Section. 

\begin{figure}[t]
\begin{picture}(300,220)(0,20)
\put(210,130){\includegraphics[angle=-90,width=0.40\textwidth]{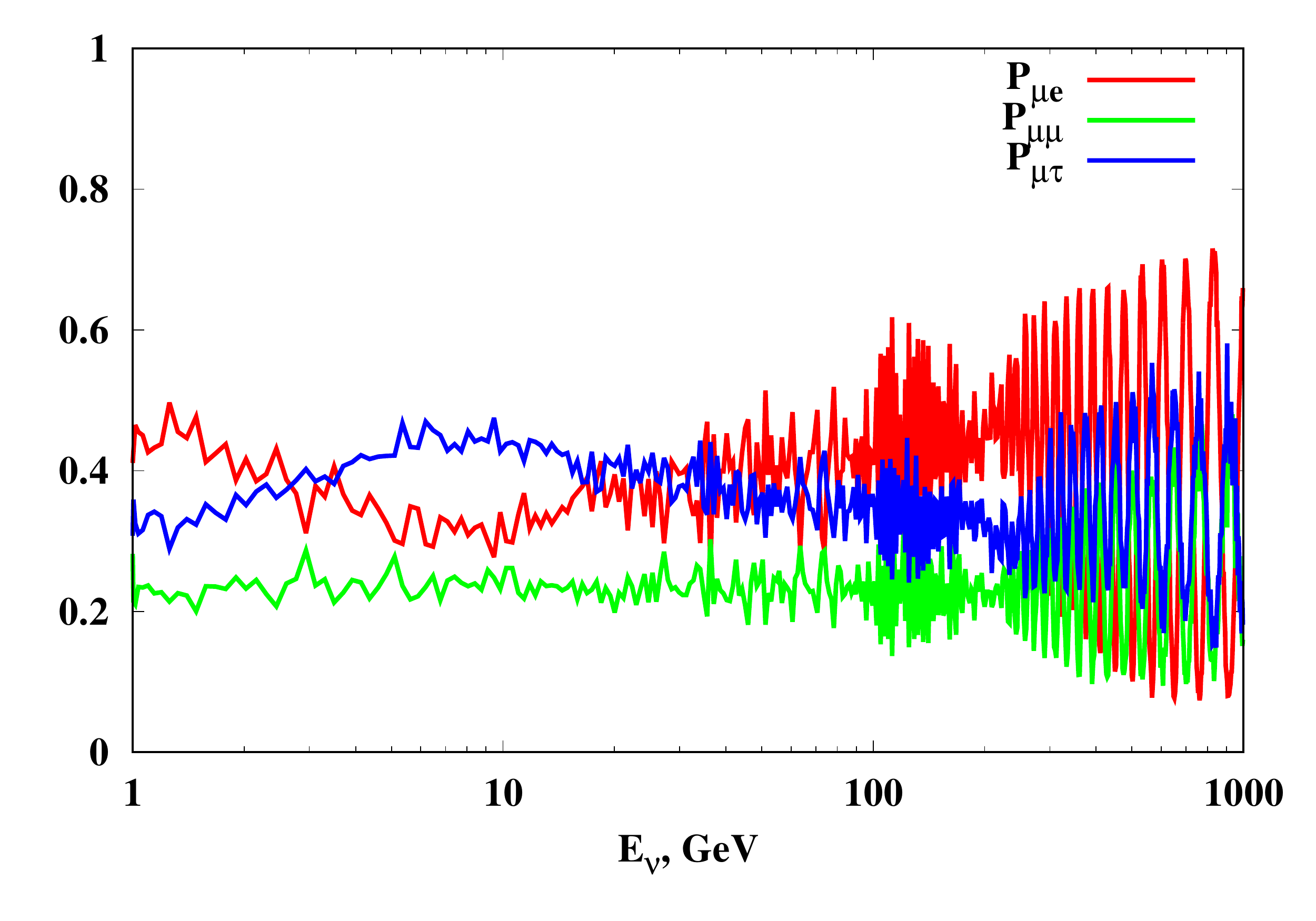}}
\put(210,250){\includegraphics[angle=-90,width=0.40\textwidth]{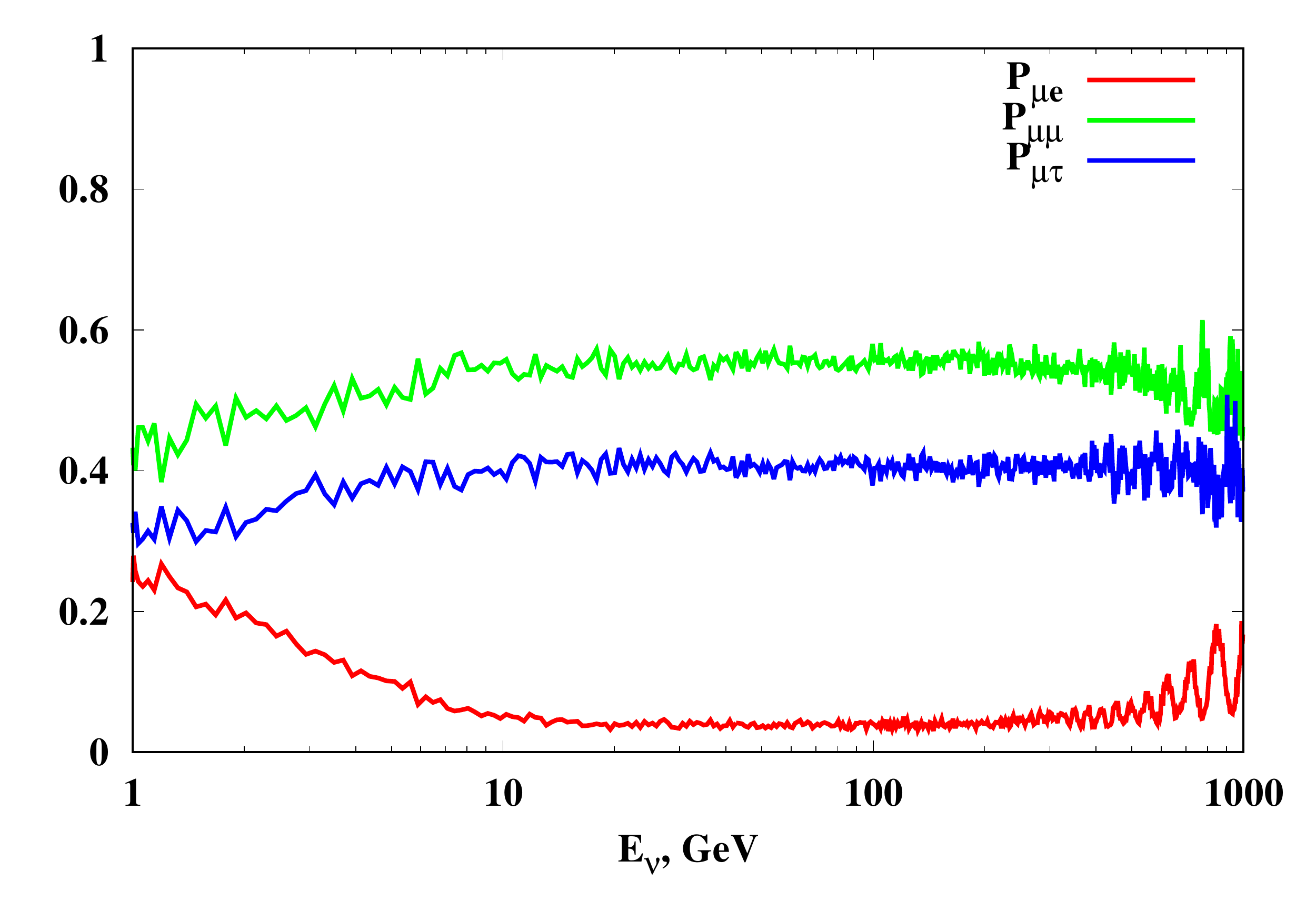}}
\put(30,130){\includegraphics[angle=-90,width=0.40\textwidth]{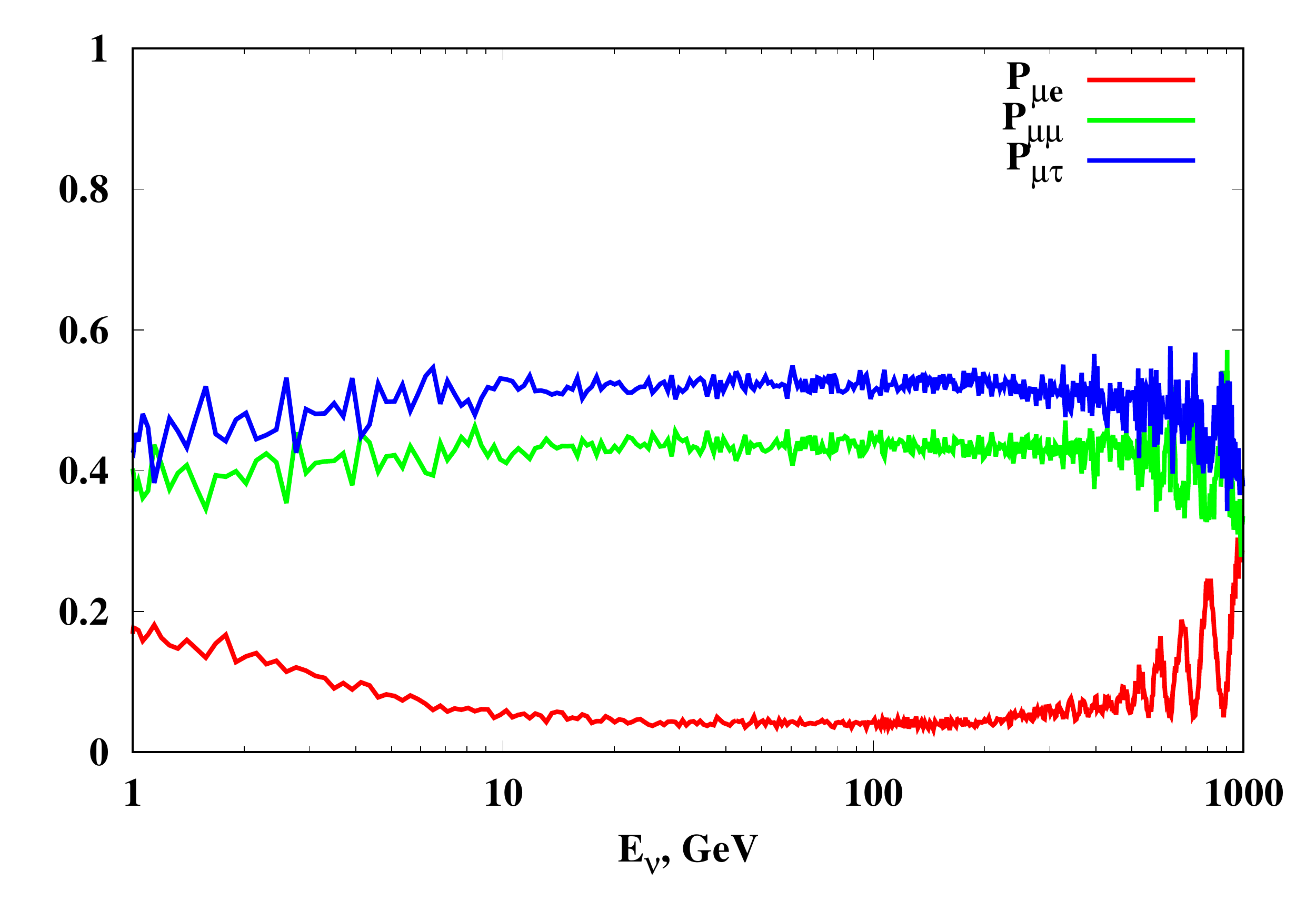}}
\put(30,250){\includegraphics[angle=-90,width=0.40\textwidth]{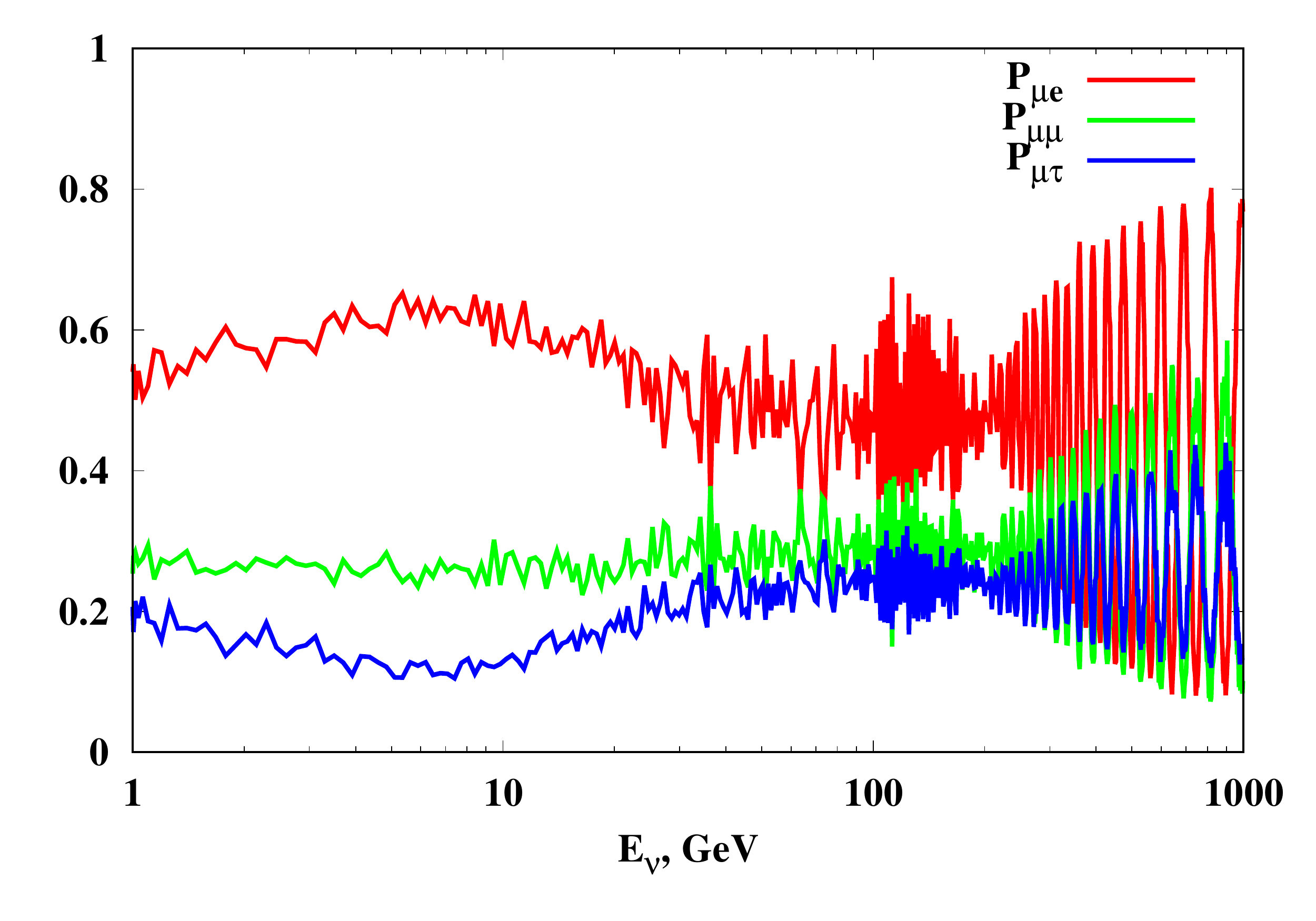}}
\end{picture}
\caption{\label{numu_emu} The same as in Fig.~\ref{numu_sm} but for $\e_{e\mu}=0.2$.}
\end{figure}

\begin{figure}[hb]
\begin{picture}(300,220)(0,20)
\put(210,130){\includegraphics[angle=-90,width=0.40\textwidth]{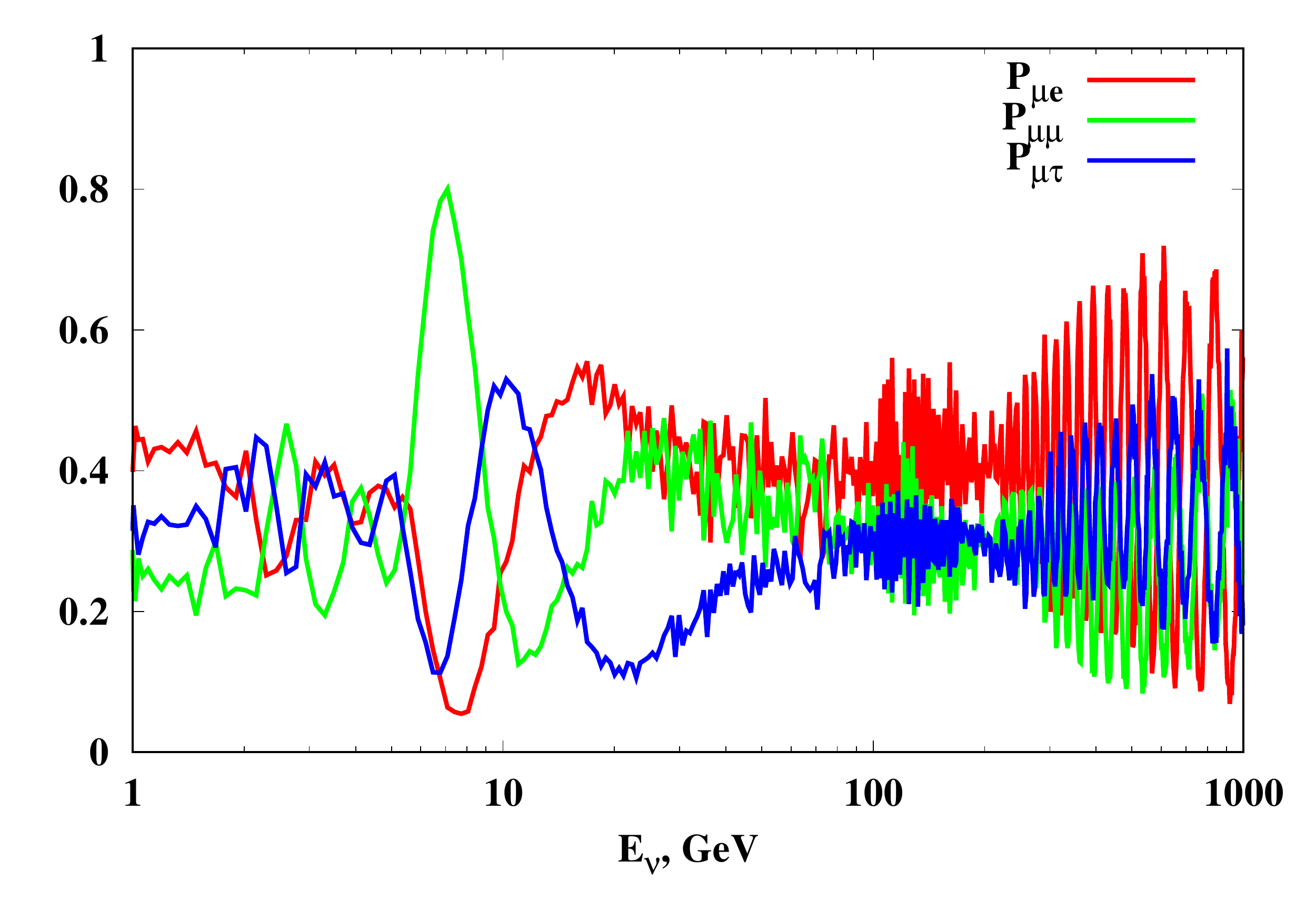}}
\put(210,250){\includegraphics[angle=-90,width=0.40\textwidth]{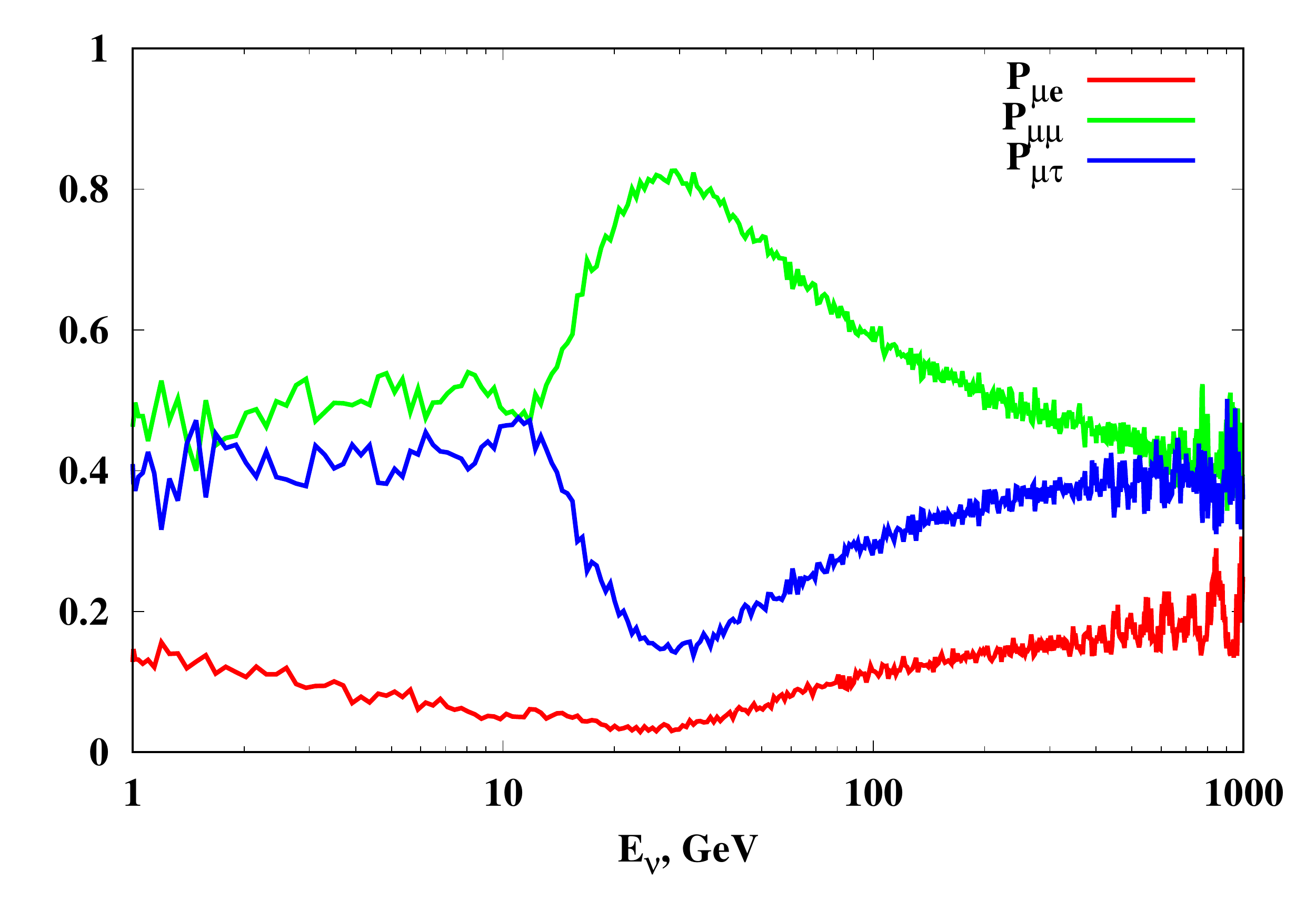}}
\put(30,130){\includegraphics[angle=-90,width=0.40\textwidth]{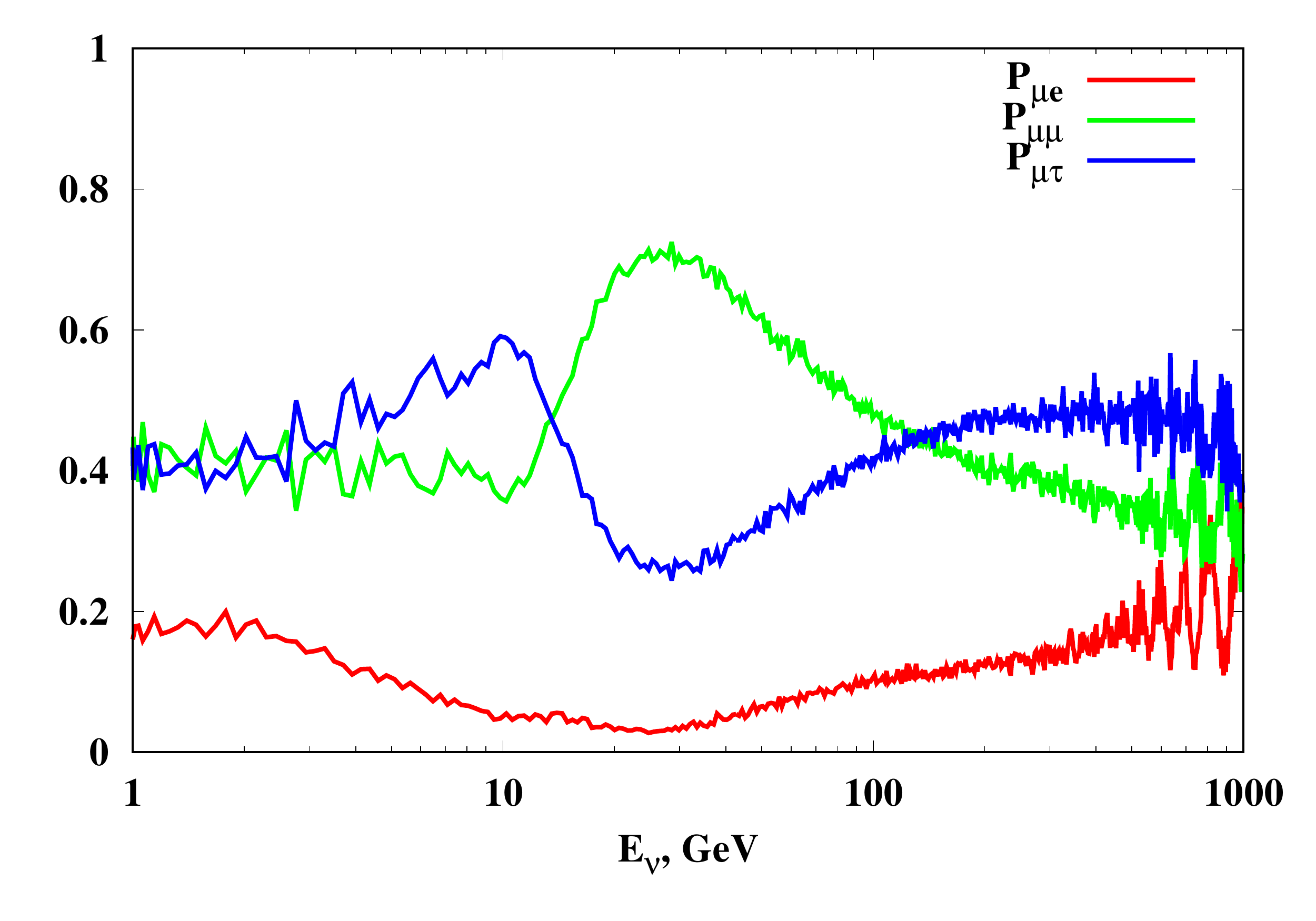}}
\put(30,250){\includegraphics[angle=-90,width=0.40\textwidth]{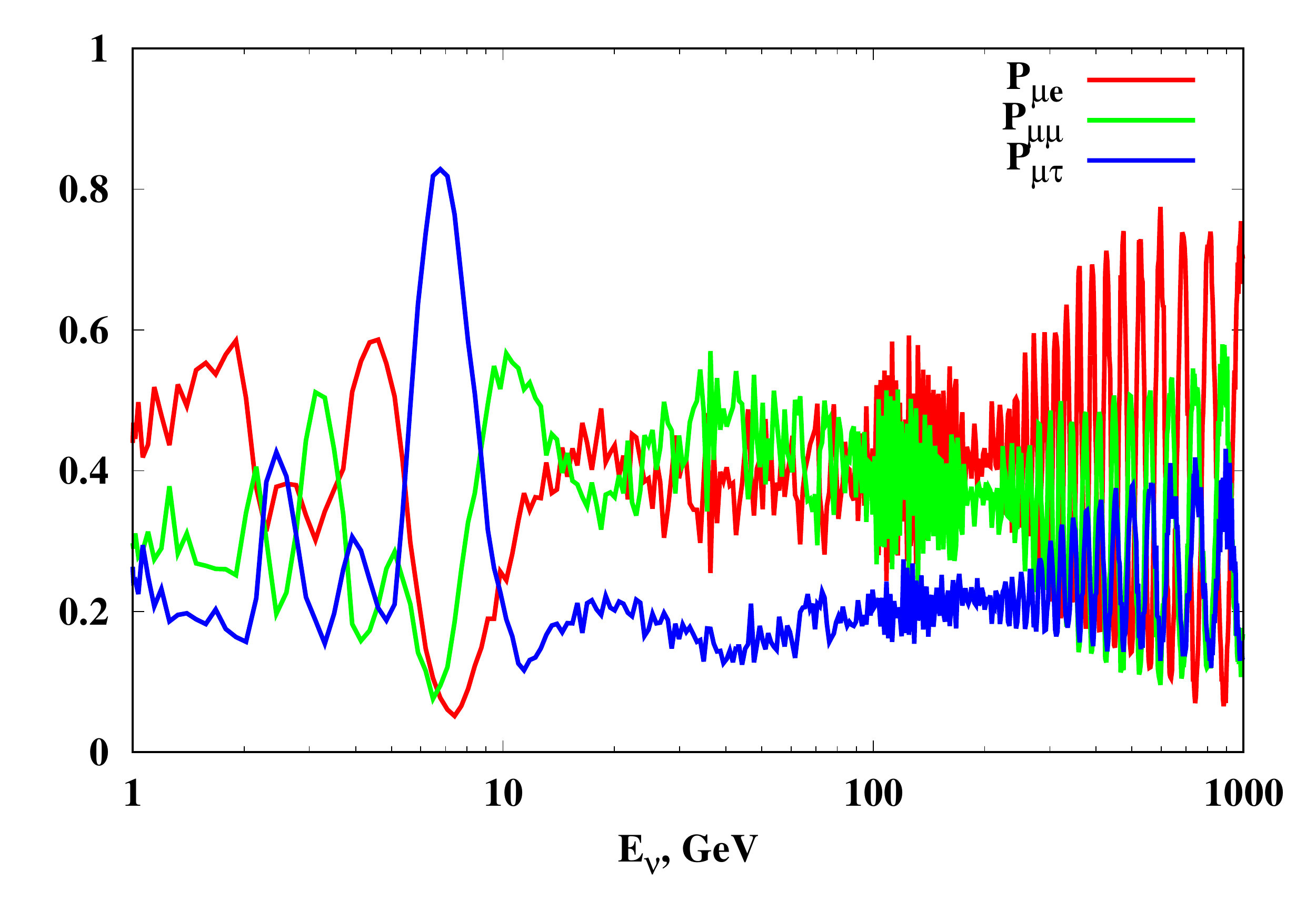}}
\end{picture}
\caption{\label{numu_emu_earth} The same as in Fig.~\ref{numu_sm_earth} but for $\e_{e\mu}=0.2$.}
\end{figure}

\begin{figure}[t]
\begin{picture}(300,220)(0,20)
\put(210,130){\includegraphics[angle=-90,width=0.40\textwidth]{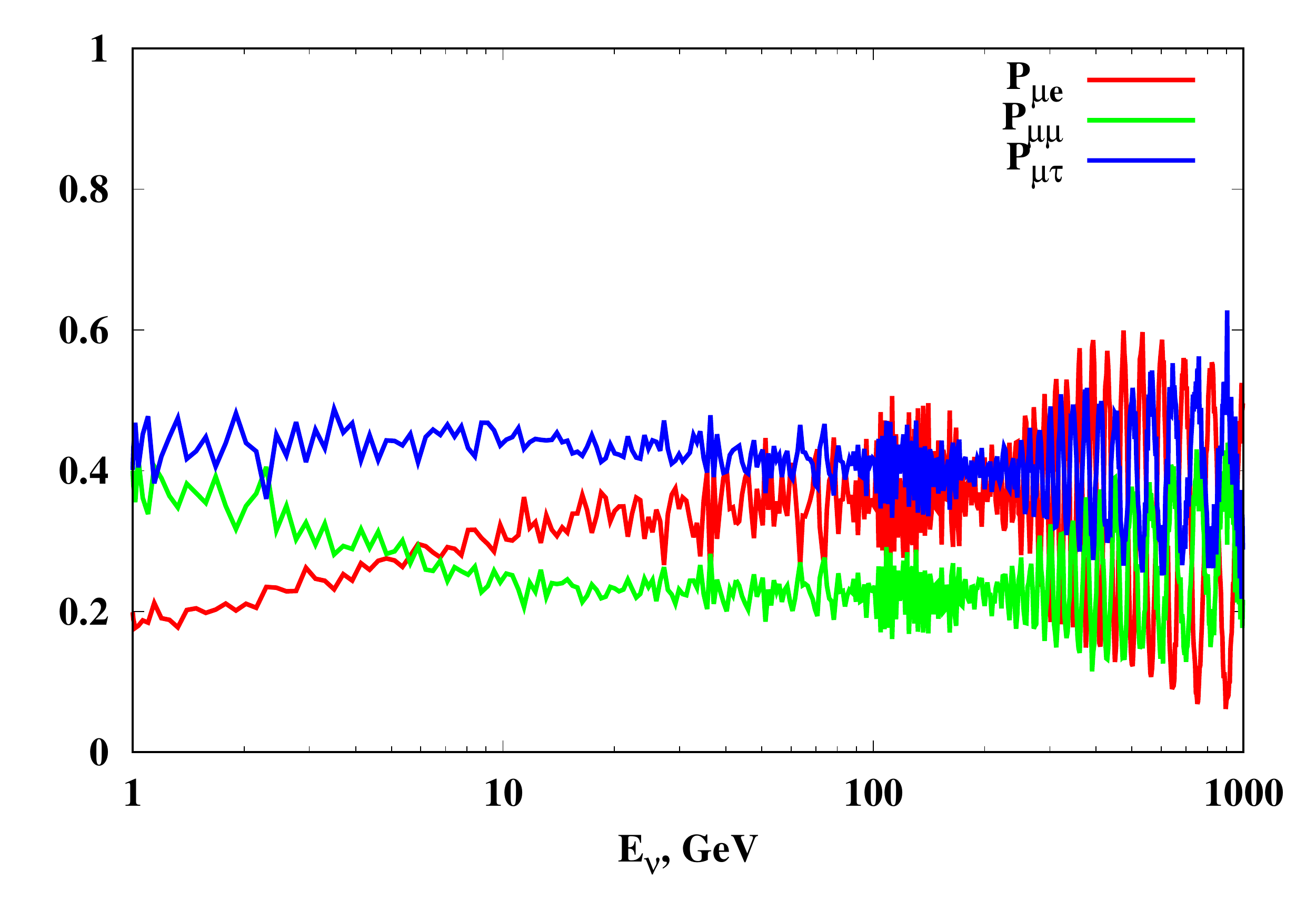}}
\put(210,250){\includegraphics[angle=-90,width=0.40\textwidth]{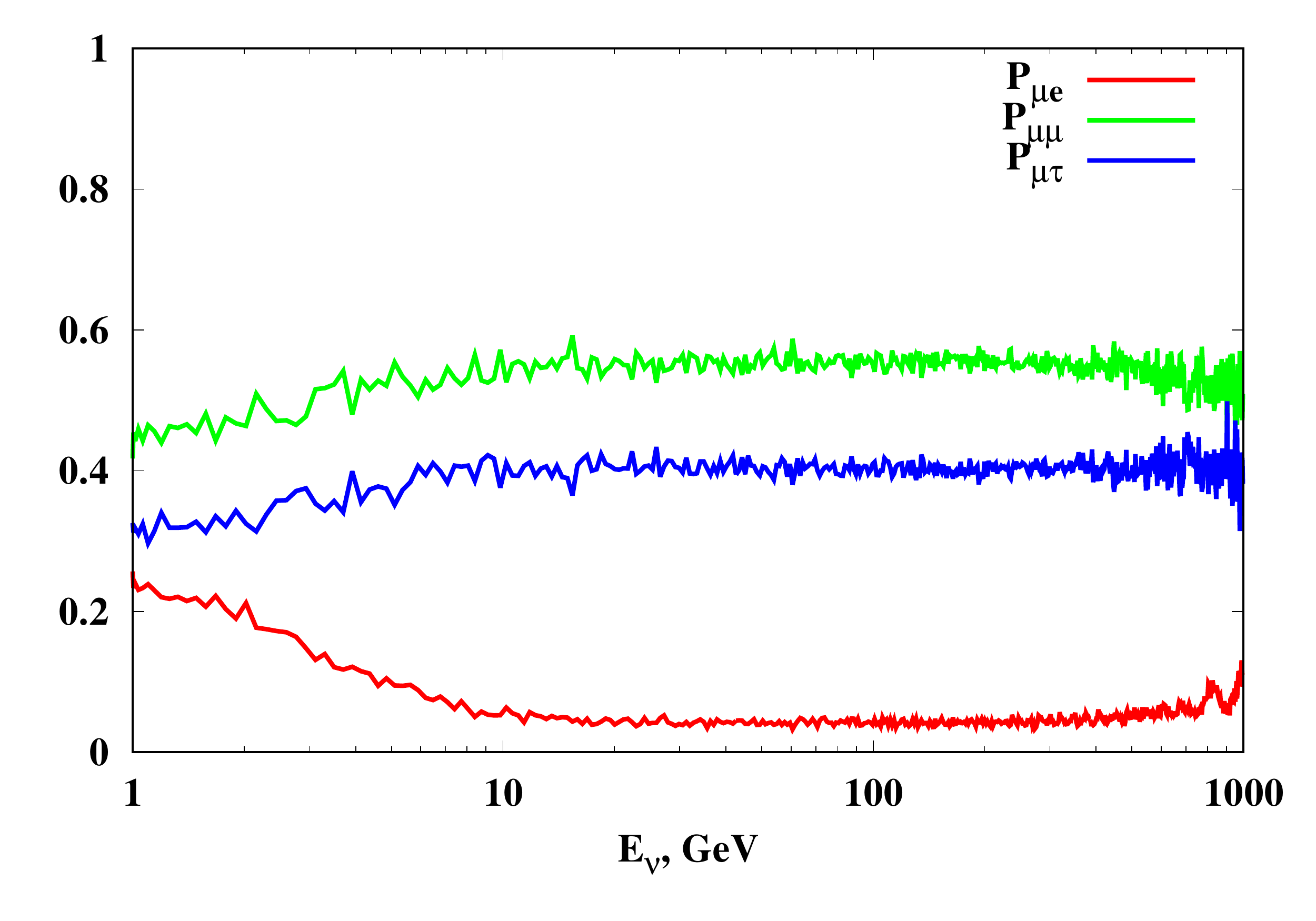}}
\put(30,130){\includegraphics[angle=-90,width=0.40\textwidth]{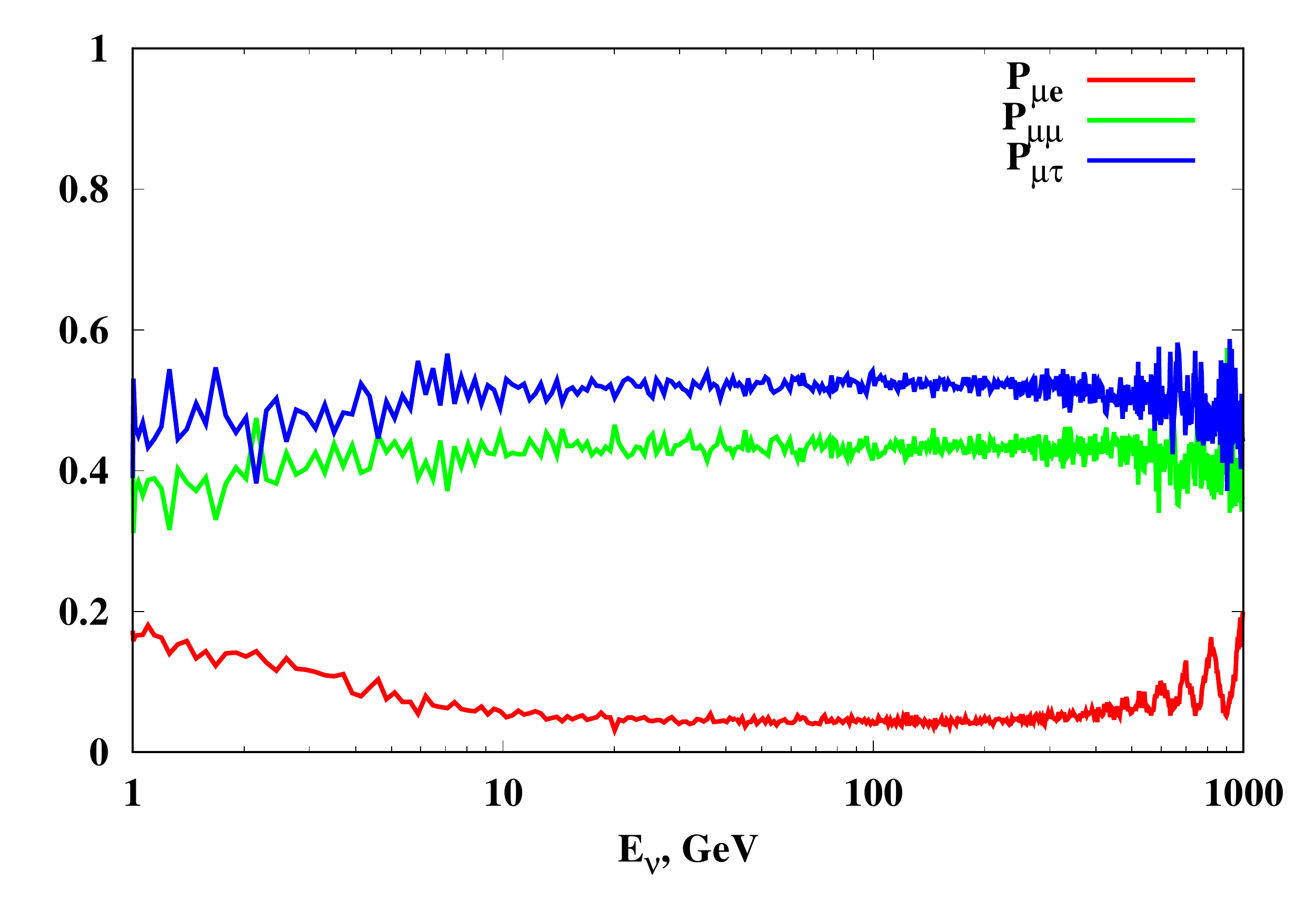}}
\put(30,250){\includegraphics[angle=-90,width=0.40\textwidth]{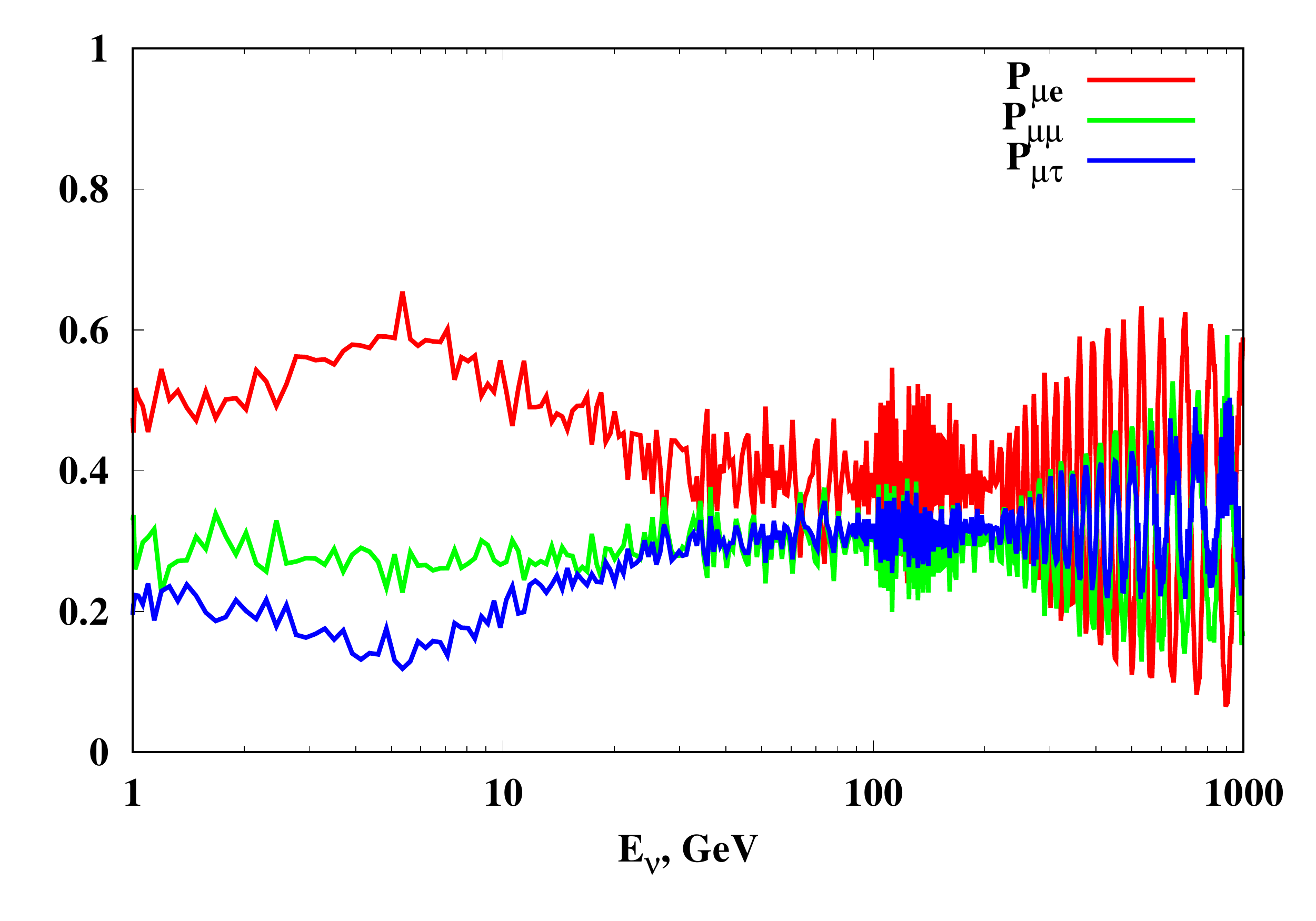}}
\end{picture}
\caption{\label{numu_emu_sign} The same as in Fig.~\ref{numu_sm} but for $\e_{e\mu}=-0.2$.}
\end{figure}

\begin{figure}[hb]
\begin{picture}(300,220)(0,20)
\put(210,130){\includegraphics[angle=-90,width=0.40\textwidth]{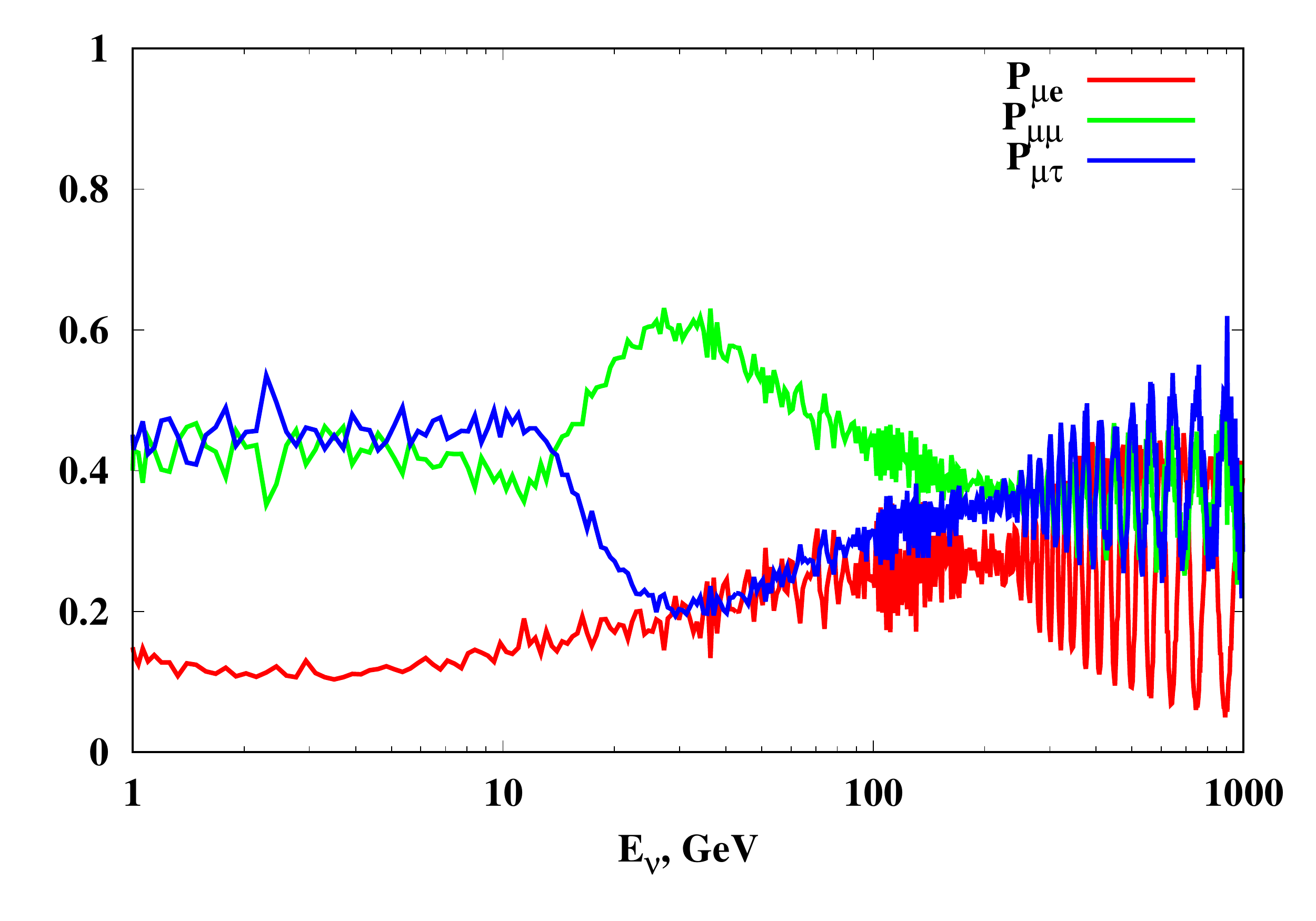}}
\put(210,250){\includegraphics[angle=-90,width=0.40\textwidth]{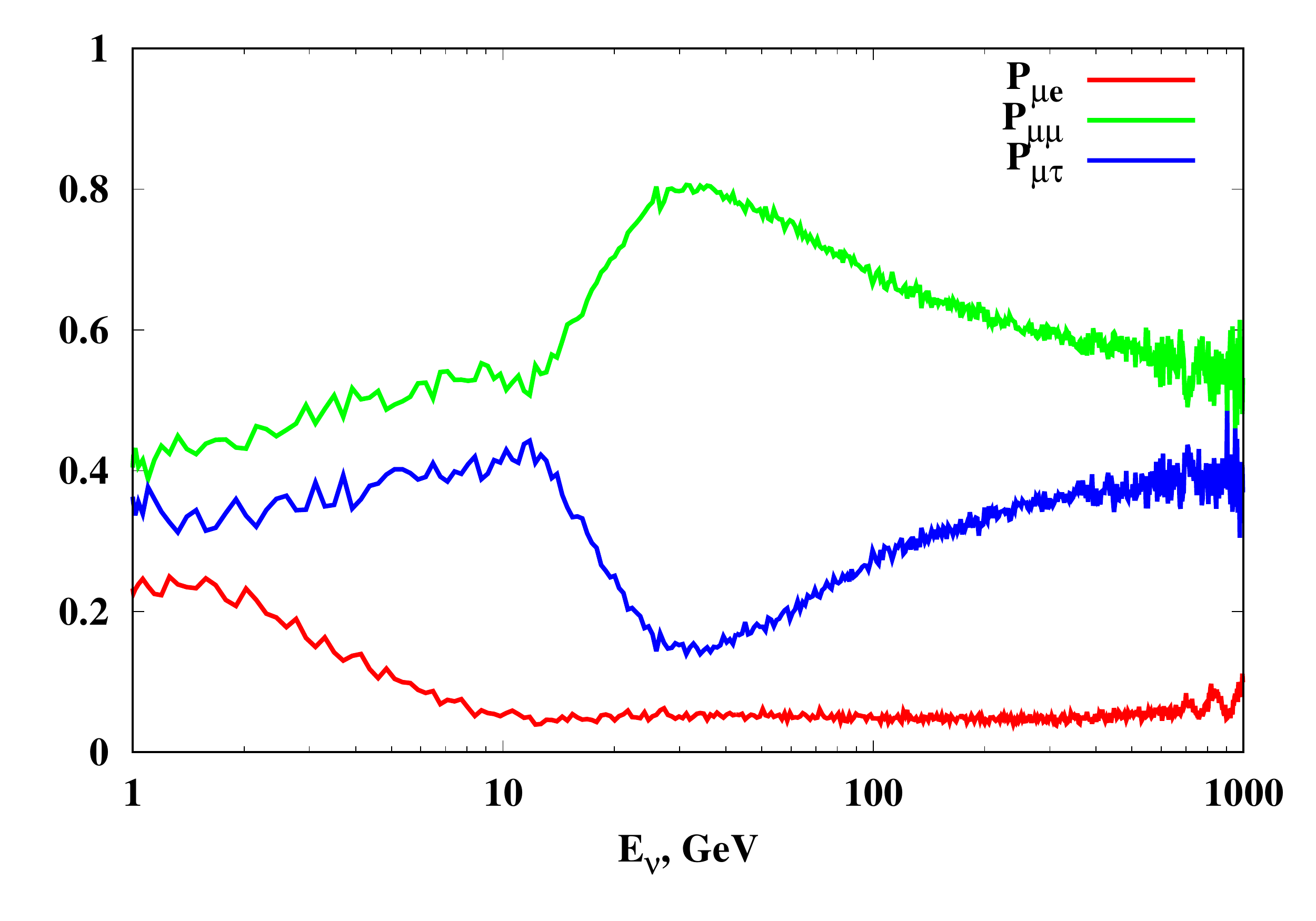}}
\put(30,130){\includegraphics[angle=-90,width=0.40\textwidth]{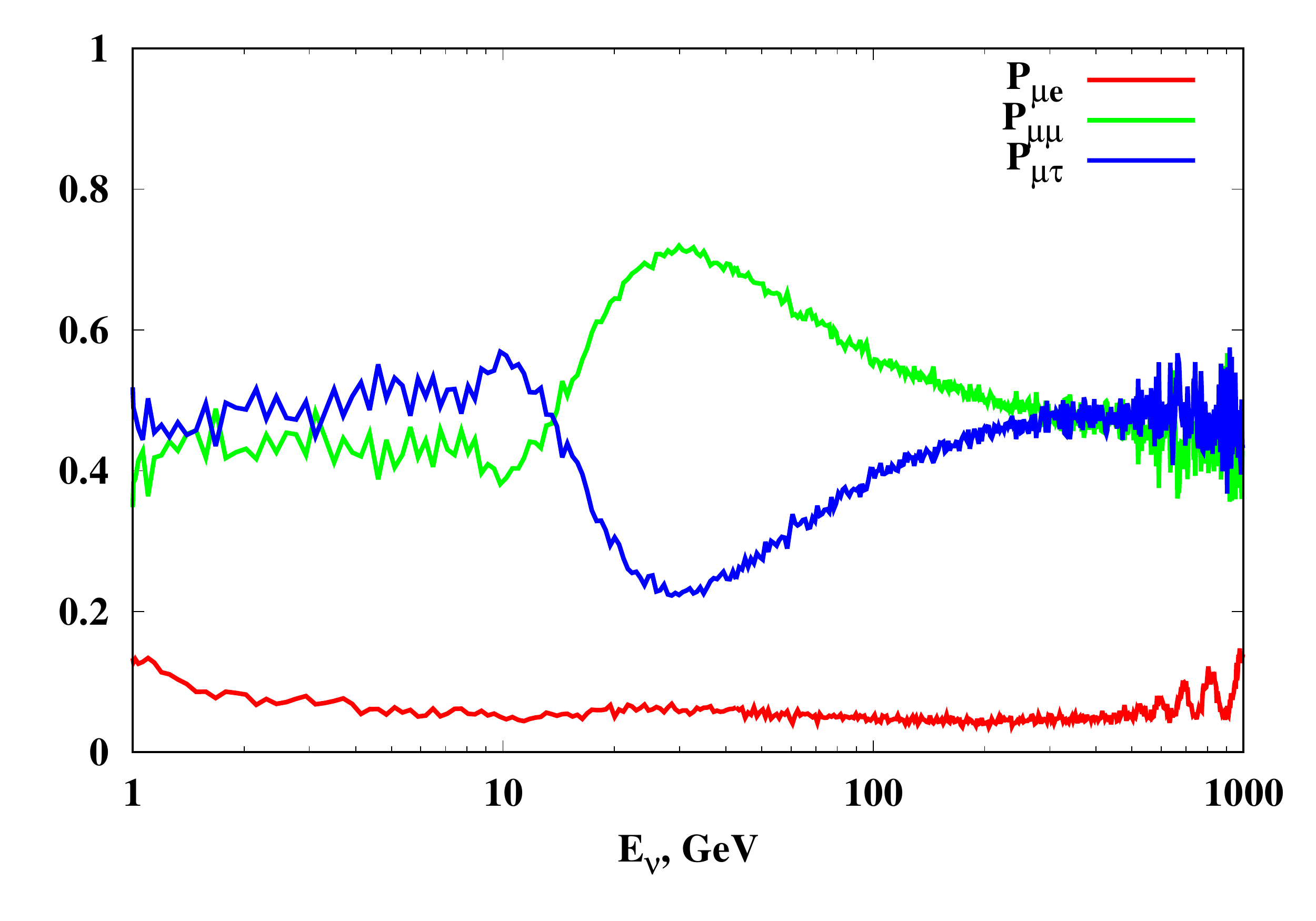}}
\put(30,250){\includegraphics[angle=-90,width=0.40\textwidth]{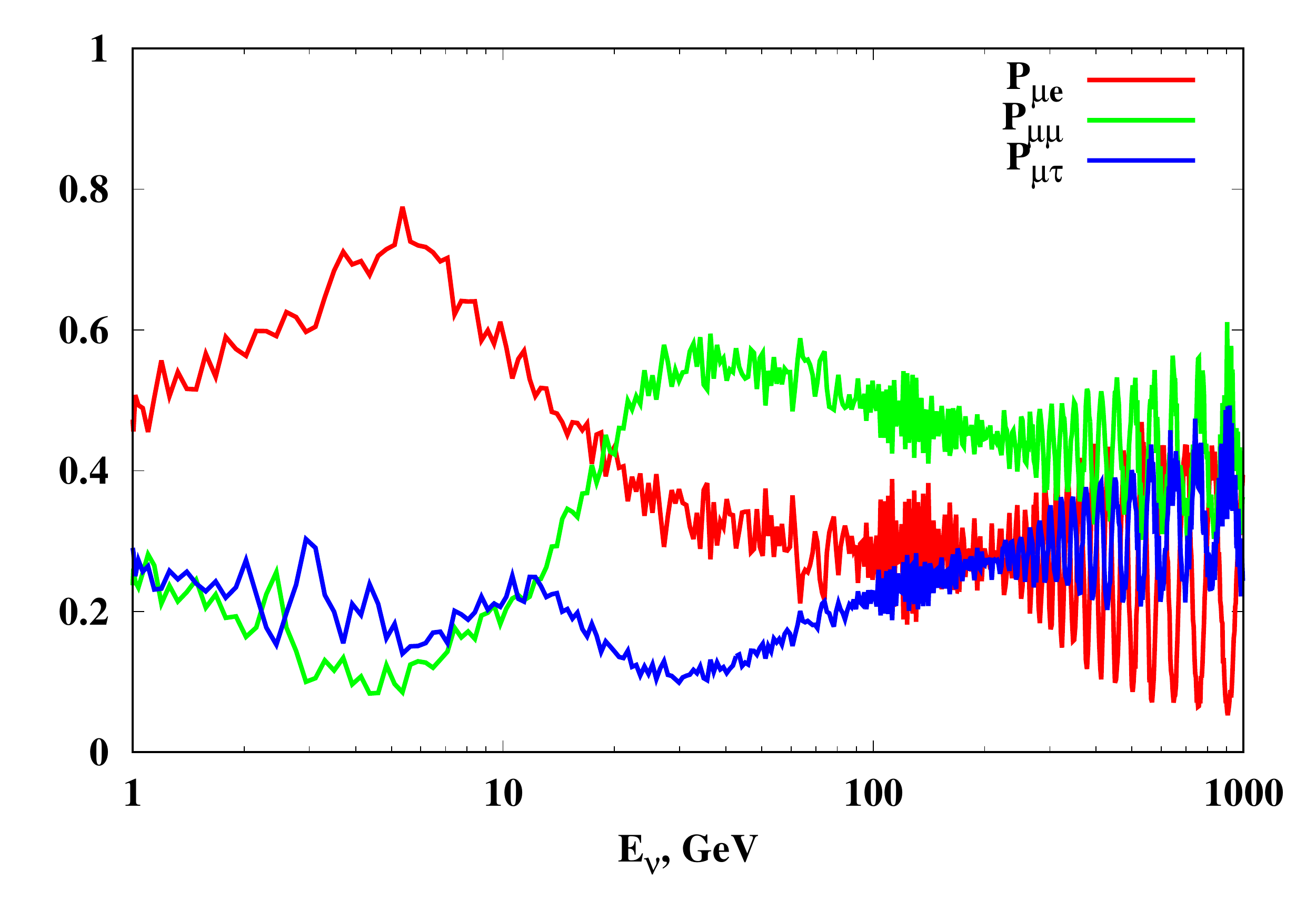}}
\end{picture}
\caption{\label{numu_emu_sign_earth}The same as in Fig.~\ref{numu_sm_earth} but for $\e_{e\mu}=-0.2$.}
\end{figure}

In Figs.~\ref{numu_emu} and~\ref{numu_emu_earth}
we plot the probabilities $P_{\mu\alpha}$, $\alpha=e,\mu, \tau$
calculated for $\e_{e\mu}=0.2$ before and after neutrino passing
through the Earth, respectively. The same probabilities but for
$\e_{e\mu}=-0.2$ are shown in Figs.~\ref{numu_emu_sign}
and~\ref{numu_emu_sign_earth}. 
We observe a similarity of the energy
dependence of these probabilities with the case $\e_{\tau\tau}=0.03$,
c.f. Figs.~\ref{numu_tautau} and~\ref{numu_tautau_earth}.
Modified mixing angles corresponding to different resonance
transitions can be found similarly to the case of non-zero
$\e_{e\tau}$ and below we present corresponding approximate expressions
\be
\tan{\theta^{e\mu}_{23}} =
\frac{\hat{s}_{12}s_{13}+\hat{c}_{12}s_{23}c_{13}}{c_{23}c_{13}}\;,
\;\;\;
\theta_{13}^{e\mu} = \theta_{13}-\hat{\theta}_{13}\;,\;\;\;
\theta_{12}^{e\mu} = \theta_{12} - \hat{\theta}_{12}\;,
\ee
where $\tan{2\hat{\theta}_{13}} = -2s_{23}\e_{e\mu}$ and
$\tan{2\hat{\theta}_{12}} = -\frac{2\e_{e\mu}c_{13}c_{23}}{c_{13}^2 -
  2\e_{e\mu}s_{13}c_{13}s_{23}}$.
For non-zero $\e_{e\mu}$ the Hamiltonian
eigenstates in the center of the Sun  (if matter term dominates) are
\be
\label{eq:3.2:4}
\cos{\phi}|\nu_e\rangle -
\sin{\phi}|\nu_\mu\rangle,\;\;
\sin{\phi}|\nu_e\rangle +
\cos{\phi}|\nu_\mu\rangle,\;\;
|\nu_\tau\rangle,
\ee
where $\tan{2\phi} = -2\e_{e\mu}$.
Numerically, for $\e_{e\mu}=\pm0.2$ one obtains $\sin^2{\phi}
\approx 0.07$ and thus muon neutrino approximately coincides with
the first among the eigenstates in~\eqref{eq:3.2:4}, which is the third
(first) energy level of the matter Hamiltonian in the solar center
for neutrino (antineutrino). For neutrino (IH) and
antineutrino (NH) $\nu_\mu$ 
produced in the center of the Sun escapes it approximately in
$|3\rangle$ 
state, see lower left and upper right panels in Figs.~\ref{numu_emu} 
and~\ref{numu_emu_sign}. In these cases 1--2 and 1--3 resonances do
not have any influence on neutrino propagation. For the case of
neutrino (NH), $\nu_\mu$ state which is the approximately the lowest
energy 
eigenstate of the Hamiltonian in the center of the Sun evolves into
$|1\rangle$ at solar surface in the adiabatic regime. At higher
energies nonadiabaticity in transition through 1--2  resonance is
important and in the limit of maximal adiabaticity violation one
obtains the mixed state $\sin^2{\theta^{e\mu}_{12}}|1\rangle\langle 1|
+ \cos^2{\theta^{e\mu}_{12}}|2\rangle\langle 2|$. Numerically,
$\sin^2{\theta^{e\mu}_{12}}\approx 0.45$ for $\e_{e\mu}=0.3$ and
$\sin^2{\theta^{e\mu}_{12}}\approx 0.18$ for $\e_{e\mu}=-0.3$ which
results in different behaviour of $P_{\mu\alpha}$ in upper
left panels in Figs.~\ref{numu_emu} and~\ref{numu_emu_sign}. 
Similarly, for the
case of antineutrino (IH) in adiabatic regime $\nu_\mu$ evolves into
$|2\rangle$ at the Earth orbit.  At energies less than about 5 GeV the
vacuum contribution to the Hamiltonian becomes comparable with the
matter term in the center of the Sun and this is responsible for the
energy dependence of the probabilities
in this energy region. Subsequent evolution in the Earth again is
similar to the case of positive $\e_{\tau\tau}$. In the upper right
and lower left panels on Fig.~\ref{numu_emu_earth} one can see the
bump-like features at energies around 30~GeV which are related 
to the oscillations of $|3\rangle$ state in the matter of the Earth. 
They result in an increase of $P_{\mu\mu}$ and decrease of
$P_{\mu\tau}$ in the intermediate neutrino energy range. In the case $\e_{e\mu}=-0.2$ behaviour of the
probabilities $P_{\mu\alpha}$ is qualitatively the same as for positive
$\e_{e\mu}$, see Figs.~\ref{numu_emu_sign} and~\ref{numu_emu_sign_earth}.

Finally, let us turn to the case of non-zero $\e_{\mu\tau}$.
\begin{figure}[hb]
\begin{picture}(300,220)(0,20)
\put(210,130){\includegraphics[angle=-90,width=0.40\textwidth]{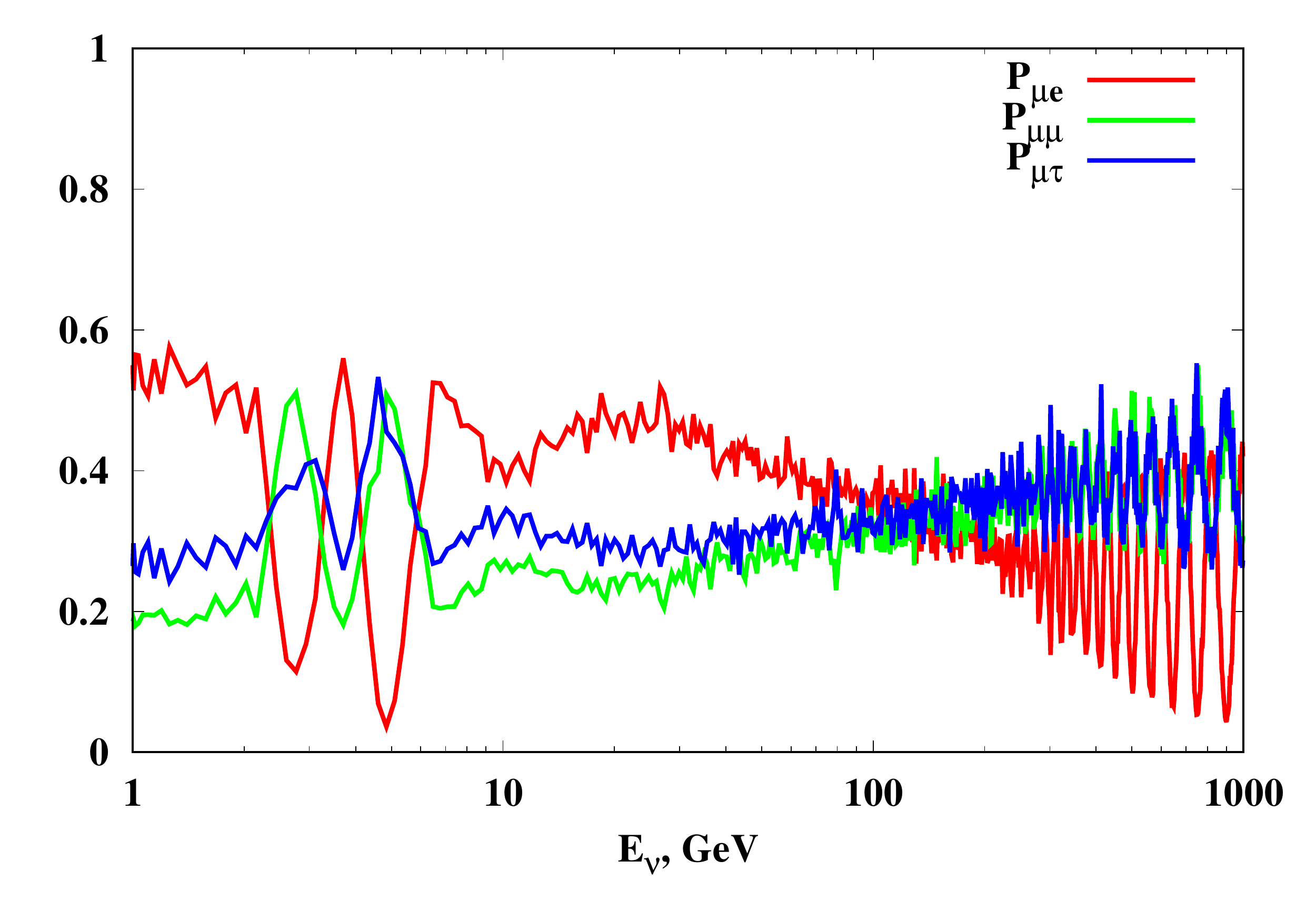}}
\put(210,250){\includegraphics[angle=-90,width=0.40\textwidth]{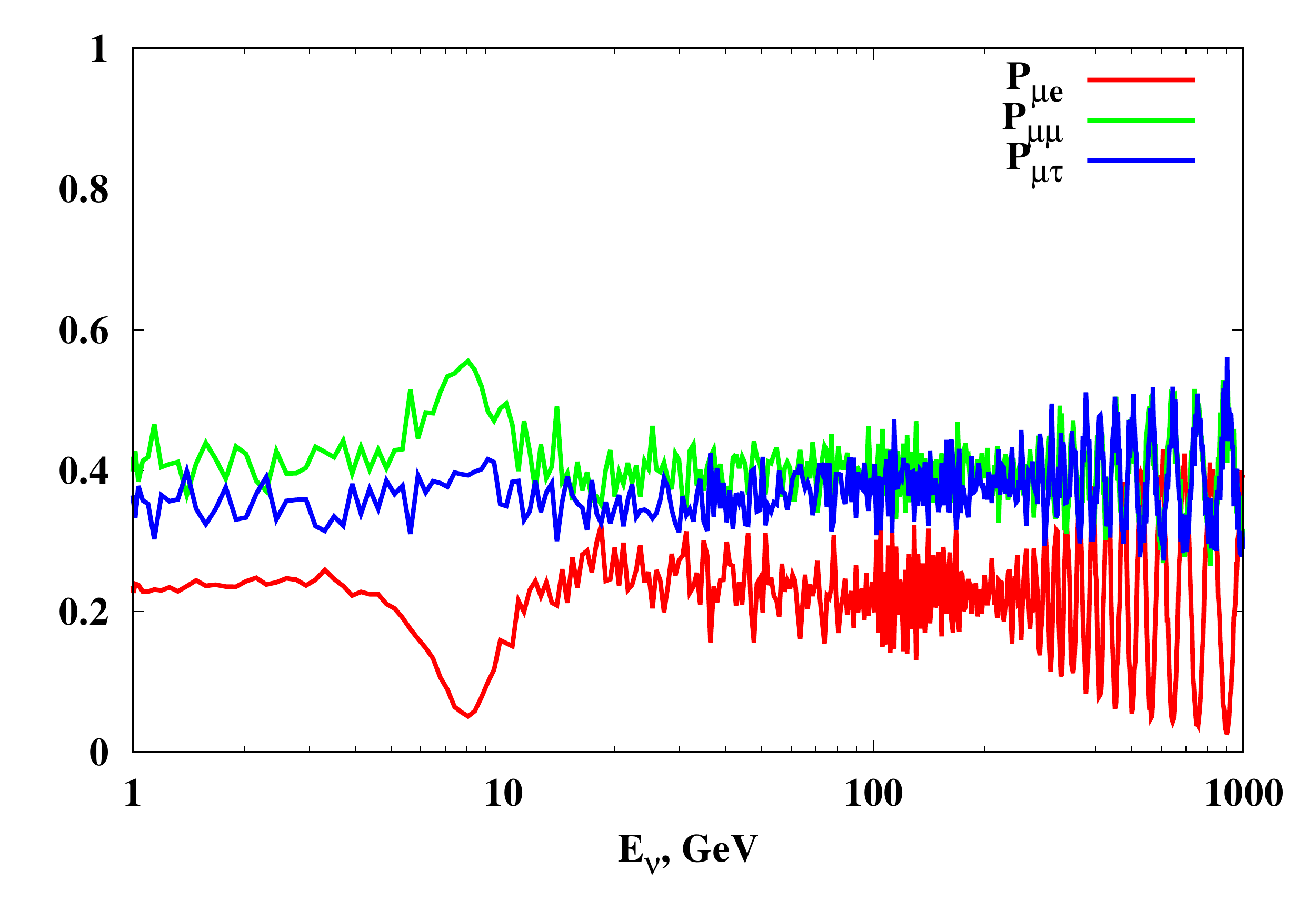}}
\put(30,130){\includegraphics[angle=-90,width=0.40\textwidth]{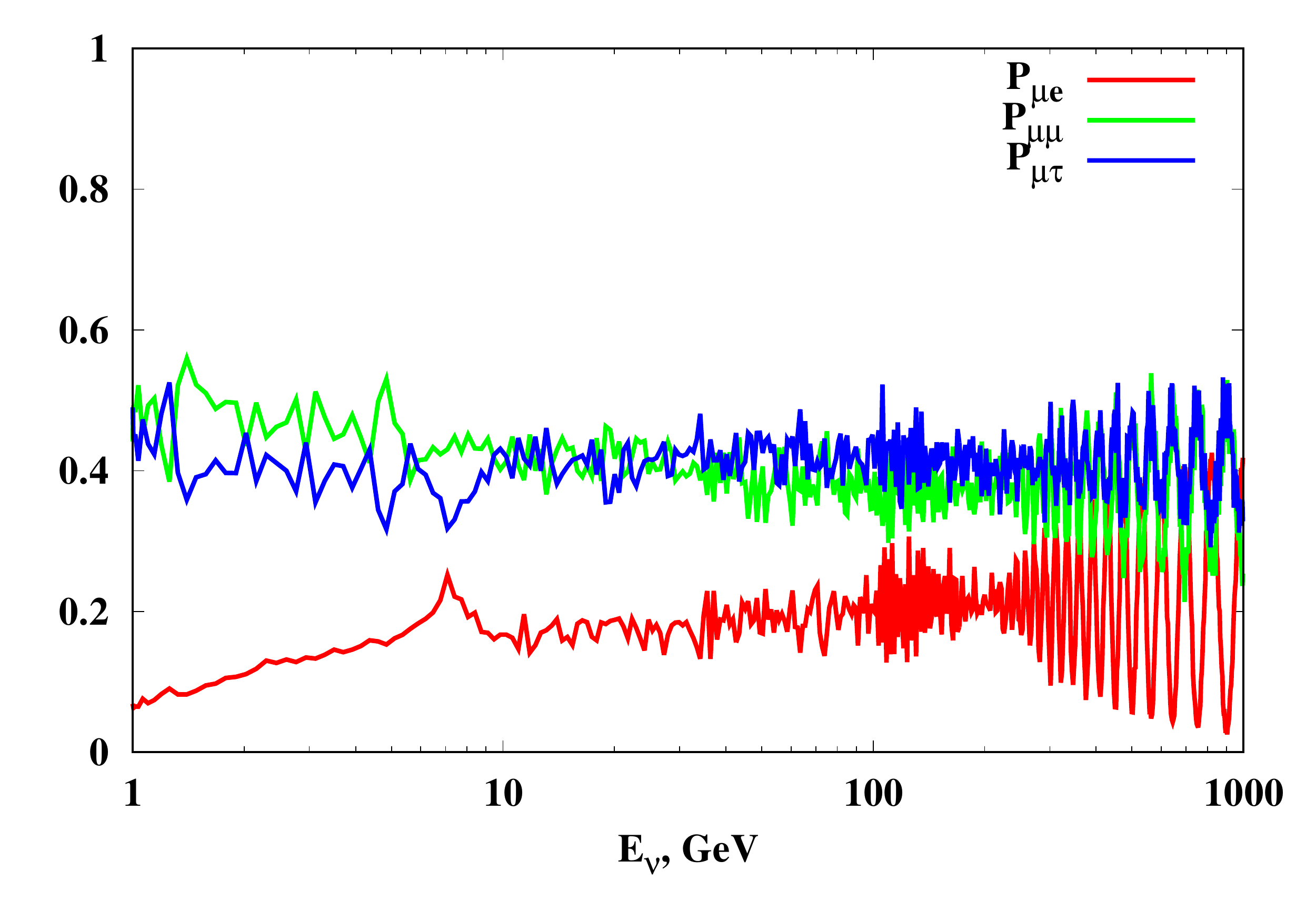}}
\put(30,250){\includegraphics[angle=-90,width=0.40\textwidth]{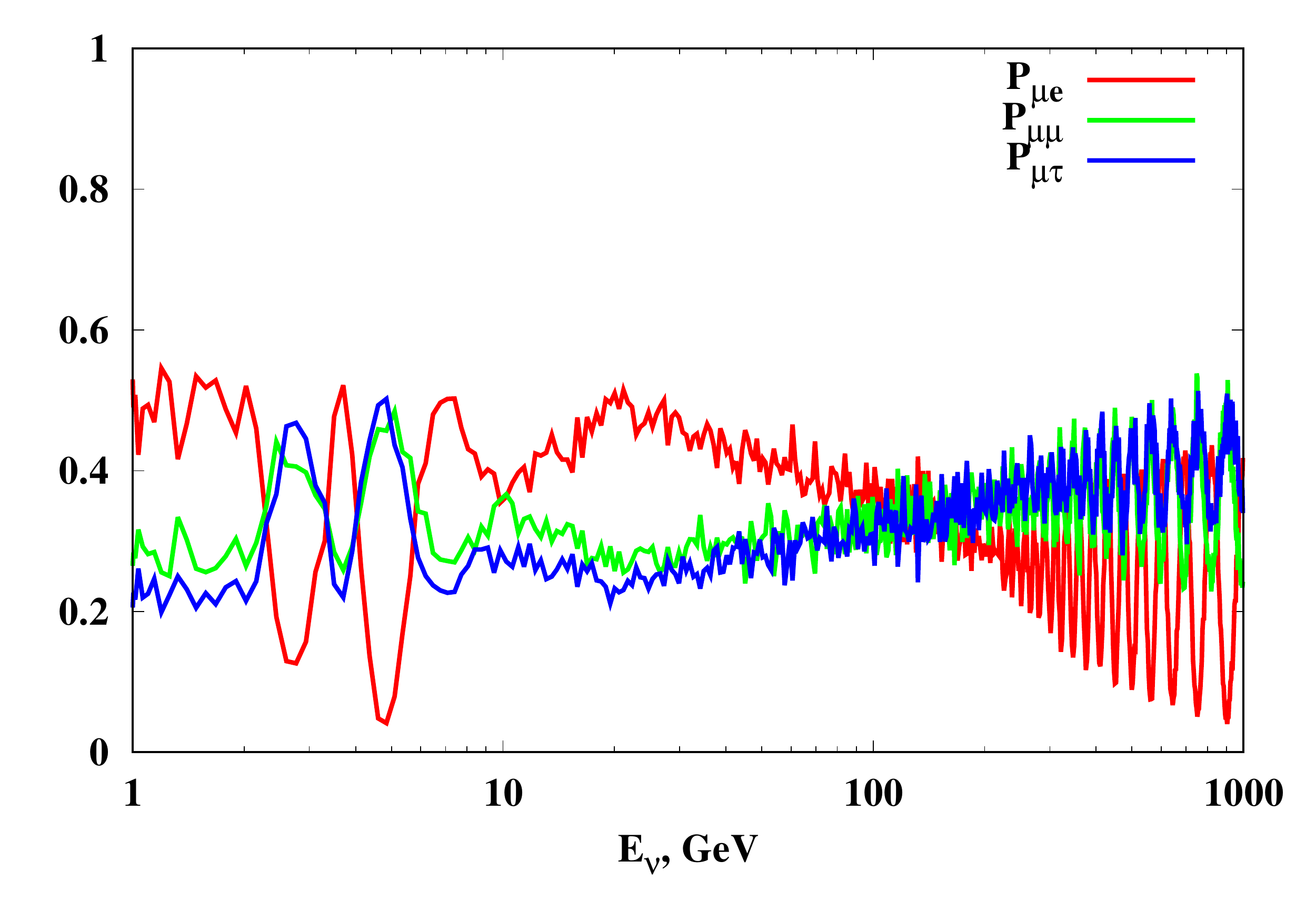}}
\end{picture}
\caption{\label{numu_mutau_earth} The same as in Fig.~\ref{numu_sm_earth} but for $\e_{\mu\tau}=0.01$.}
\end{figure}
\begin{figure}[hb]
\begin{picture}(300,220)(0,20)
\put(210,130){\includegraphics[angle=-90,width=0.40\textwidth]{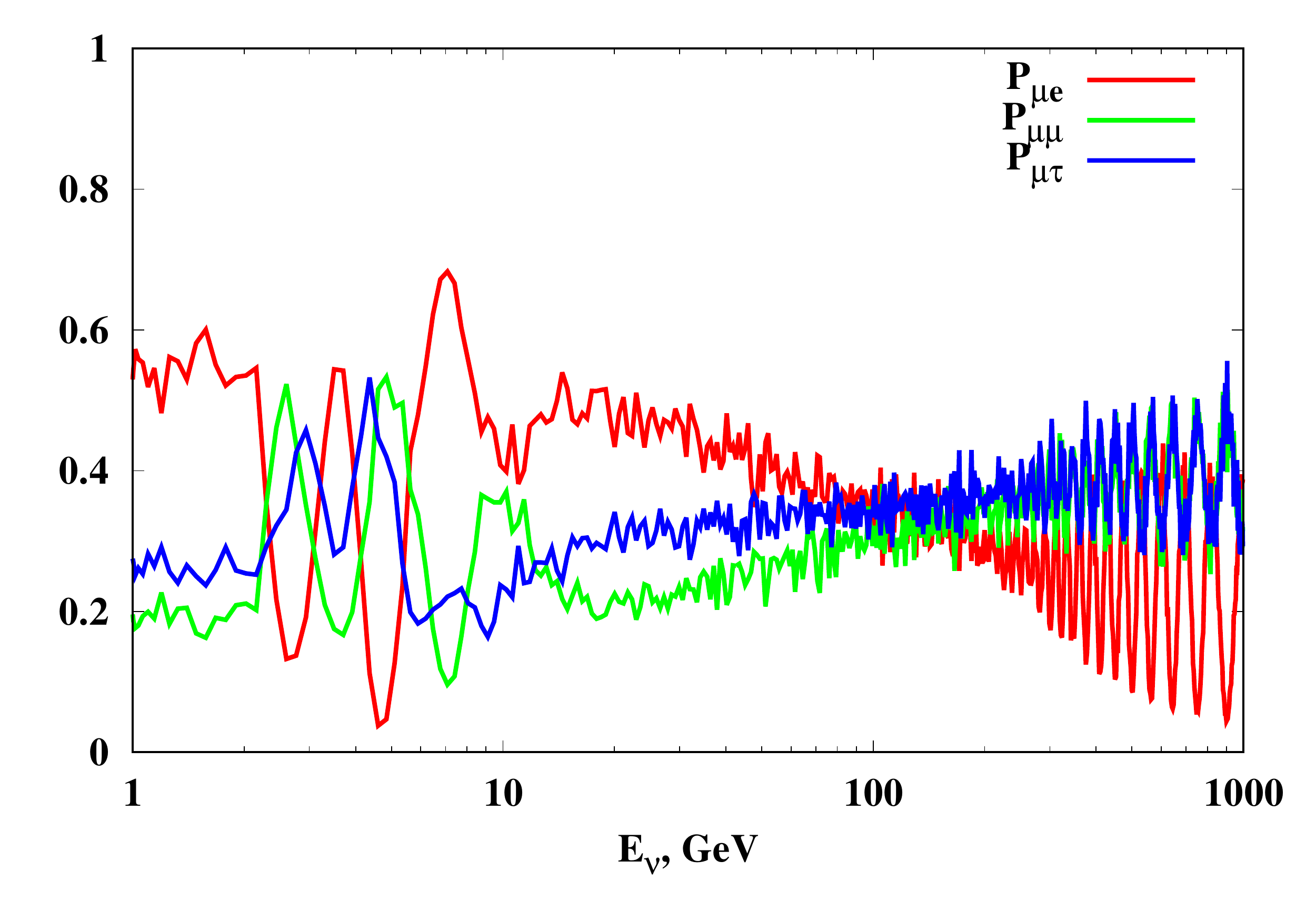}}
\put(210,250){\includegraphics[angle=-90,width=0.40\textwidth]{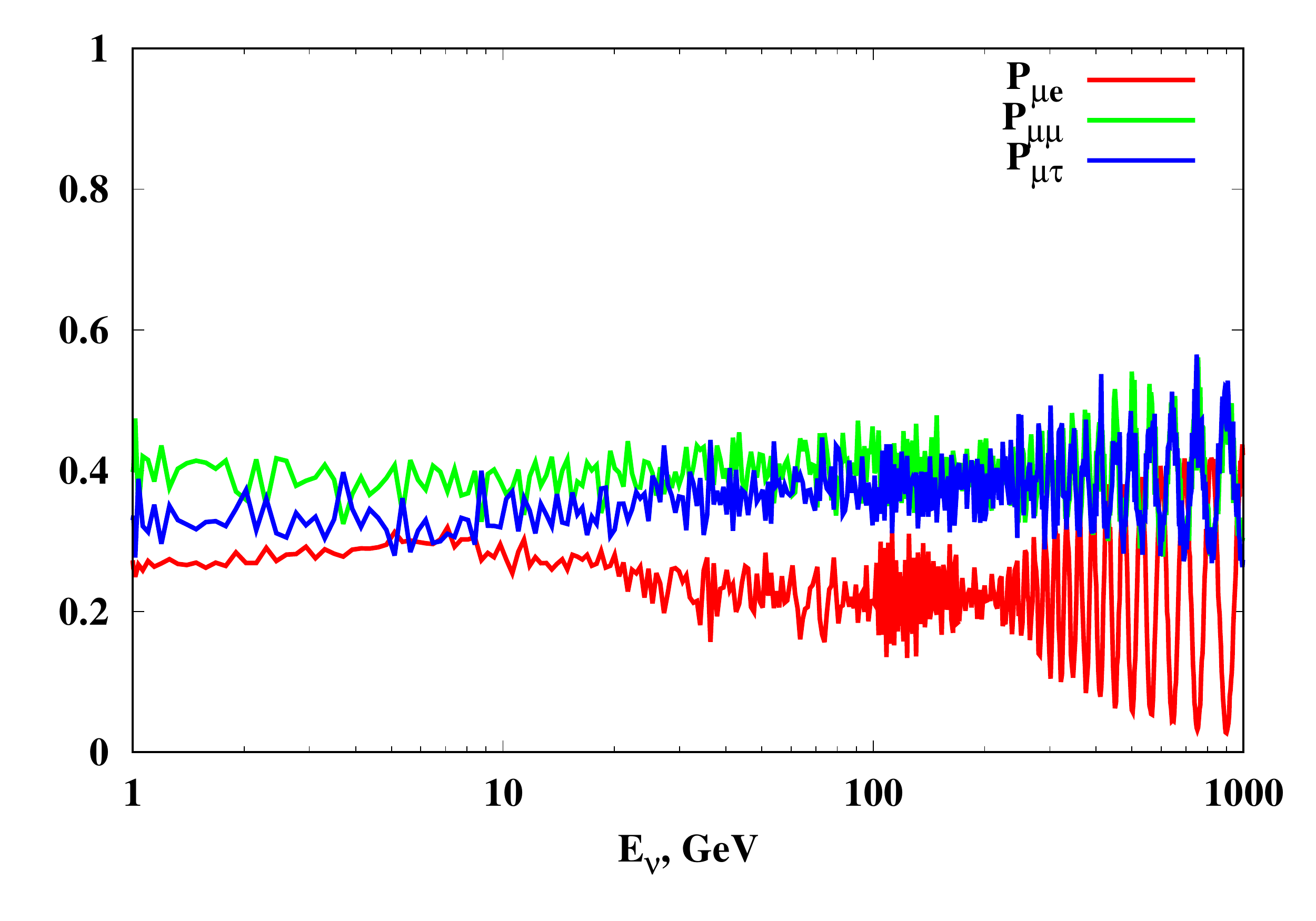}}
\put(30,130){\includegraphics[angle=-90,width=0.40\textwidth]{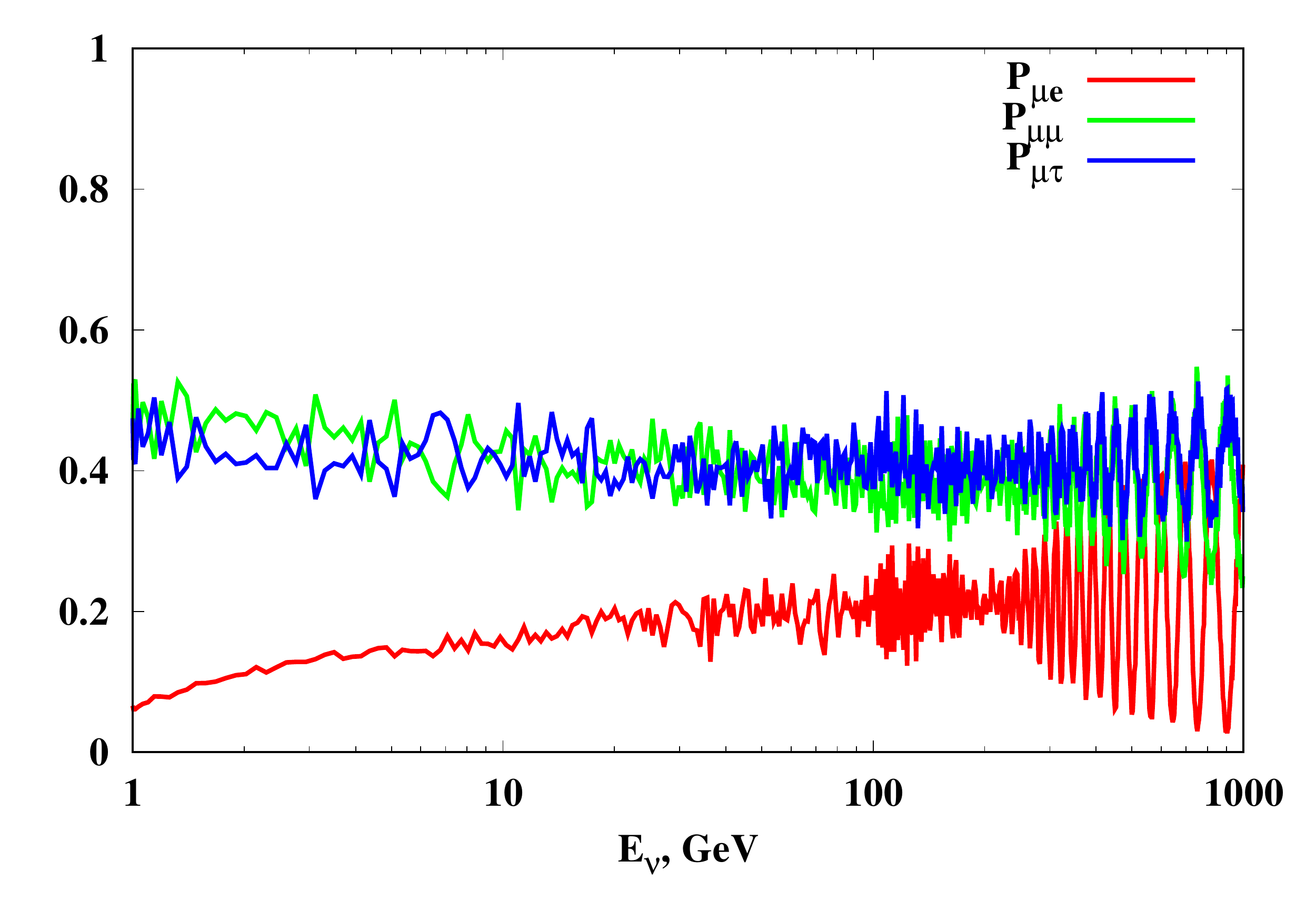}}
\put(30,250){\includegraphics[angle=-90,width=0.40\textwidth]{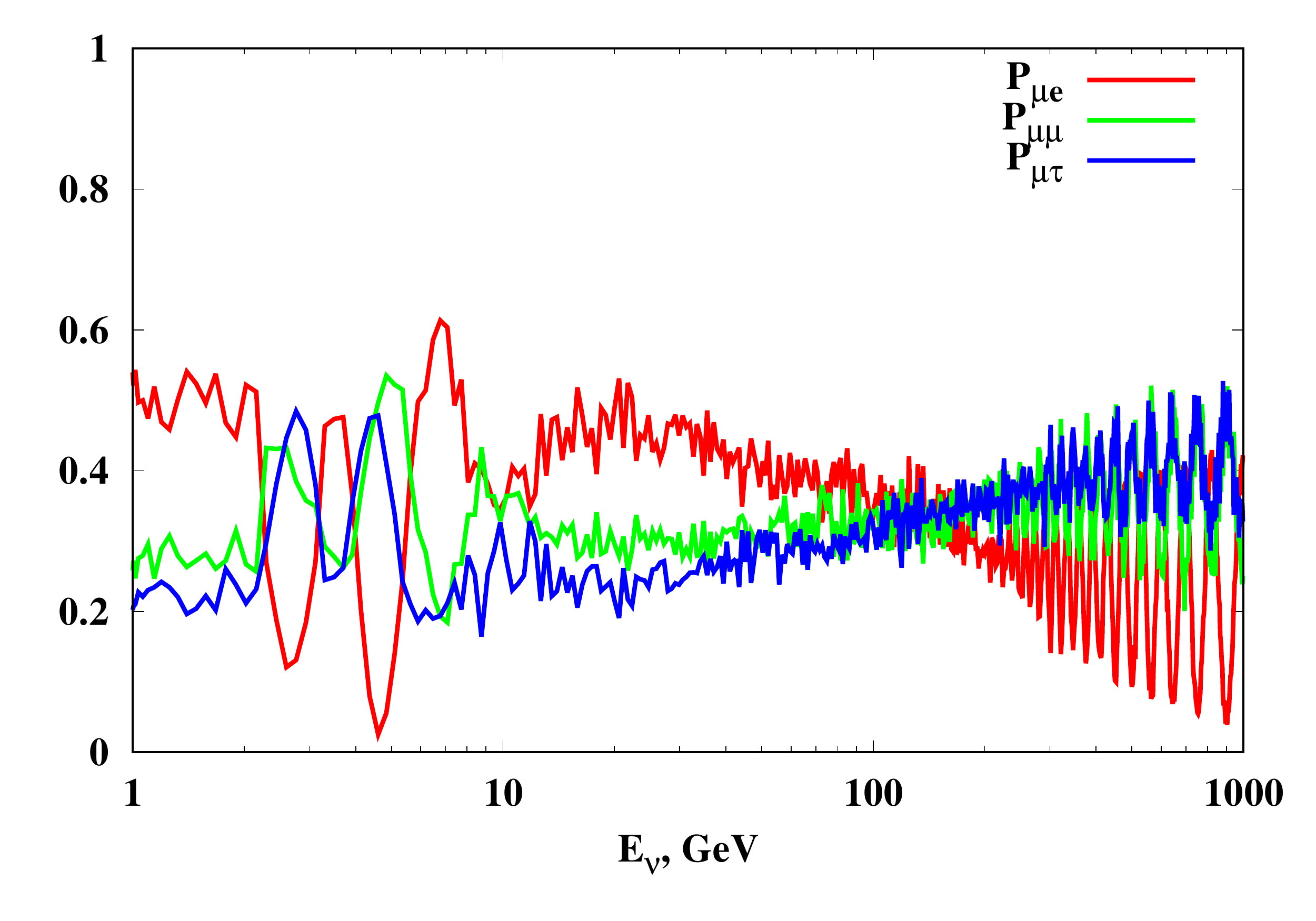}}
\end{picture}
\caption{\label{numu_mutau_sign_earth} The same as in Fig.~\ref{numu_sm_earth} but for $\e_{\mu\tau}=-0.01$.}
\end{figure}
Now the
matter term in the Hamiltonian can be diagonalized by rotation
$\nu=R_{23}^\dagger(\frac{\pi}{4})\nu^\prime$. New Hamiltonian takes the form
\be
H^\prime = \frac{1}{2E_\nu}U^\prime{\rm
  diag}(m_1^2,m_2^2.m_3^2)U^{\prime\dagger} + V_e\left(
\begin{array}{ccc}
  1 & & \\
  & -\e_{\mu\tau} &\\
  & & \e_{\mu\tau}
\end{array}
\right)\;,
\ee
where $U^\prime =
R_{23}\left(\theta_{23}-\frac{\pi}{4}\right)R_{13}(\theta_{13})R_{12}(\theta_{12})$. Numerically  
the value of $\theta_{23}-\frac{\pi}{4}$ is rather small with
$\sin^2{\left(\theta_{23}-\frac{\pi}{4}\right)} \lsim 0.01$ and 
2--3 transition appears to be in non-adiabatic regime for $E_\nu\gsim
1$~GeV, see Eq.~\eqref{eq:3.1:5}. Thus muon neutrino (as well as
tau neutrino) which is superposition of the Hamiltonian eigenstates
undergoes fast oscillations and leaves its central part approximately as
$\frac{1}{2}\left(|2_m\rangle\langle 
2_m| + |3_m\rangle \langle 3_m|\right)$, where $|2_m\rangle$ and
$|3_m\rangle$ are defined in Eq.~\eqref{eq:2:3}. This is almost
coincide with the result obtained for the no-NSI case, see discussion after
Eq.~\eqref{eq:2:3}. Thus, subsequent evolution in the Sun and
respective probabilities $P_{\mu\alpha}$ will be the same as in
Fig.~\ref{numu_sm}. We check this numerically for $\e_{\mu\tau}=\pm
0.01$. The effect of propagation through the Earth
is presented in Figs.~\ref{numu_mutau_earth}
and~\ref{numu_mutau_sign_earth} for different signs of
$\e_{\mu\tau}=\pm0.01$. We observe that effect of non-zero
$\e_{\mu\tau}$ is considerably milder as compared with that of
non-zero $\e_{\tau\tau}$,
$\e_{\mu\mu}$, $\e_{e\tau}$ or $\e_{e\mu}$. Still as we will see in the
next Section interactions of neutrino with the matter of the Sun 
can result in nontrivial dependence on non-zero $\e_{\mu\tau}$.

\section{Full Monte-Carlo analysis}
In the previous Section we study evolution of monochromatic
neutrino from the center of the Sun to the Earth completely neglecting  
of CC and NC neutrino scatterings. Here we present results of the full
Monte-Carlo simulation of the neutrino propagation from the Sun to the
Earth. Detailed description of our numerical code and in particular
comparison with WimpSim
package~\cite{Blennow:2007tw,wimpsim,Edsjo:2017kjk} had been presented
in Ref.~\cite{Boliev:2013ai}. Here we briefly sketch its main features. 
For initial neutrino energy spectra of chosen annihilation channels
$\tau^+\tau^-$, $W^+W^-$ and $b\bar{b}$ we use those obtained with
WimpSim. Annihilation of dark matter into $\tau^+\tau^-$ results in
$\nu_\tau(\bar{\nu}_\tau)$-dominated neutrino flux at production,
while for $b\bar{b}$ channel this flux is saturated by electron and
muon neutrino flavors. As for the case $W^+W^-$ all neutrino flavors
present almost in equal parts. We simulate  
annihilation point near the center of the Sun according to 
space dark matter distribution~\eqref{eq:2:1}. Apart from oscillations
we take into account CC and NC 
interactions of neutrinos. Corresponding cross section have
been calculated including tau-mass effects using formulas presented in
Ref.~\cite{Paschos:2001np}. NC interactions result in change of 
neutrino energy leaving its flavor content intact while CC
interactions in case of electron and muon neutrino result in their
disappearance from the flux. At the same time for tau-neutrino
regeneration in CC interactions takes place because produced
tau-lepton decays into tau-neutrino of lower energy. This process is important in
particular for $\tau^+\tau^-$ annihilation channel.  
We simulate time distribution for position of the Sun on the sky for a
detector placed at 52$^\circ$ North latitude which corresponds to
the position of Baikal-GVD
project~\cite{Avrorin:2015skm,Avrorin:2016zna}. Rather close results 
are expected for the positions of
KM3NeT~\cite{Katz:2006wv,Adrian-Martinez:2016fdl} and  
Super-Kamiokande~\cite{Abe:2011ts}. As for IceCube-Gen2
detector~\cite{Aartsen:2014njl} we expect that the effect of 
neutrino propagation through the Earth will be negligible because the
Sun there is always close to the horizon. 

\begin{figure}[!htb]
\begin{picture}(300,190)(0,40)
\put(260,130){\includegraphics[angle=-90,width=0.31\textwidth]{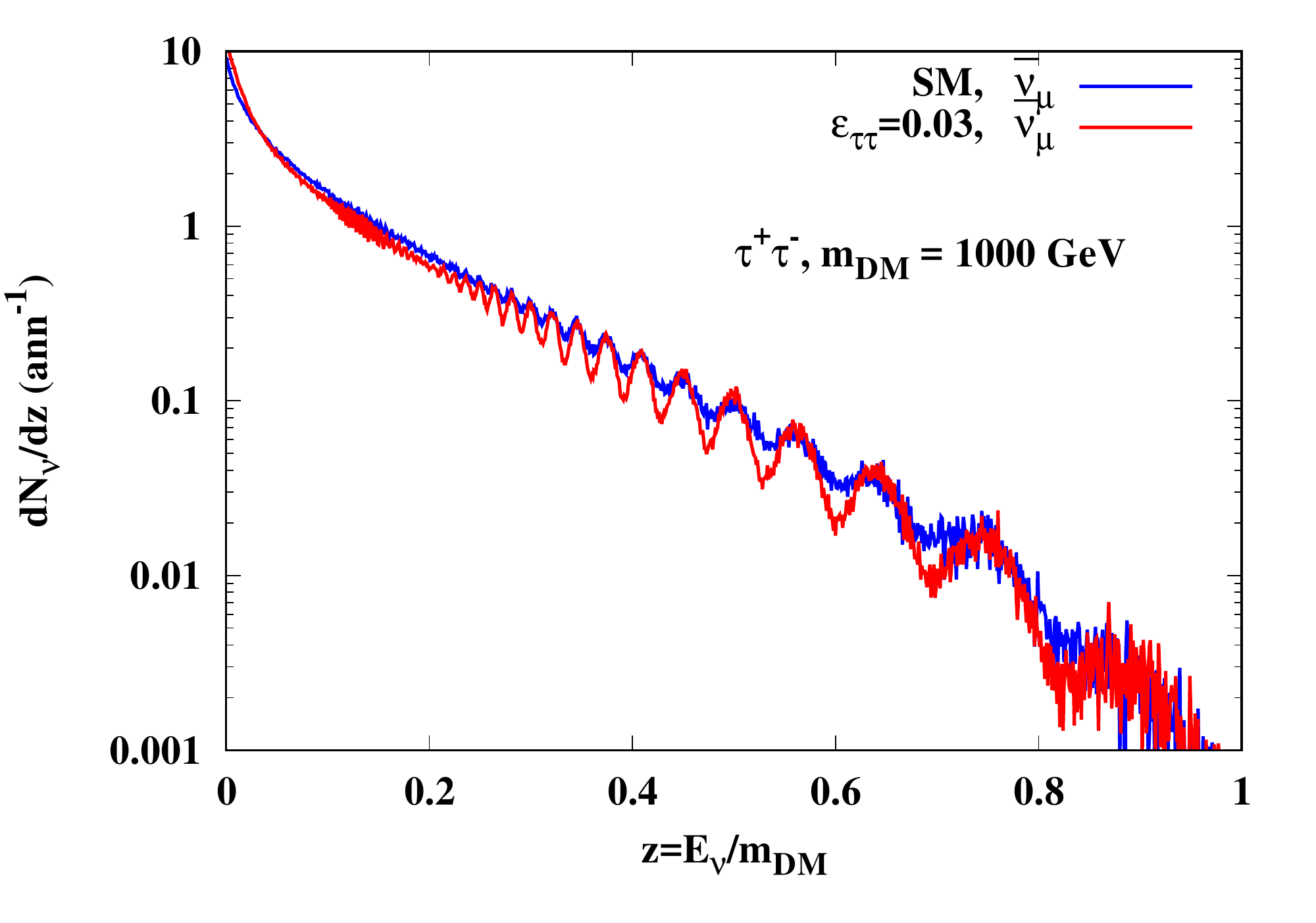}}
\put(260,240){\includegraphics[angle=-90,width=0.31\textwidth]{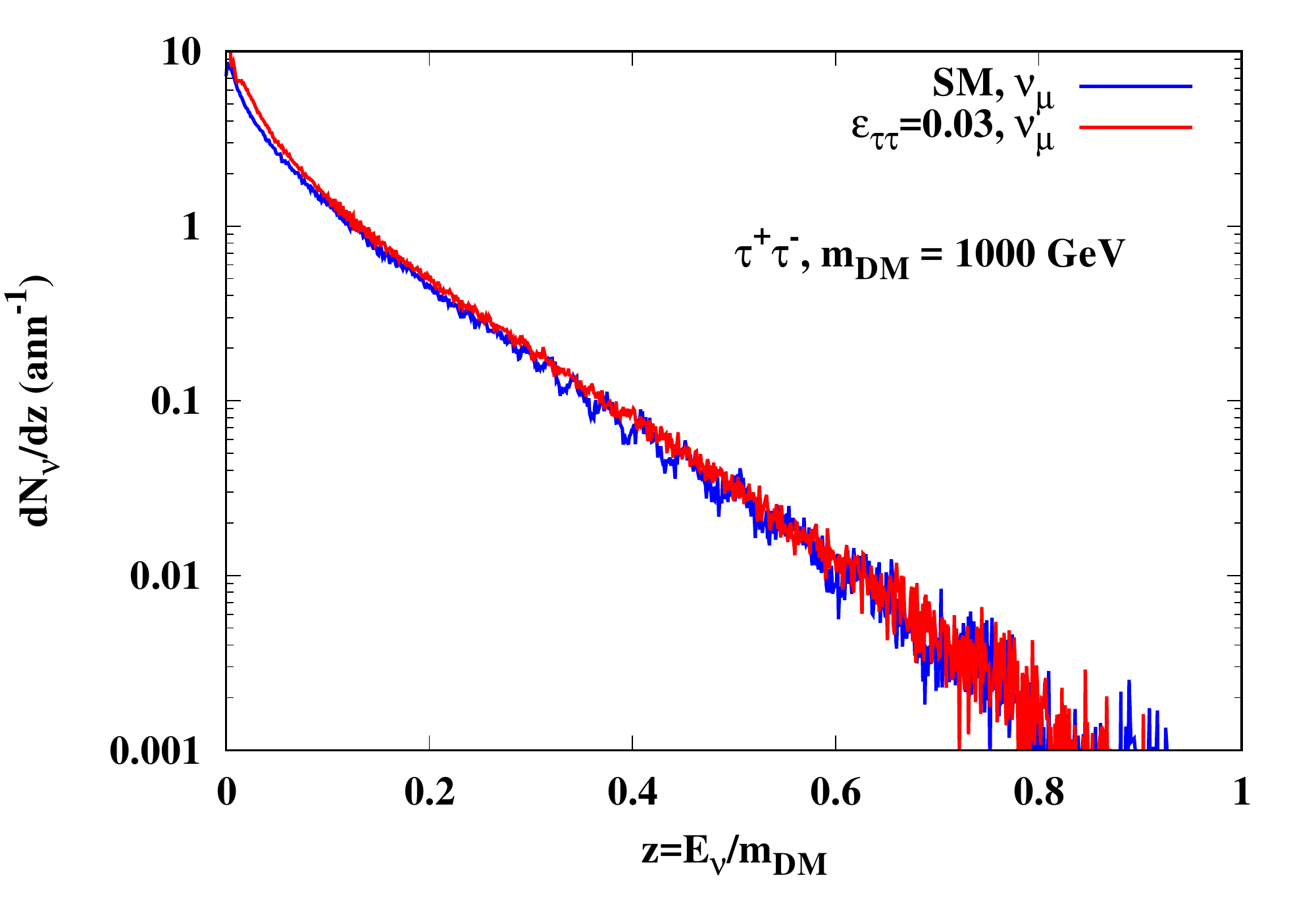}}
\put(130,130){\includegraphics[angle=-90,width=0.31\textwidth]{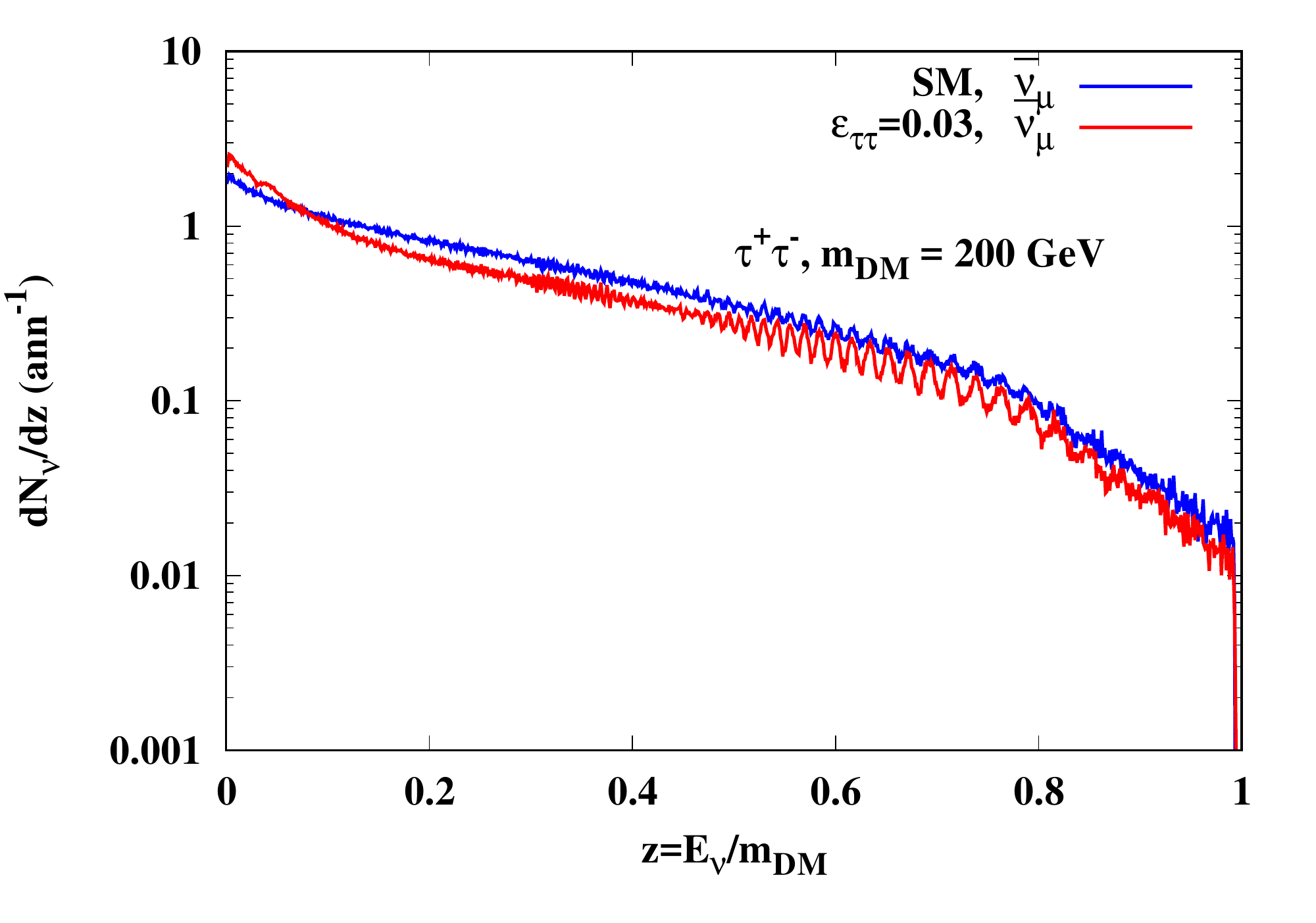}}
\put(130,240){\includegraphics[angle=-90,width=0.31\textwidth]{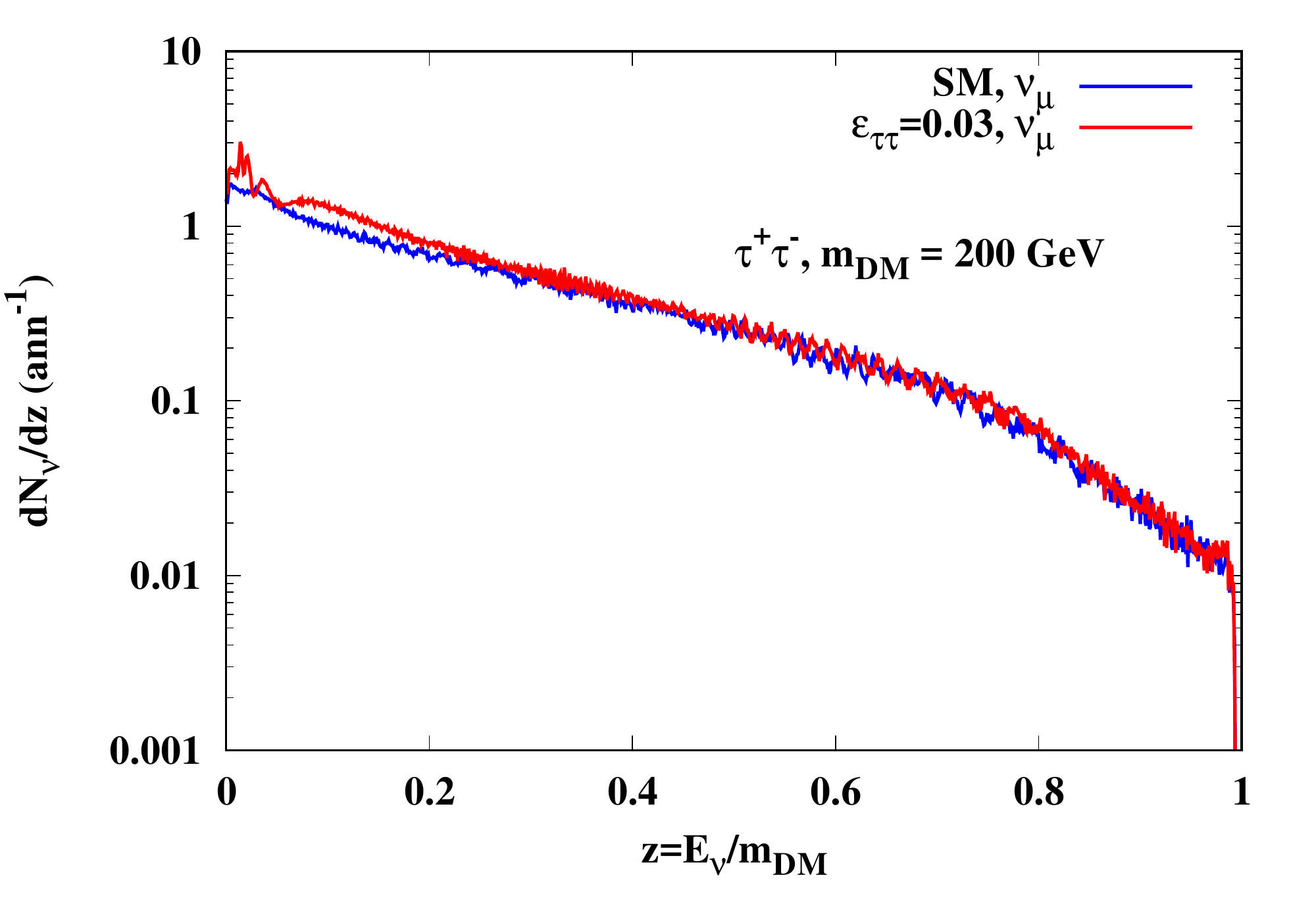}}
\put(0,130){\includegraphics[angle=-90,width=0.31\textwidth]{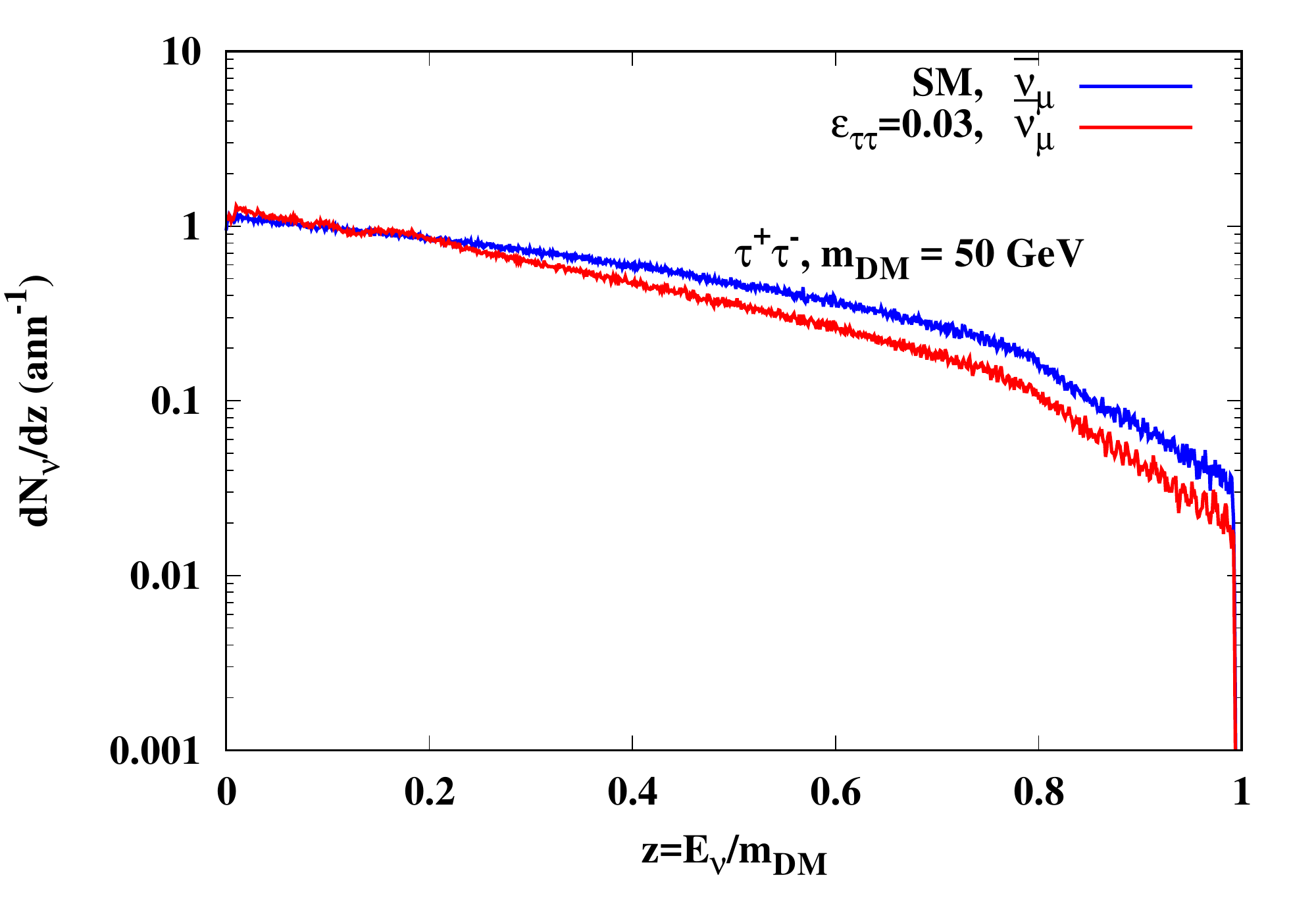}}
\put(0,240){\includegraphics[angle=-90,width=0.31\textwidth]{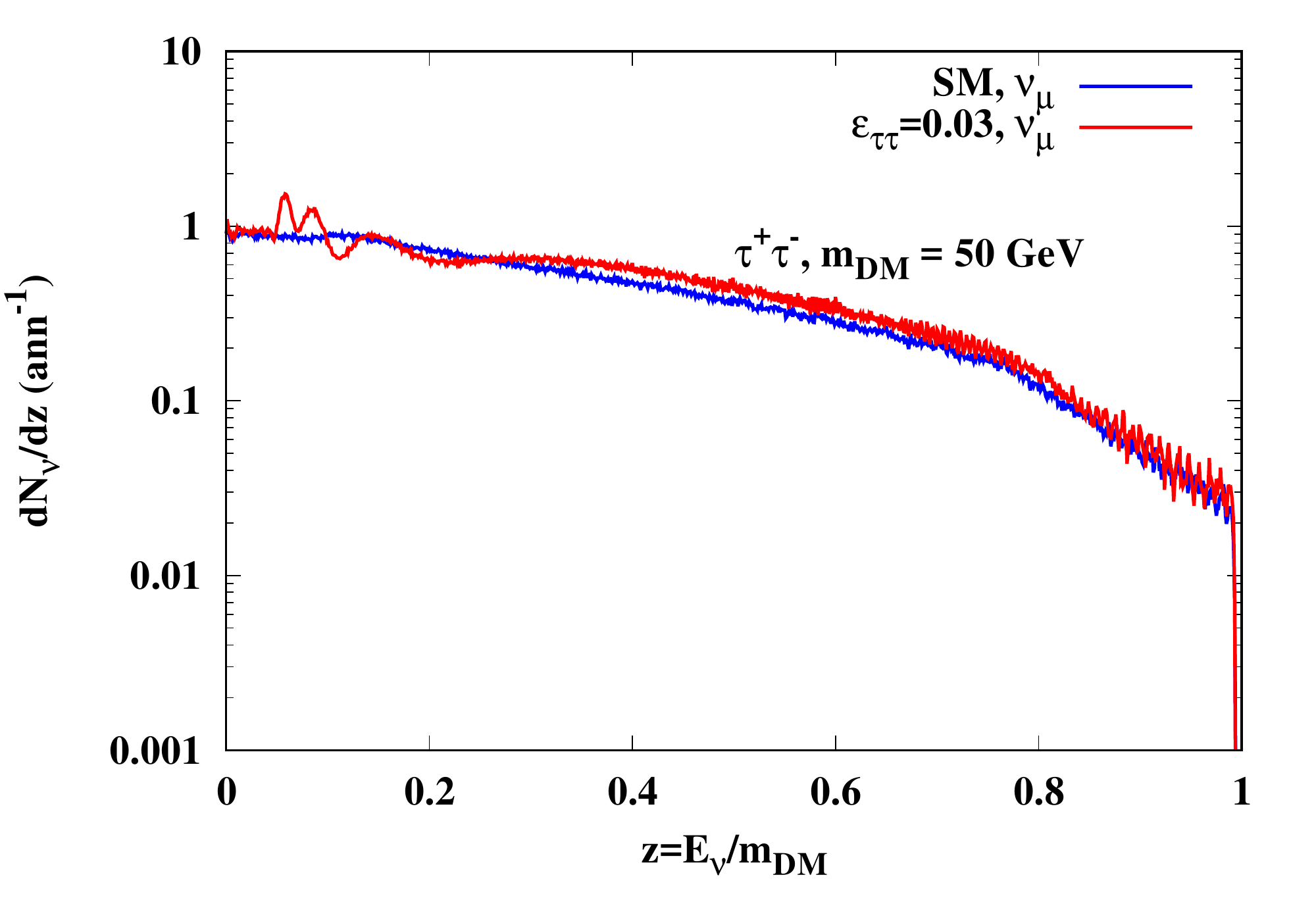}}
\end{picture}
\caption{\label{tau_tautau} Muon neutrino (upper panels) and
  antineutrino (lower panels) energy spectra at the Earth for
  $\tau^+\tau^-$ dark matter annihilation channel for $m_{DM}=50, 200$
  and 1000~GeV calculated with $\e_{\tau\tau}=0.03$ (red curve) and
  without NSI (blue curve). Normal mass ordering is assumed.}
\end{figure}

\begin{figure}[!htb]
\begin{picture}(300,190)(0,40)
\put(260,130){\includegraphics[angle=-90,width=0.31\textwidth]{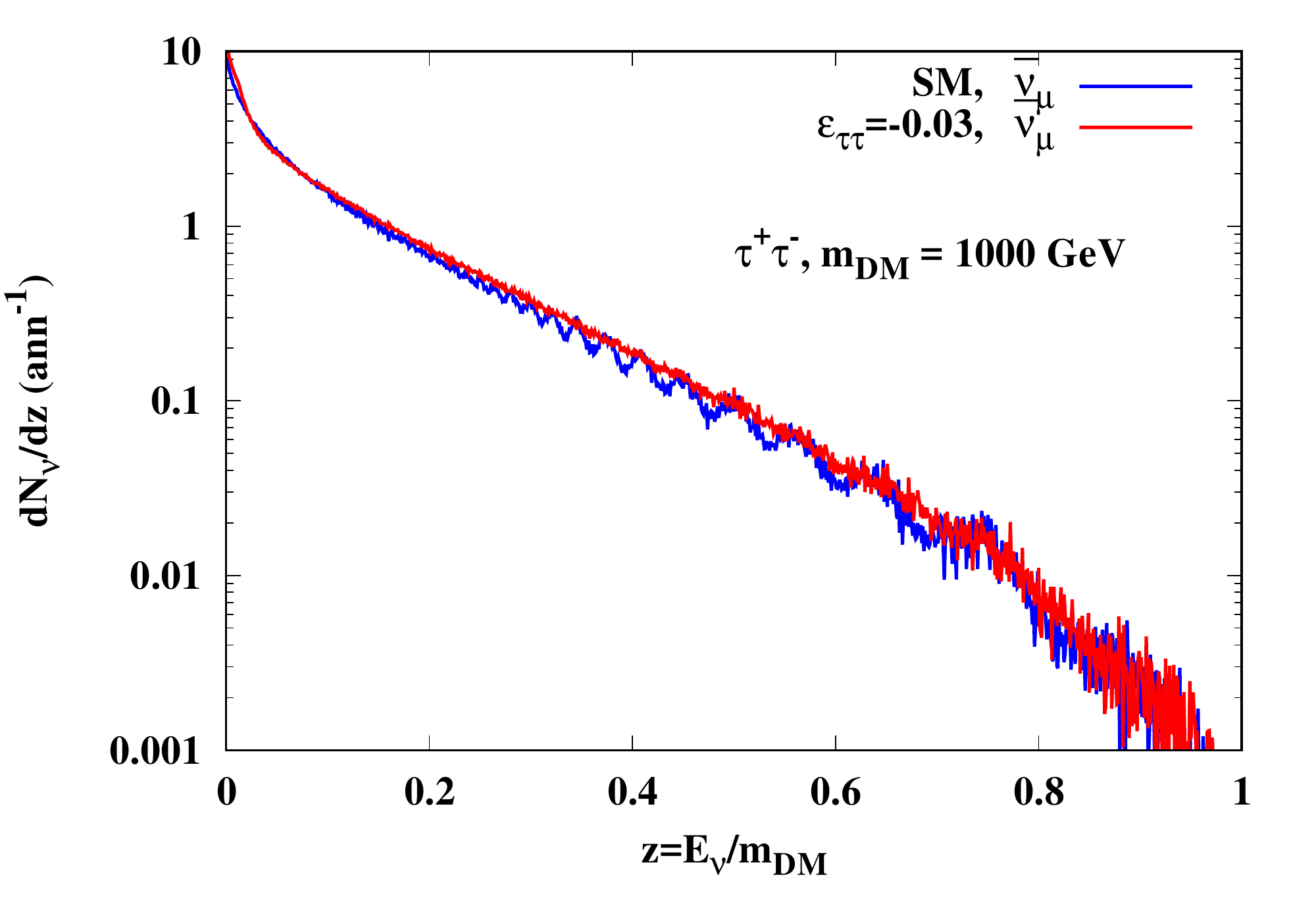}}
\put(260,240){\includegraphics[angle=-90,width=0.31\textwidth]{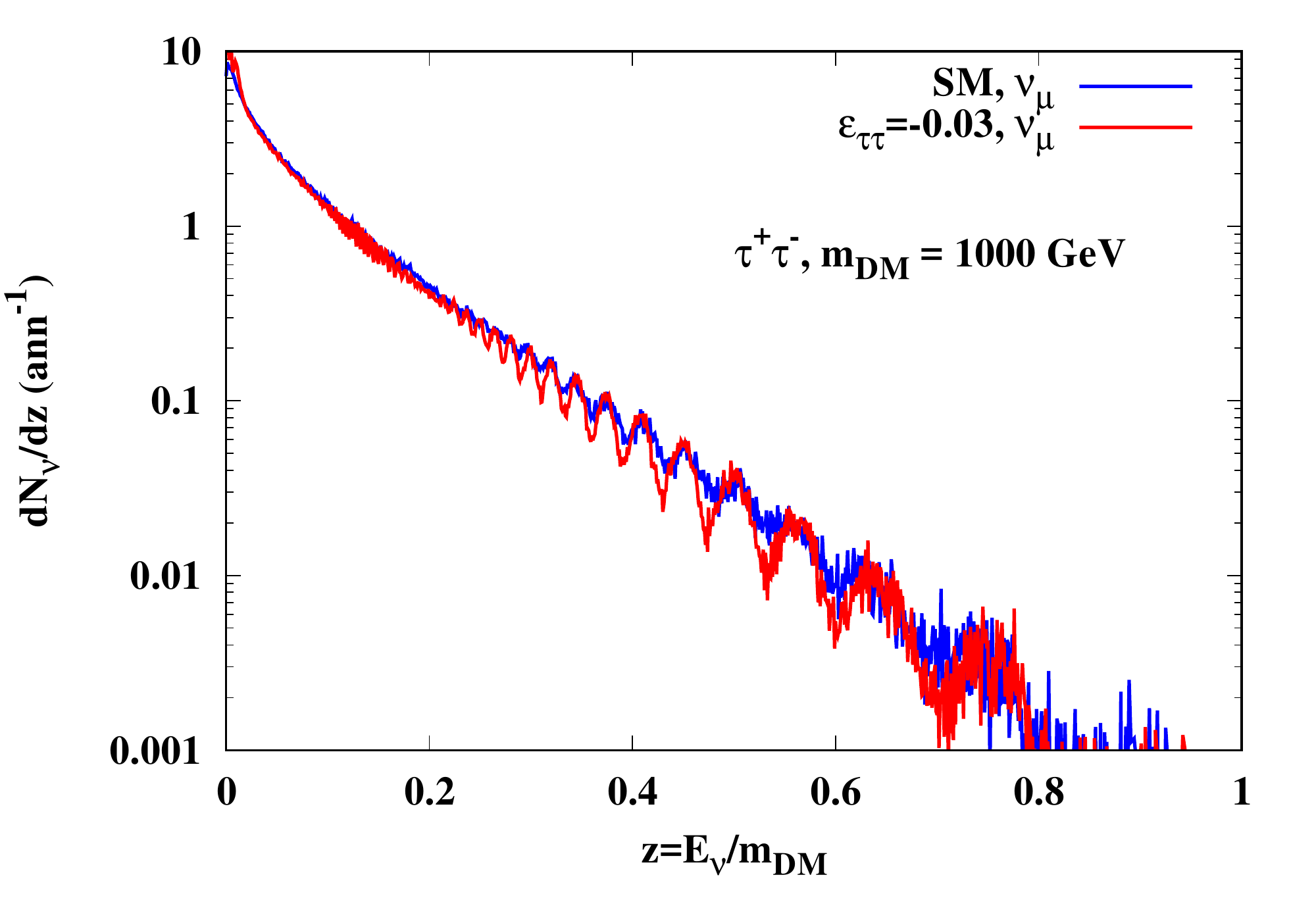}}
\put(130,130){\includegraphics[angle=-90,width=0.31\textwidth]{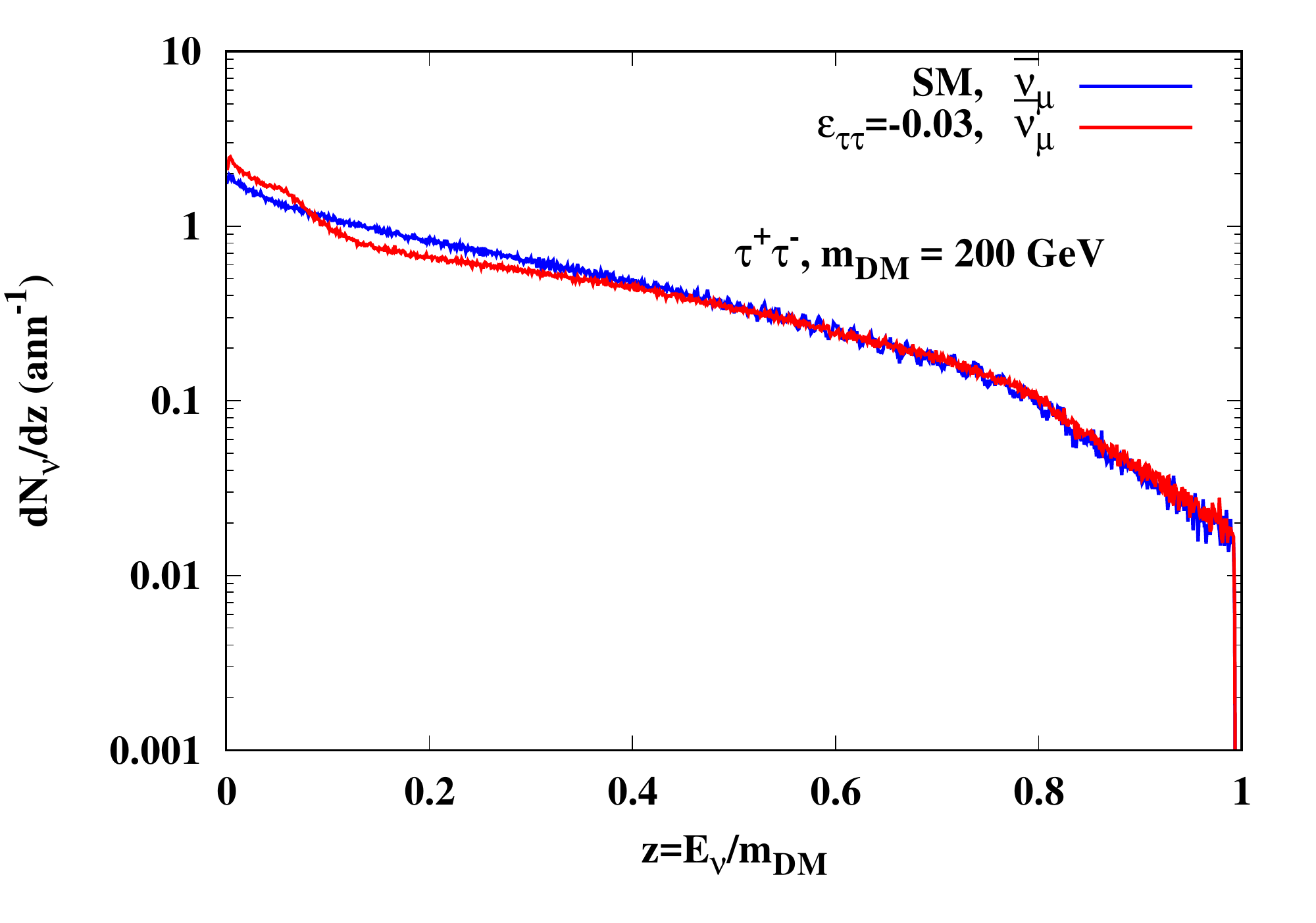}}
\put(130,240){\includegraphics[angle=-90,width=0.31\textwidth]{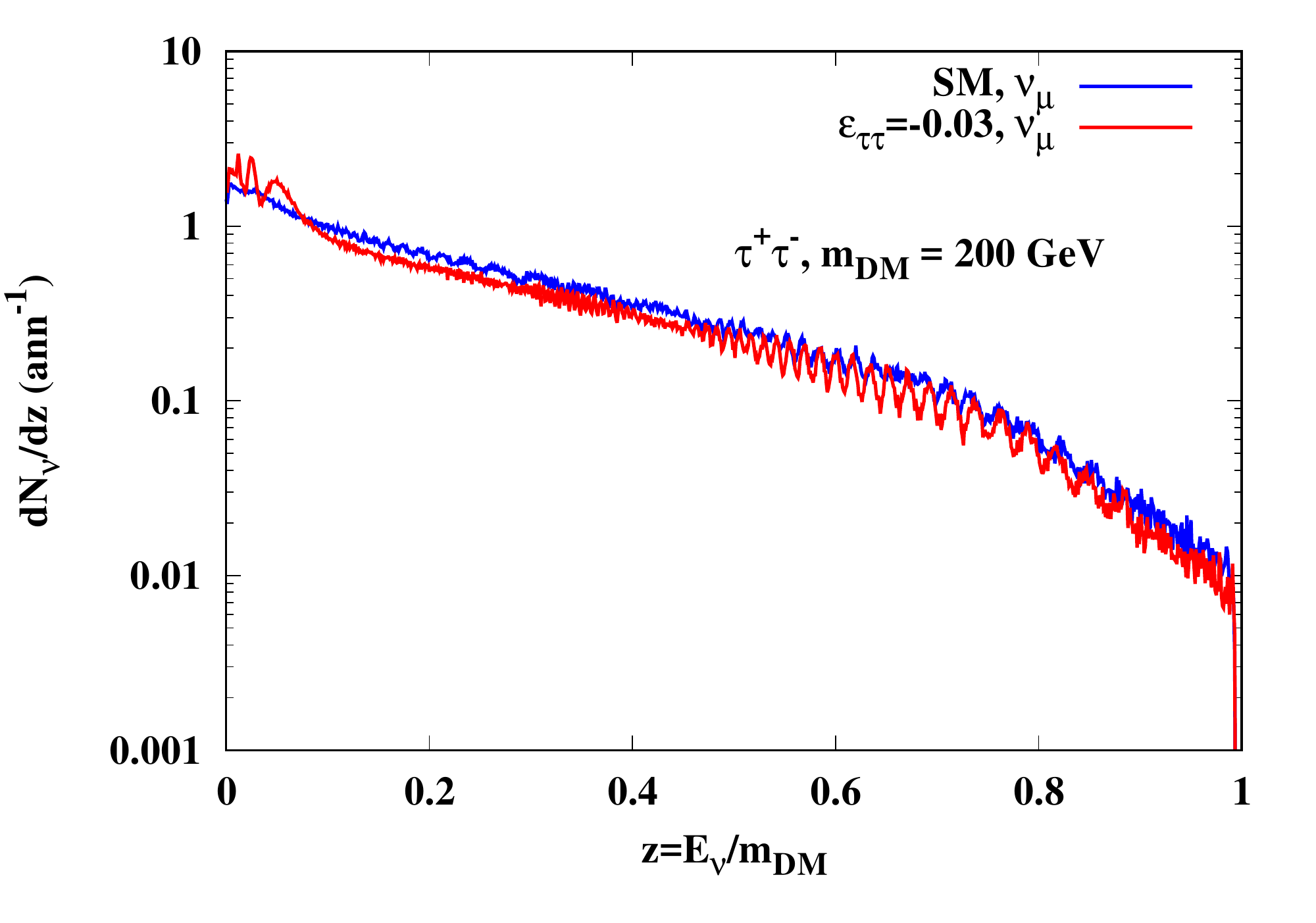}}
\put(0,130){\includegraphics[angle=-90,width=0.31\textwidth]{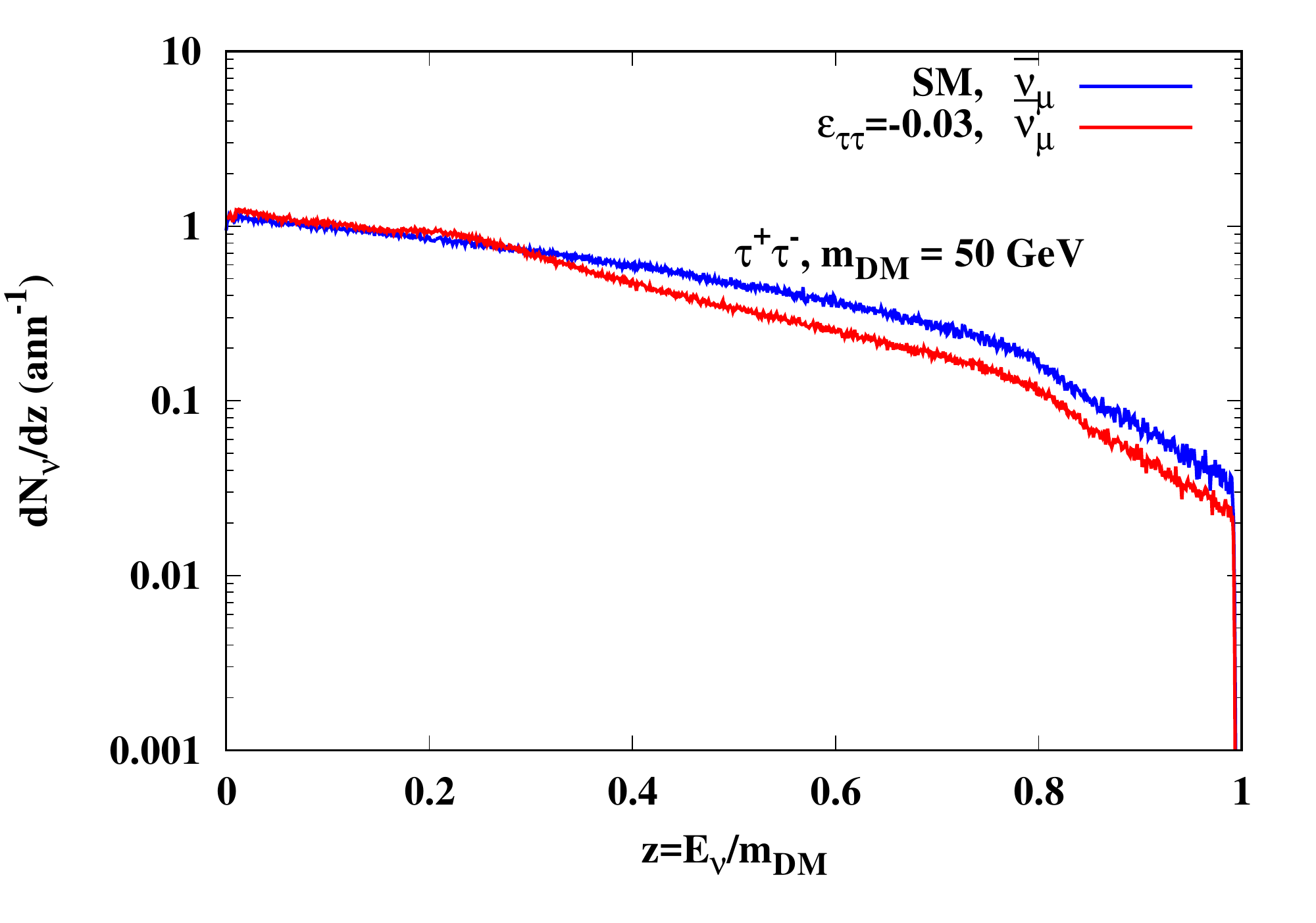}}
\put(0,240){\includegraphics[angle=-90,width=0.31\textwidth]{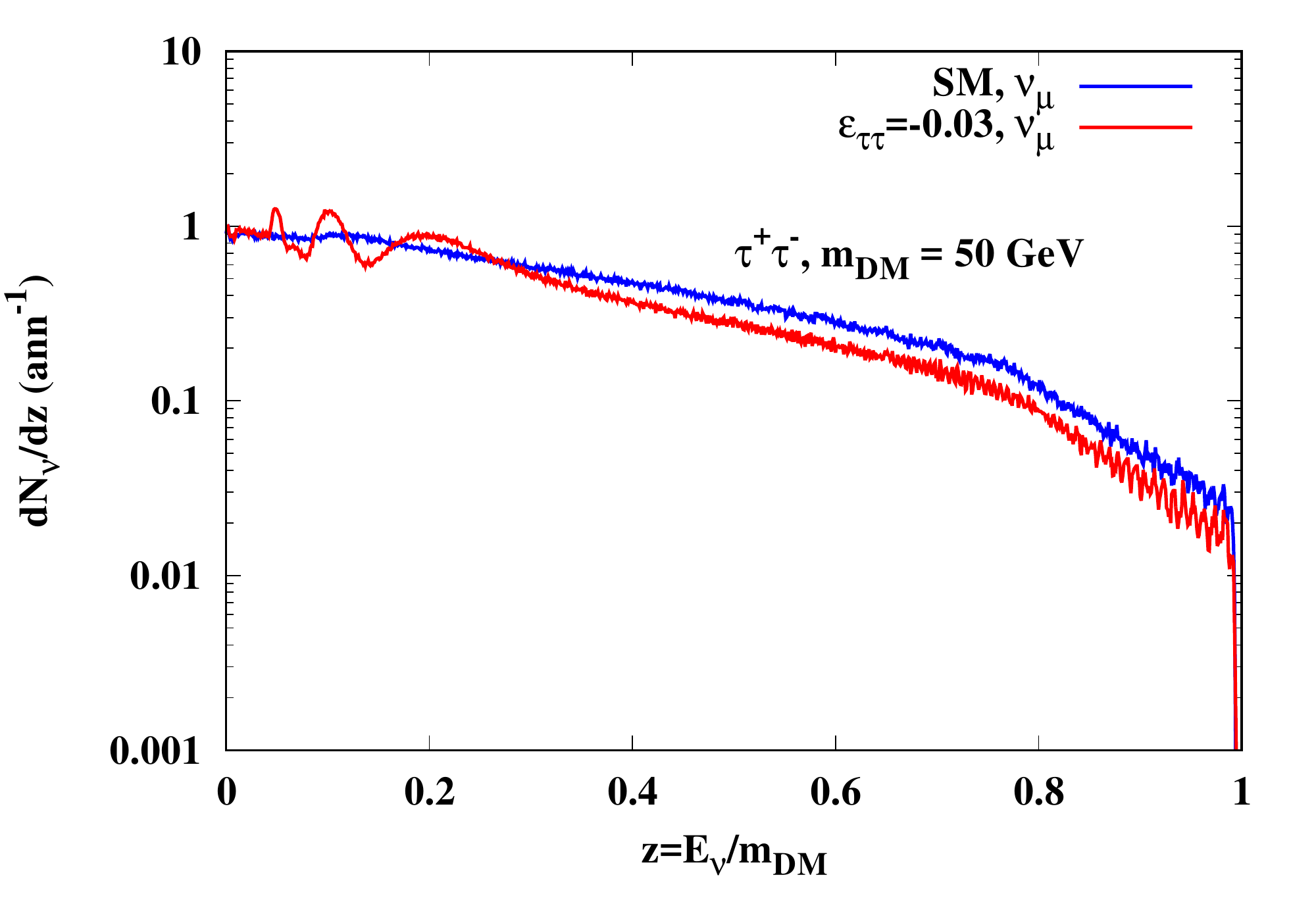}}
\end{picture}
\caption{\label{tau_tautau_sign} The same as in Fig.~\ref{tau_tautau}
  but for $\e_{\tau\tau}=-0.03$ and NH.}
\end{figure}

In the Figures presented below we show in red lines final muon
neutrino and antineutrino energy spectra for $\tau^+\tau^-$
annihilation channel and non-zero NSI parameters. For
comparison we show also muon neutrino energy spectra without NSI in
blue lines. In Fig.~\ref{tau_tautau} and~\ref{tau_tautau_sign} we
present 
muon neutrino and antineutrino energy spectra for $\e_{\tau\tau}=0.03$
and $\e_{\tau\tau}=-0.03$, respectively. Here we choose a
representative set of masses of dark matter particles 50, 200 and
1000~GeV and assume 
normal neutrino mass ordering for simulation of neutrino
oscillations. The spectra are normalized to single act of dark matter 
annihilation per second. As compared to the no-NSI case,  the largest
difference appears for smaller masses of dark matter particles. The
similar energy spectra 
\begin{figure}[!htb]
\begin{picture}(300,190)(0,40)
\put(260,130){\includegraphics[angle=-90,width=0.31\textwidth]{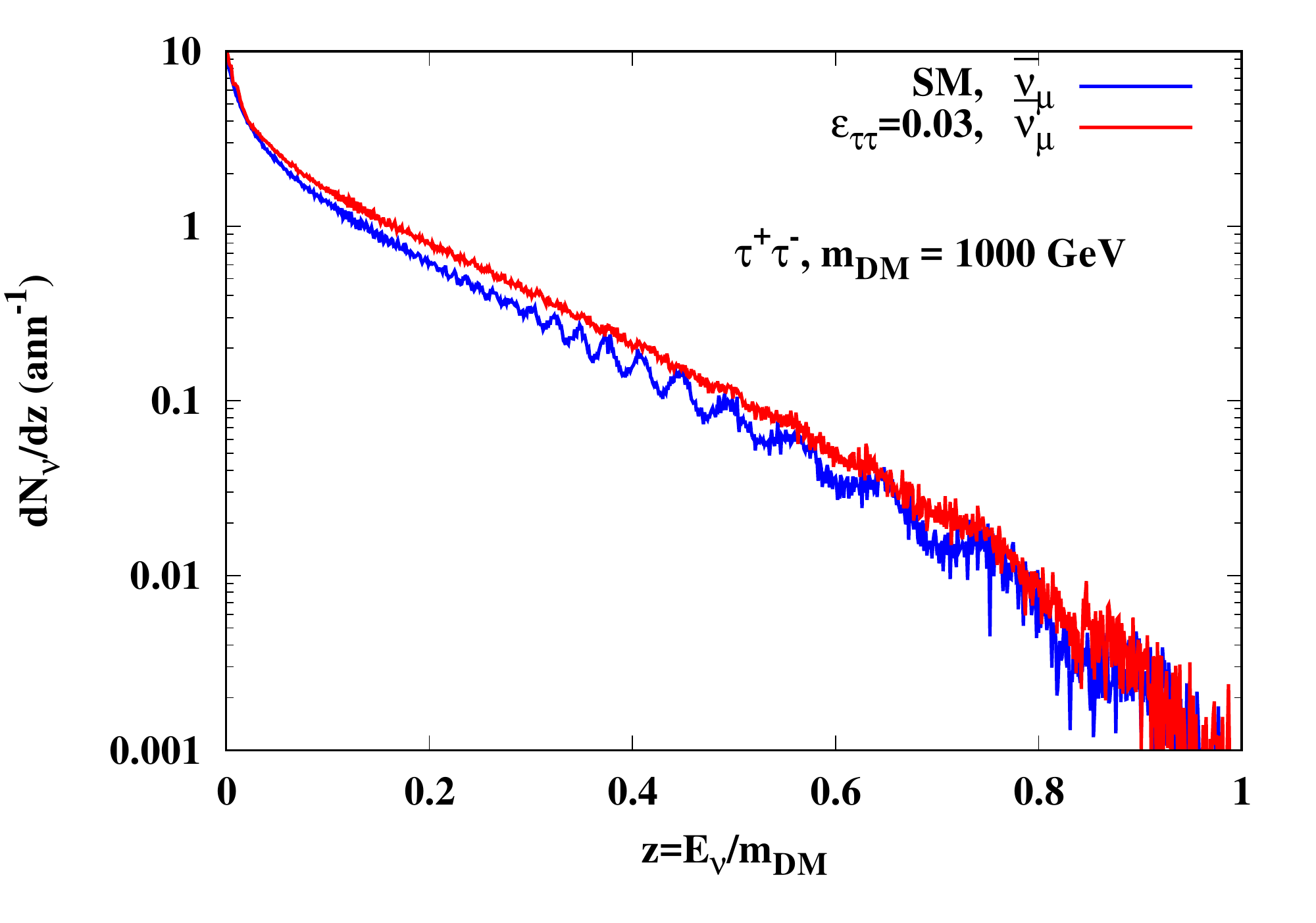}}
\put(260,240){\includegraphics[angle=-90,width=0.31\textwidth]{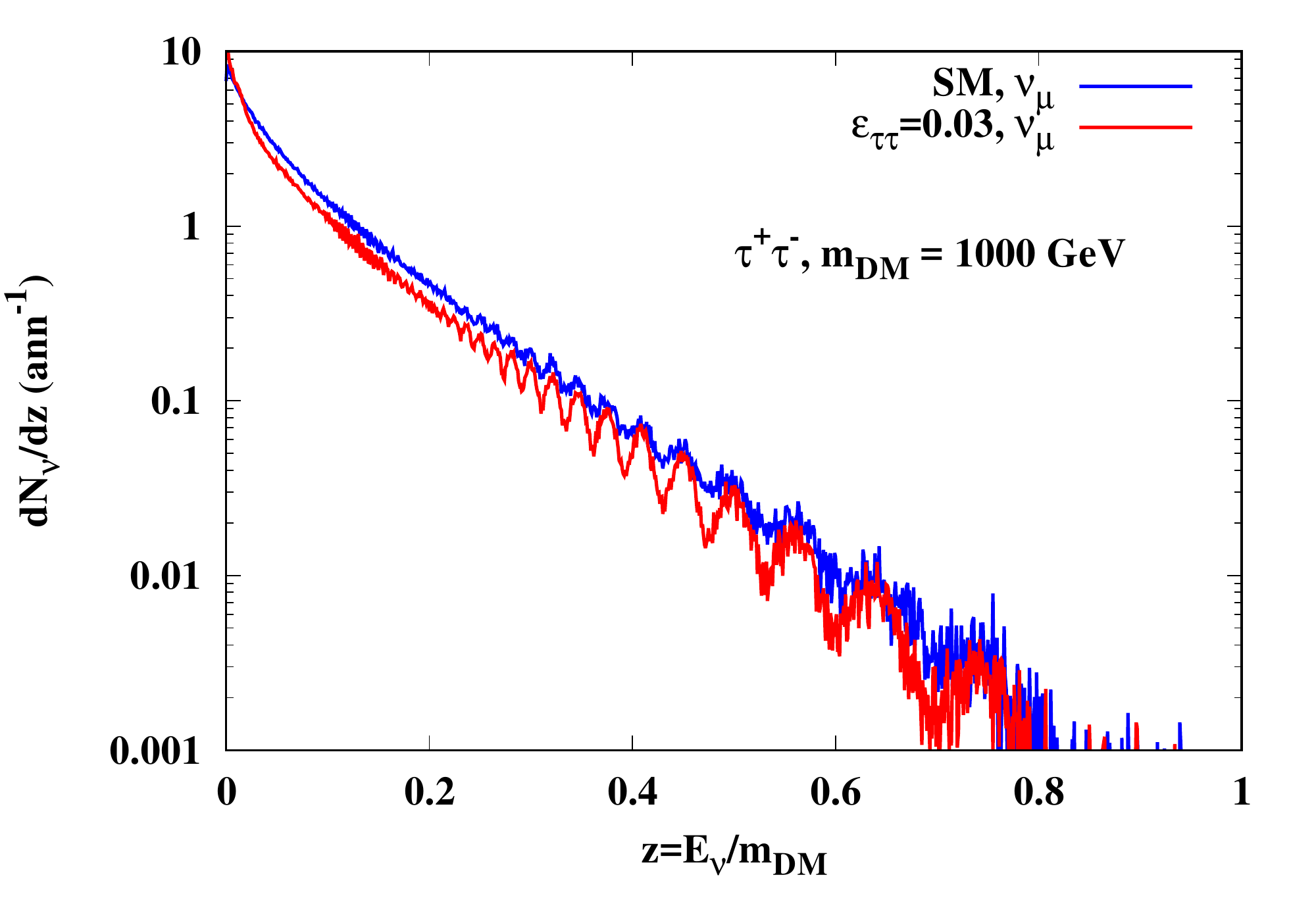}}
\put(130,130){\includegraphics[angle=-90,width=0.31\textwidth]{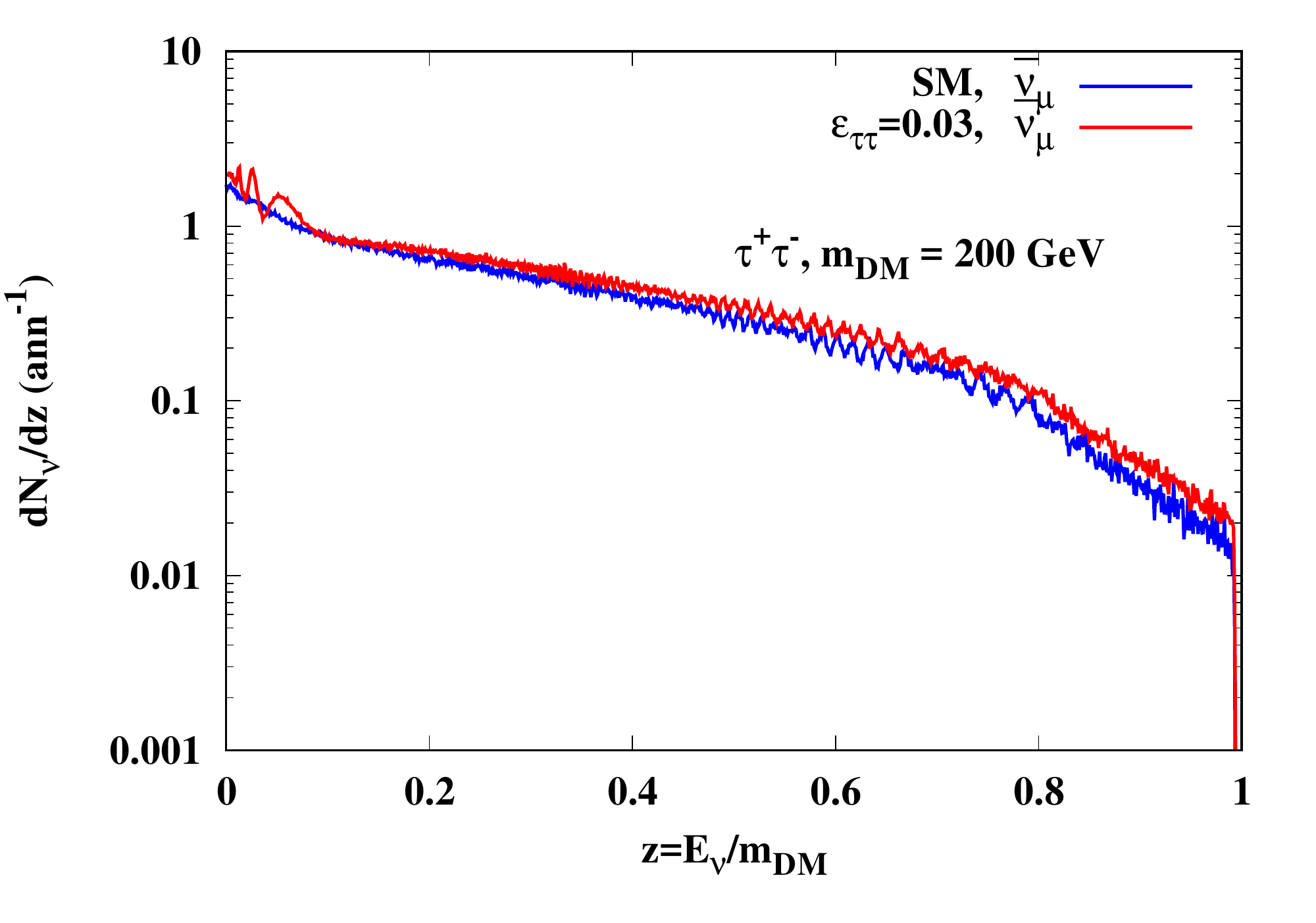}}
\put(130,240){\includegraphics[angle=-90,width=0.31\textwidth]{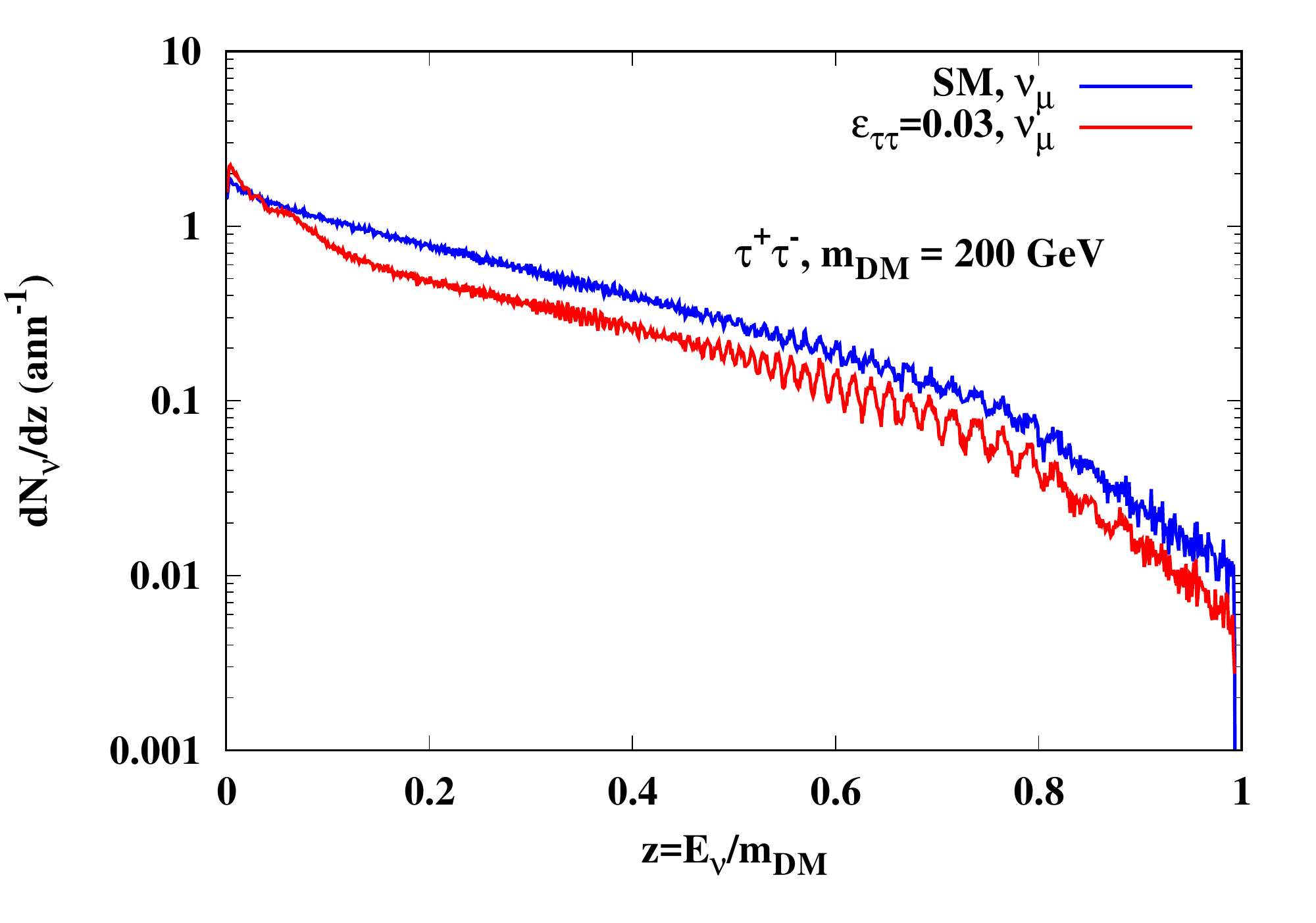}}
\put(0,130){\includegraphics[angle=-90,width=0.31\textwidth]{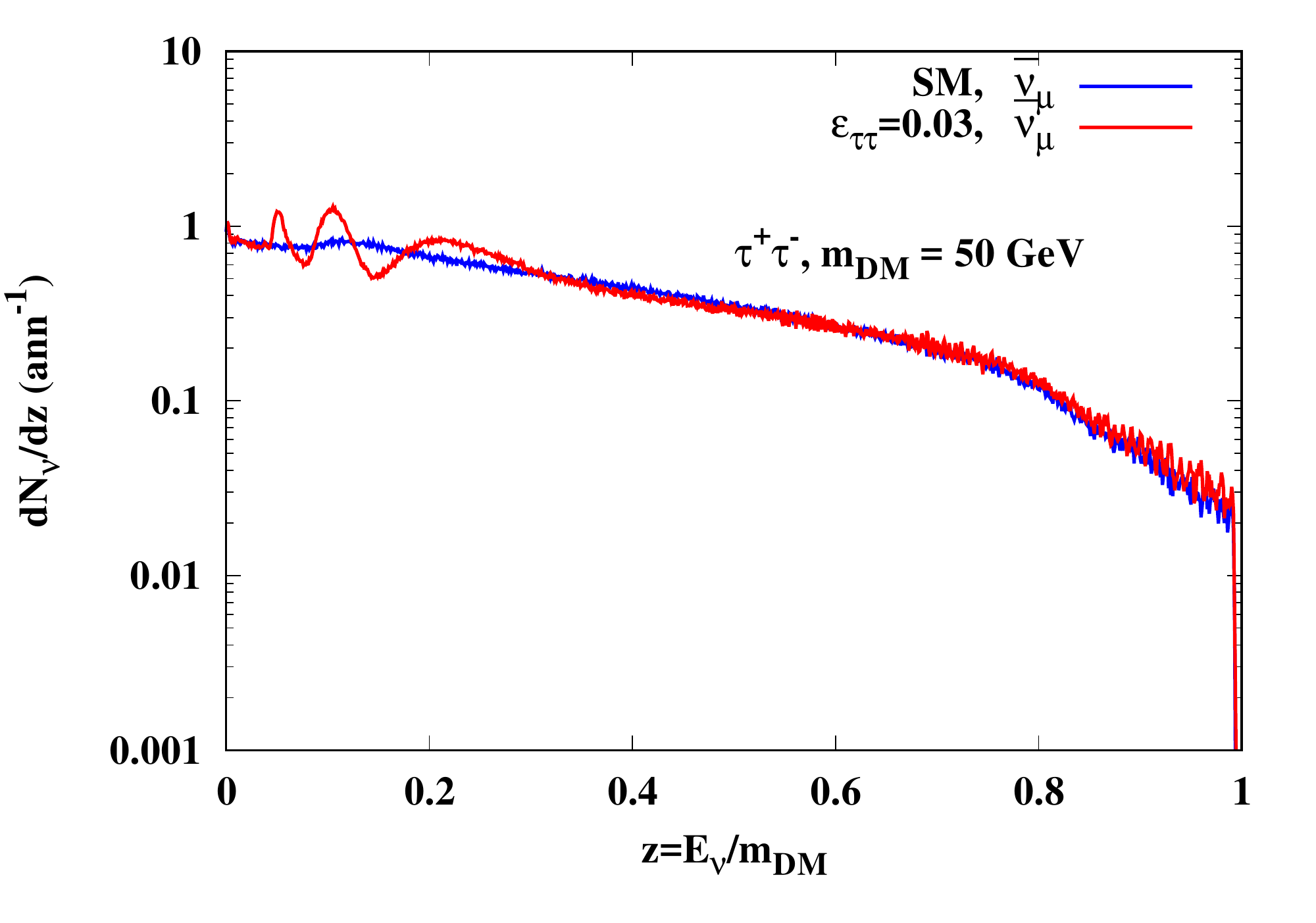}}
\put(0,240){\includegraphics[angle=-90,width=0.31\textwidth]{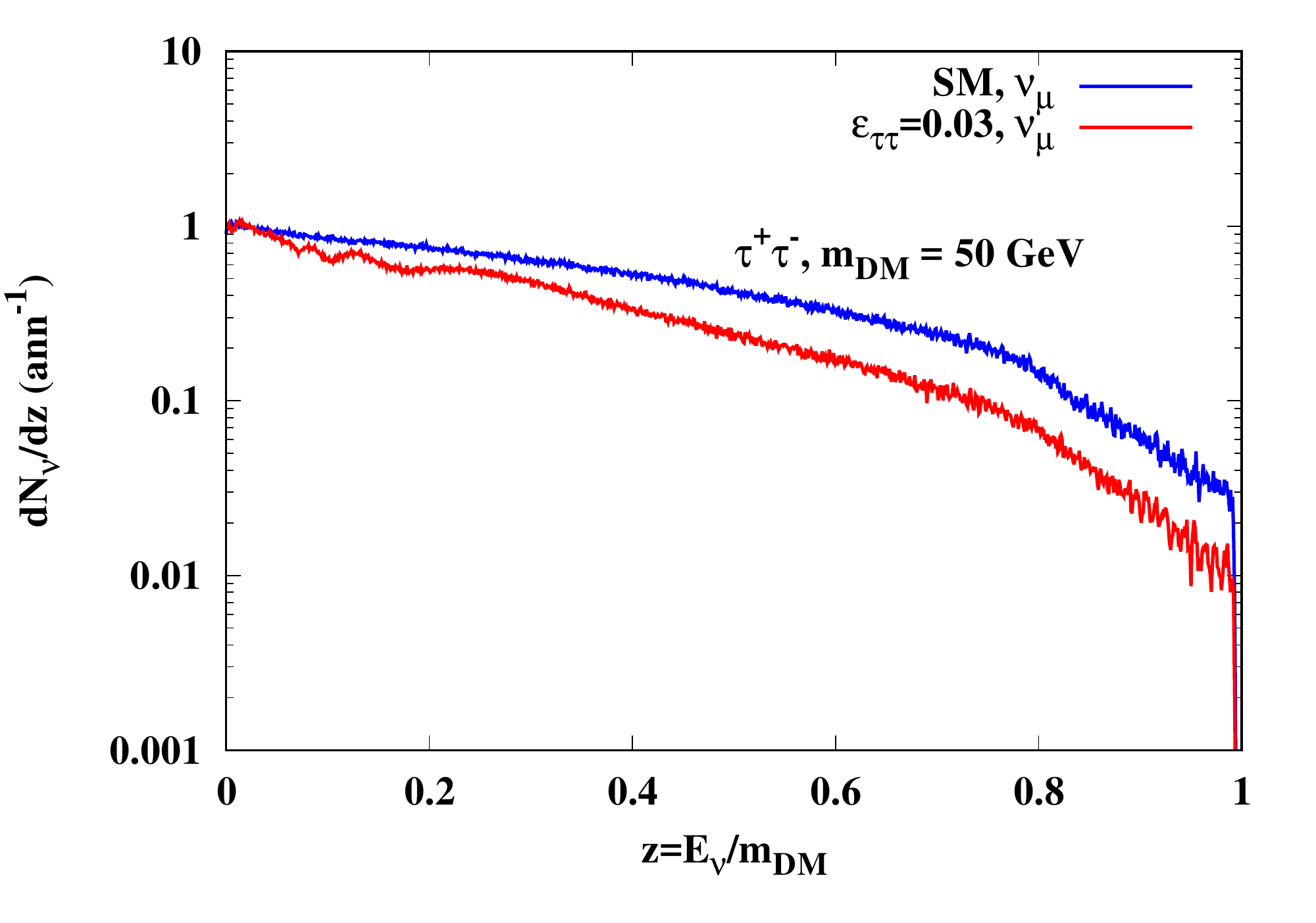}}
\end{picture}
\caption{\label{tau_tautau_inv} The same as in Fig.~\ref{tau_tautau}
  but for $\e_{\tau\tau}=0.03$ and IH.}
\end{figure}
\begin{figure}[!htb]
\begin{picture}(300,190)(0,40)
\put(260,130){\includegraphics[angle=-90,width=0.31\textwidth]{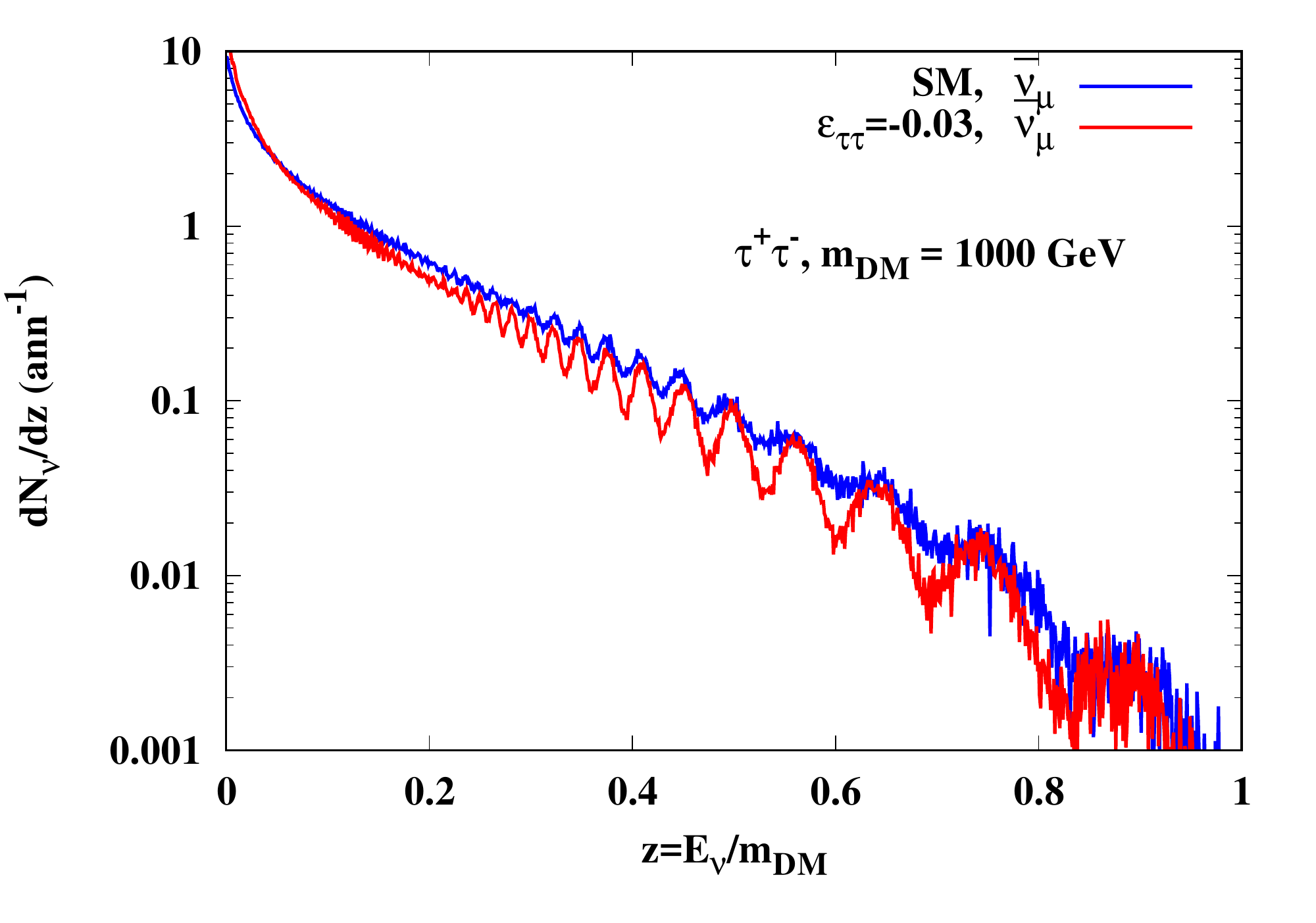}}
\put(260,240){\includegraphics[angle=-90,width=0.31\textwidth]{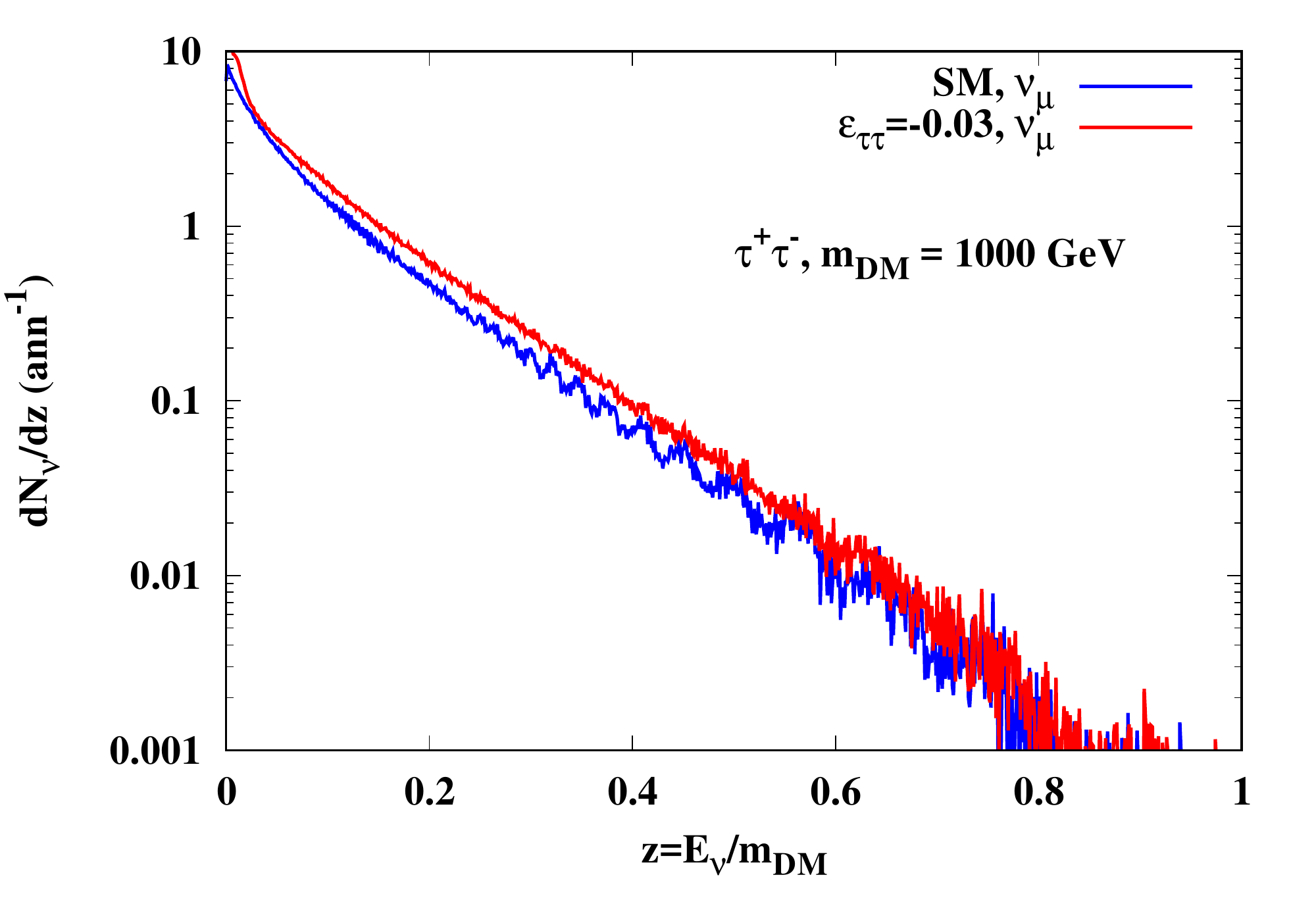}}
\put(130,130){\includegraphics[angle=-90,width=0.31\textwidth]{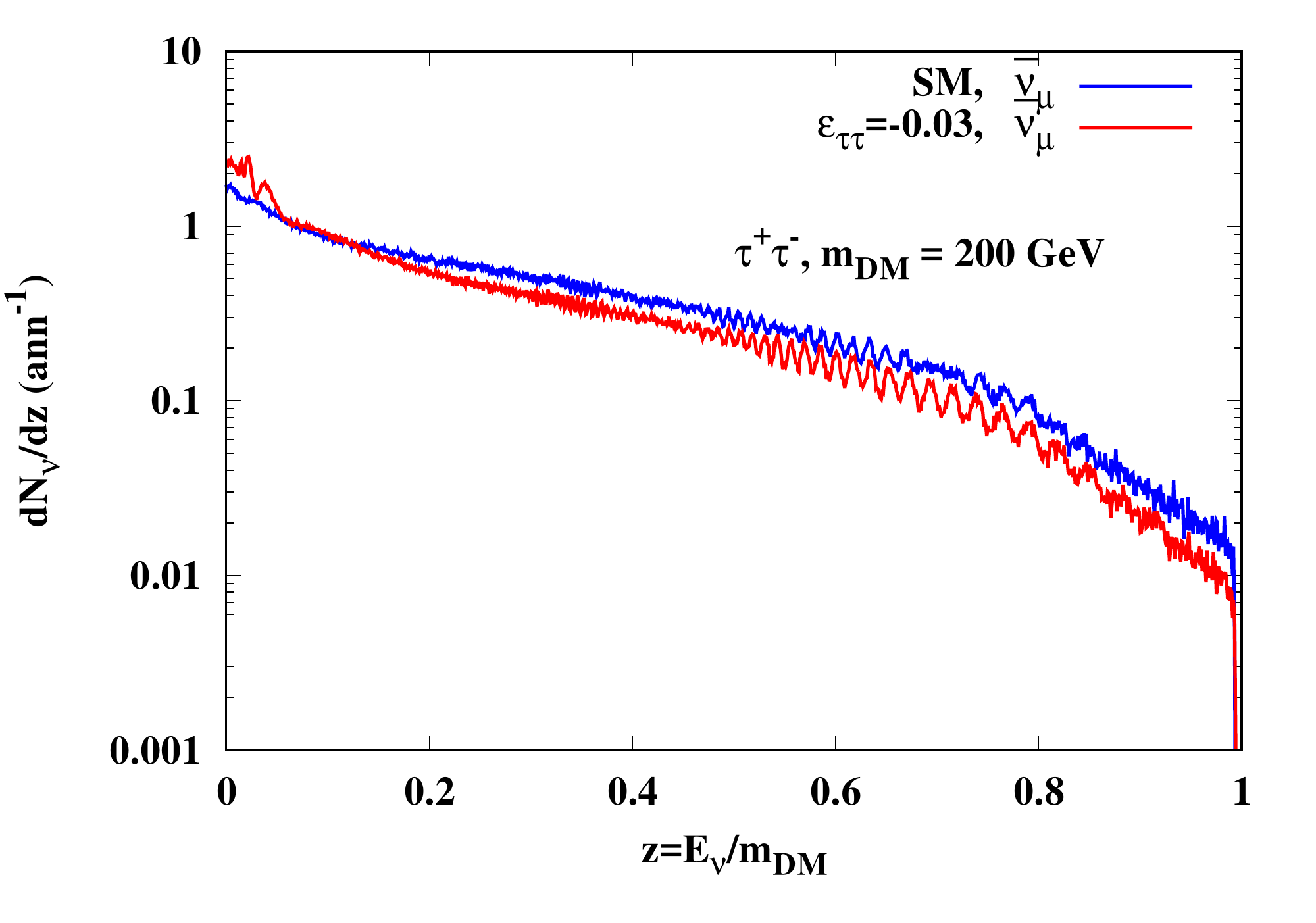}}
\put(130,240){\includegraphics[angle=-90,width=0.31\textwidth]{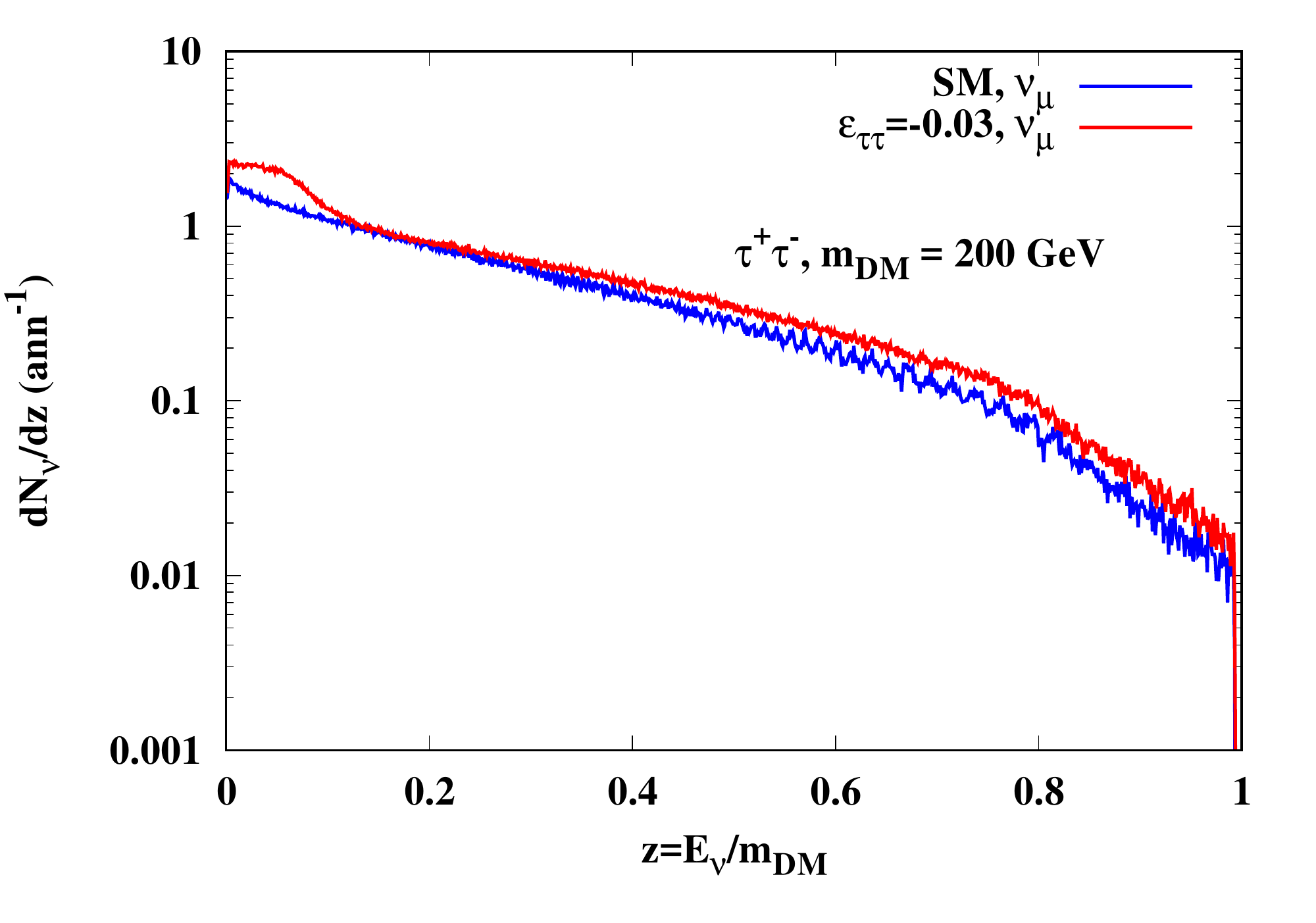}}
\put(0,130){\includegraphics[angle=-90,width=0.31\textwidth]{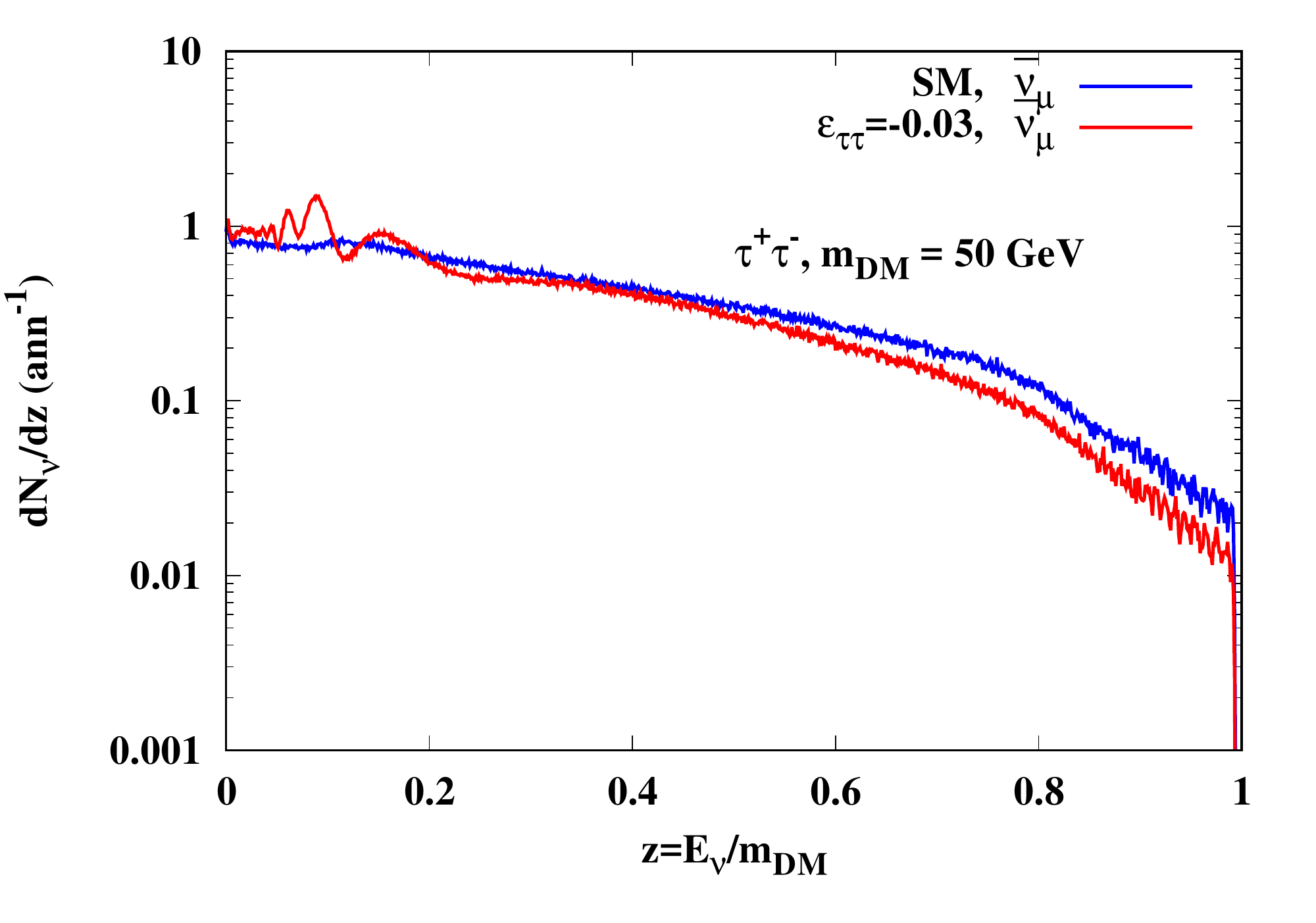}}
\put(0,240){\includegraphics[angle=-90,width=0.31\textwidth]{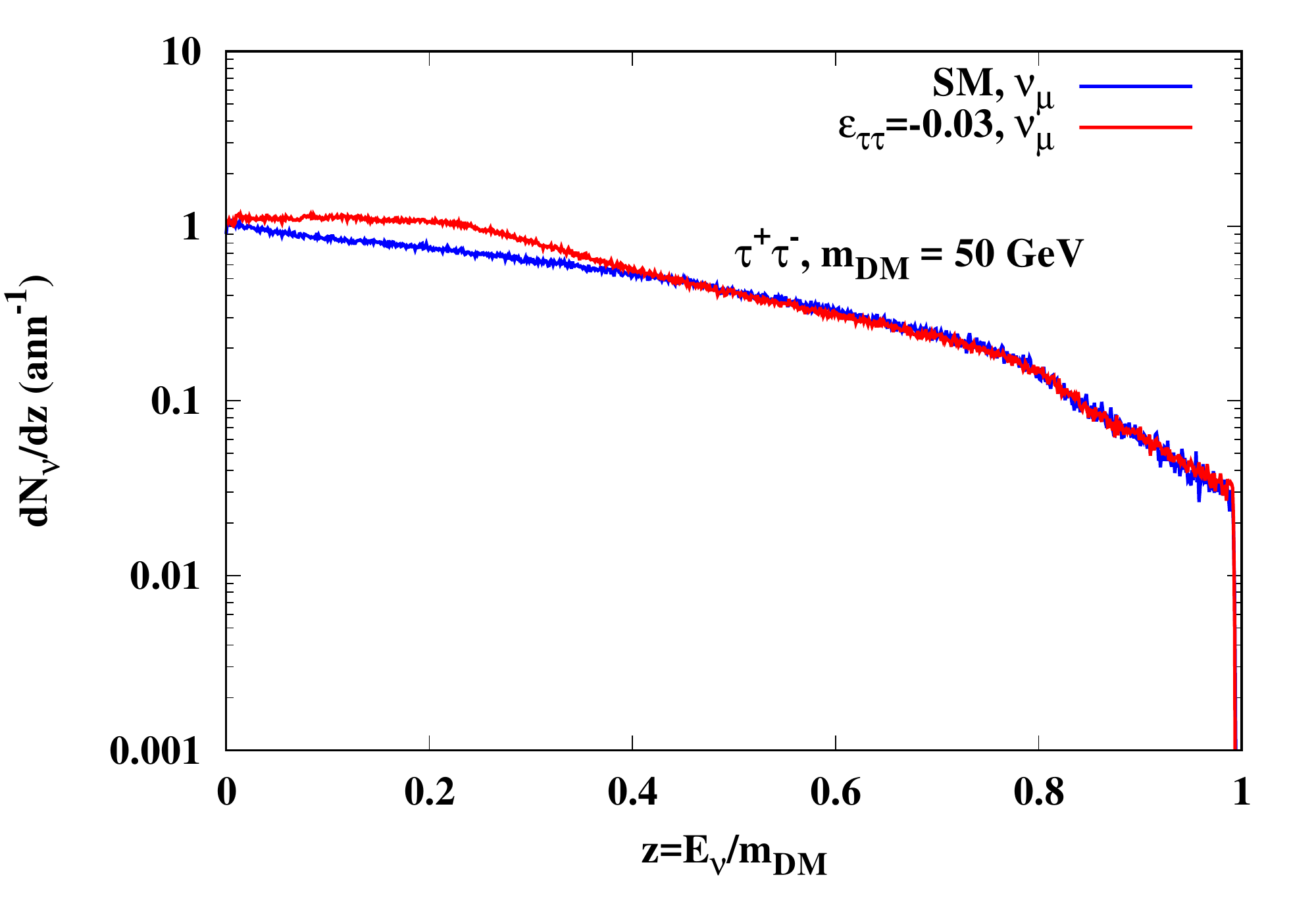}}
\end{picture}
\caption{\label{tau_tautau_inv_sign} The same as in Fig.~\ref{tau_tautau}
  but for $\e_{\tau\tau} = -0.03$ and IH.}
\end{figure}
but for inverted mass ordering are shown in Figs.~\ref{tau_tautau_inv}
and~\ref{tau_tautau_inv_sign}. Here the difference in the spectra is
somewhat larger and considerable even for $m_{DM}=1000$~GeV.
The case of $\e_{e\tau}=\pm0.4$ presented in Figs.~\ref{tau_etau}
and~\ref{tau_etau_sign} 
\begin{figure}[!htb]
\begin{picture}(300,190)(0,40)
\put(260,130){\includegraphics[angle=-90,width=0.31\textwidth]{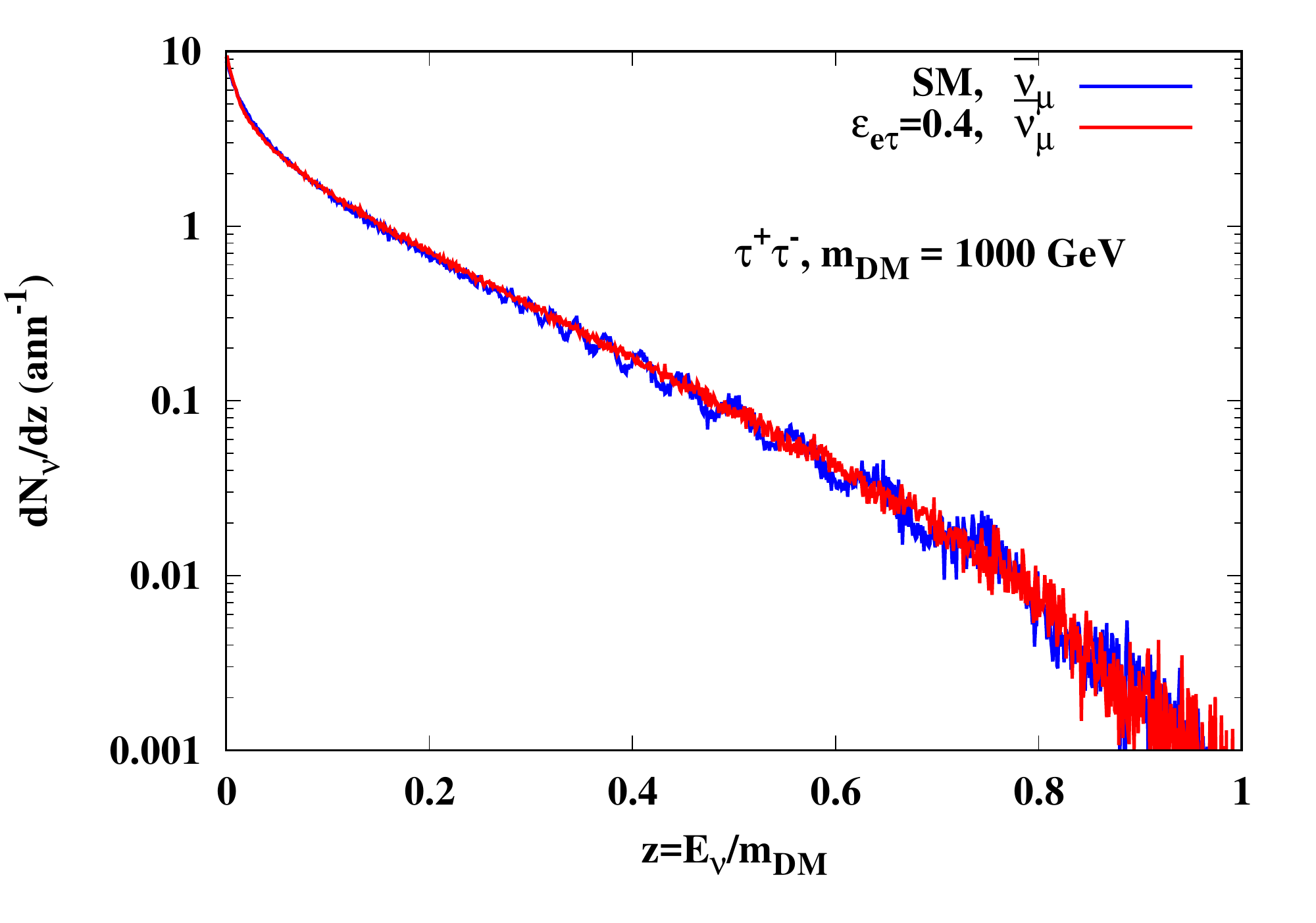}}
\put(260,240){\includegraphics[angle=-90,width=0.31\textwidth]{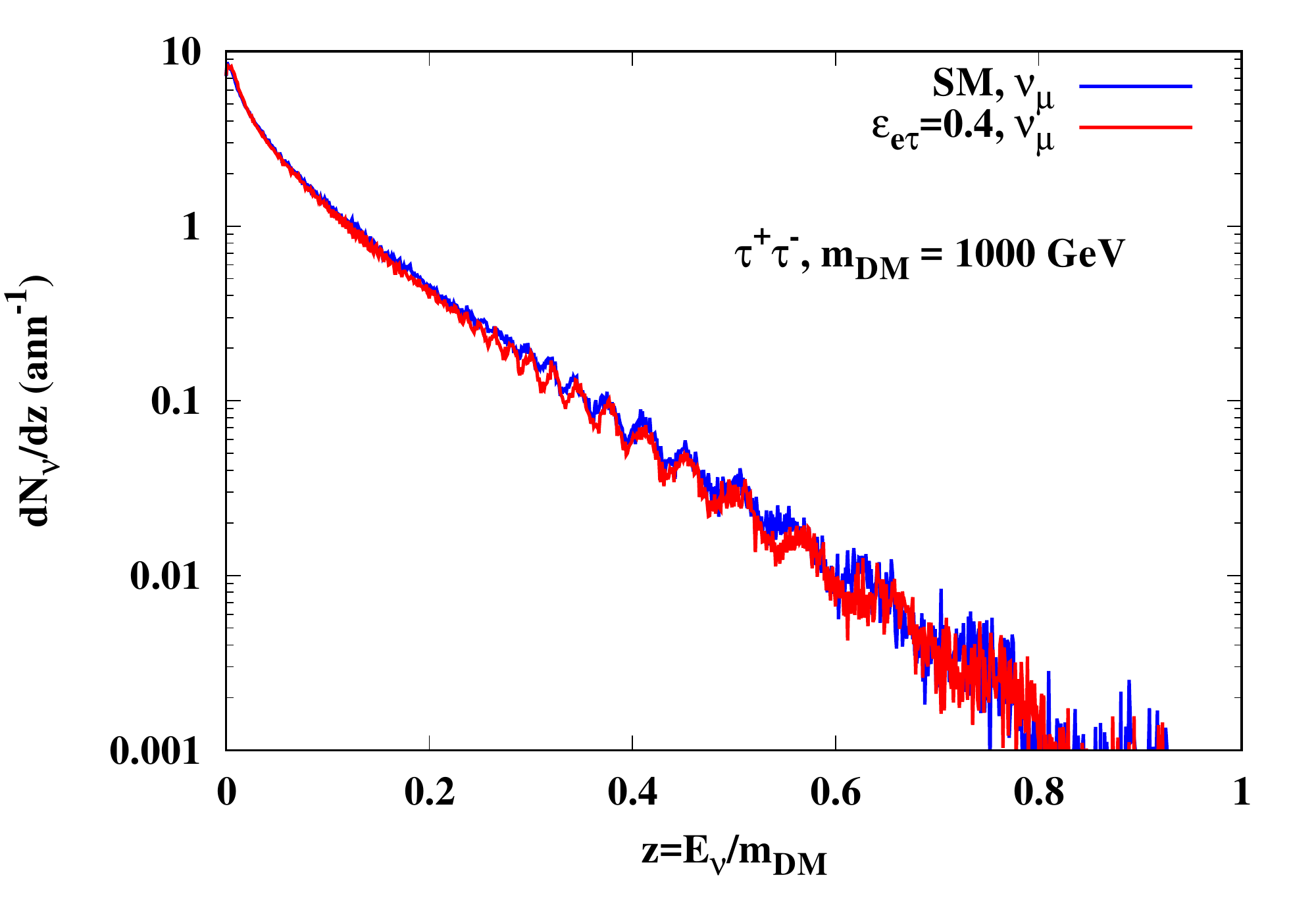}}
\put(130,130){\includegraphics[angle=-90,width=0.31\textwidth]{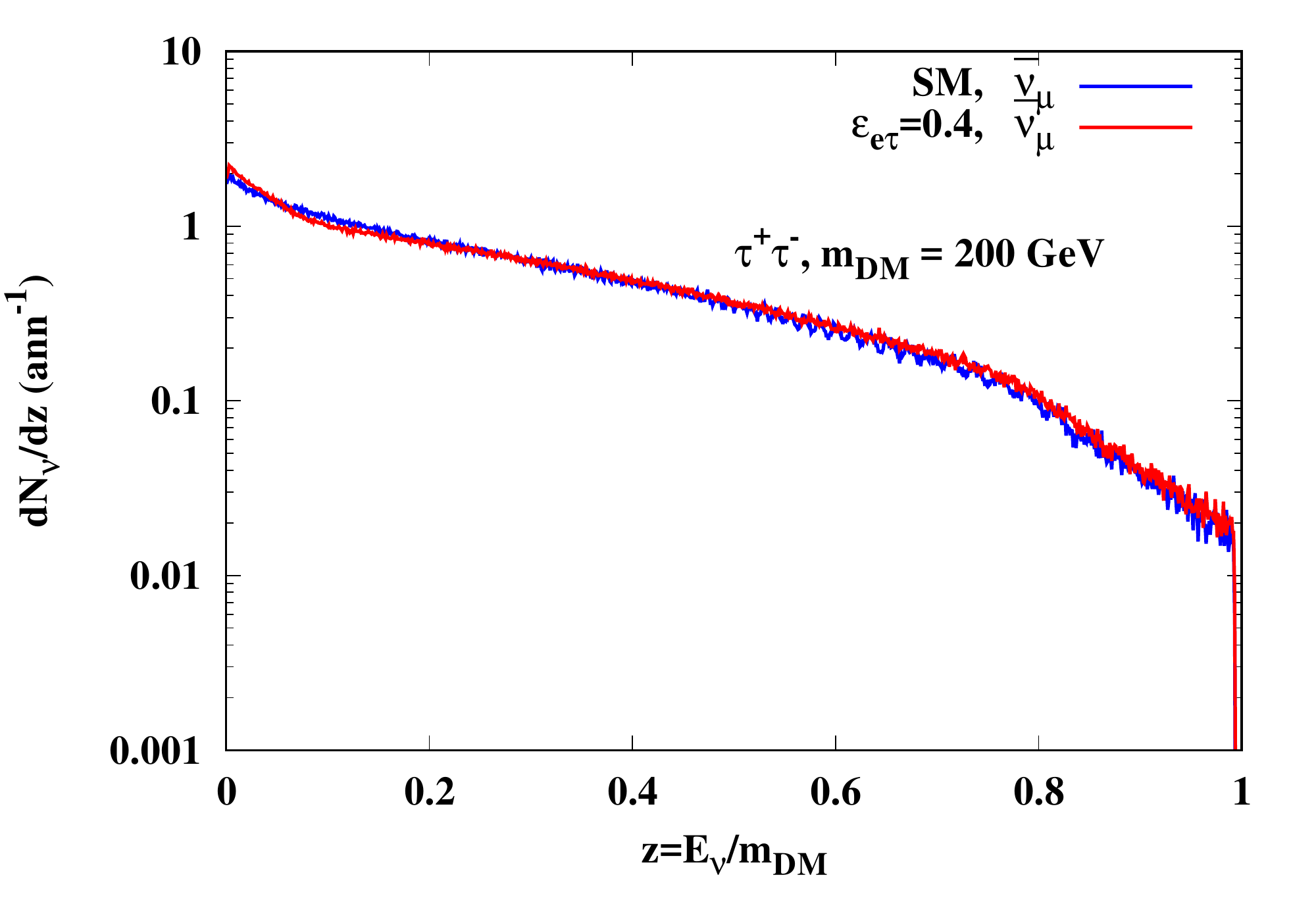}}
\put(130,240){\includegraphics[angle=-90,width=0.31\textwidth]{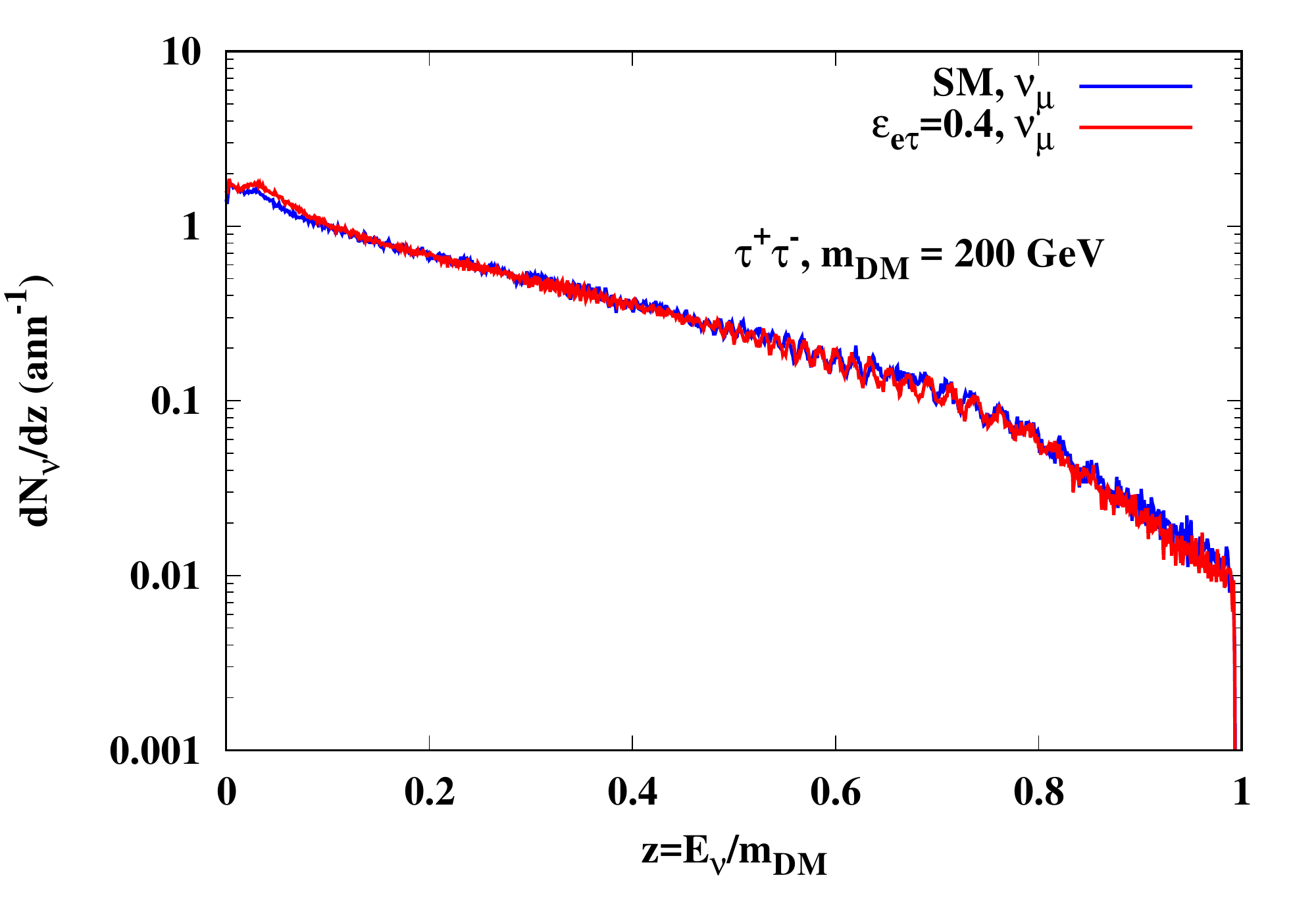}}
\put(0,130){\includegraphics[angle=-90,width=0.31\textwidth]{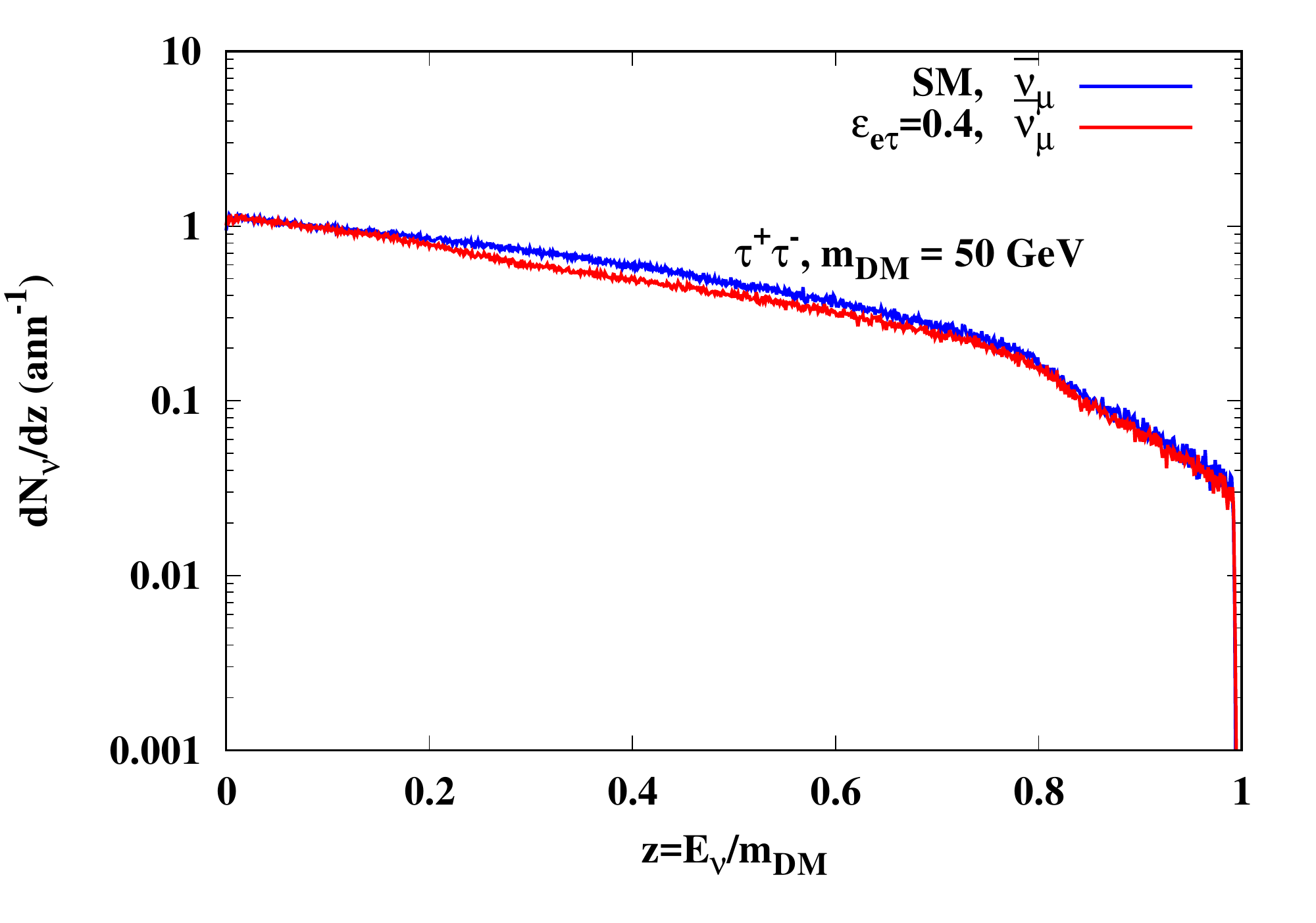}}
\put(0,240){\includegraphics[angle=-90,width=0.31\textwidth]{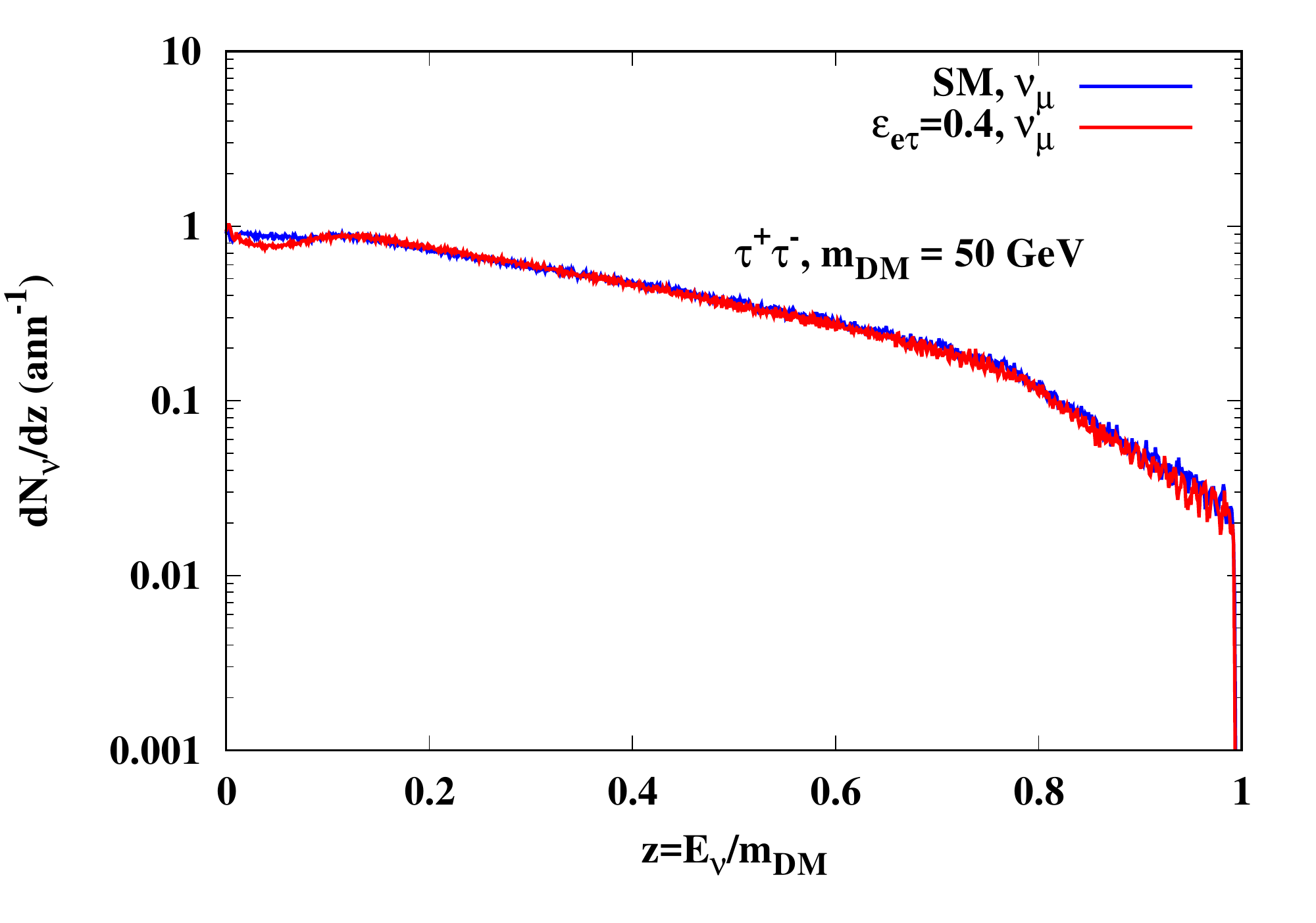}}
\end{picture}
\caption{\label{tau_etau} The same as in Fig.~\ref{tau_tautau}
  but for $\e_{e\tau} = 0.4$ and NH.}
\end{figure}
\begin{figure}[!htb]
\begin{picture}(300,190)(0,40)
\put(260,130){\includegraphics[angle=-90,width=0.31\textwidth]{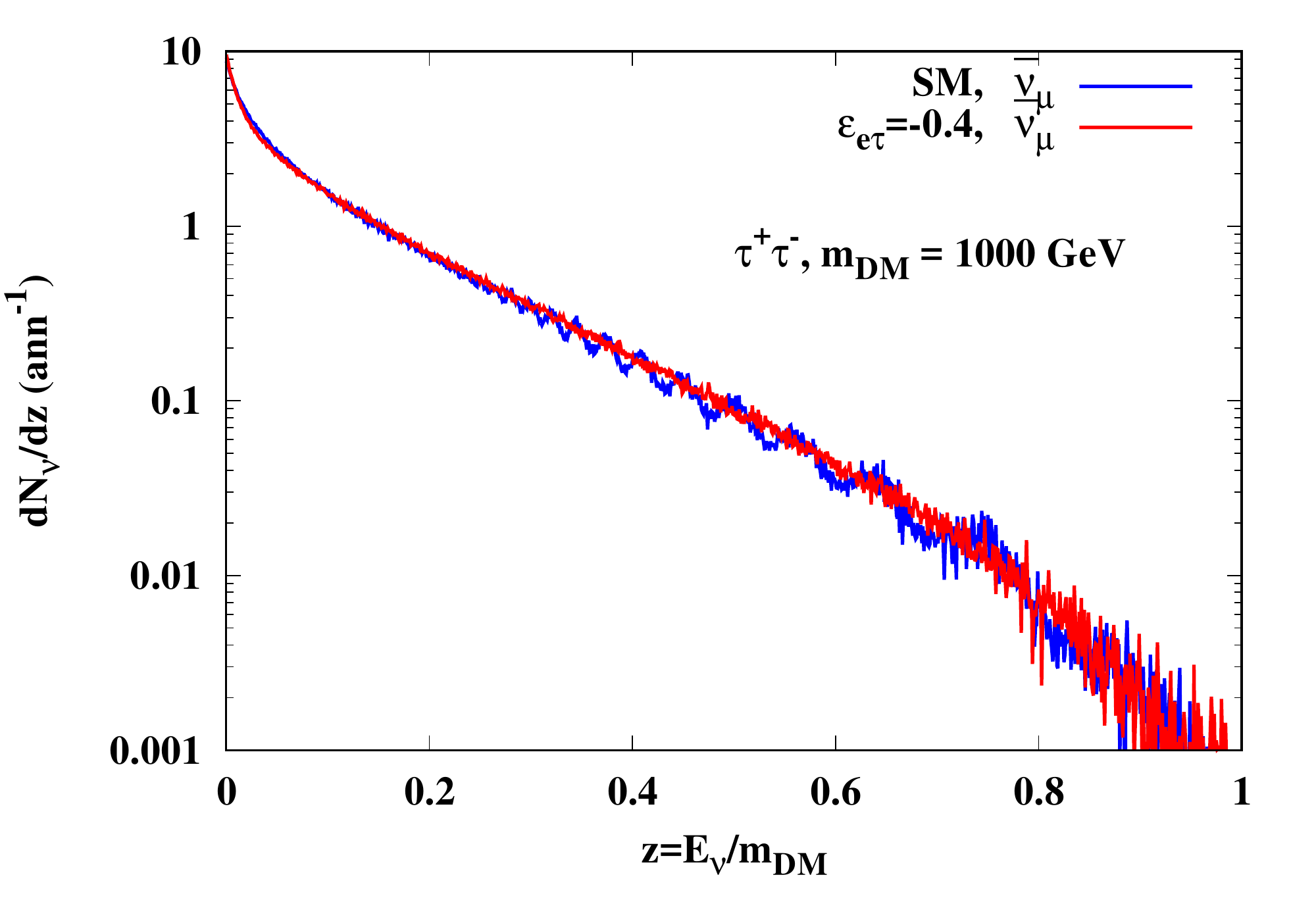}}
\put(260,240){\includegraphics[angle=-90,width=0.31\textwidth]{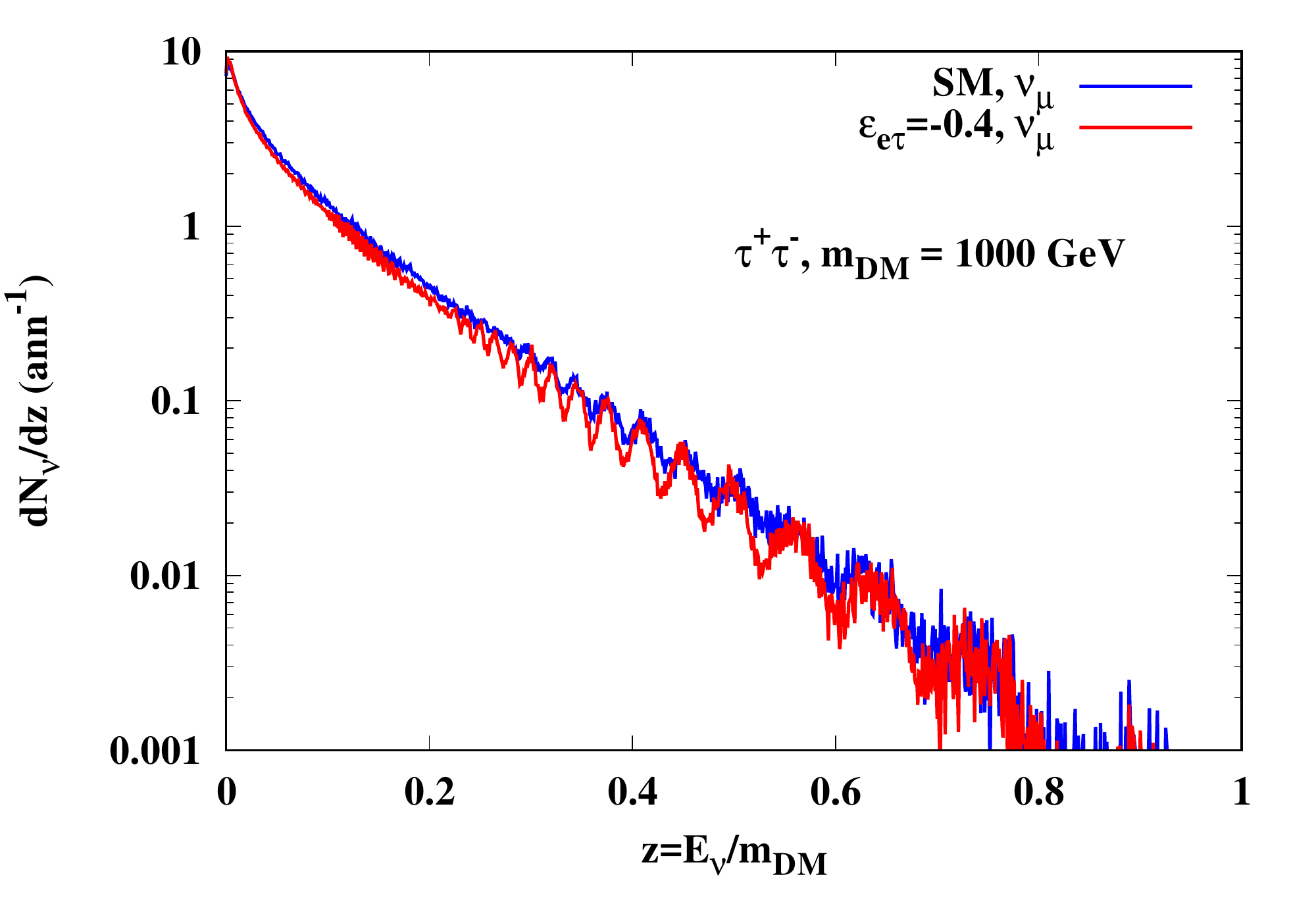}}
\put(130,130){\includegraphics[angle=-90,width=0.31\textwidth]{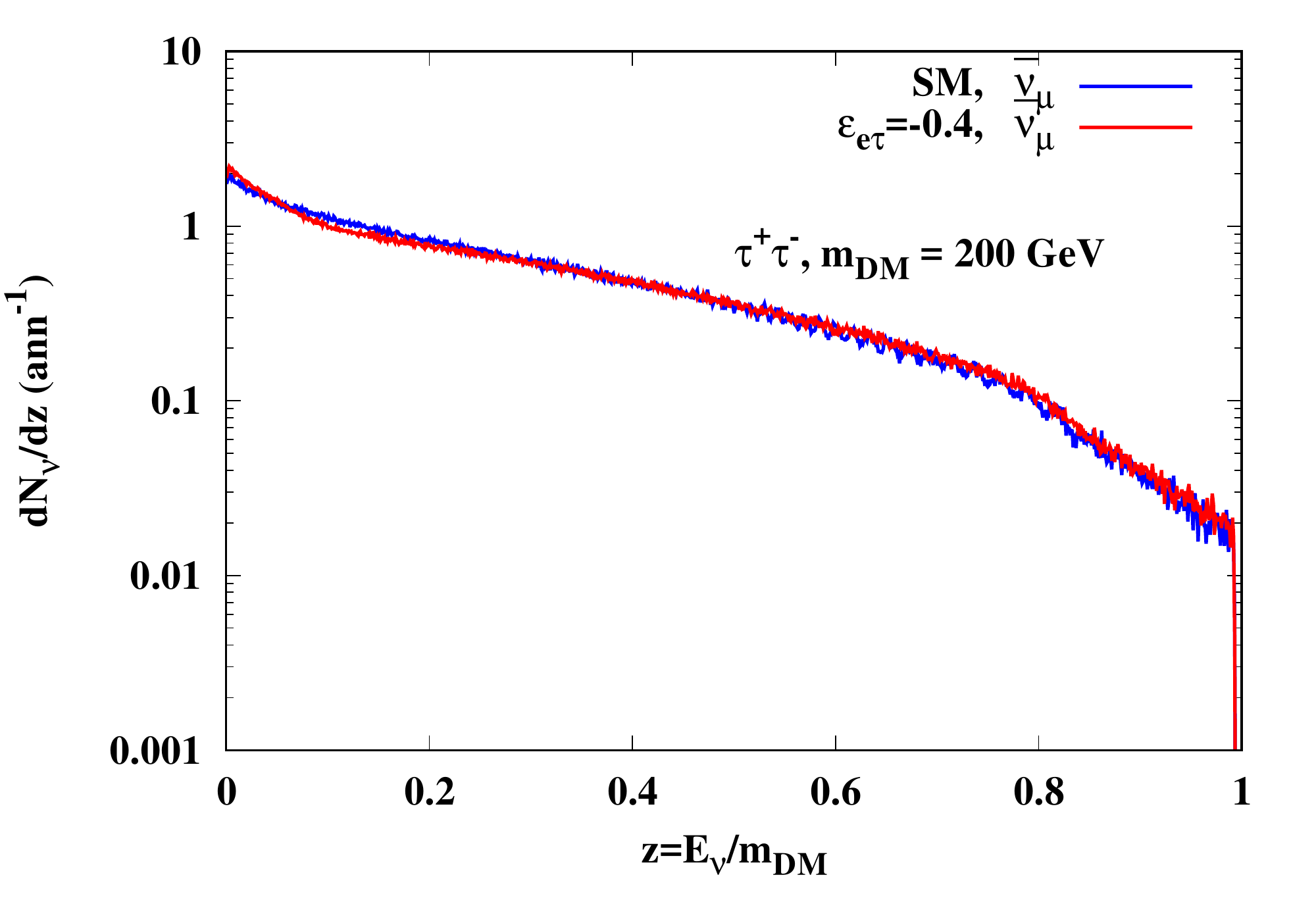}}
\put(130,240){\includegraphics[angle=-90,width=0.31\textwidth]{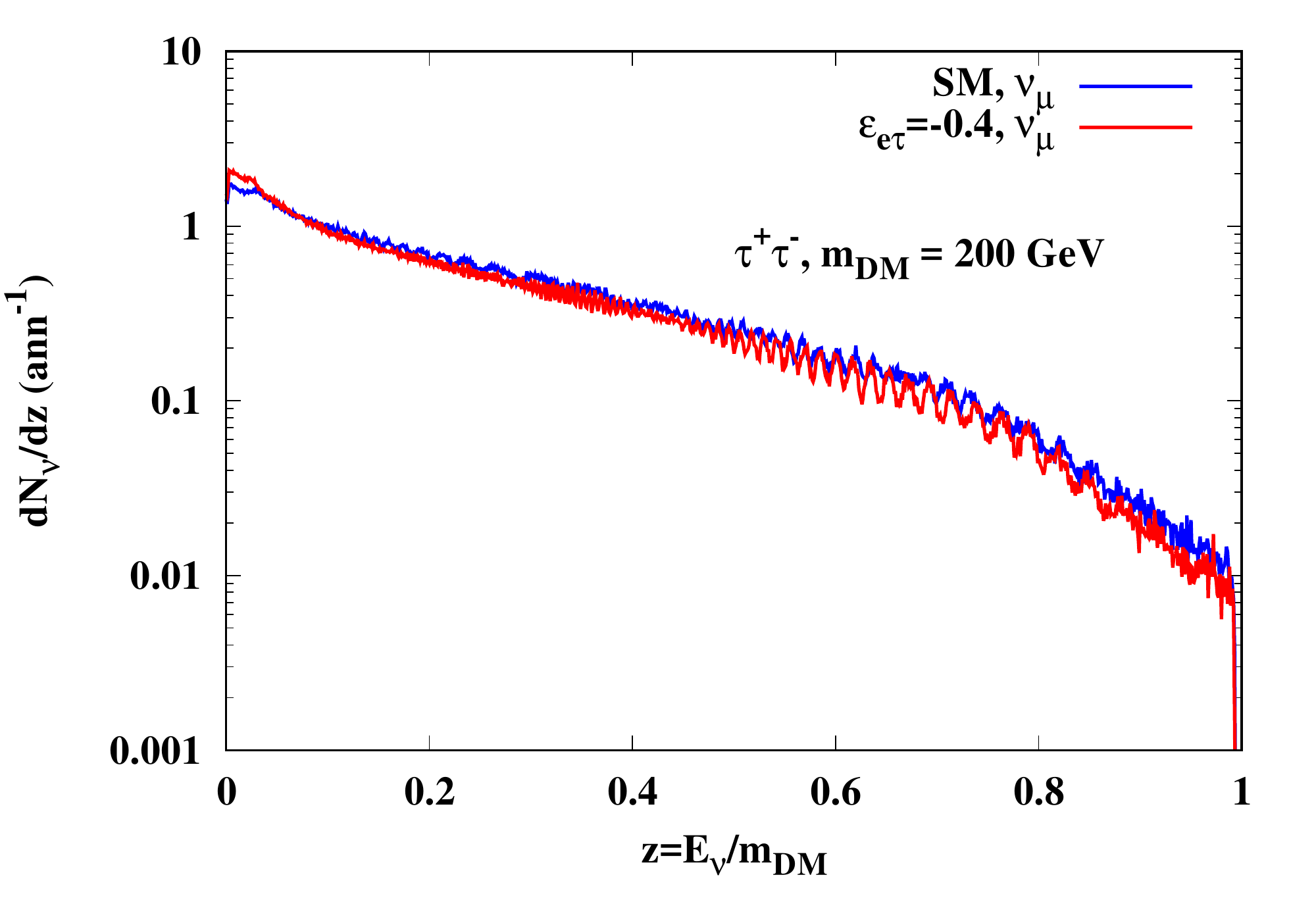}}
\put(0,130){\includegraphics[angle=-90,width=0.31\textwidth]{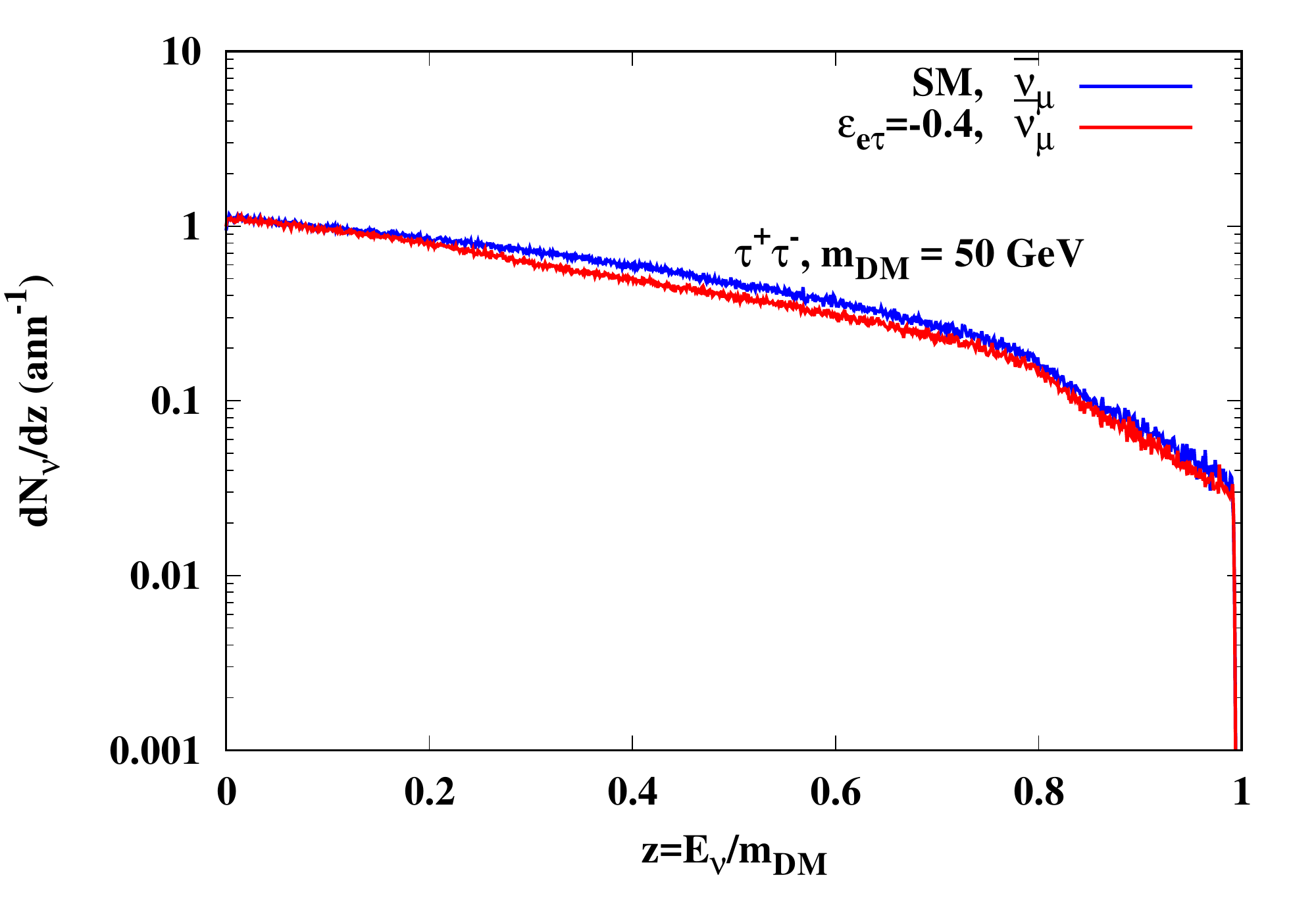}}
\put(0,240){\includegraphics[angle=-90,width=0.31\textwidth]{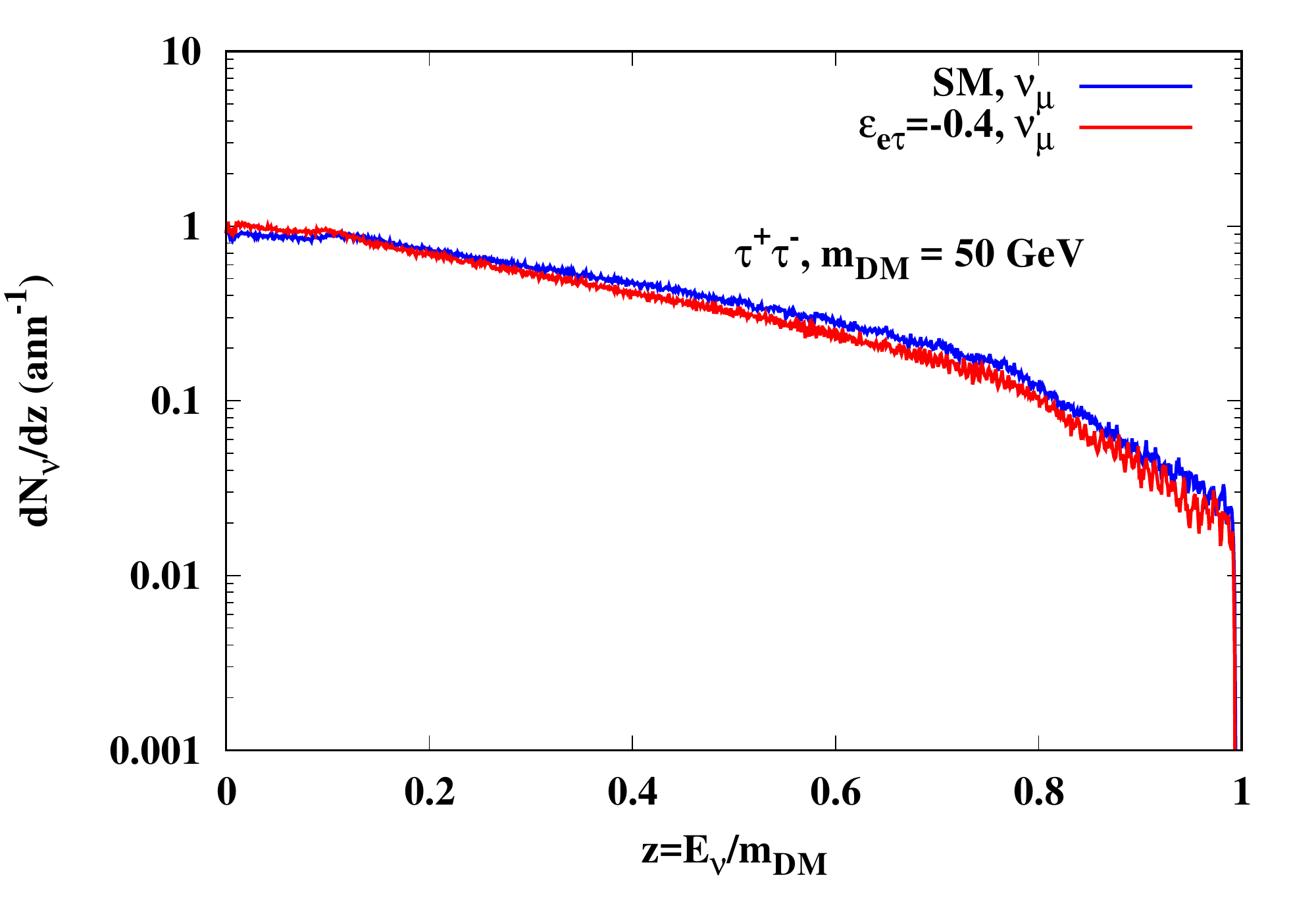}}
\end{picture}
\caption{\label{tau_etau_sign}The same as in Fig.~\ref{tau_tautau}
  but for $\e_{e\tau} = -0.4$ and NH.}
\end{figure}
assuming normal mass ordering. 
The same case but with inverted mass ordering is shown in
Figs.~\ref{tau_etau_inv} and~\ref{tau_etau_inv_sign}.
\begin{figure}[!htb]
\begin{picture}(300,190)(0,40)
\put(260,130){\includegraphics[angle=-90,width=0.31\textwidth]{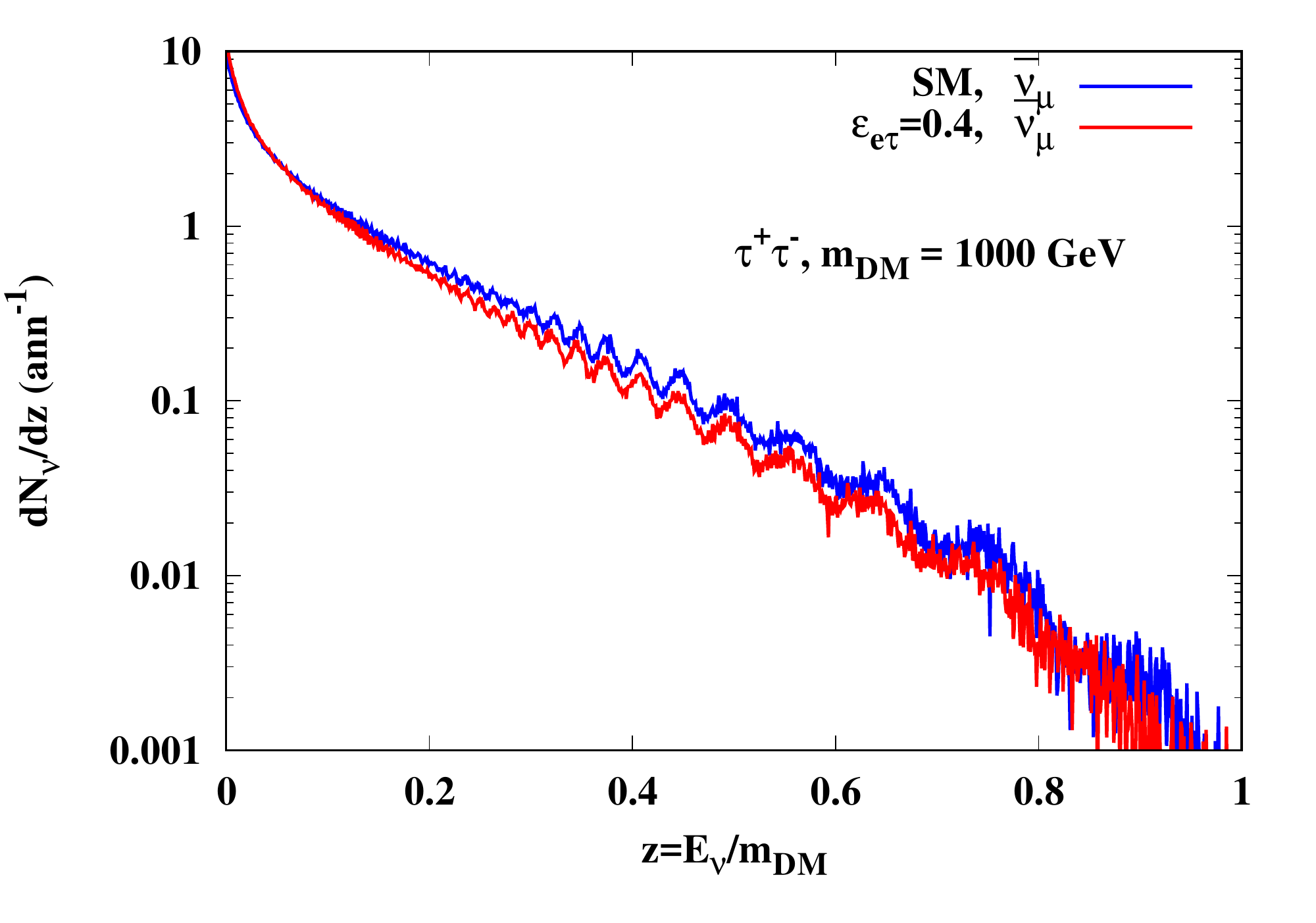}}
\put(260,240){\includegraphics[angle=-90,width=0.31\textwidth]{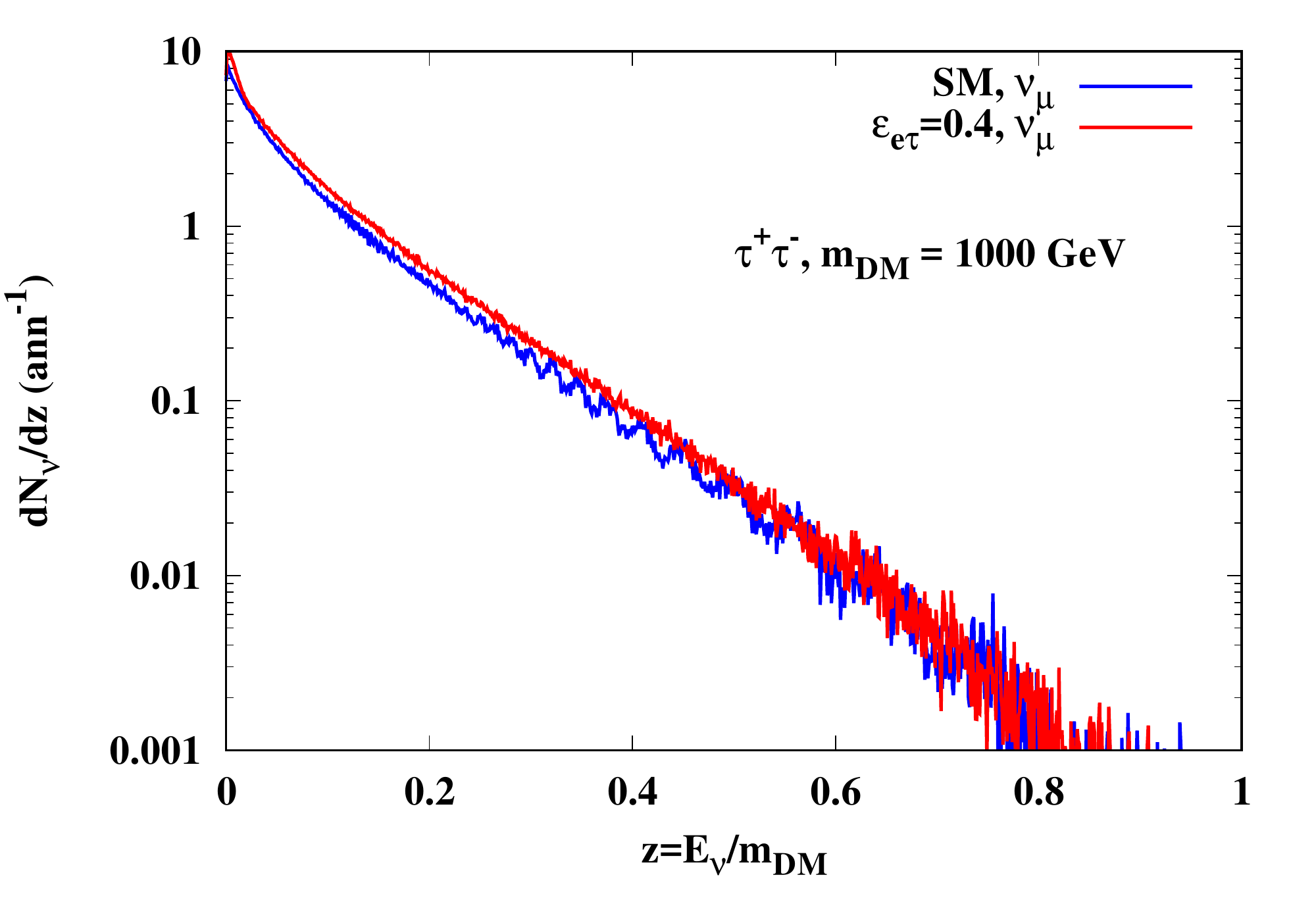}}
\put(130,130){\includegraphics[angle=-90,width=0.31\textwidth]{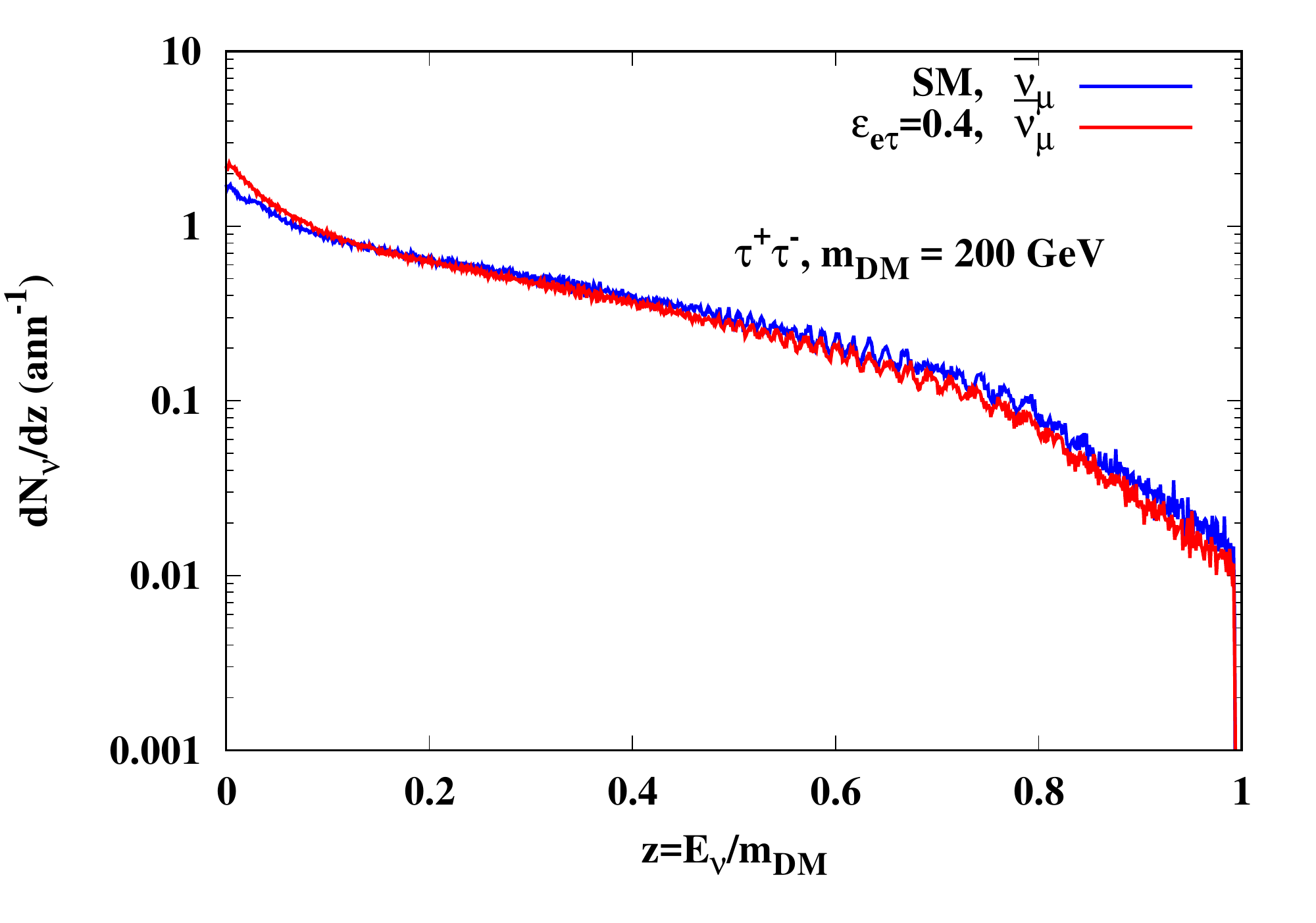}}
\put(130,240){\includegraphics[angle=-90,width=0.31\textwidth]{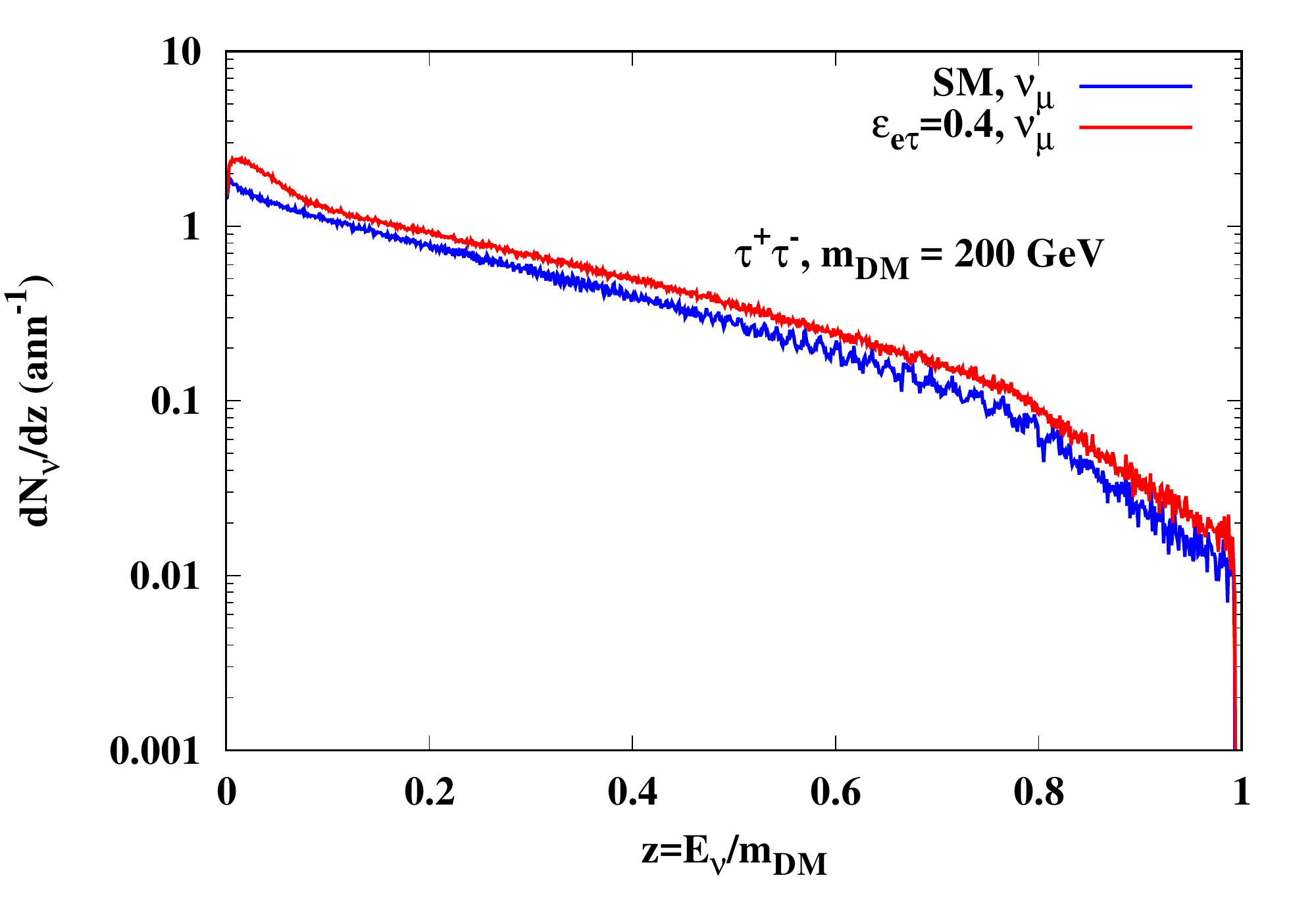}}
\put(0,130){\includegraphics[angle=-90,width=0.31\textwidth]{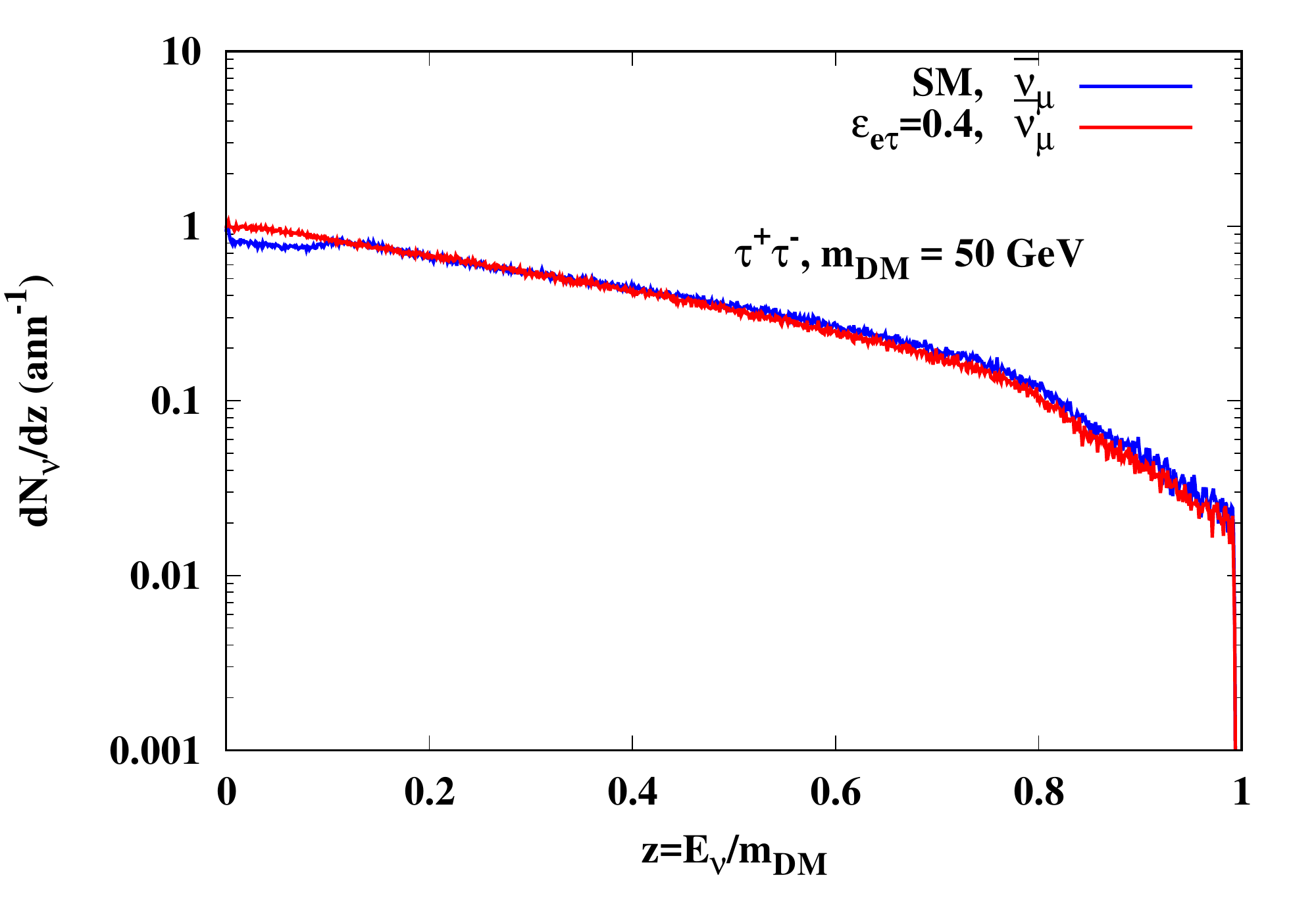}}
\put(0,240){\includegraphics[angle=-90,width=0.31\textwidth]{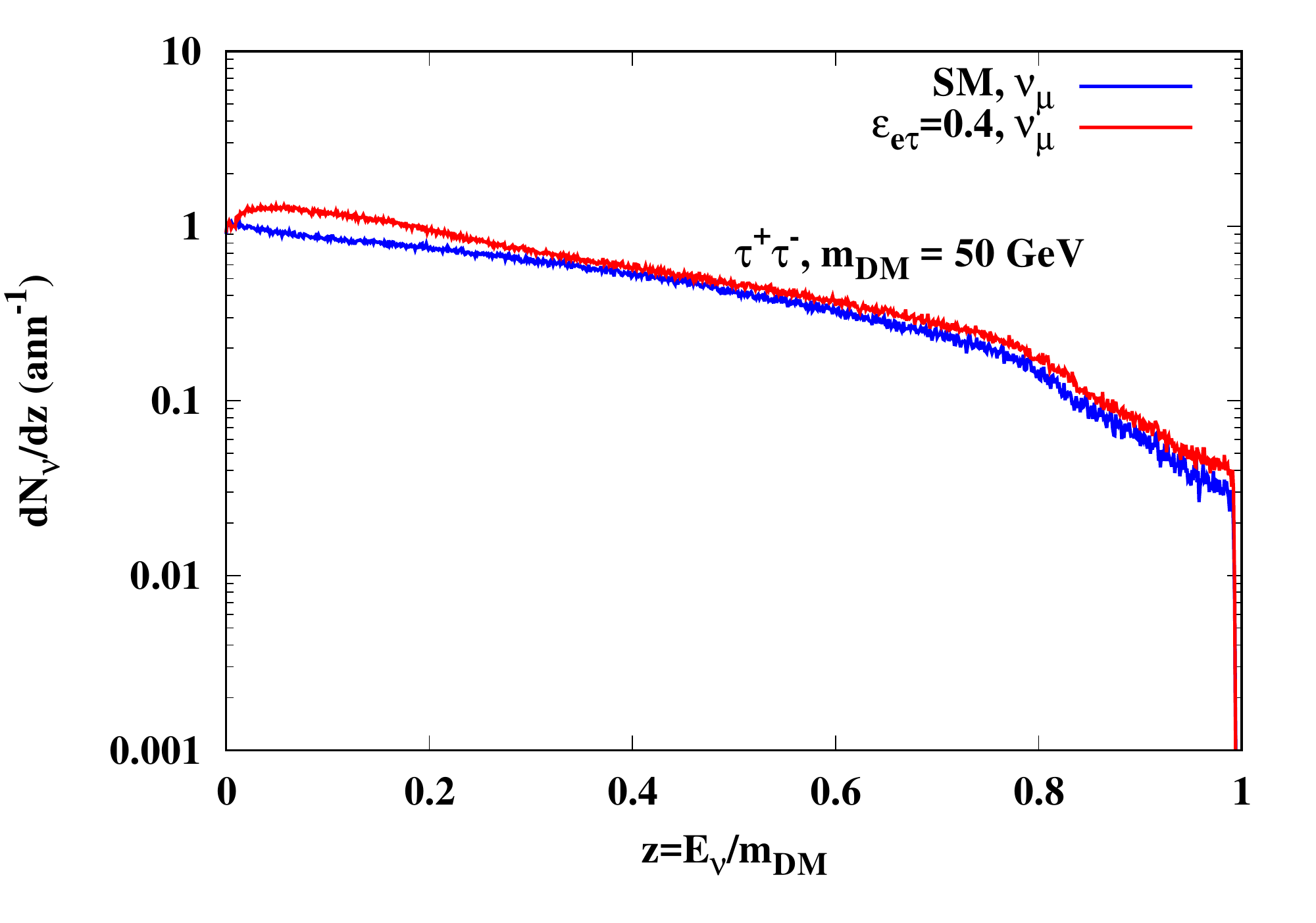}}
\end{picture}
\caption{\label{tau_etau_inv} The same as in Fig.~\ref{tau_tautau}
  but for $\e_{e\tau} = 0.4$ and IH.}
\end{figure}
\begin{figure}[!htb]
\begin{picture}(300,190)(0,40)
\put(260,130){\includegraphics[angle=-90,width=0.31\textwidth]{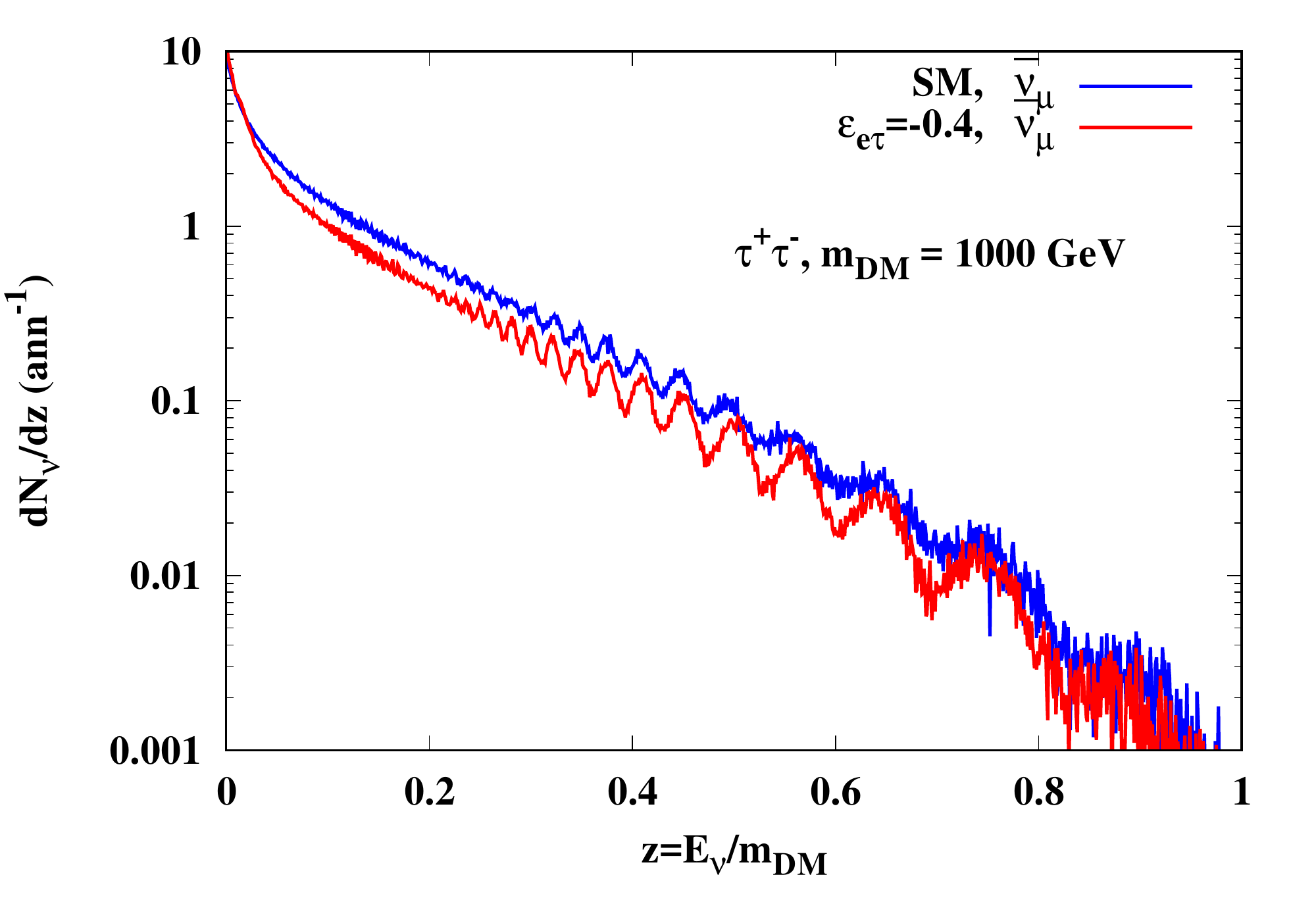}}
\put(260,240){\includegraphics[angle=-90,width=0.31\textwidth]{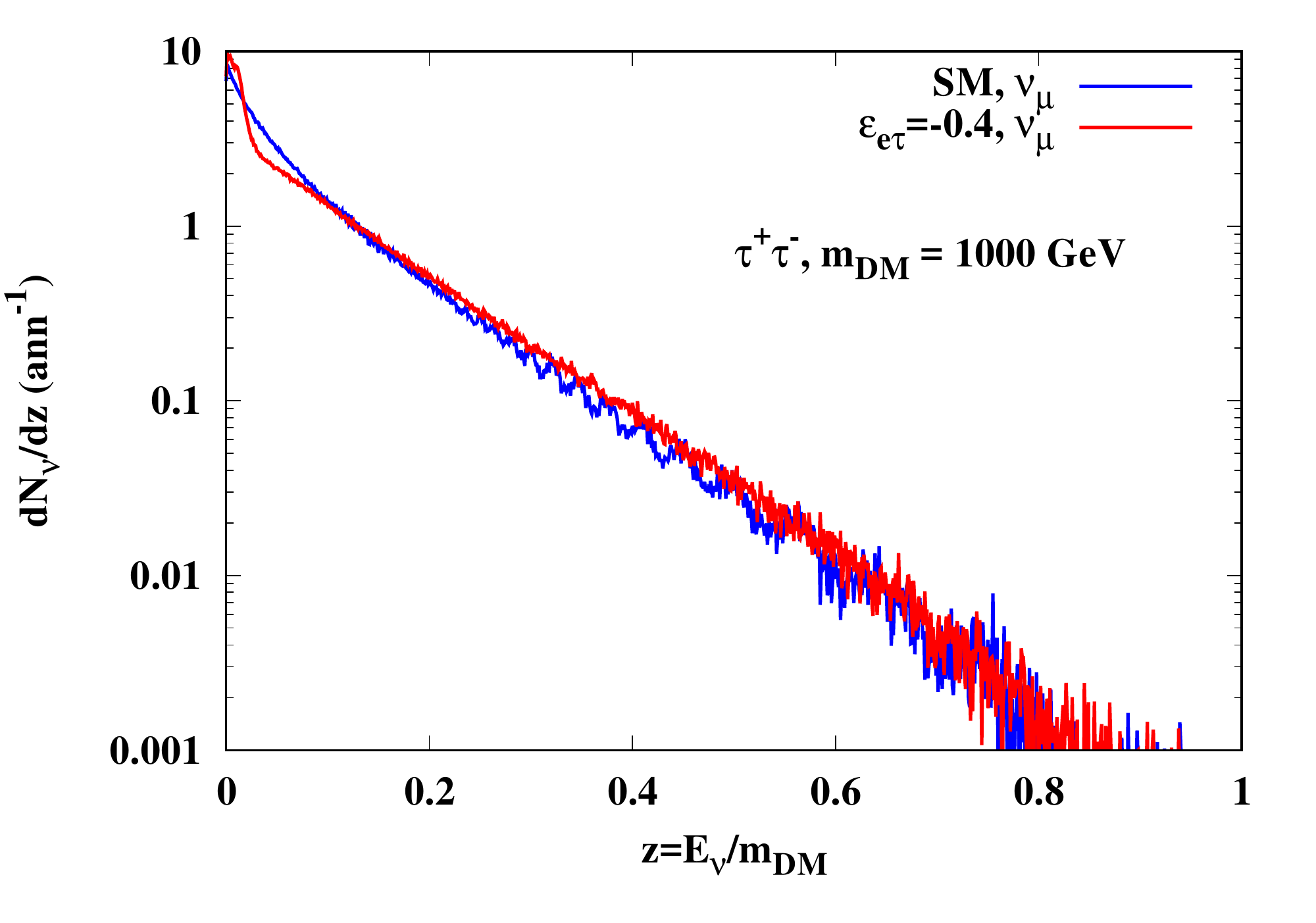}}
\put(130,130){\includegraphics[angle=-90,width=0.31\textwidth]{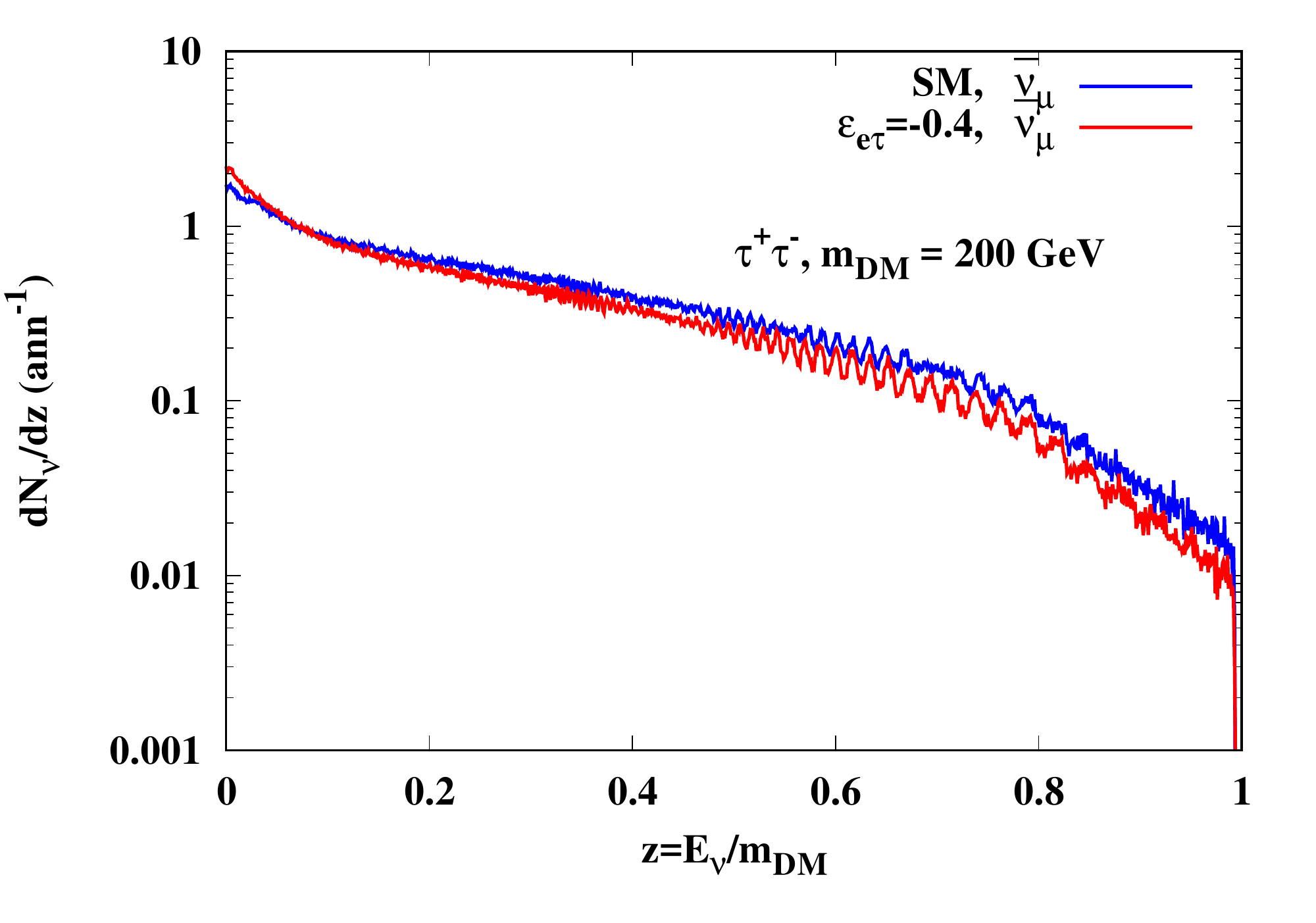}}
\put(130,240){\includegraphics[angle=-90,width=0.31\textwidth]{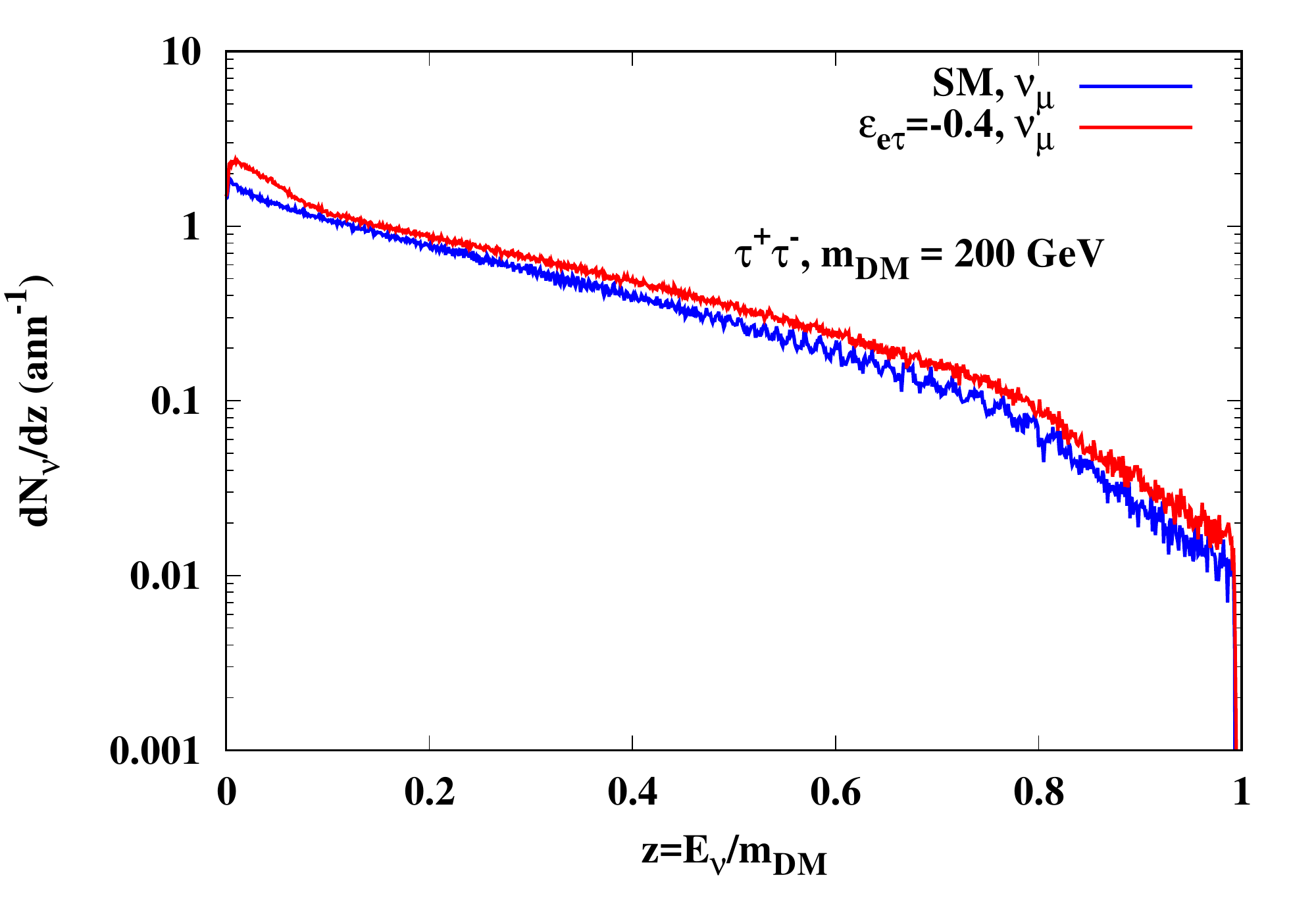}}
\put(0,130){\includegraphics[angle=-90,width=0.31\textwidth]{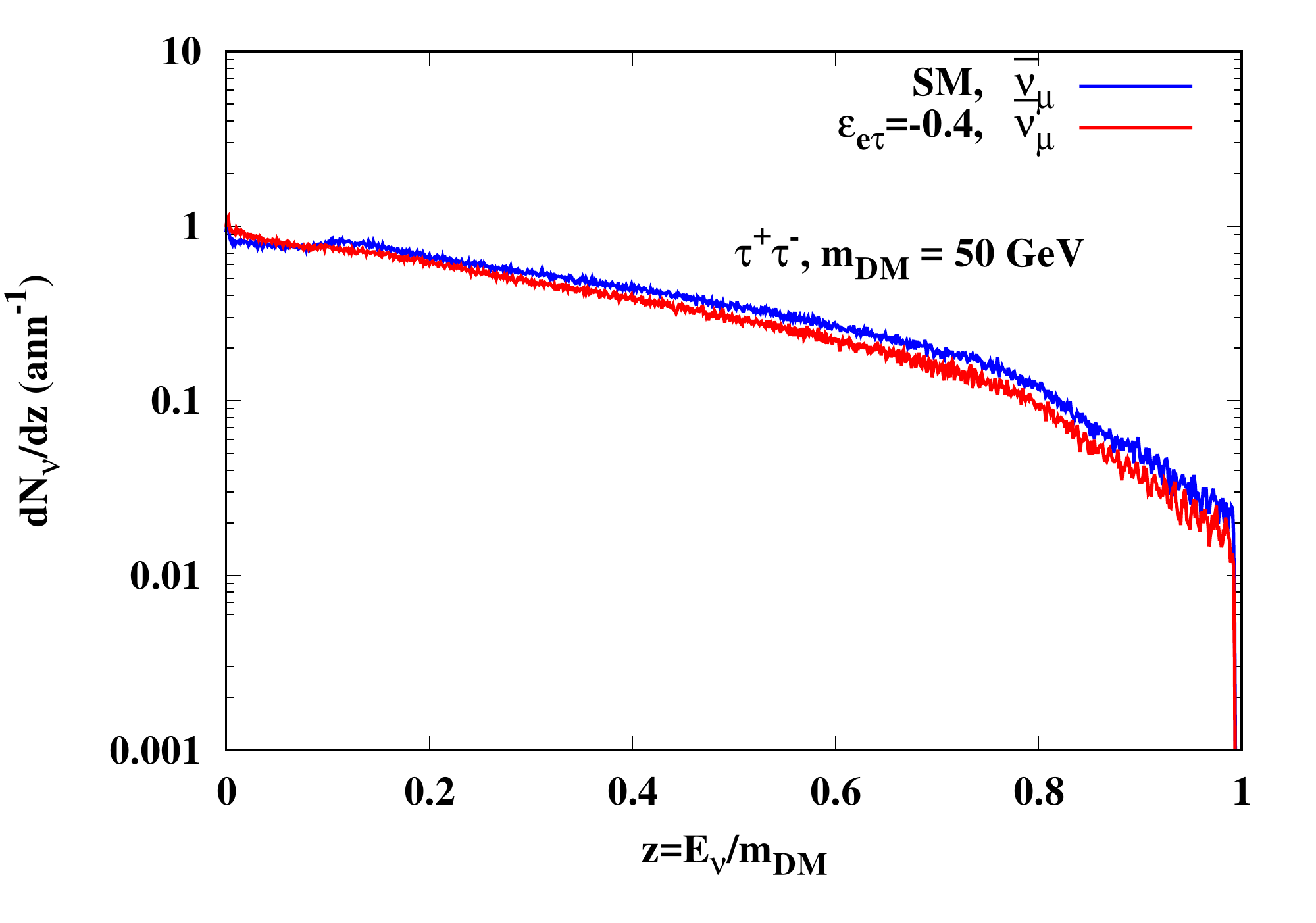}}
\put(0,240){\includegraphics[angle=-90,width=0.31\textwidth]{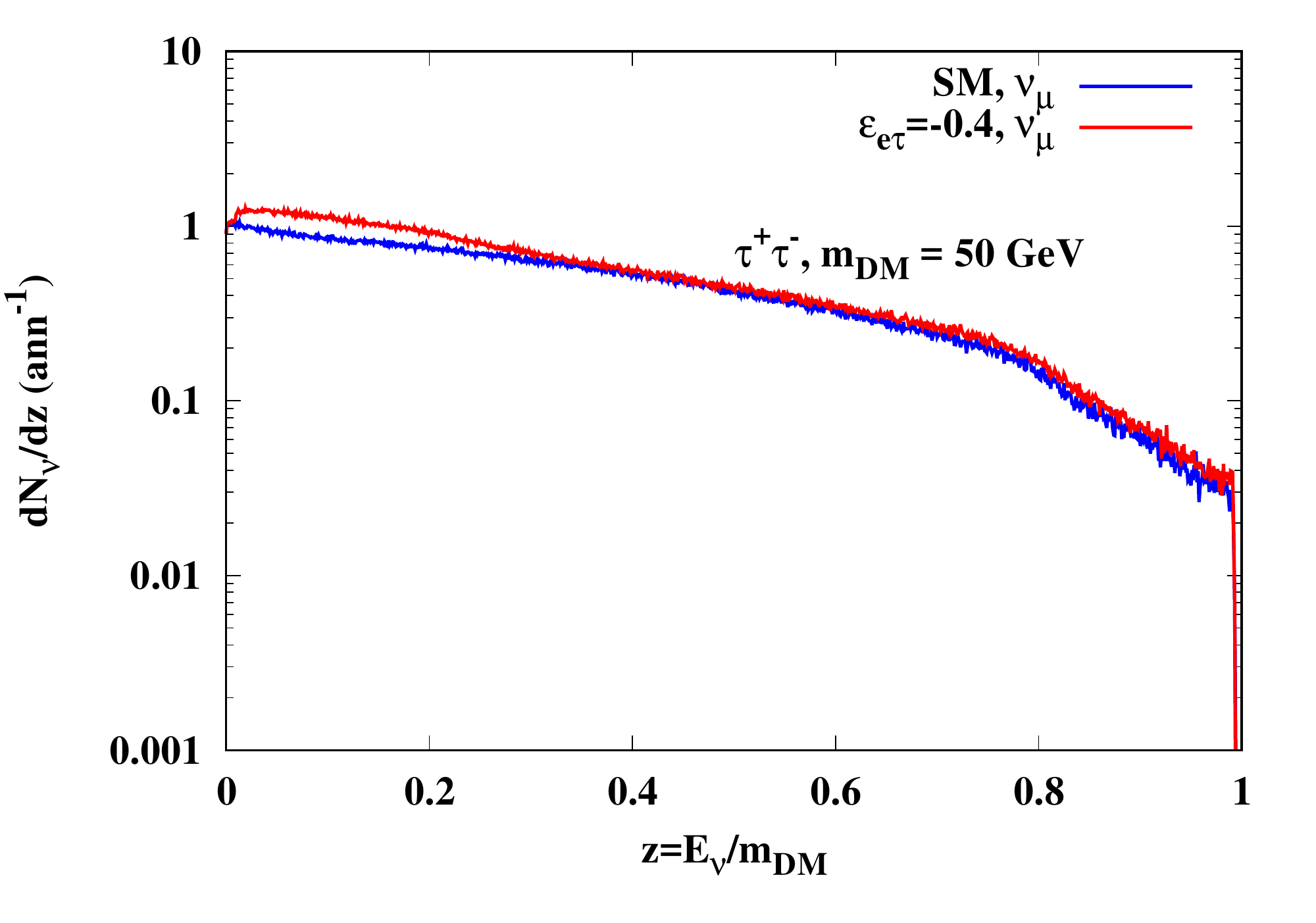}}
\end{picture}
\caption{\label{tau_etau_inv_sign} The same as in Fig.~\ref{tau_tautau}
  but for $\e_{e\tau} = -0.4$ and IH.}
\end{figure}
We also see considerable deviations of muon neutrino flux from the no-NSI
case. Next,  the spectra for $\e_{e\mu}=\pm0.2$
\begin{figure}[!htb]
\begin{picture}(300,190)(0,40)
\put(260,130){\includegraphics[angle=-90,width=0.31\textwidth]{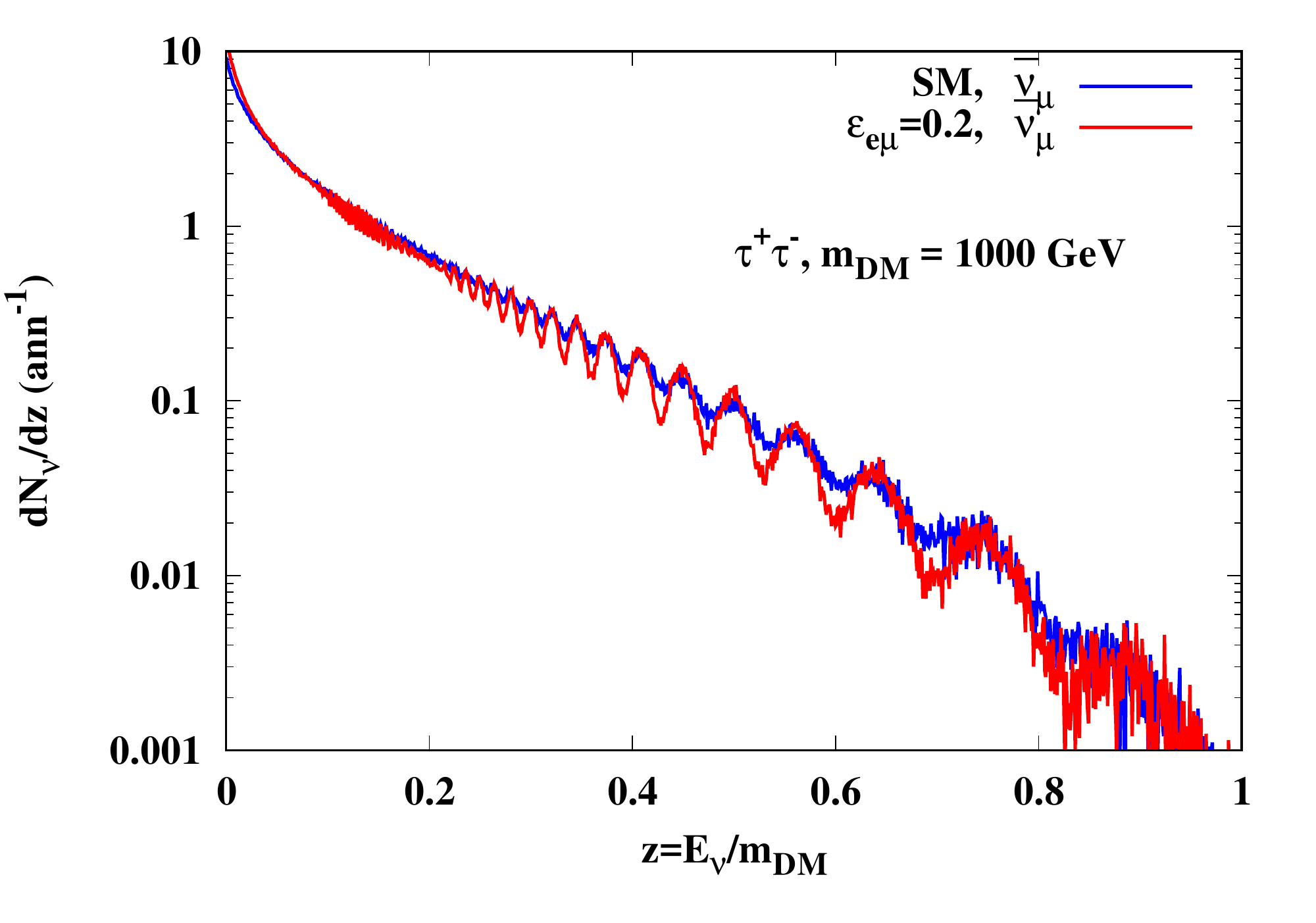}}
\put(260,240){\includegraphics[angle=-90,width=0.31\textwidth]{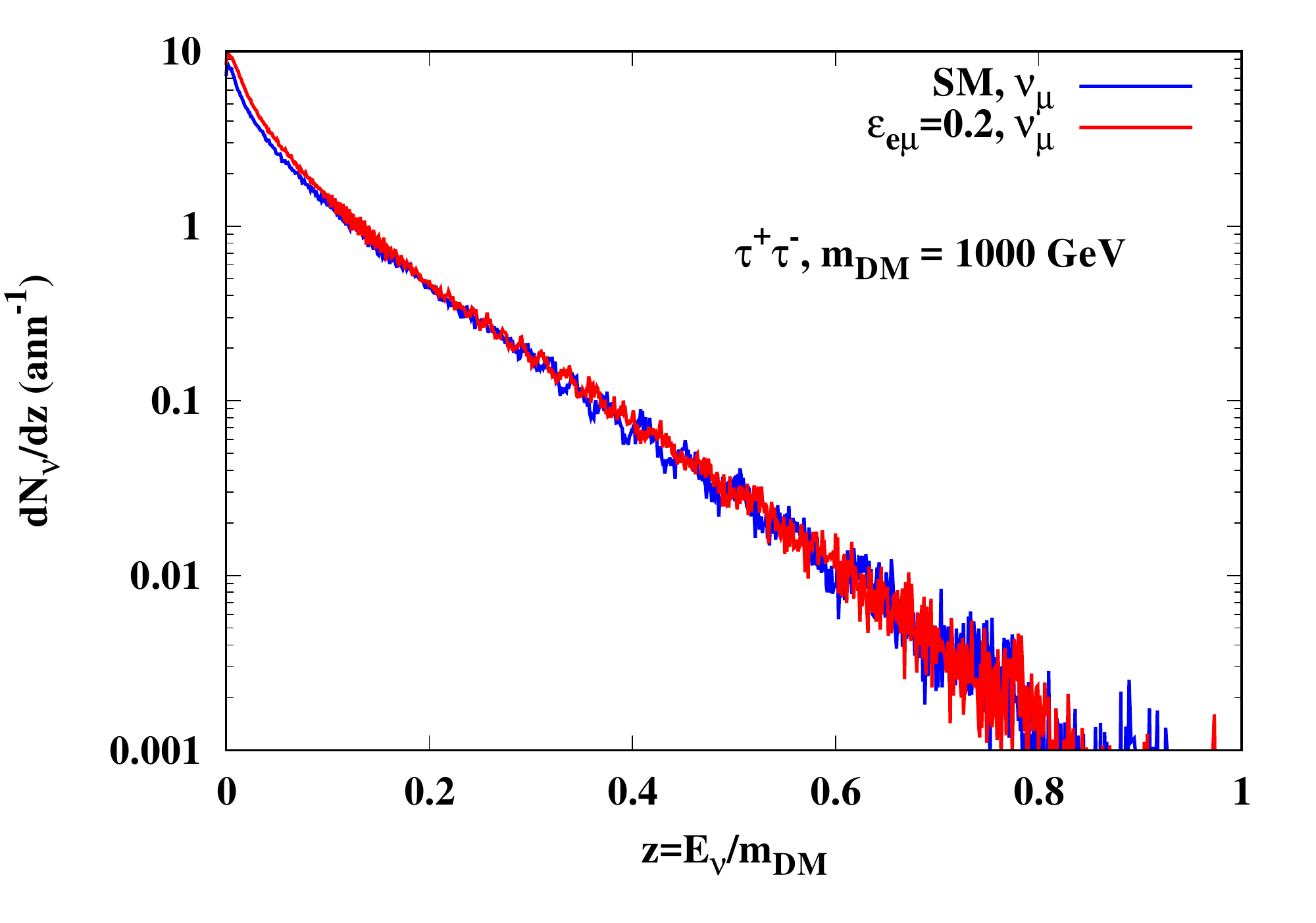}}
\put(130,130){\includegraphics[angle=-90,width=0.31\textwidth]{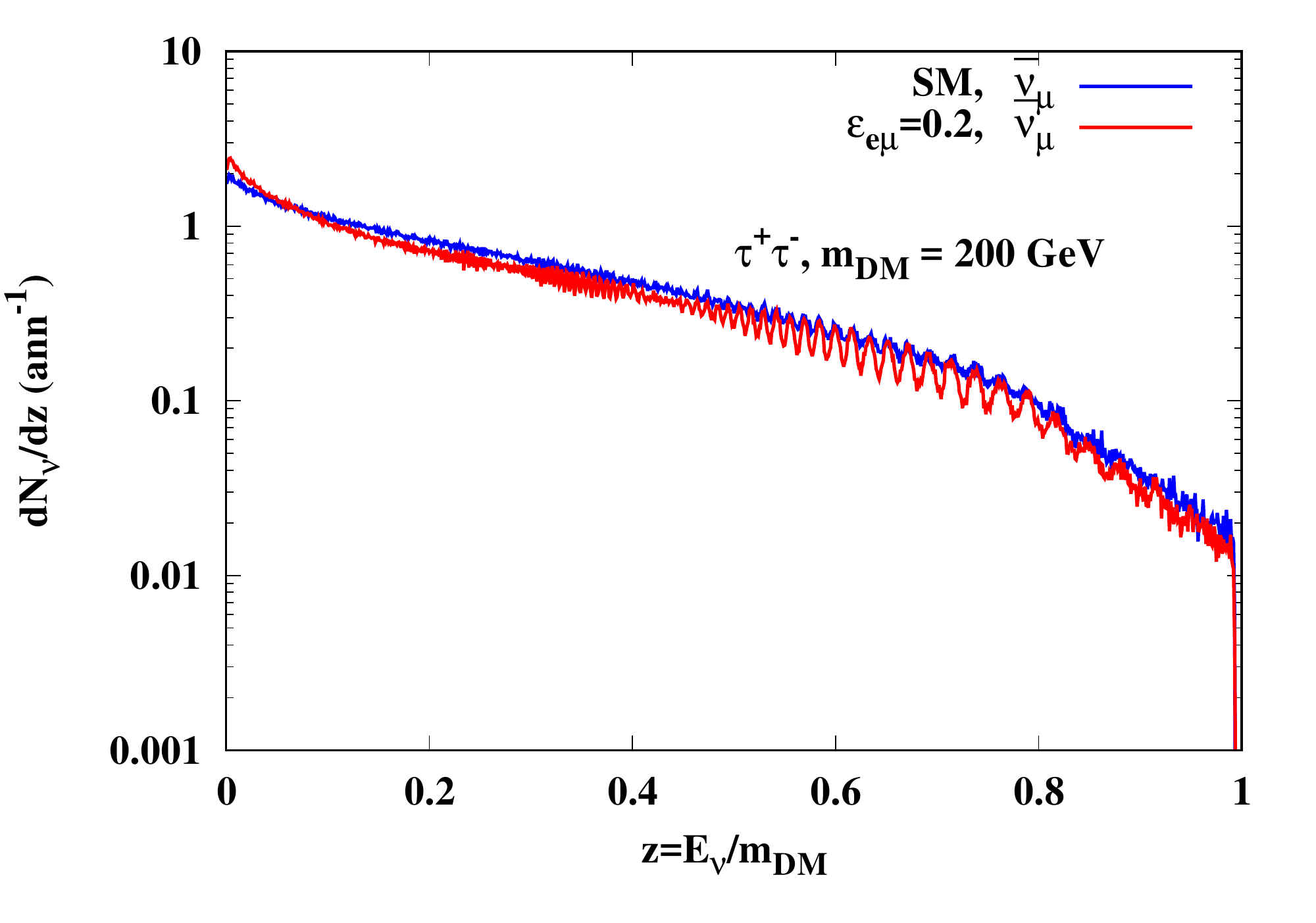}}
\put(130,240){\includegraphics[angle=-90,width=0.31\textwidth]{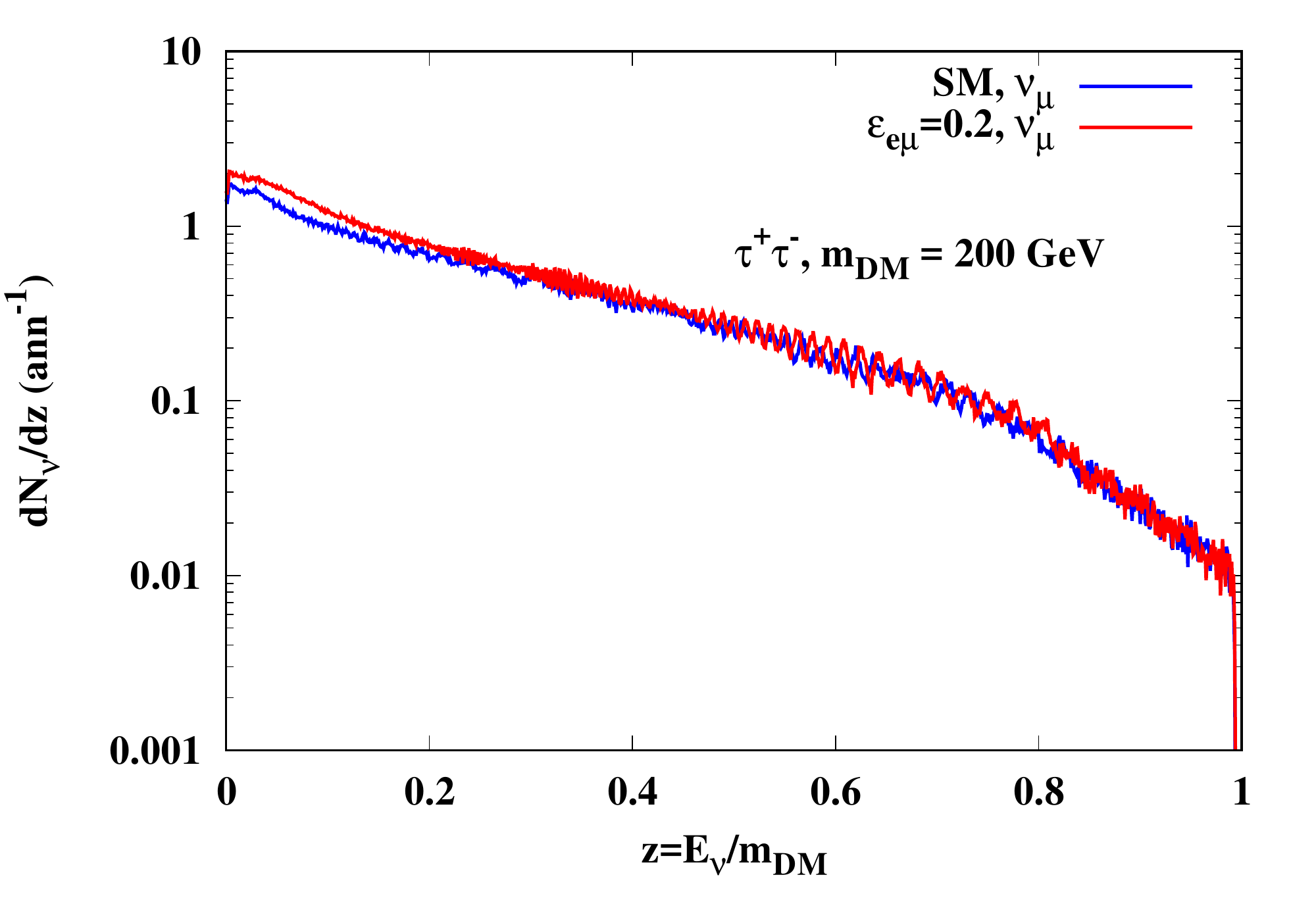}}
\put(0,130){\includegraphics[angle=-90,width=0.31\textwidth]{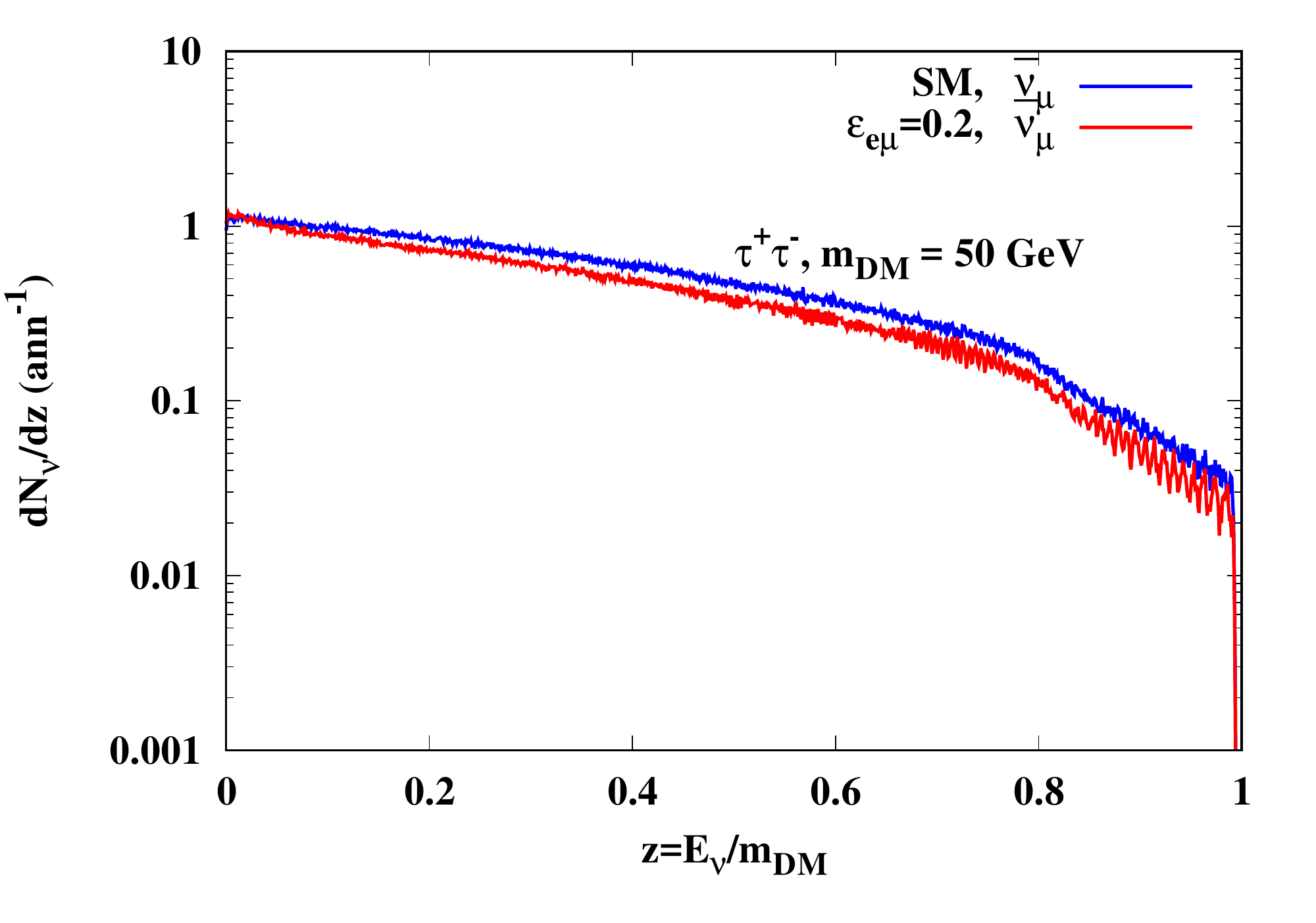}}
\put(0,240){\includegraphics[angle=-90,width=0.31\textwidth]{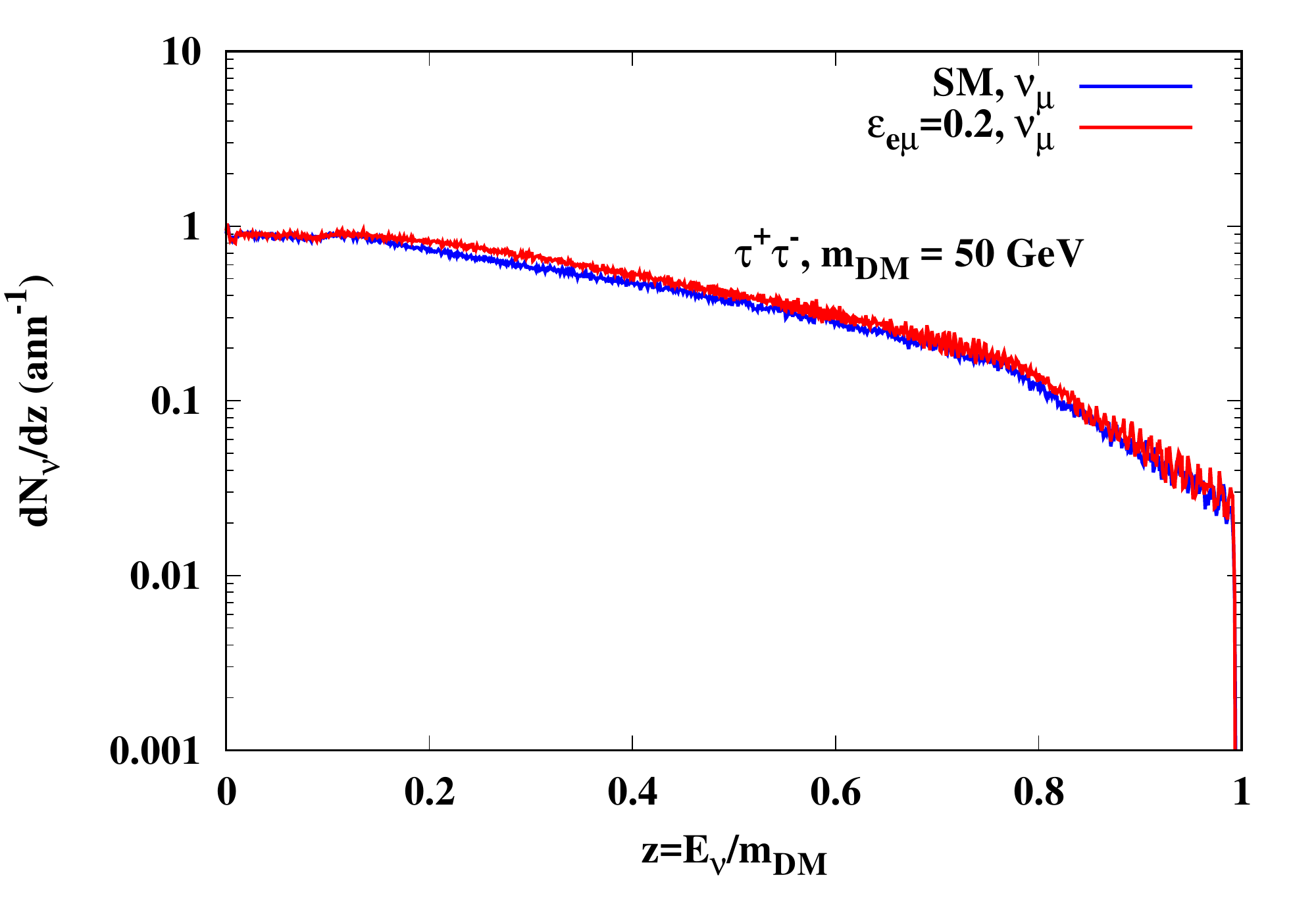}}
\end{picture}
\caption{\label{tau_emu} The same as in Fig.~\ref{tau_tautau}
  but for $\e_{e\mu} = 0.2$ and NH.}
\end{figure}
\begin{figure}[!htb]
\begin{picture}(300,190)(0,40)
\put(260,130){\includegraphics[angle=-90,width=0.31\textwidth]{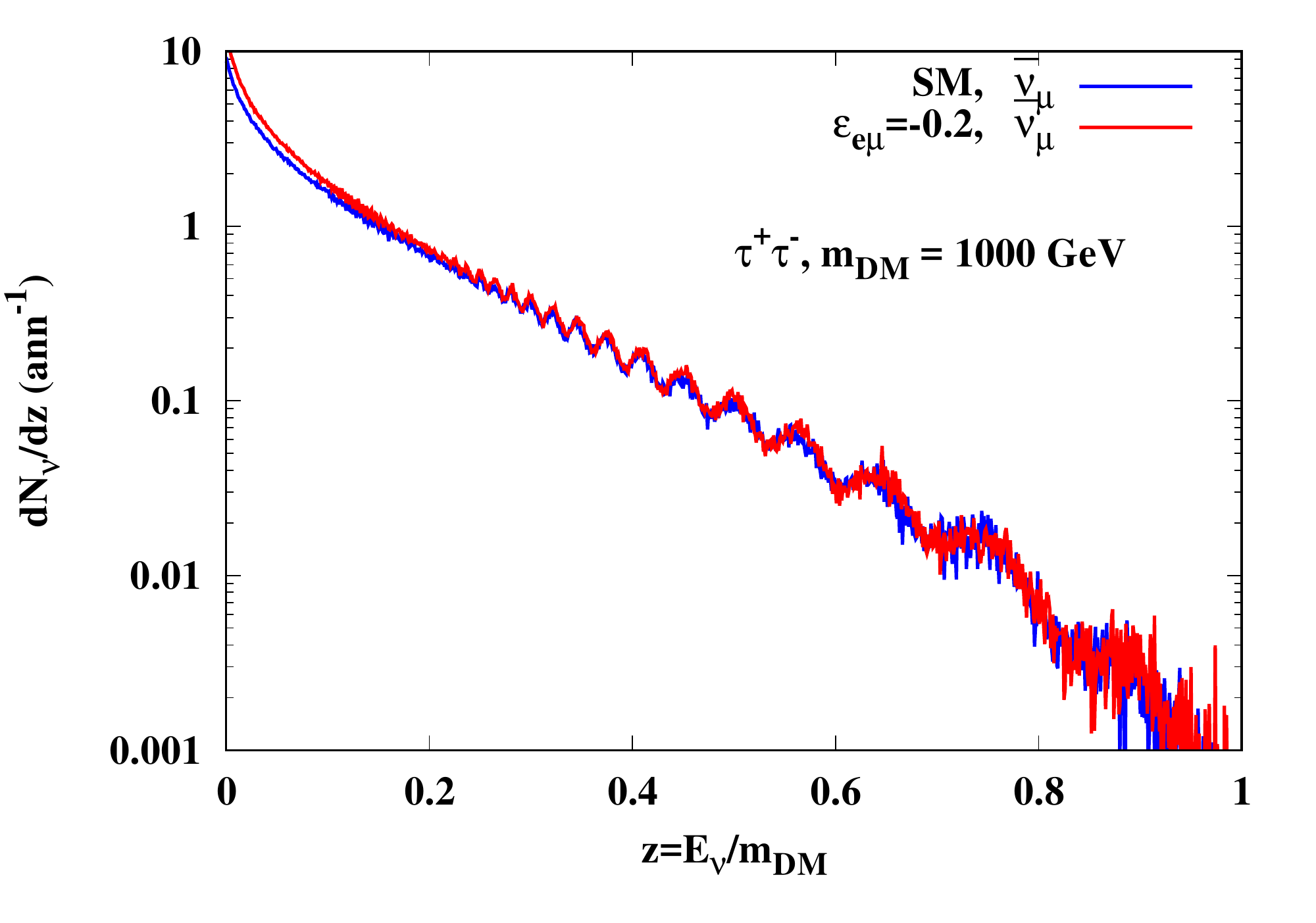}}
\put(260,240){\includegraphics[angle=-90,width=0.31\textwidth]{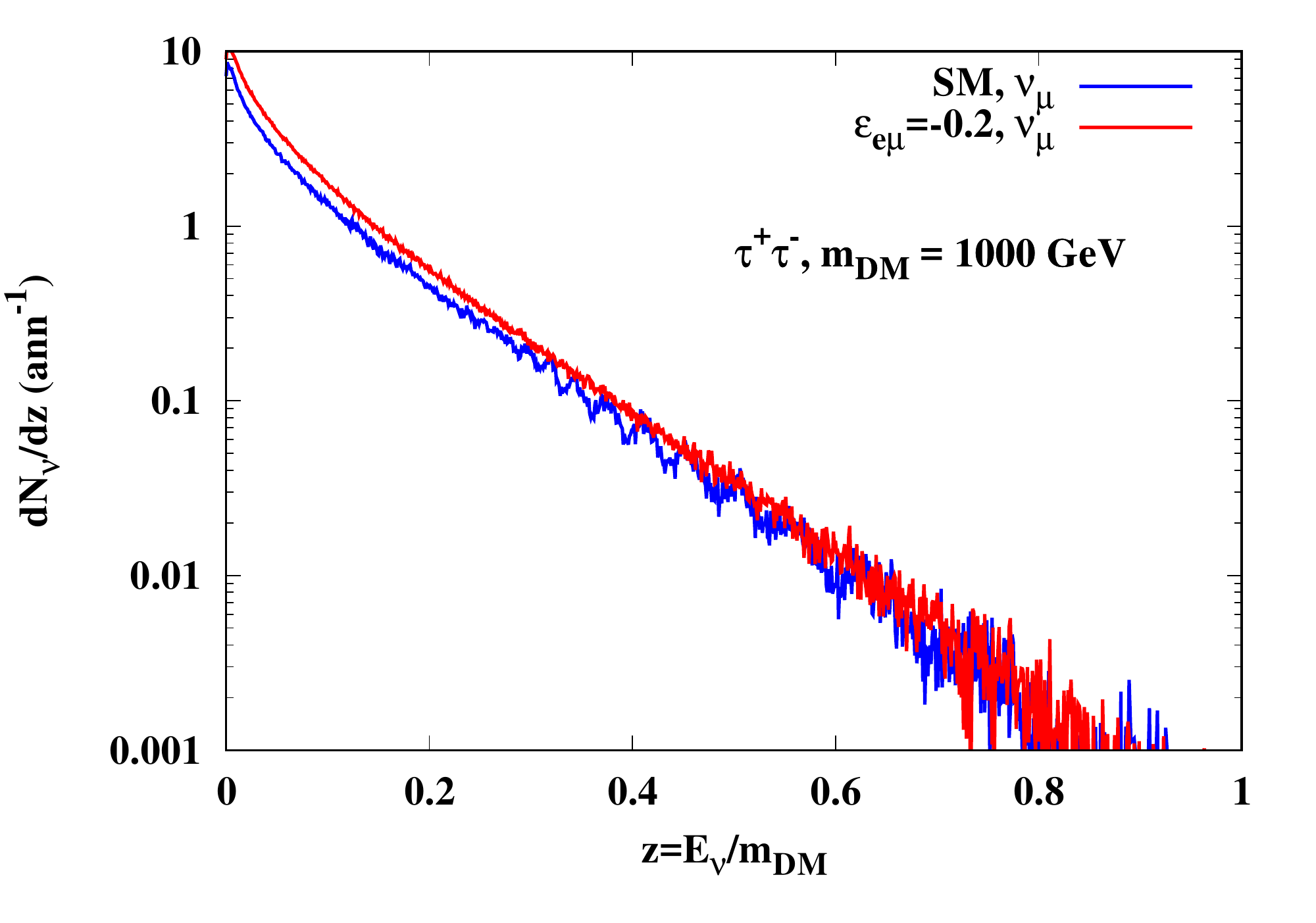}}
\put(130,130){\includegraphics[angle=-90,width=0.31\textwidth]{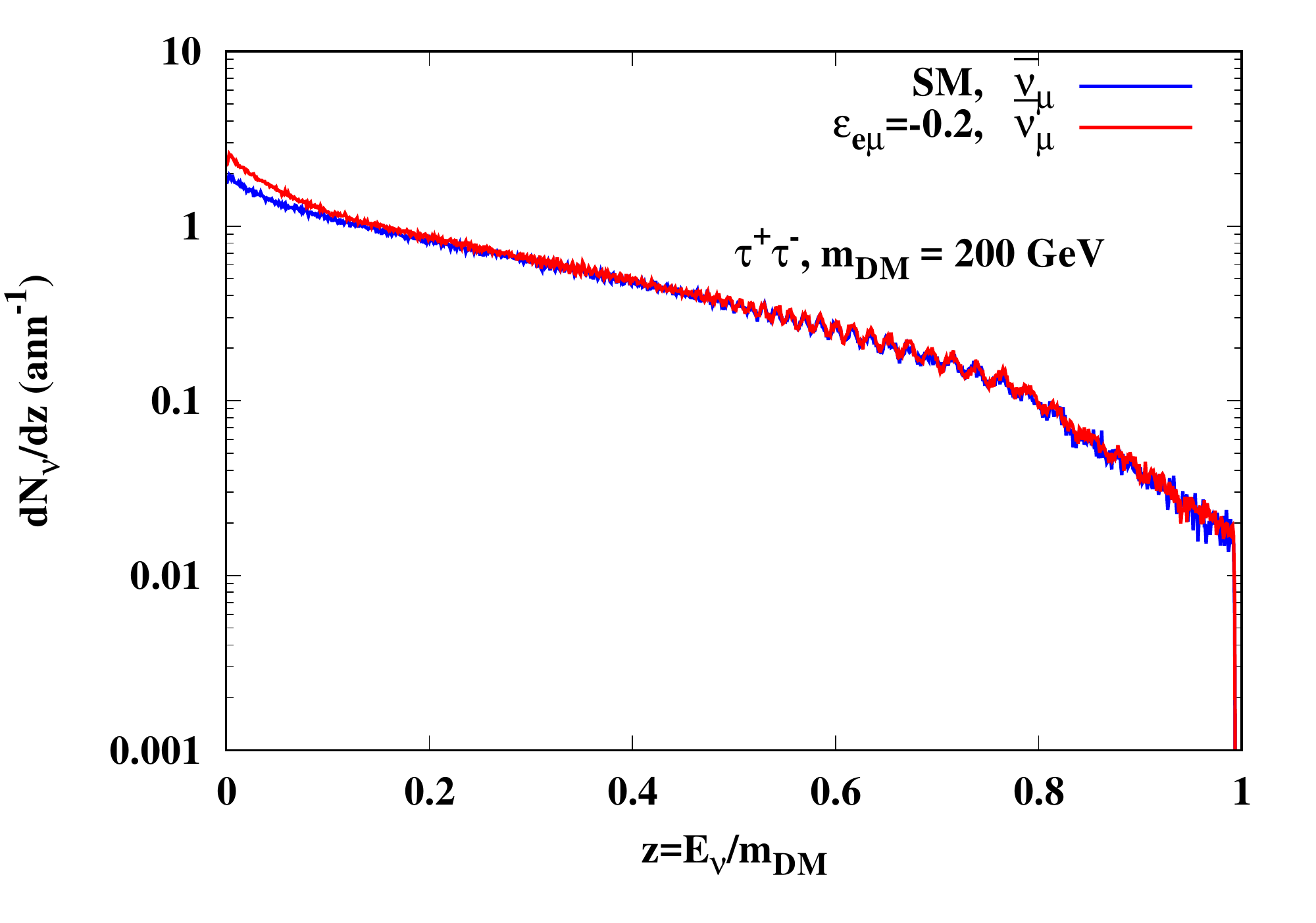}}
\put(130,240){\includegraphics[angle=-90,width=0.31\textwidth]{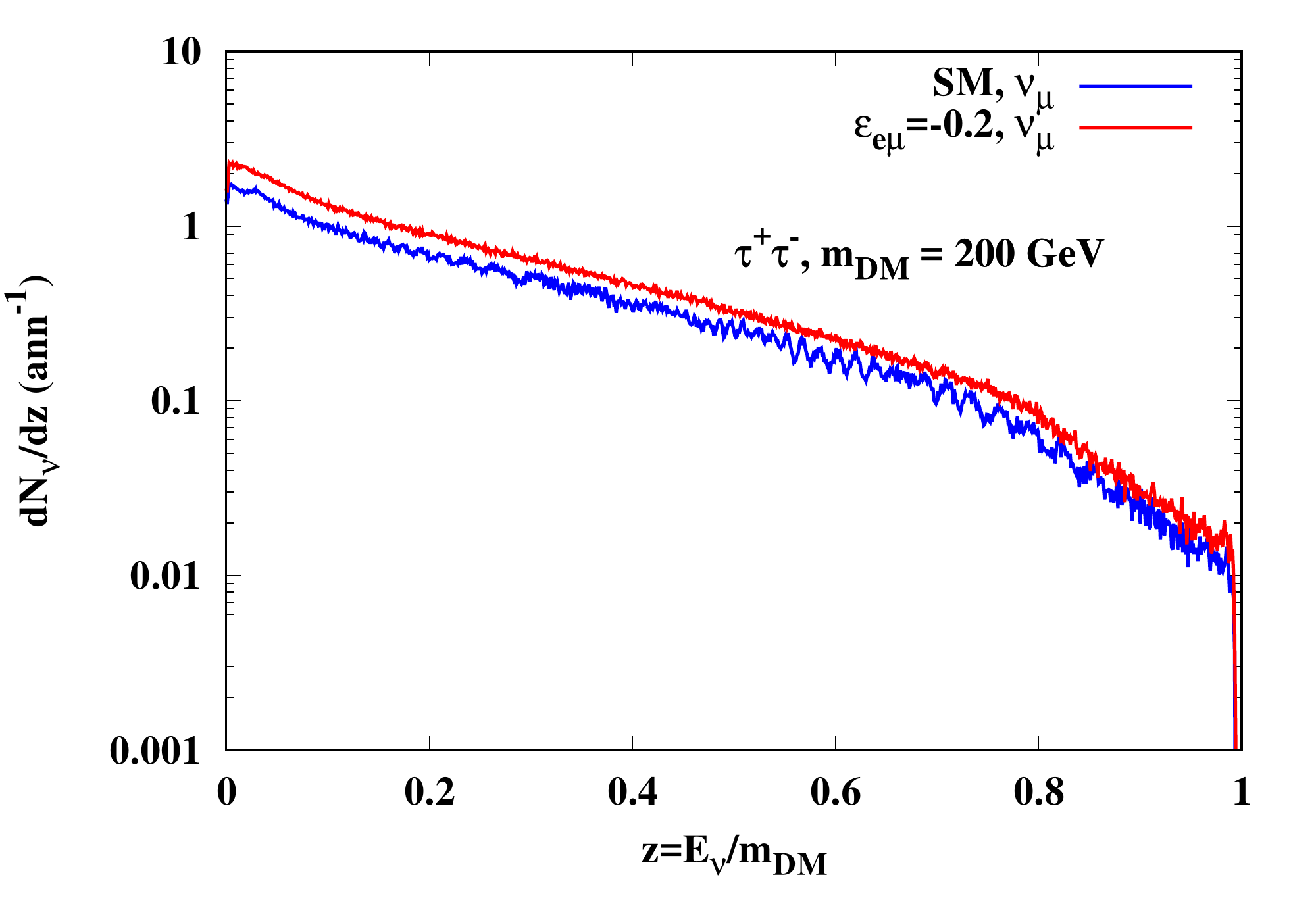}}
\put(0,130){\includegraphics[angle=-90,width=0.31\textwidth]{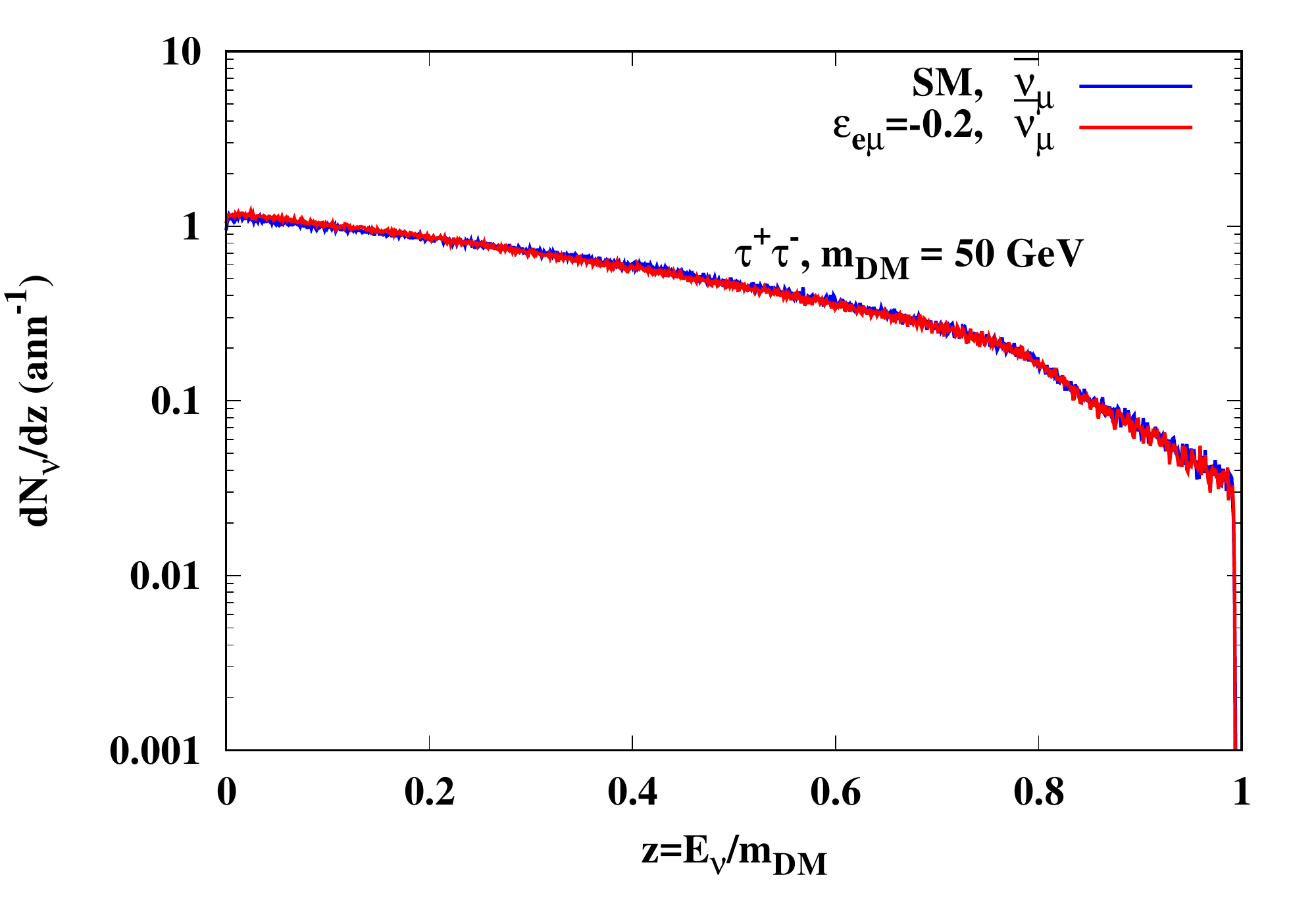}}
\put(0,240){\includegraphics[angle=-90,width=0.31\textwidth]{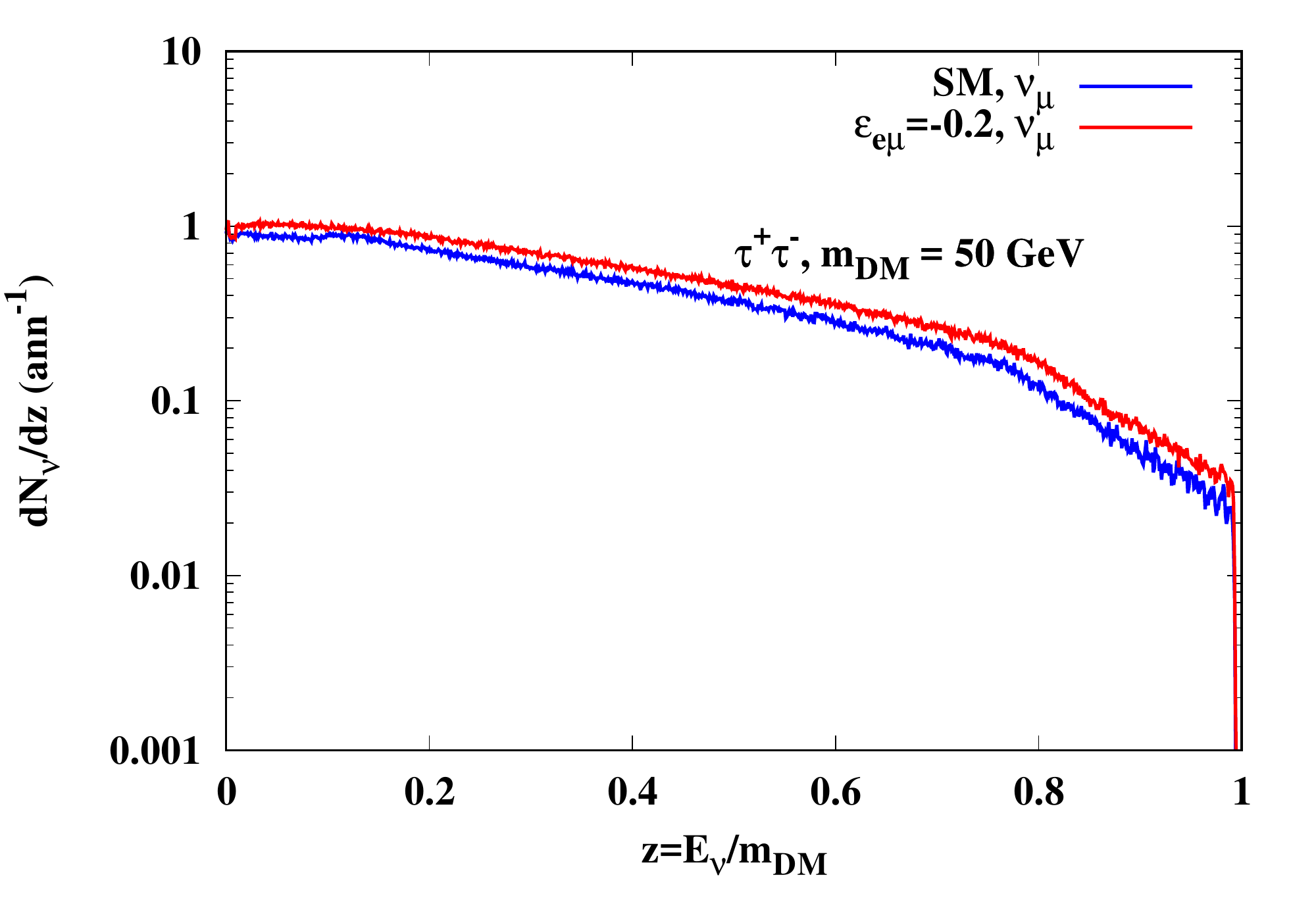}}
\end{picture}
\caption{\label{tau_emu_sign} The same as in Fig.~\ref{tau_tautau}
  but for $\e_{e\mu} = -0.2$ and NH.}
\end{figure}
are presented in Figs.~\ref{tau_emu} and~\ref{tau_emu_sign} for normal
mass hierarchy and 
\begin{figure}[!htb]
\begin{picture}(300,190)(0,40)
\put(260,130){\includegraphics[angle=-90,width=0.31\textwidth]{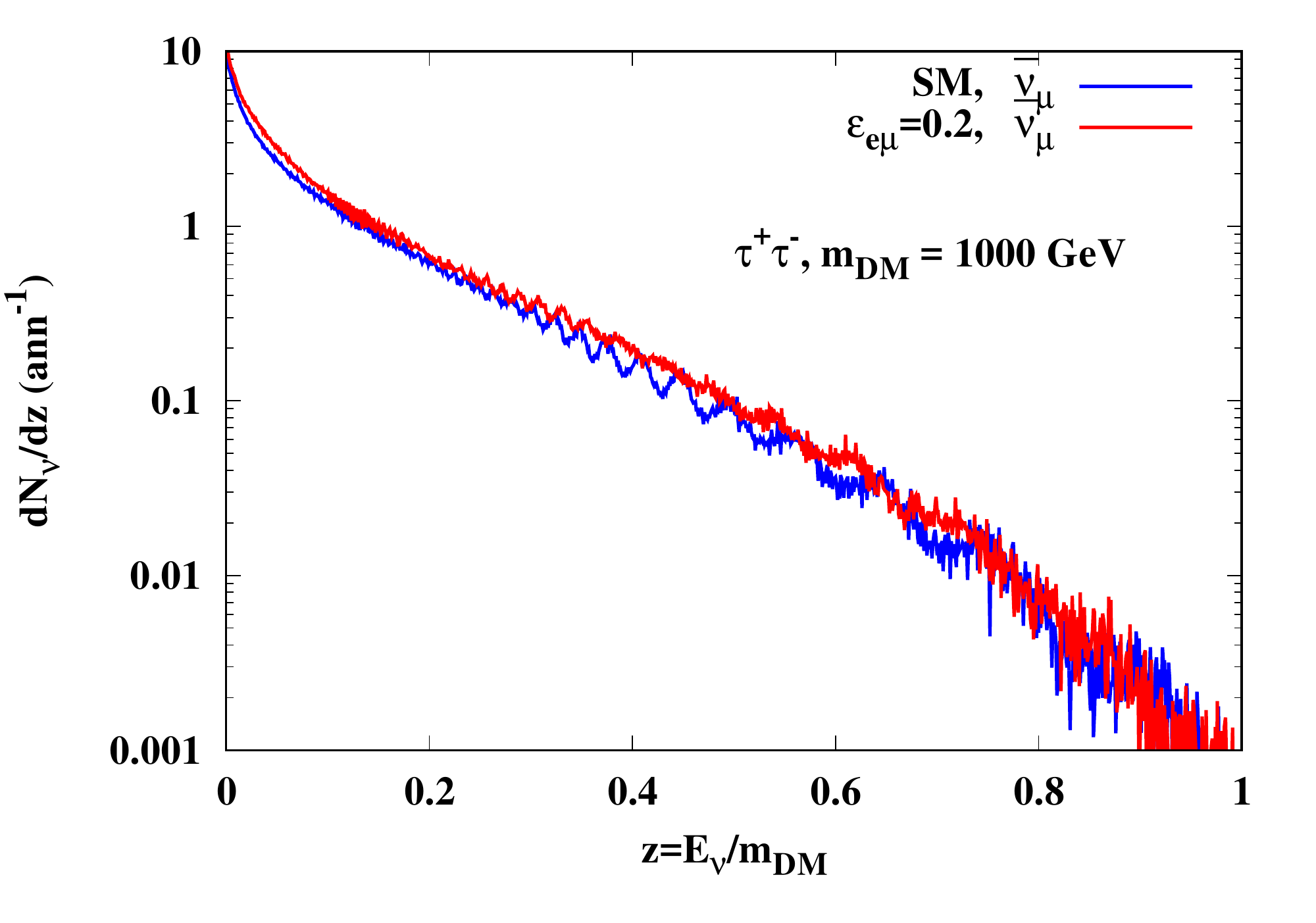}}
\put(260,240){\includegraphics[angle=-90,width=0.31\textwidth]{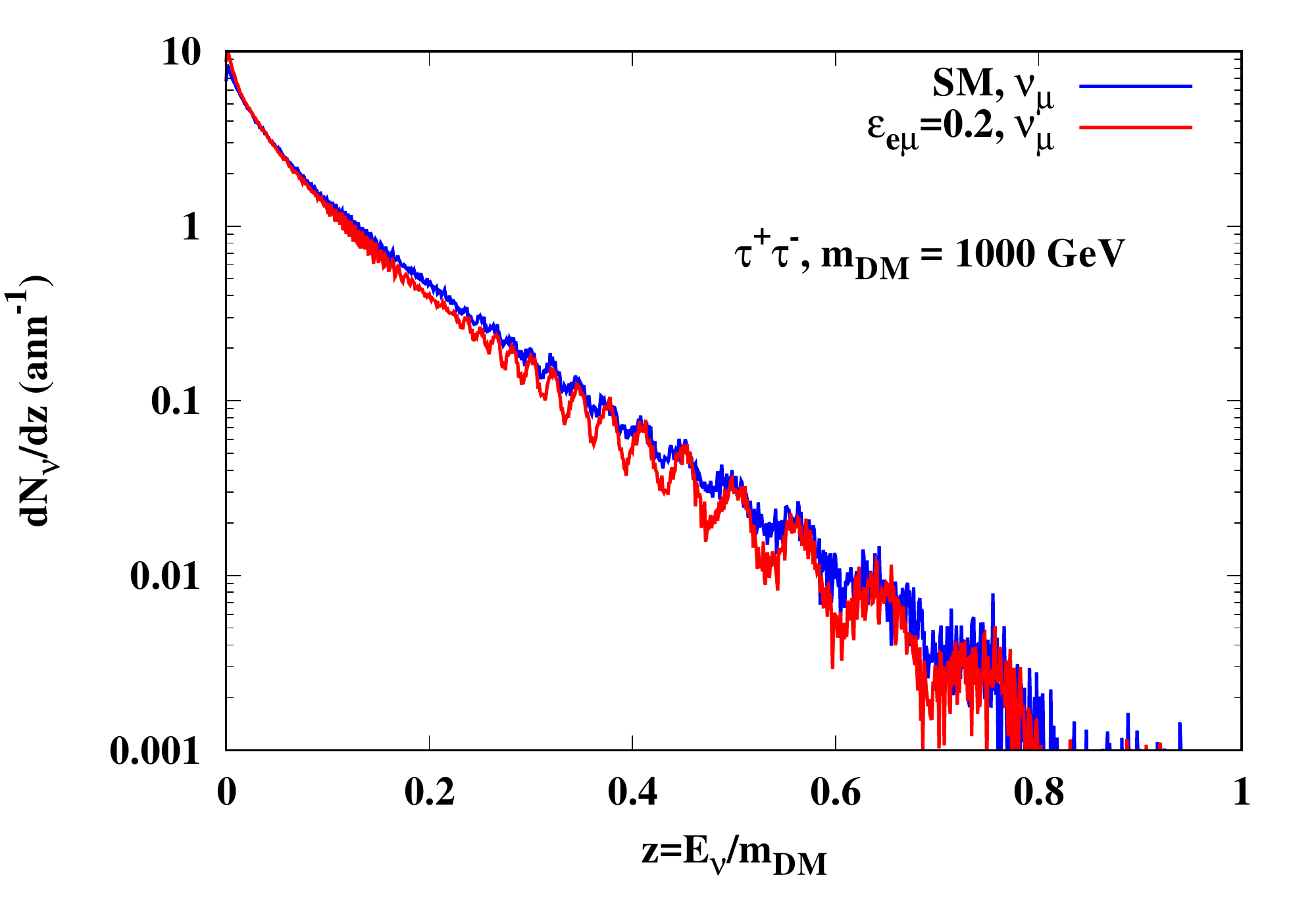}}
\put(130,130){\includegraphics[angle=-90,width=0.31\textwidth]{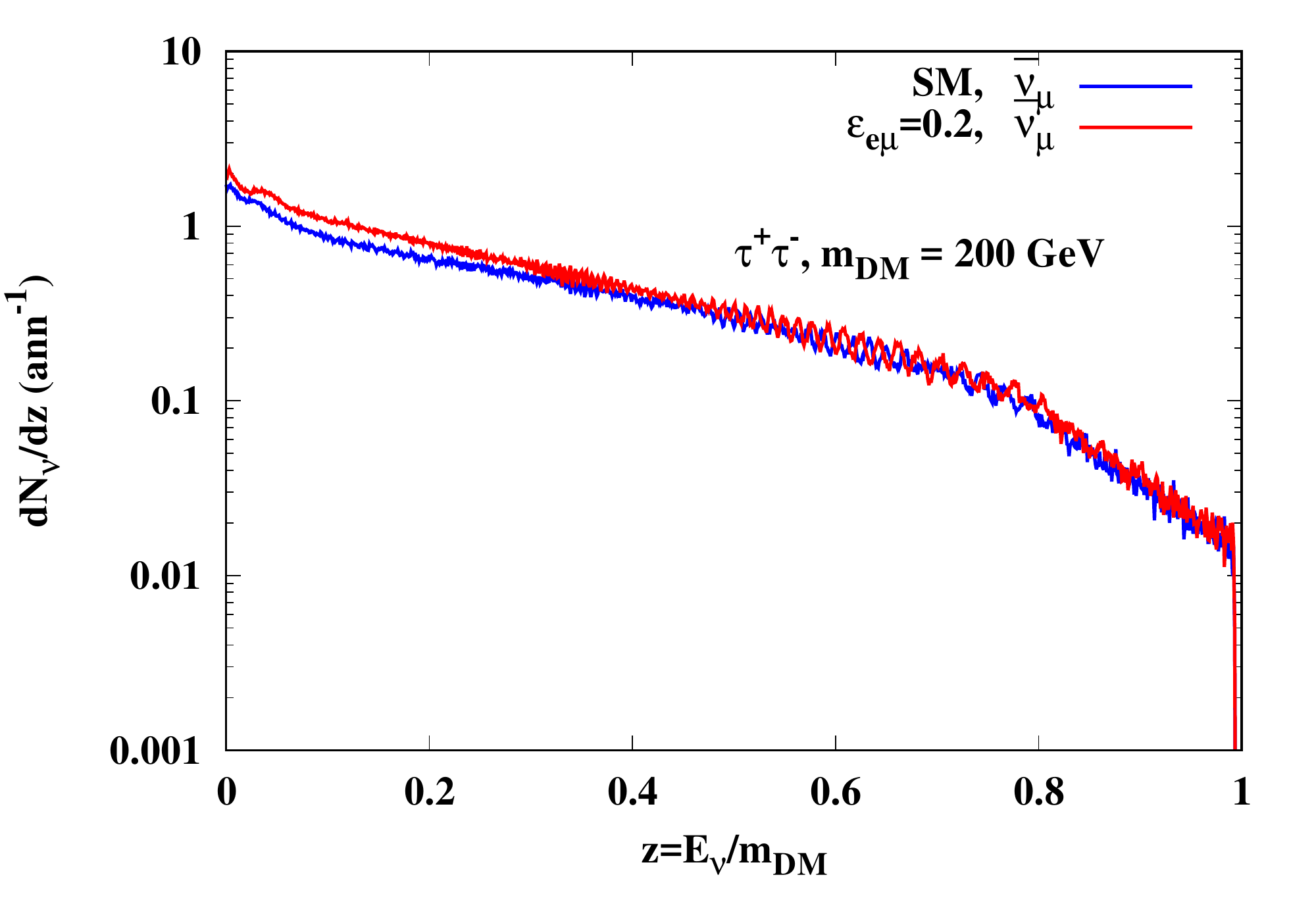}}
\put(130,240){\includegraphics[angle=-90,width=0.31\textwidth]{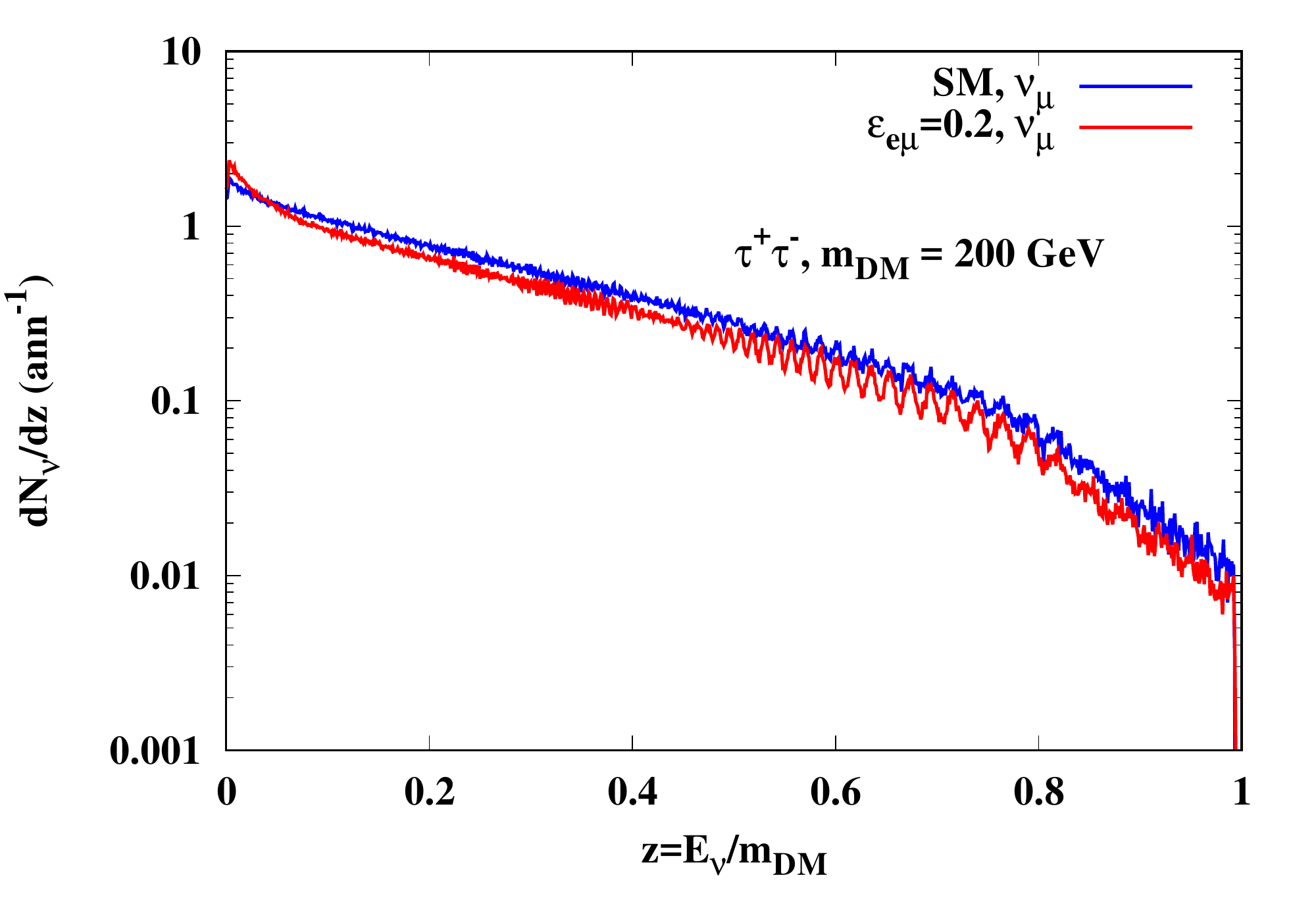}}
\put(0,130){\includegraphics[angle=-90,width=0.31\textwidth]{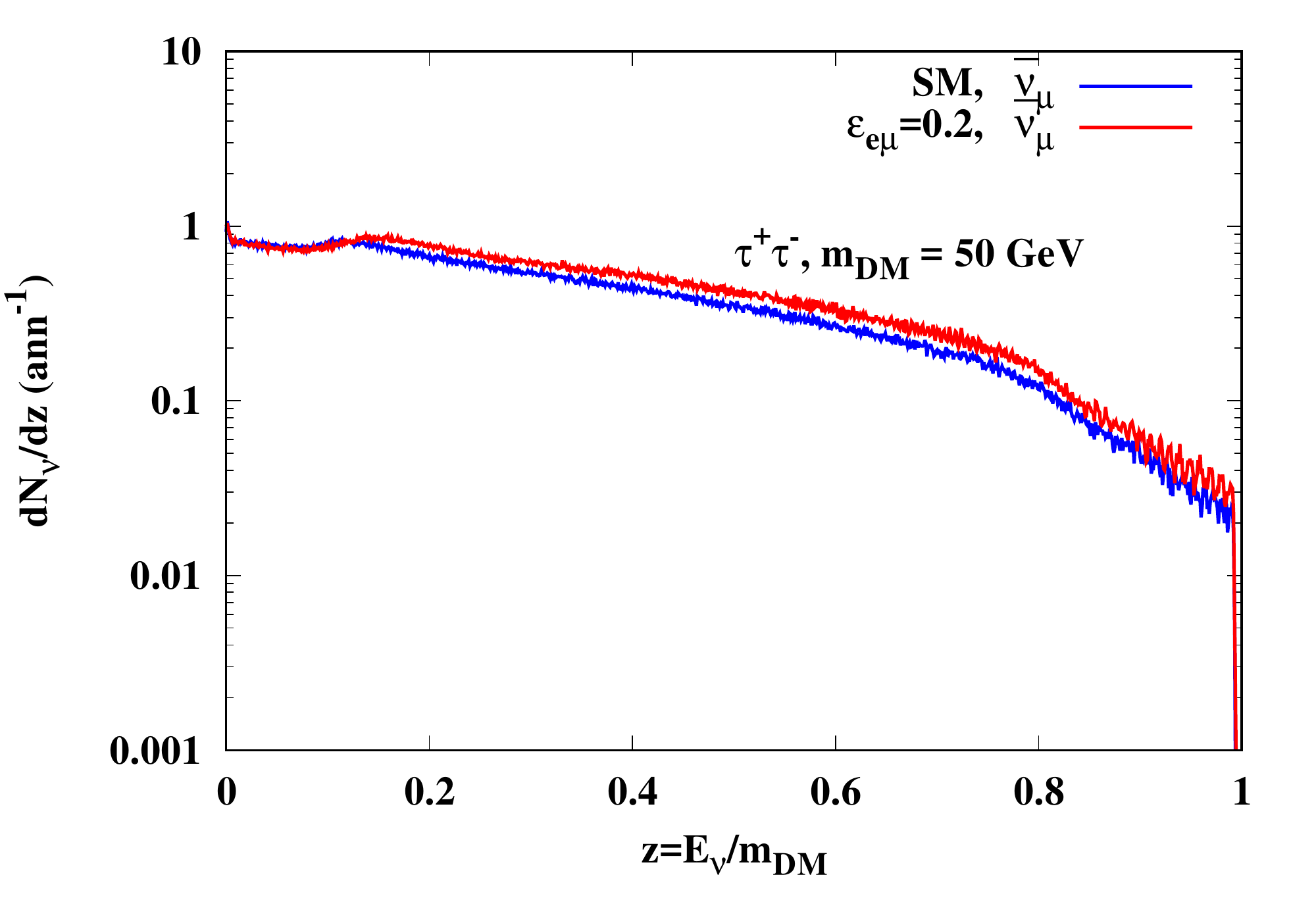}}
\put(0,240){\includegraphics[angle=-90,width=0.31\textwidth]{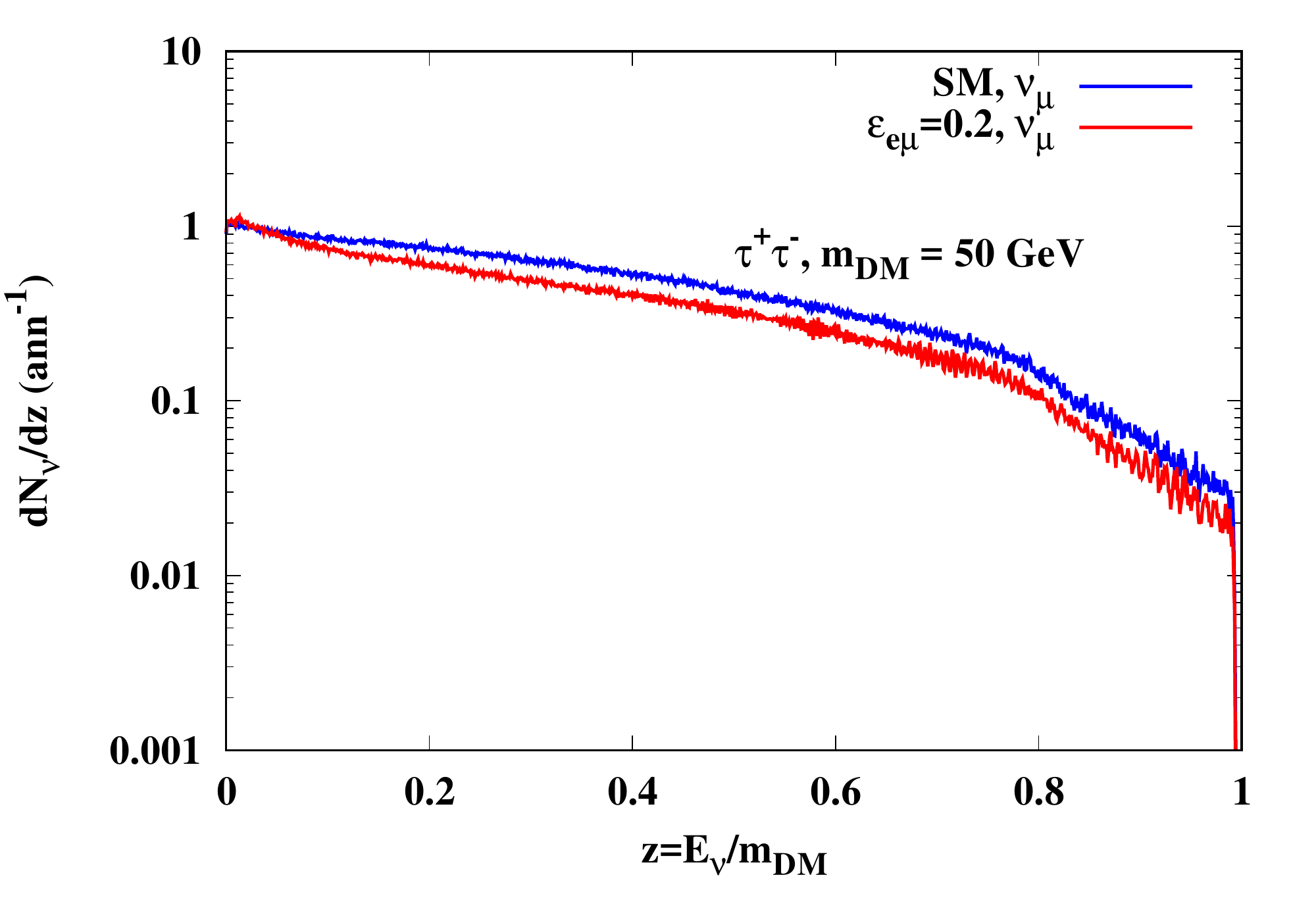}}
\end{picture}
\caption{\label{tau_emu_inv} The same as in Fig.~\ref{tau_tautau}
  but for $\e_{e\mu} = 0.2$ and IH.}
\end{figure}
\begin{figure}[!htb]
\begin{picture}(300,190)(0,40)
\put(260,130){\includegraphics[angle=-90,width=0.31\textwidth]{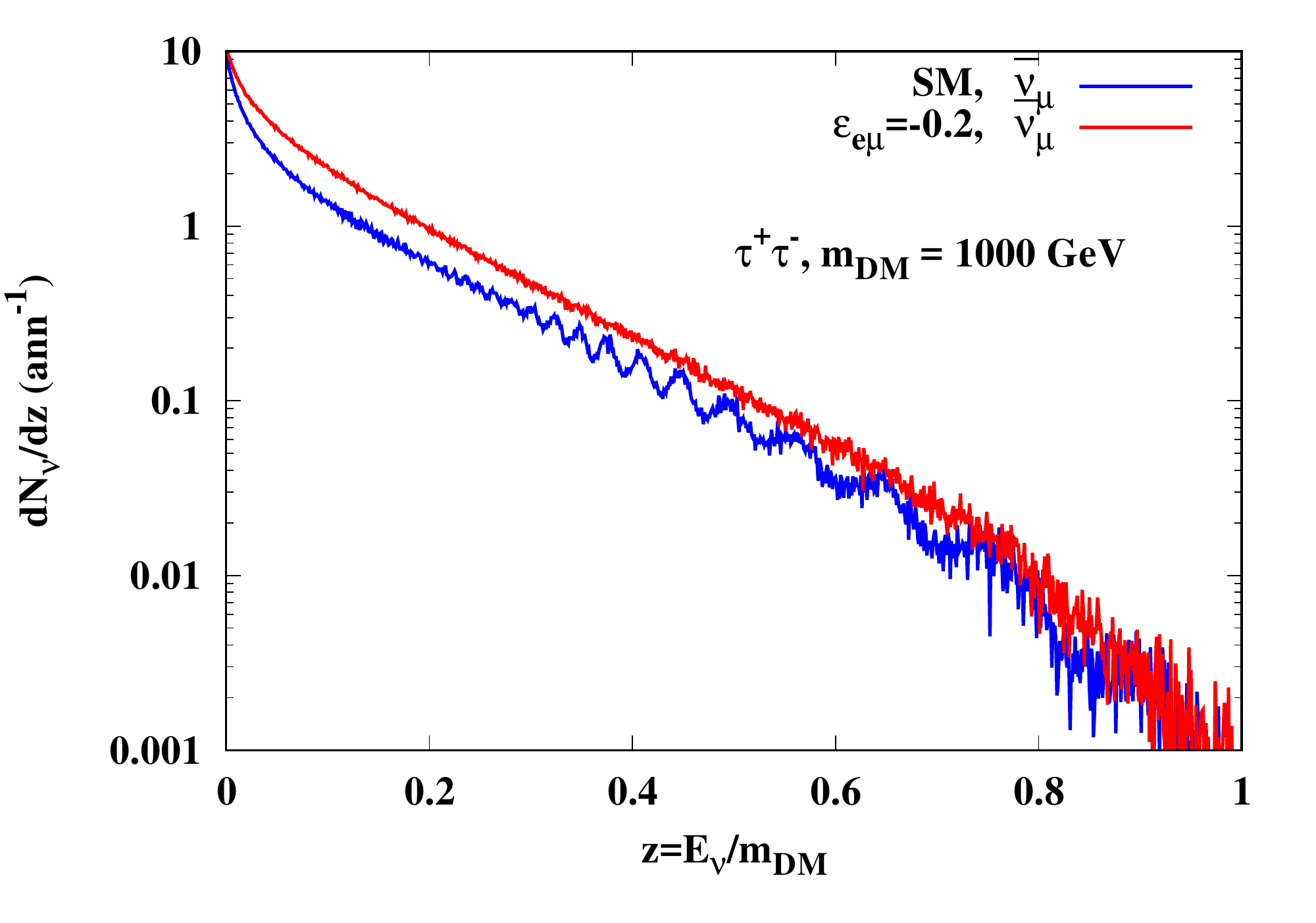}}
\put(260,240){\includegraphics[angle=-90,width=0.31\textwidth]{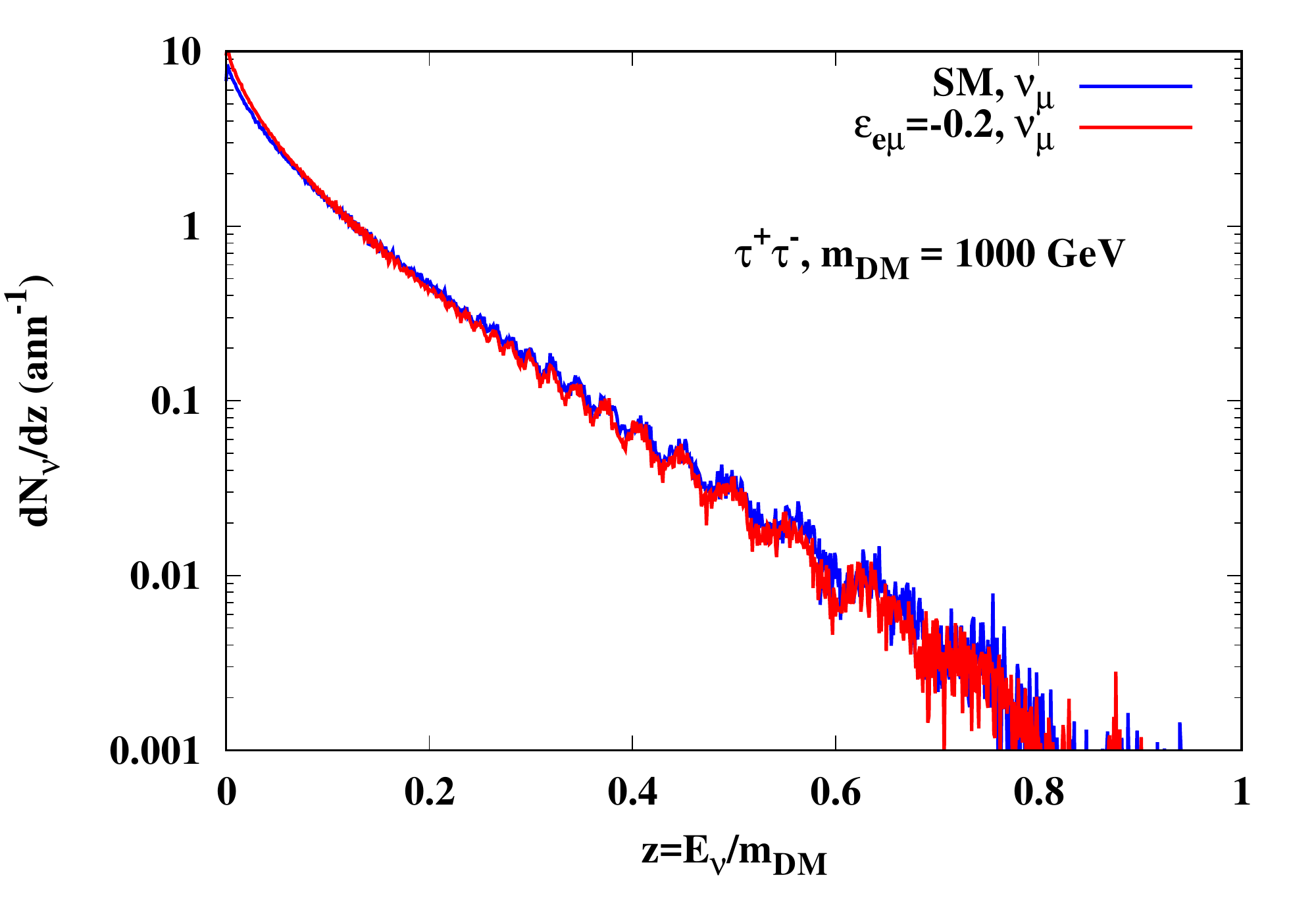}}
\put(130,130){\includegraphics[angle=-90,width=0.31\textwidth]{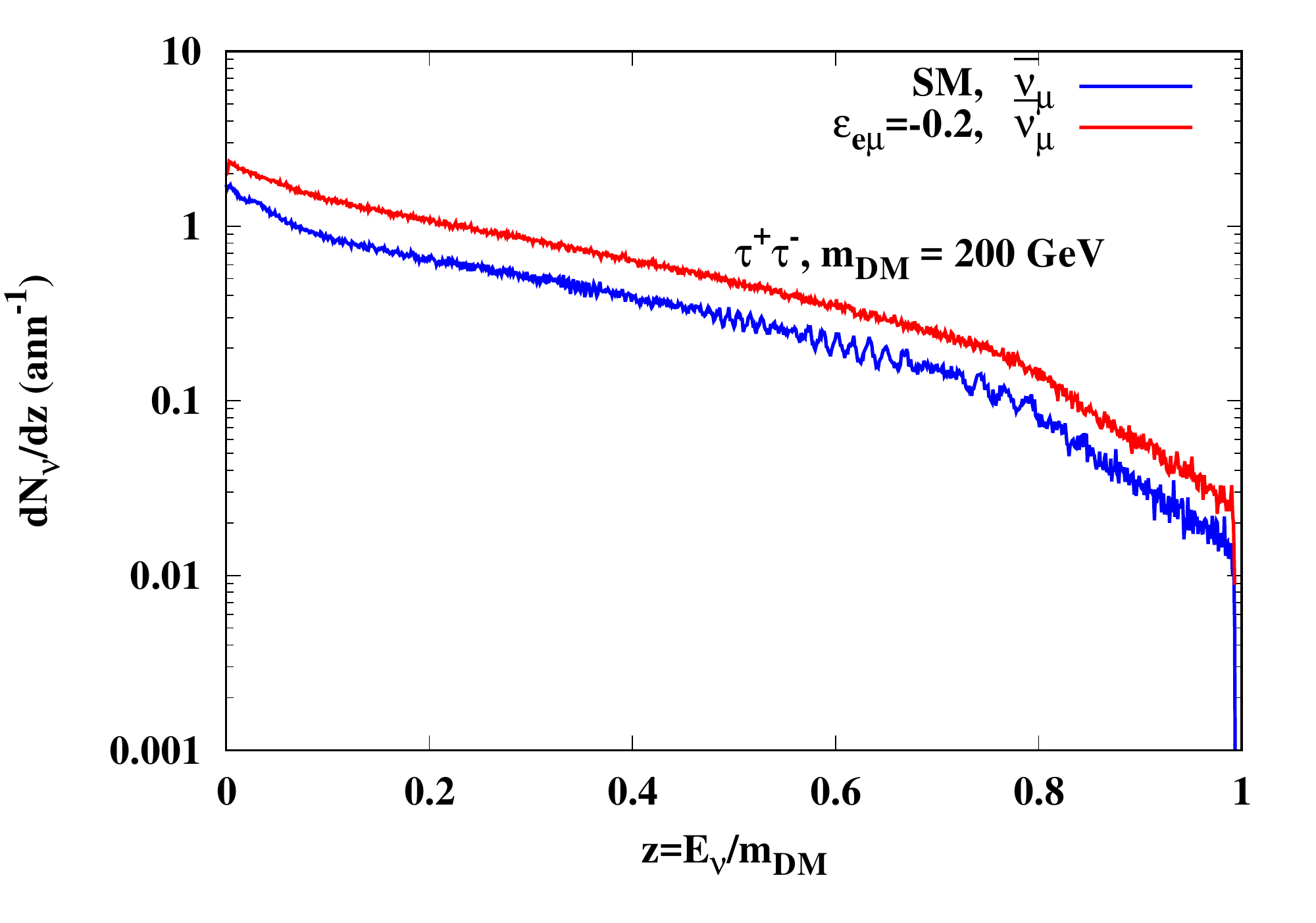}}
\put(130,240){\includegraphics[angle=-90,width=0.31\textwidth]{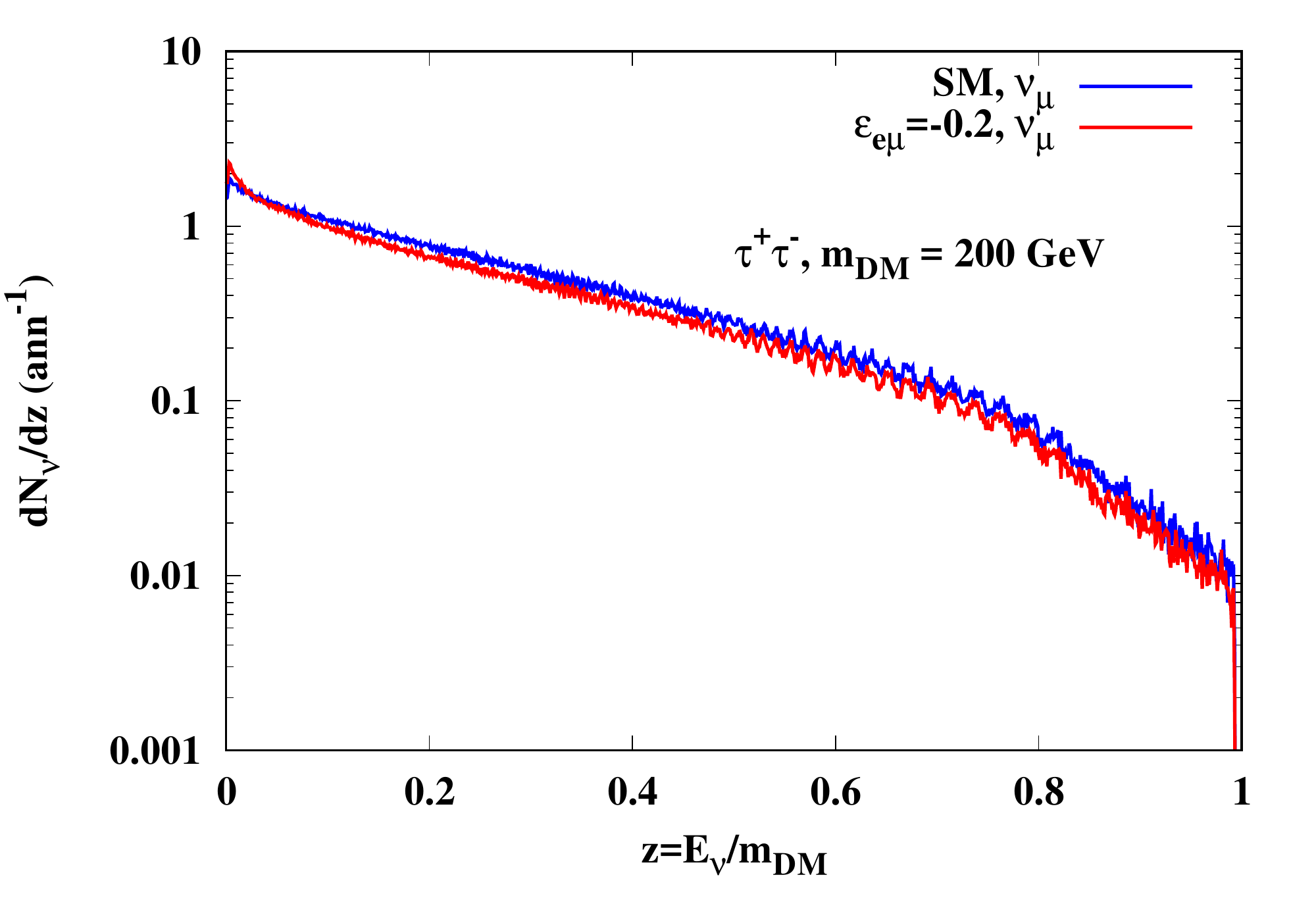}}
\put(0,130){\includegraphics[angle=-90,width=0.31\textwidth]{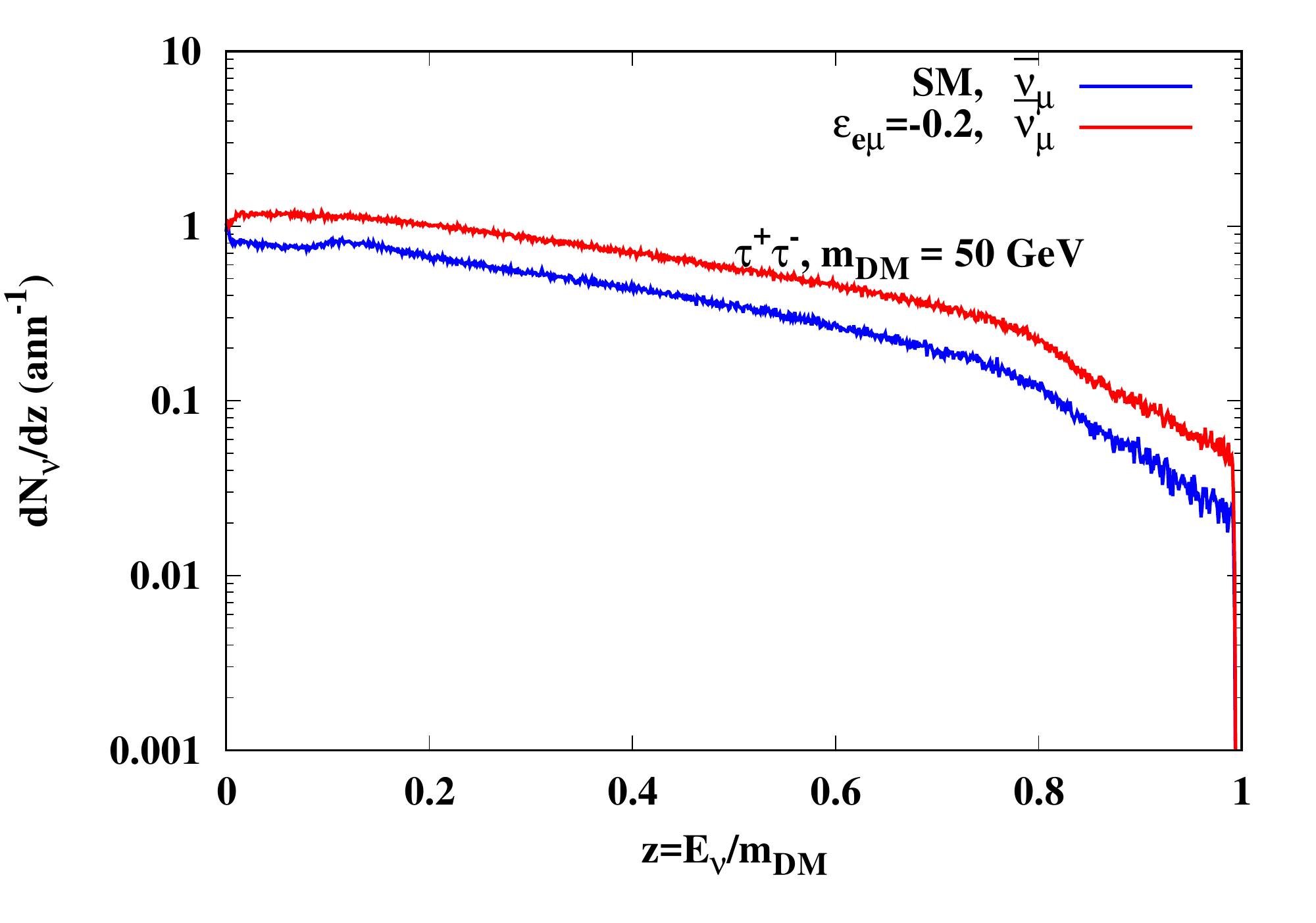}}
\put(0,240){\includegraphics[angle=-90,width=0.31\textwidth]{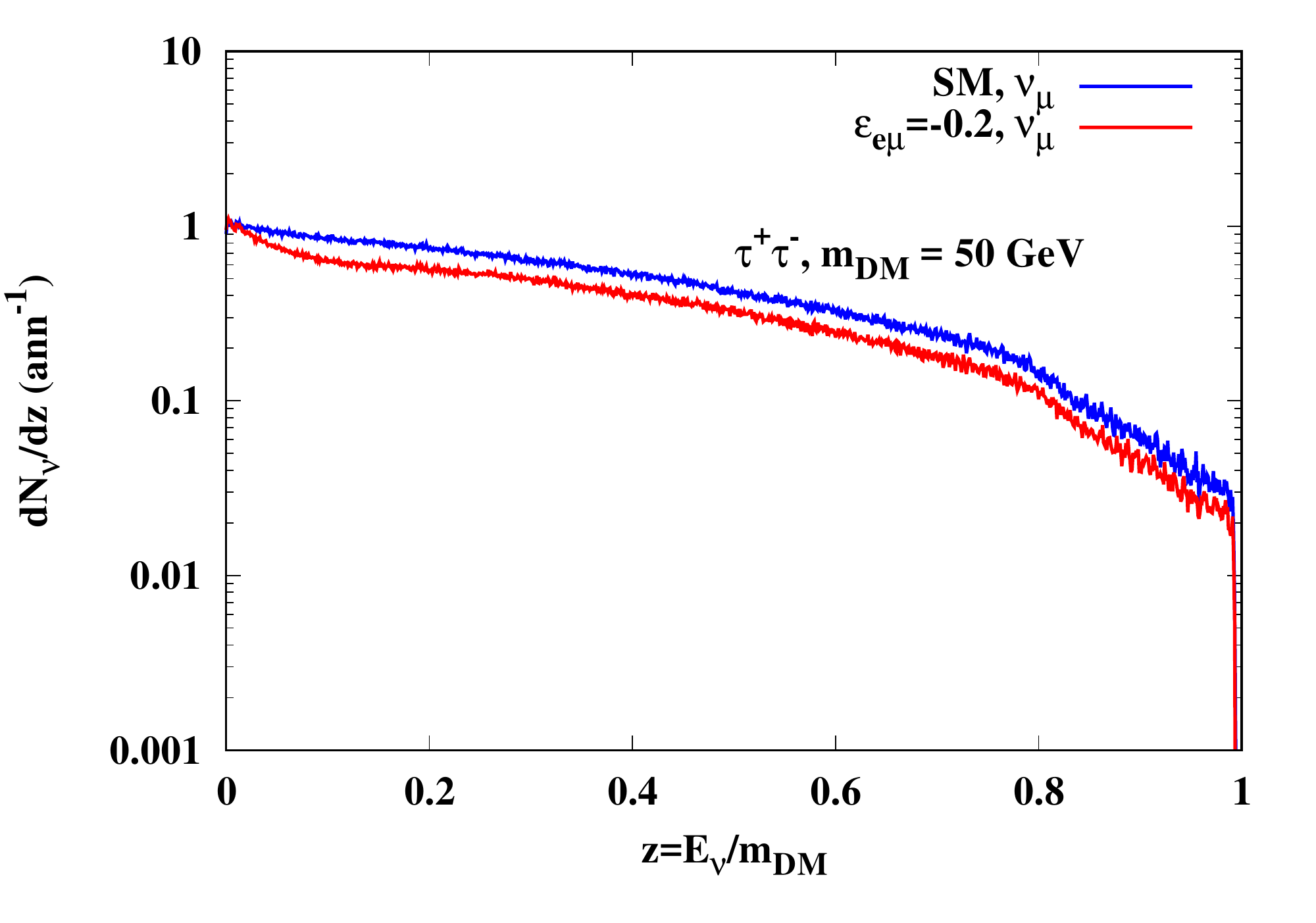}}
\end{picture}
\caption{\label{tau_emu_inv_sign} The same as in Fig.~\ref{tau_tautau}
  but for $\e_{e\mu} = -0.2$ and IH.}
\end{figure}
in Figs.~\ref{tau_emu_inv} and~\ref{tau_emu_inv_sign} for inverted
mass hierarchy. General conclusion drawn from these Figures is that
the non-standard neutrino interactions for some values of their
parameters 
can significantly change neutrino and/or antineutrino energy spectra
in particular for small masses of dark matter particles. 
Finally, in Figs.~\ref{tau_mutau}--\ref{tau_mutau_inv_sign}
\begin{figure}[!htb]
\begin{picture}(300,190)(0,40)
\put(260,130){\includegraphics[angle=-90,width=0.31\textwidth]{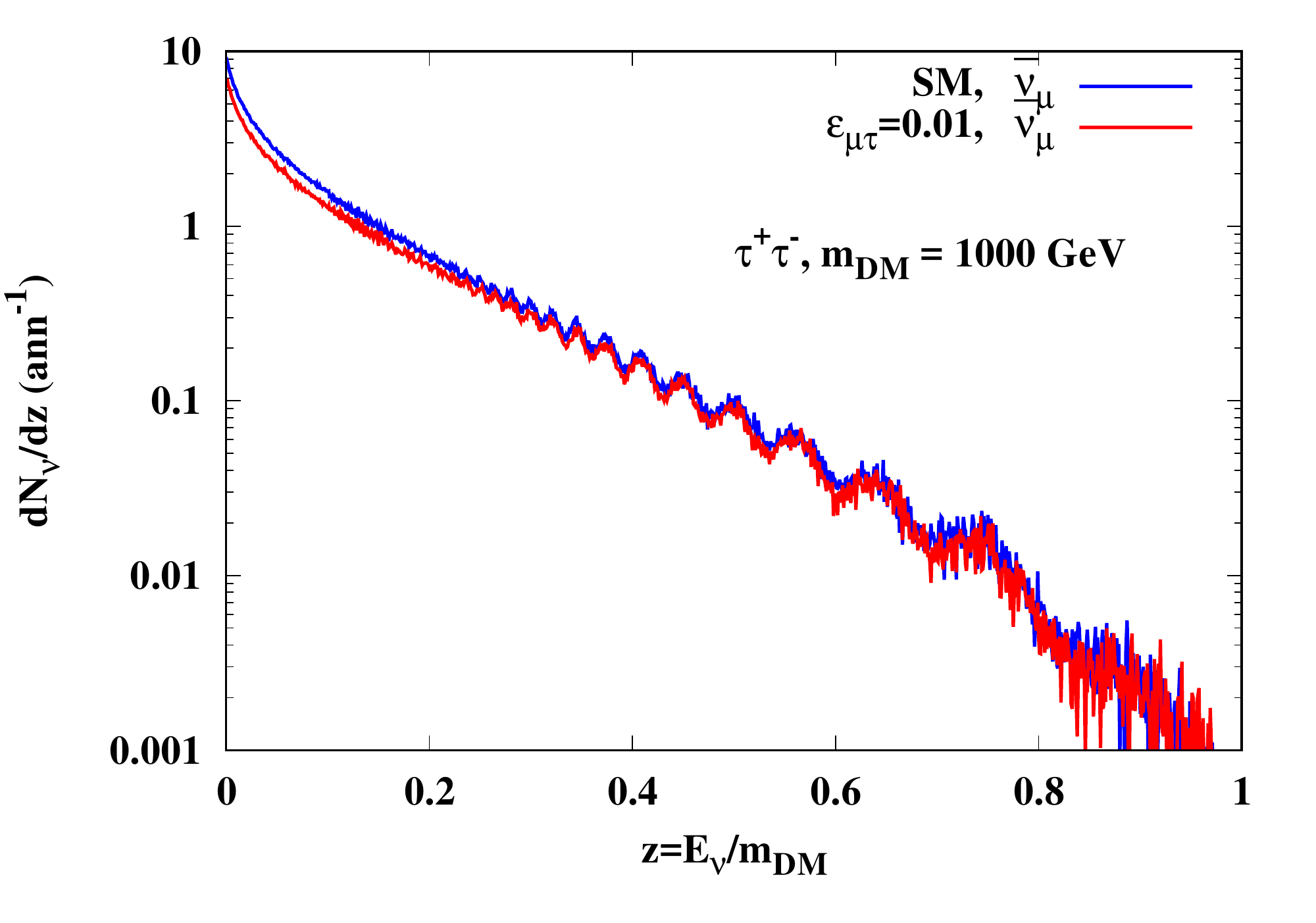}}
\put(260,240){\includegraphics[angle=-90,width=0.31\textwidth]{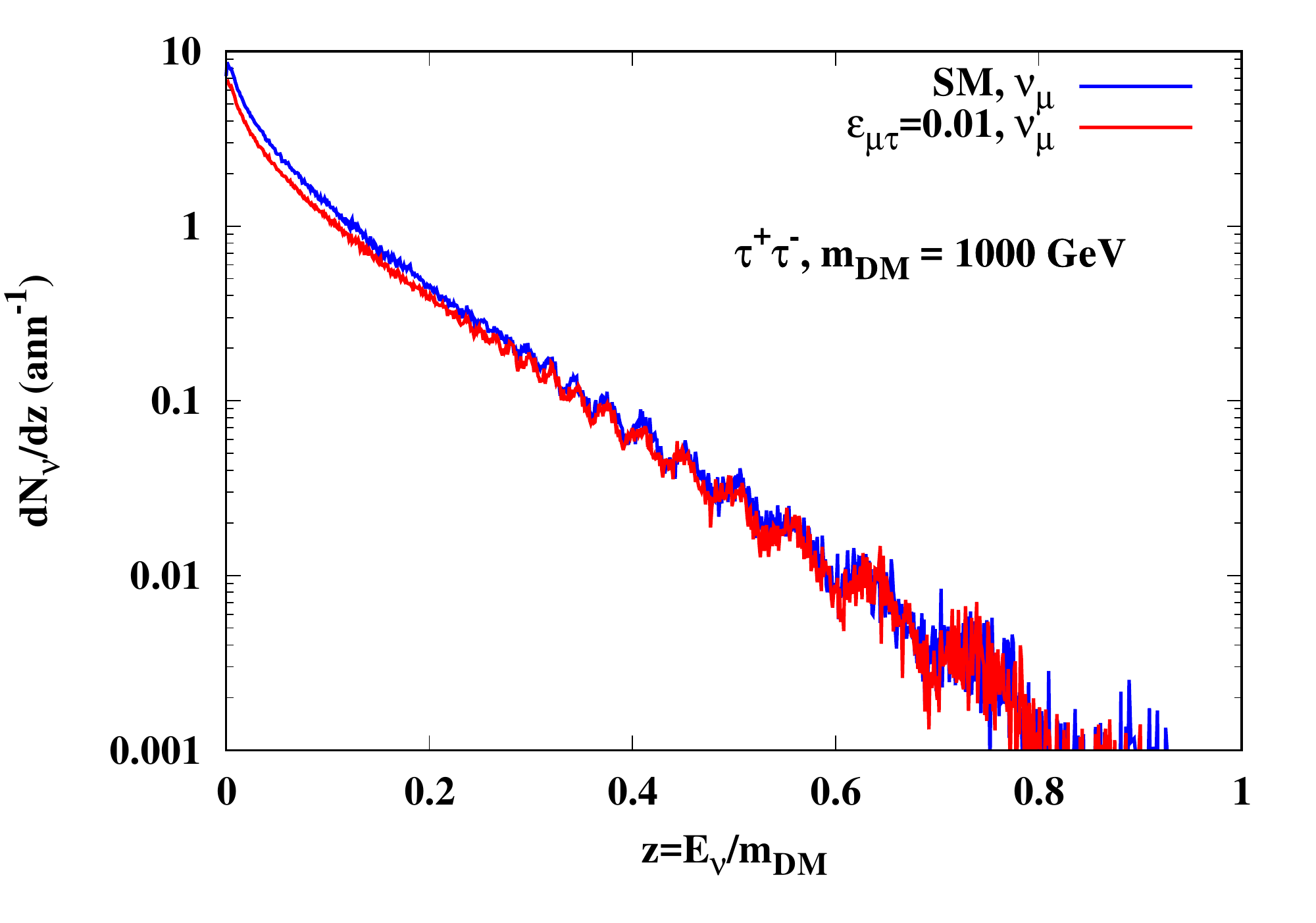}}
\put(130,130){\includegraphics[angle=-90,width=0.31\textwidth]{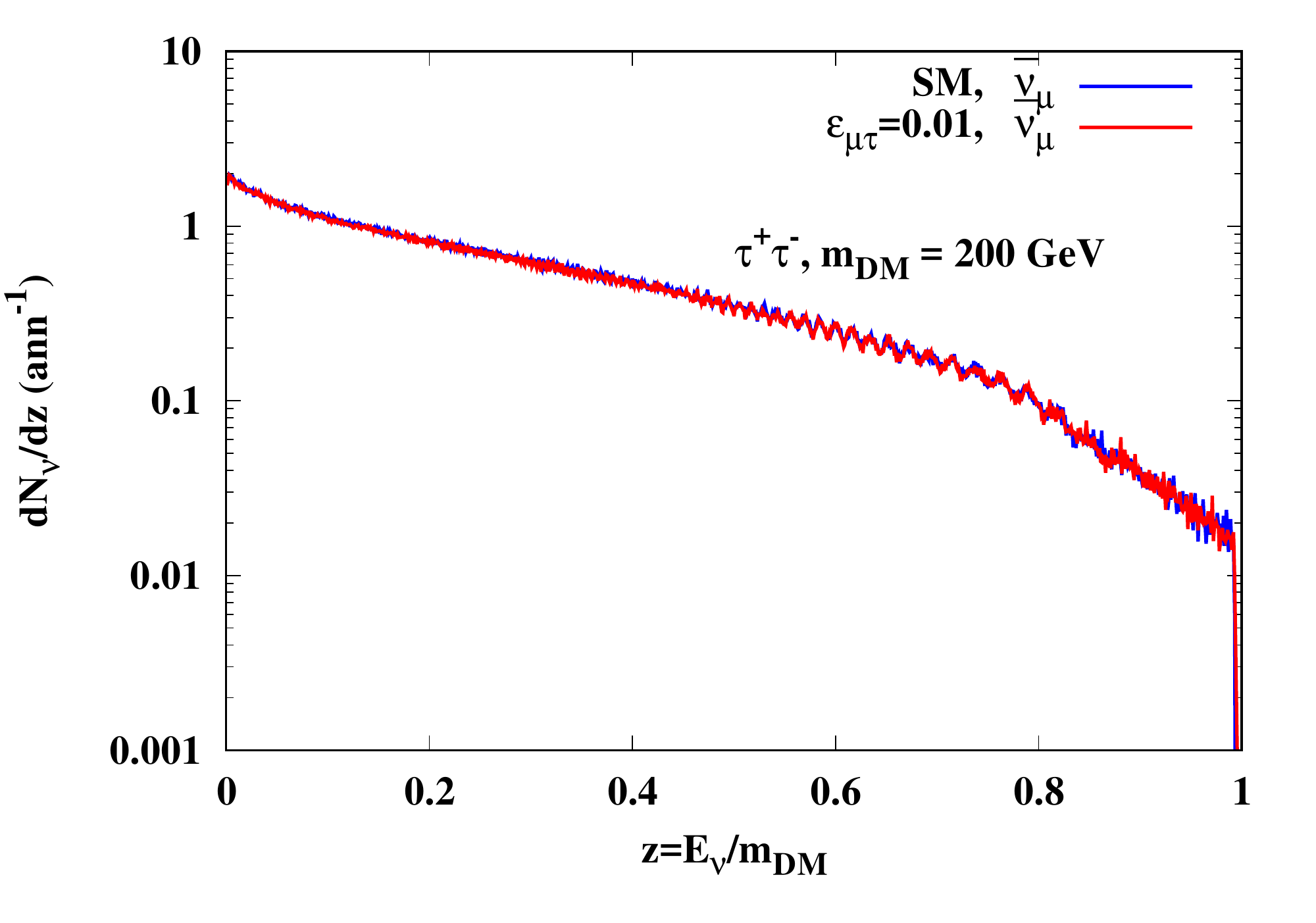}}
\put(130,240){\includegraphics[angle=-90,width=0.31\textwidth]{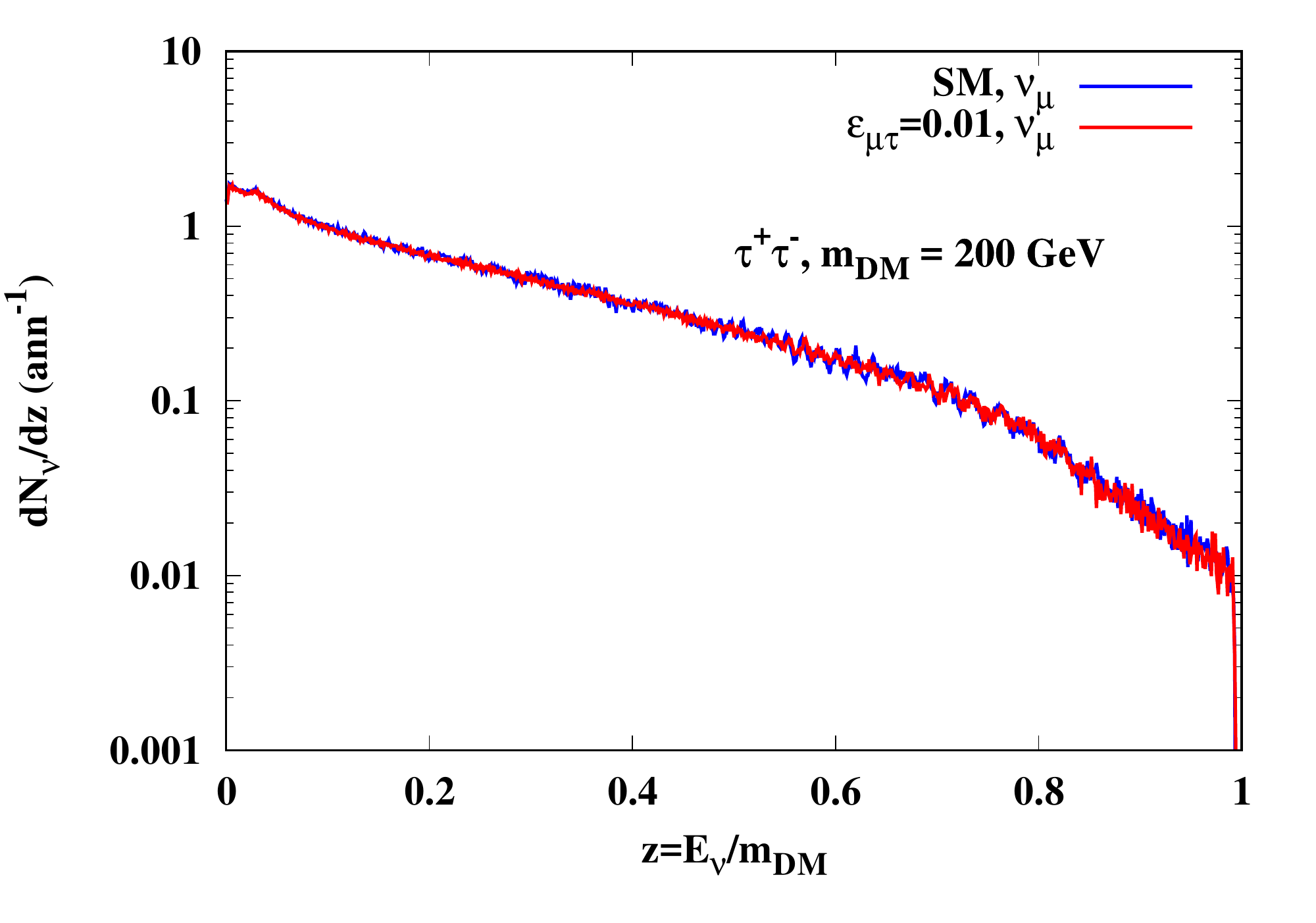}}
\put(0,130){\includegraphics[angle=-90,width=0.31\textwidth]{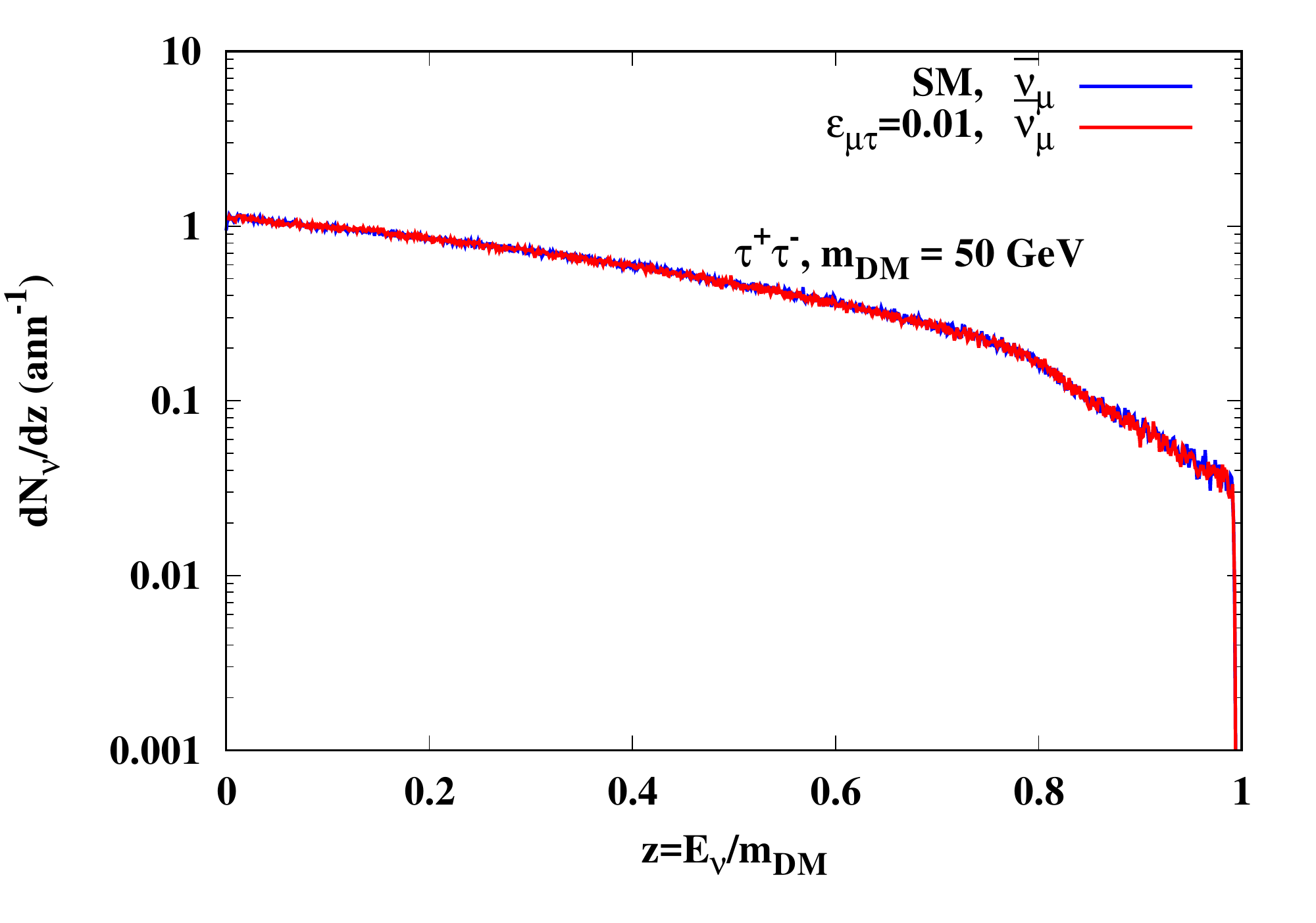}}
\put(0,240){\includegraphics[angle=-90,width=0.31\textwidth]{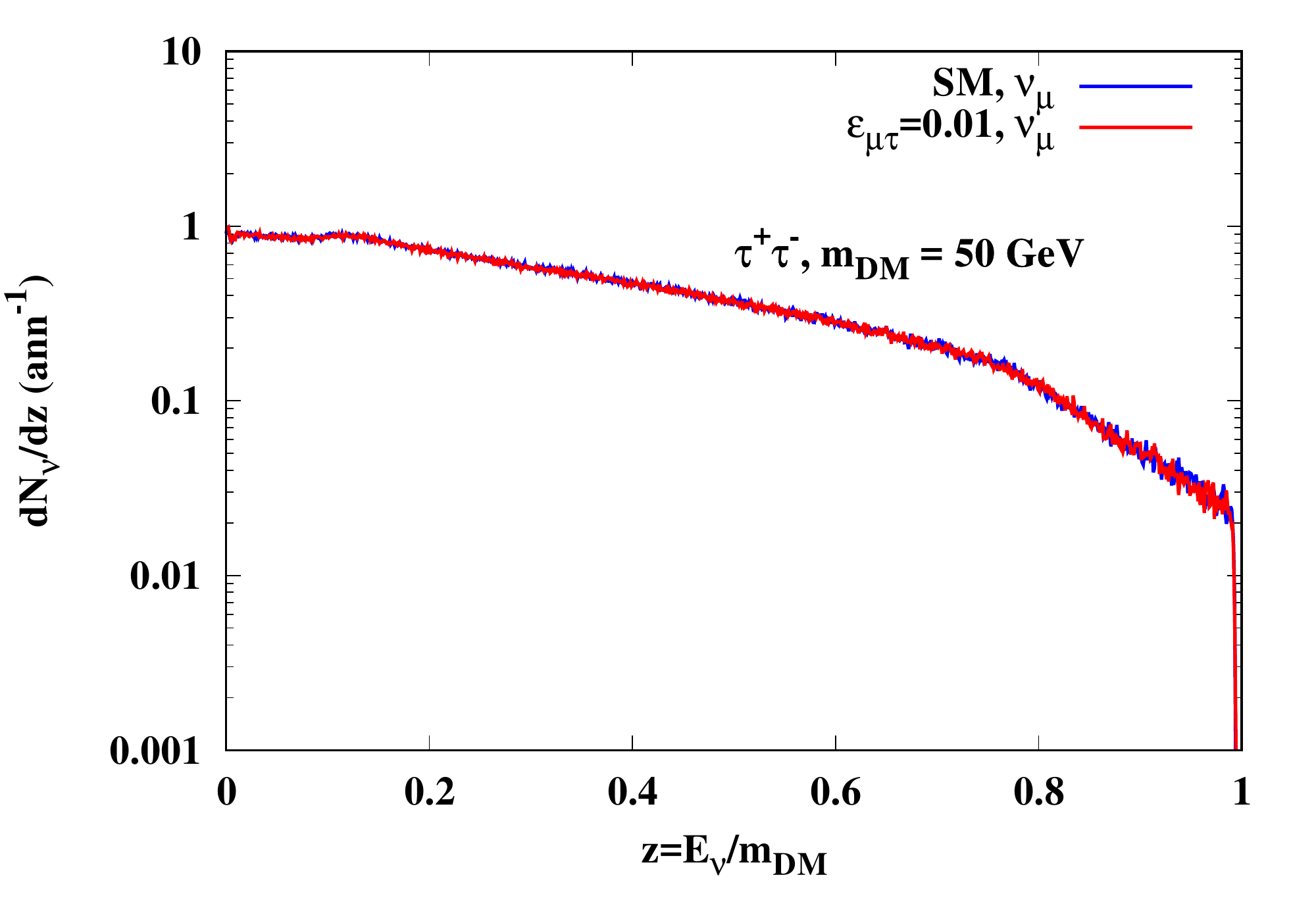}}
\end{picture}
\caption{\label{tau_mutau} The same as in Fig.~\ref{tau_tautau}
  but for $\e_{\mu\tau} = 0.01$ and NH.}
\end{figure}
\begin{figure}[!htb]
\begin{picture}(300,190)(0,40)
\put(260,130){\includegraphics[angle=-90,width=0.31\textwidth]{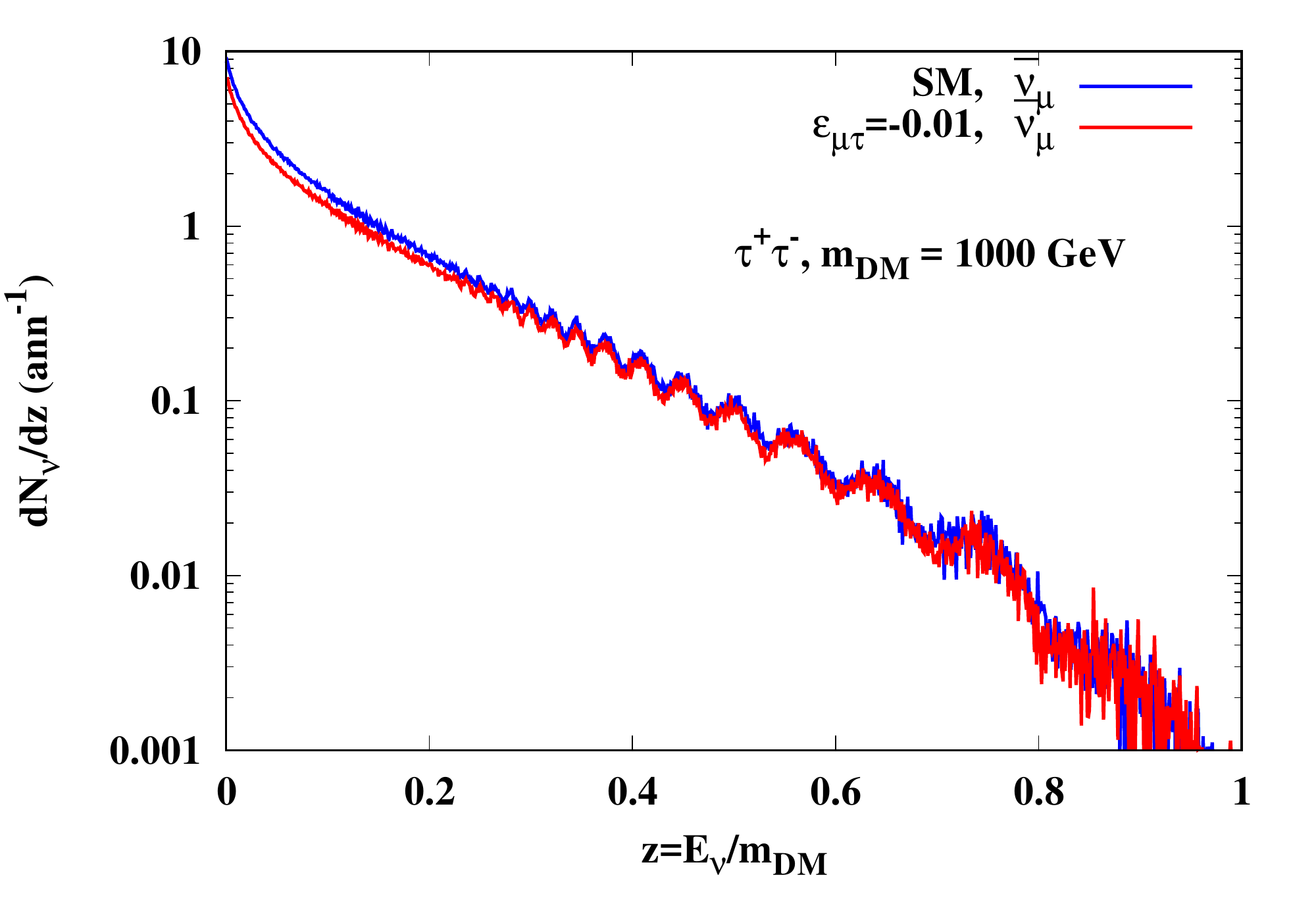}}
\put(260,240){\includegraphics[angle=-90,width=0.31\textwidth]{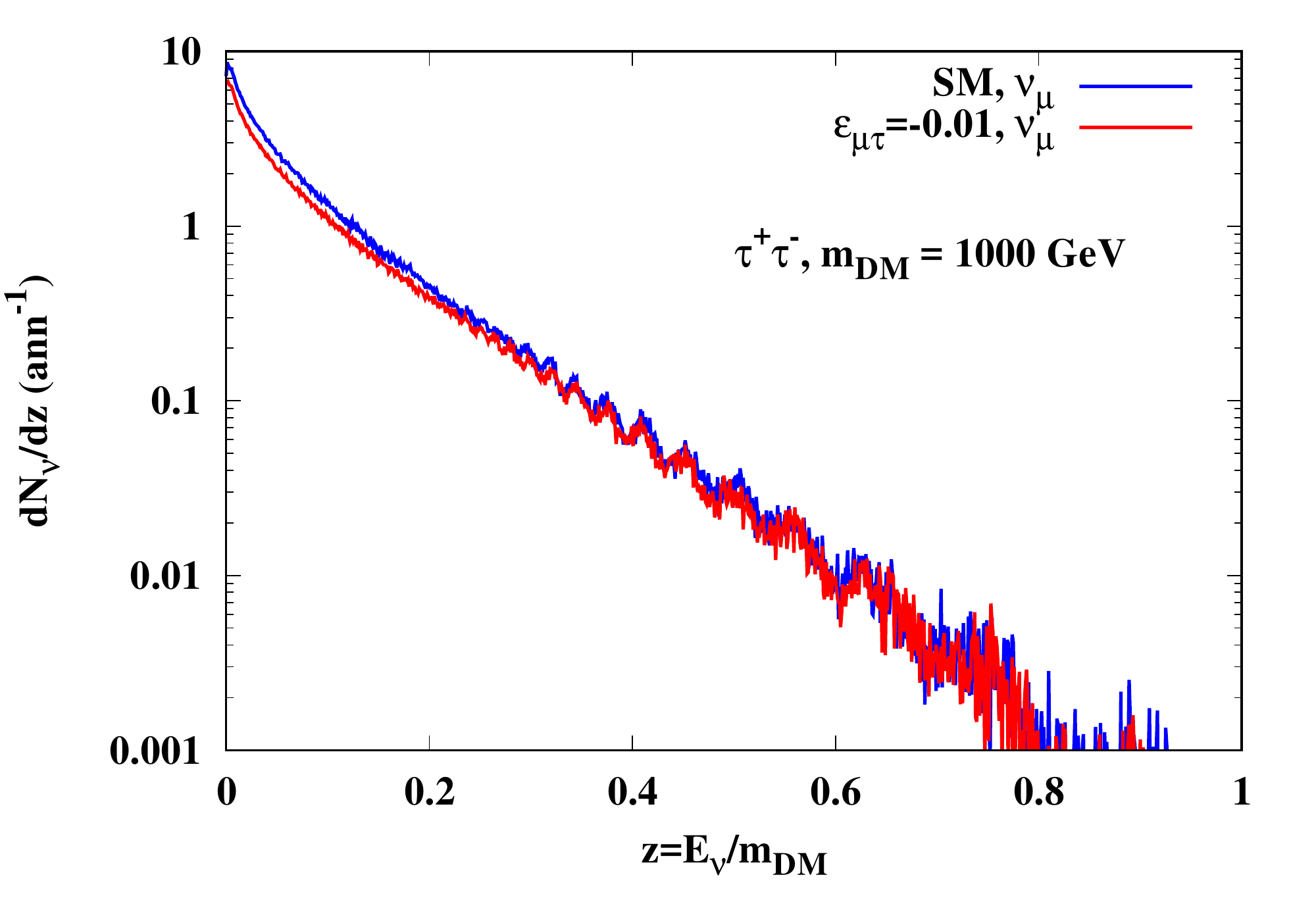}}
\put(130,130){\includegraphics[angle=-90,width=0.31\textwidth]{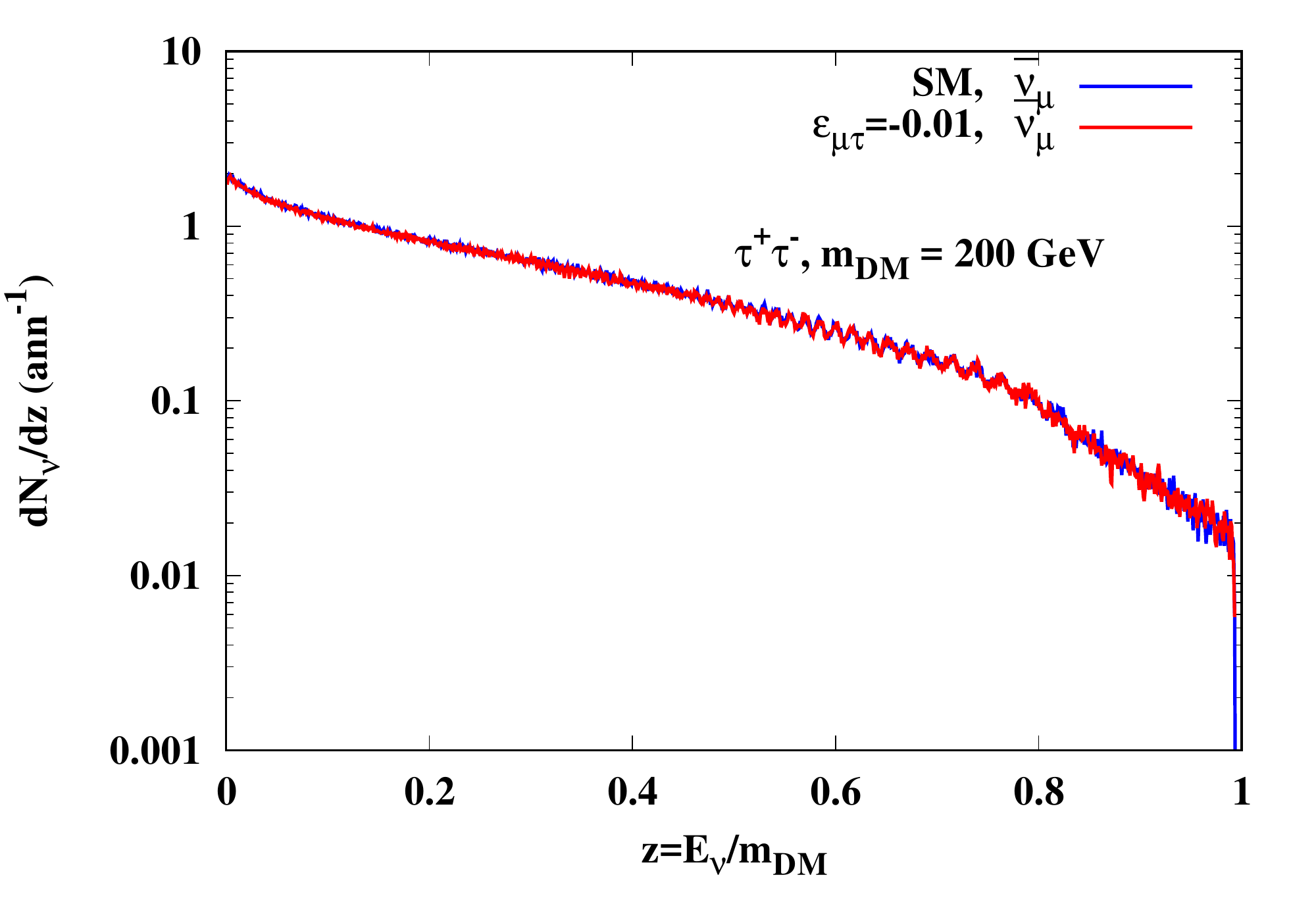}}
\put(130,240){\includegraphics[angle=-90,width=0.31\textwidth]{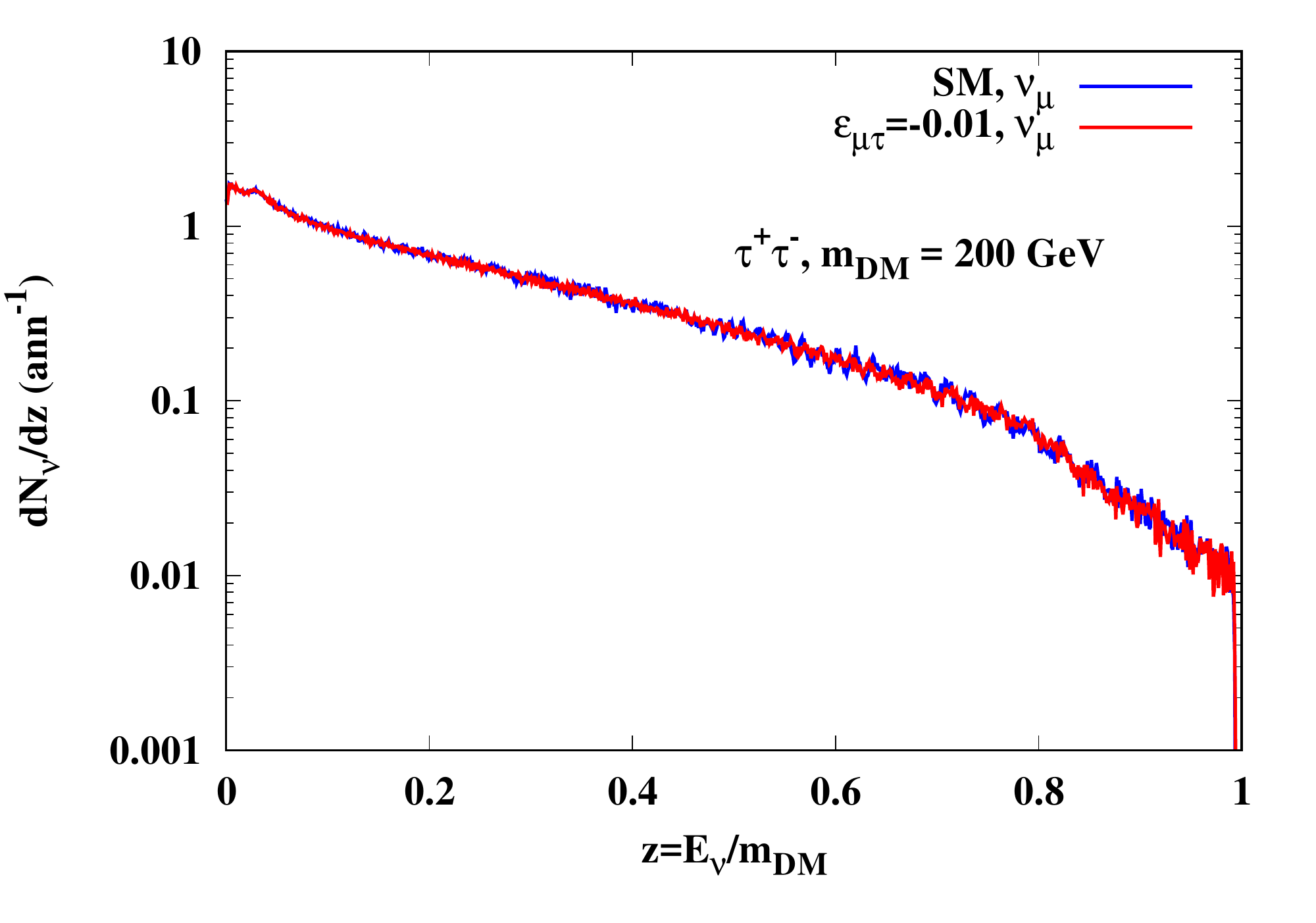}}
\put(0,130){\includegraphics[angle=-90,width=0.31\textwidth]{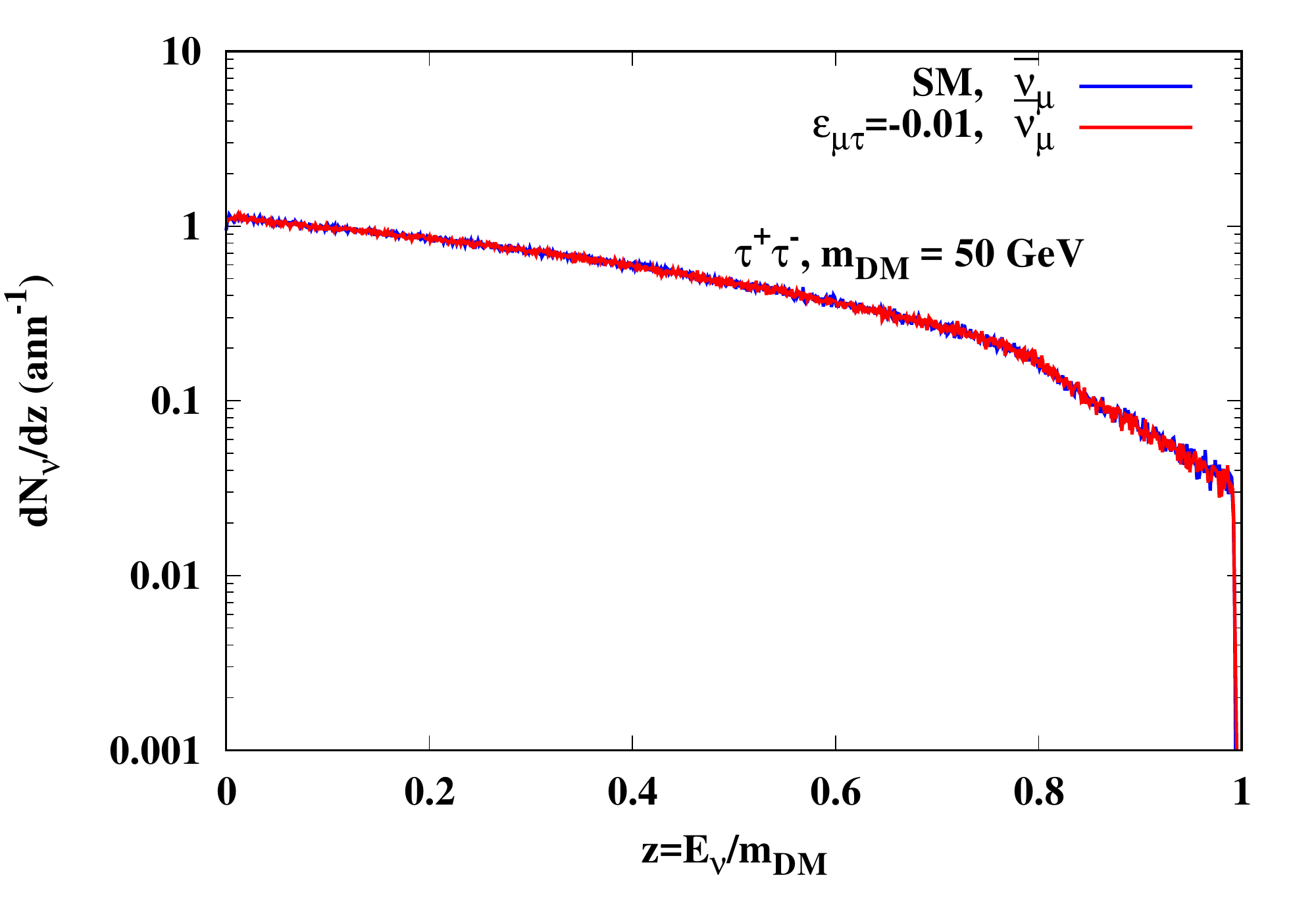}}
\put(0,240){\includegraphics[angle=-90,width=0.31\textwidth]{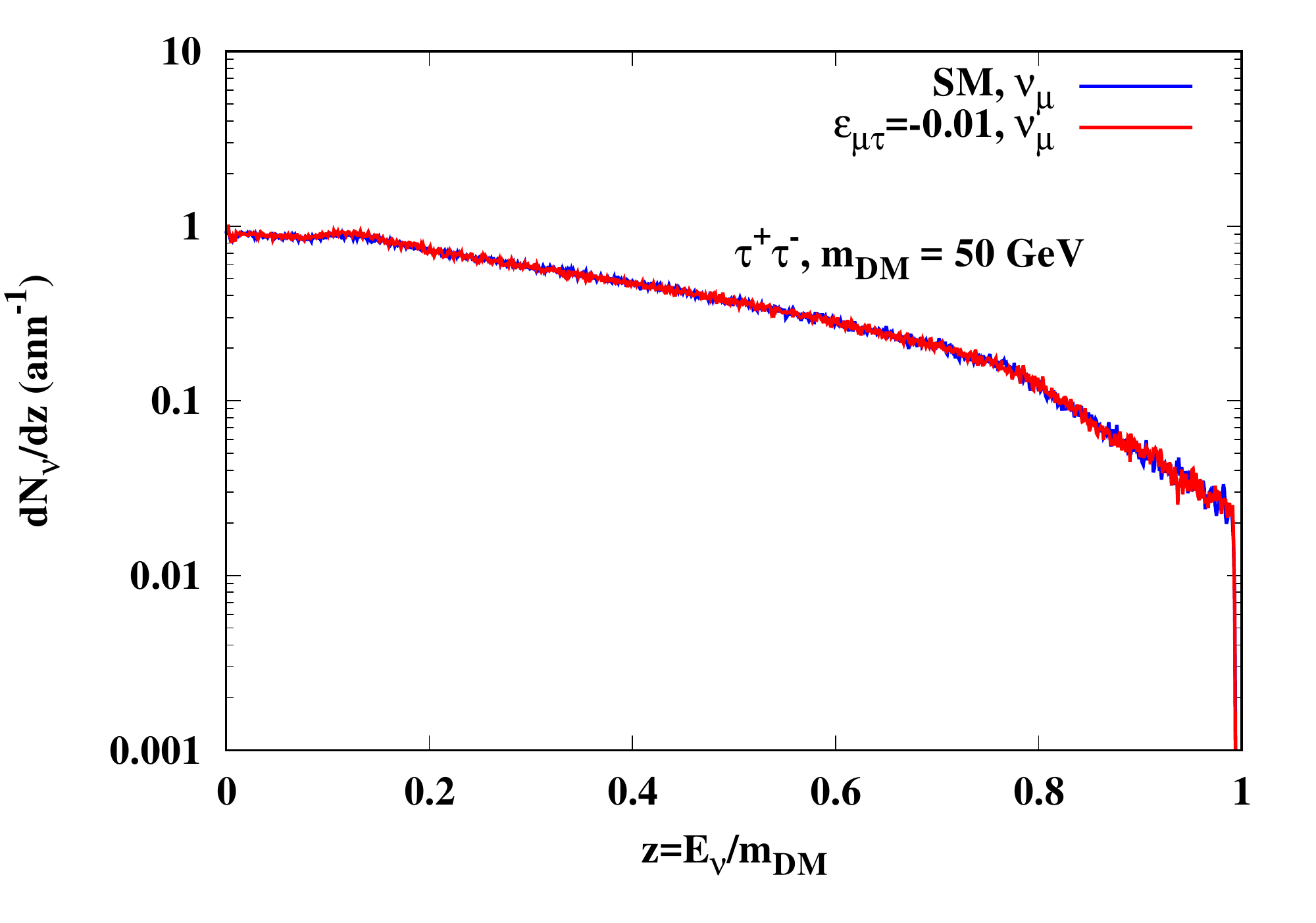}}
\end{picture}
\caption{\label{tau_mutau_sign} The same as in Fig.~\ref{tau_tautau}
  but for $\e_{\mu\tau} = -0.01$ and NH.}
\end{figure}
\begin{figure}[!htb]
\begin{picture}(300,190)(0,40)
\put(260,130){\includegraphics[angle=-90,width=0.31\textwidth]{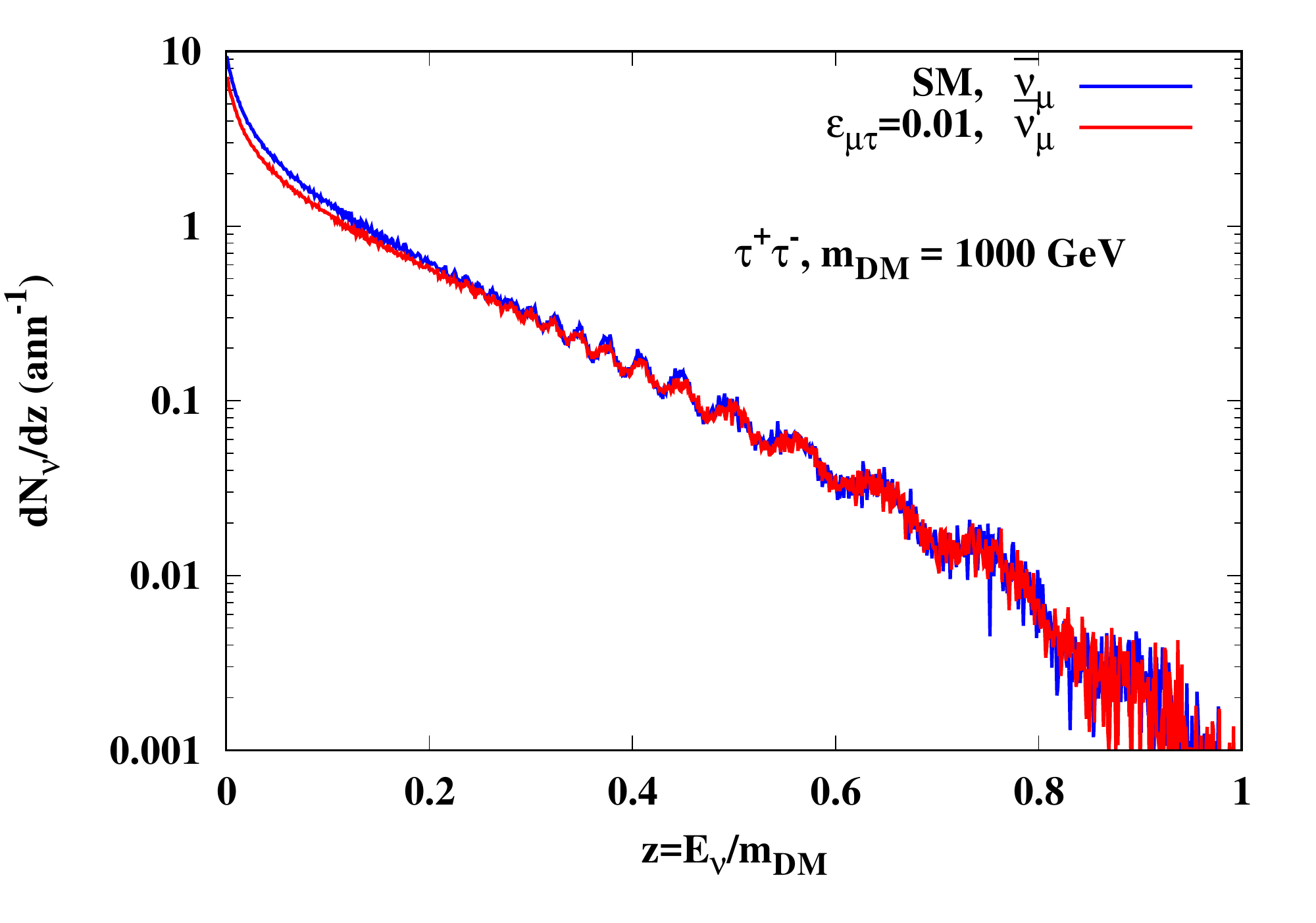}}
\put(260,240){\includegraphics[angle=-90,width=0.31\textwidth]{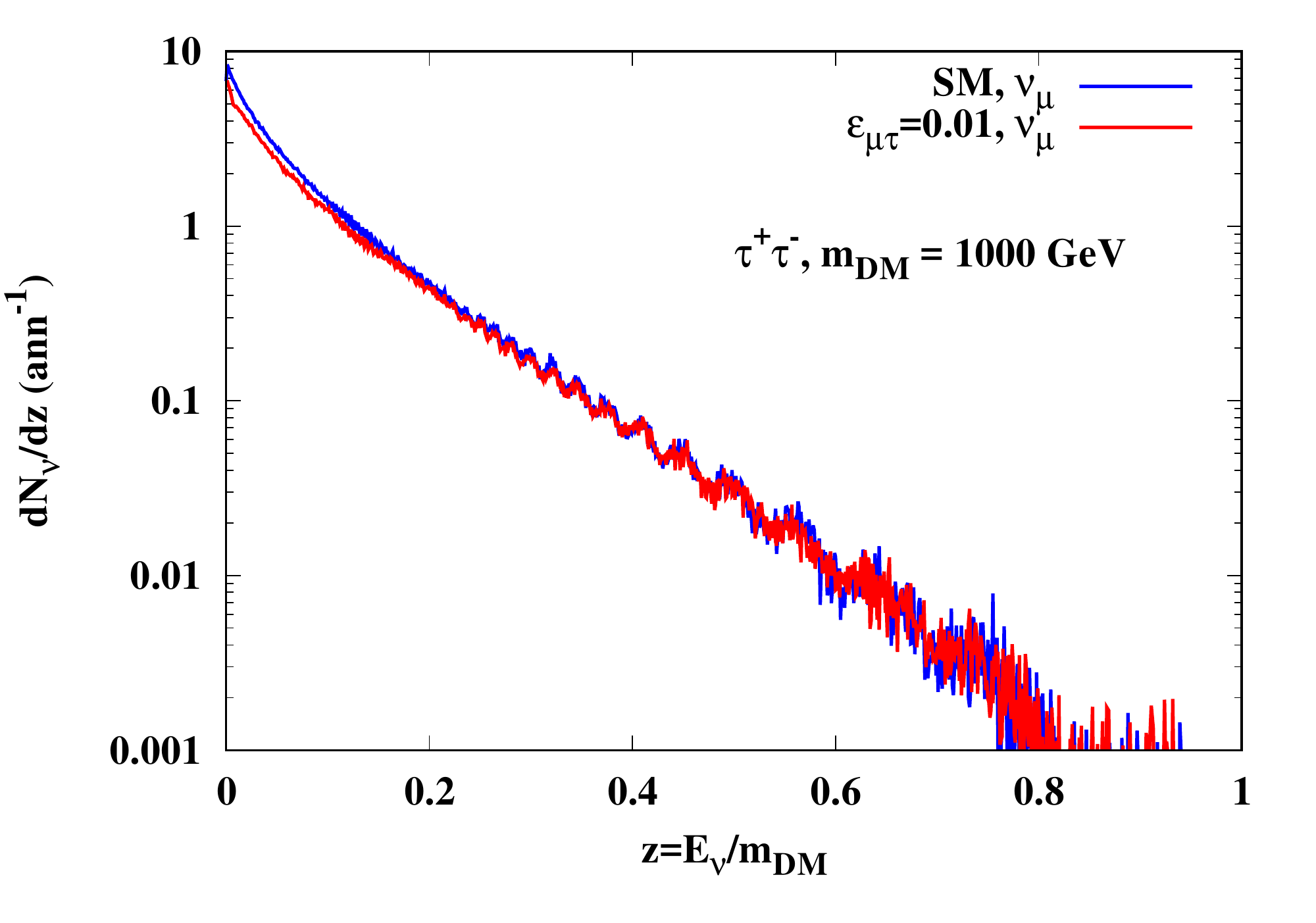}}
\put(130,130){\includegraphics[angle=-90,width=0.31\textwidth]{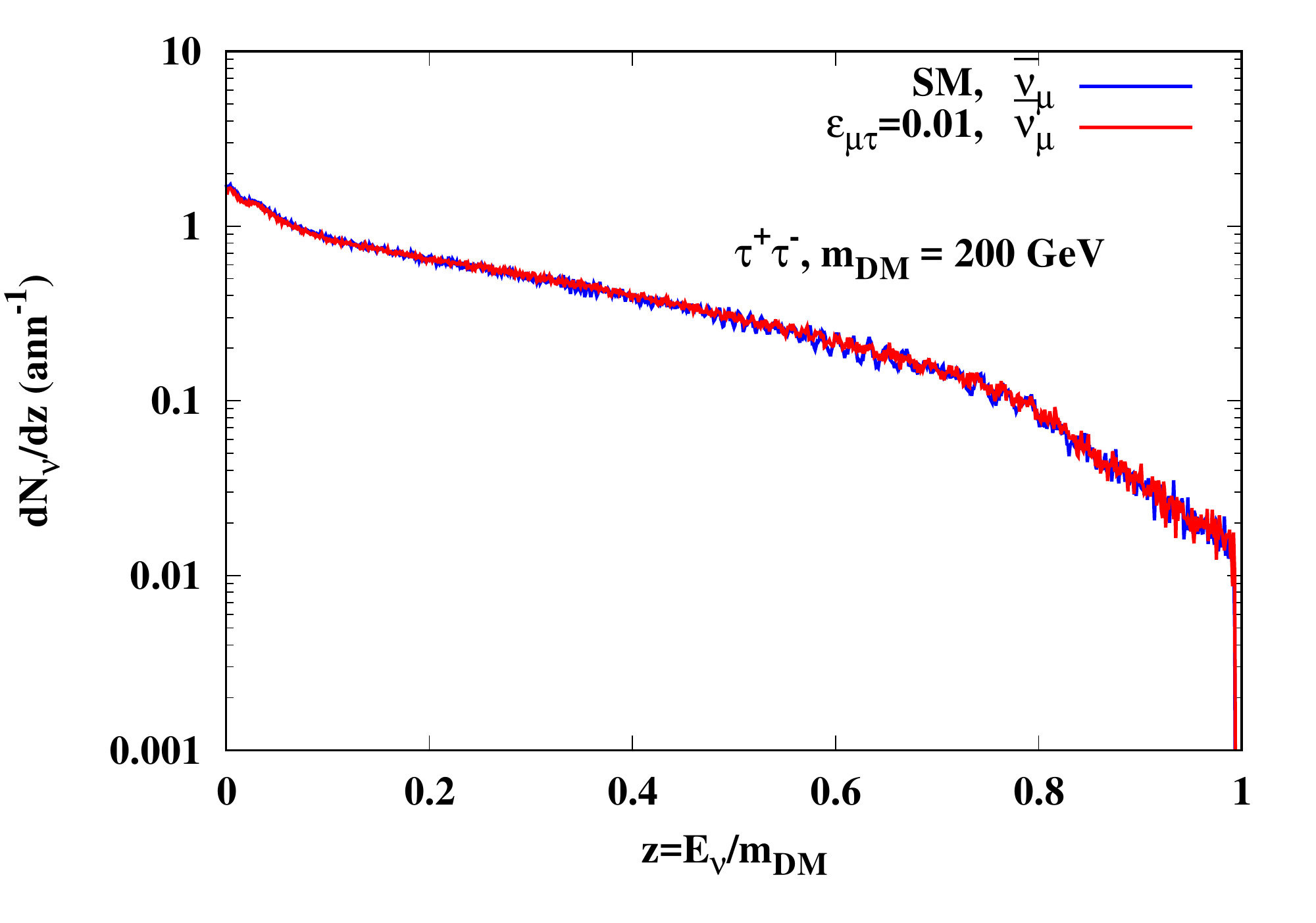}}
\put(130,240){\includegraphics[angle=-90,width=0.31\textwidth]{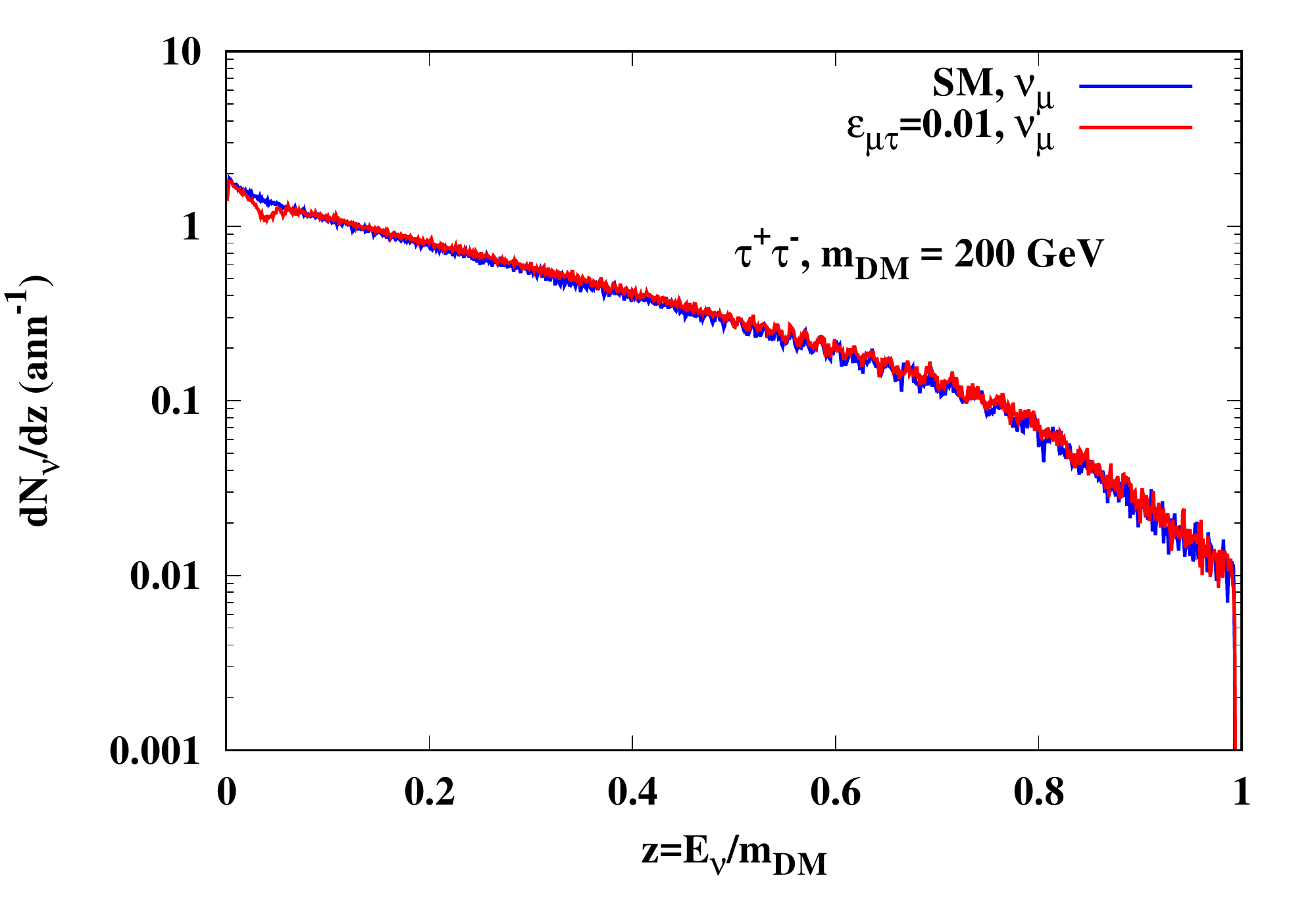}}
\put(0,130){\includegraphics[angle=-90,width=0.31\textwidth]{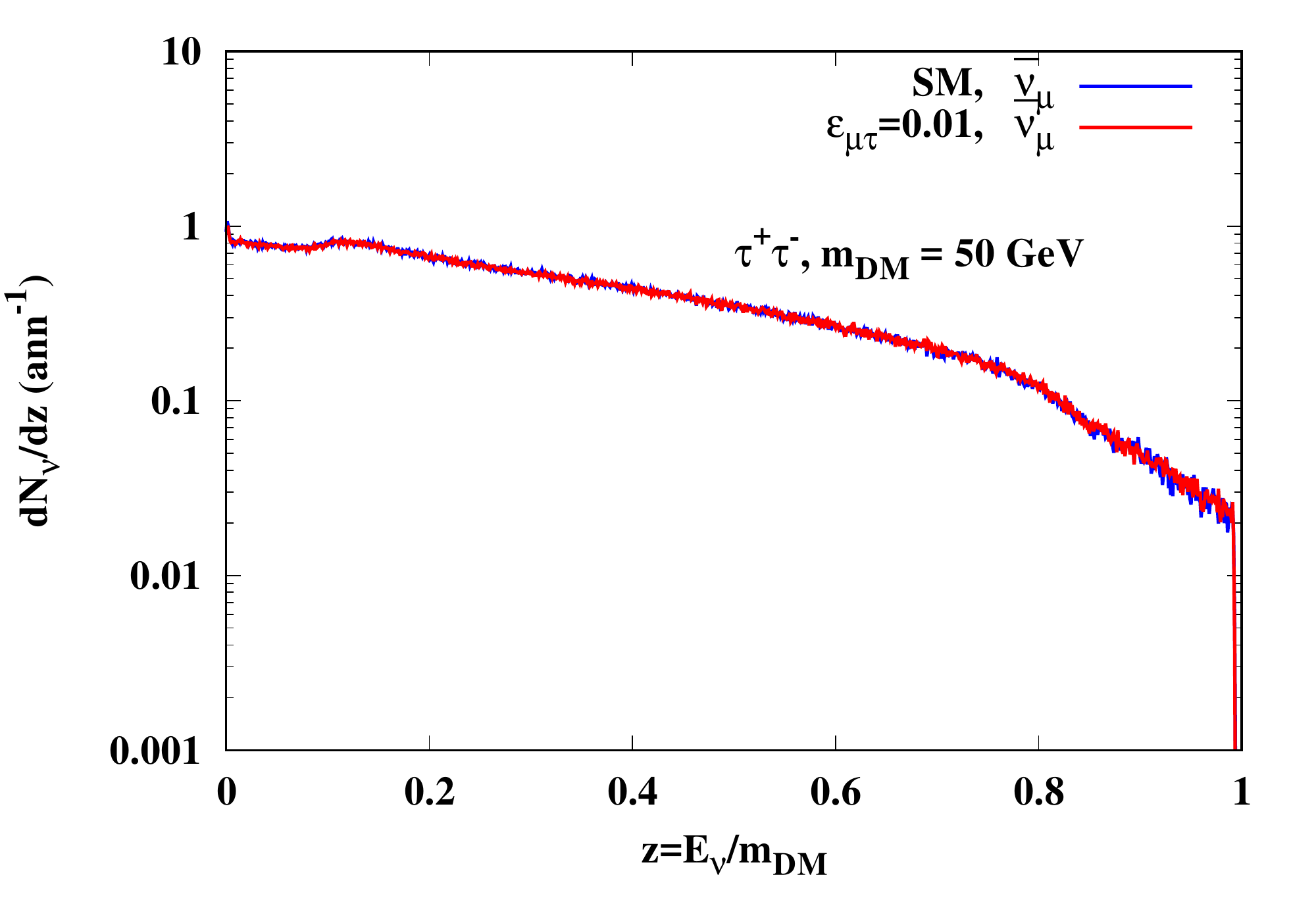}}
\put(0,240){\includegraphics[angle=-90,width=0.31\textwidth]{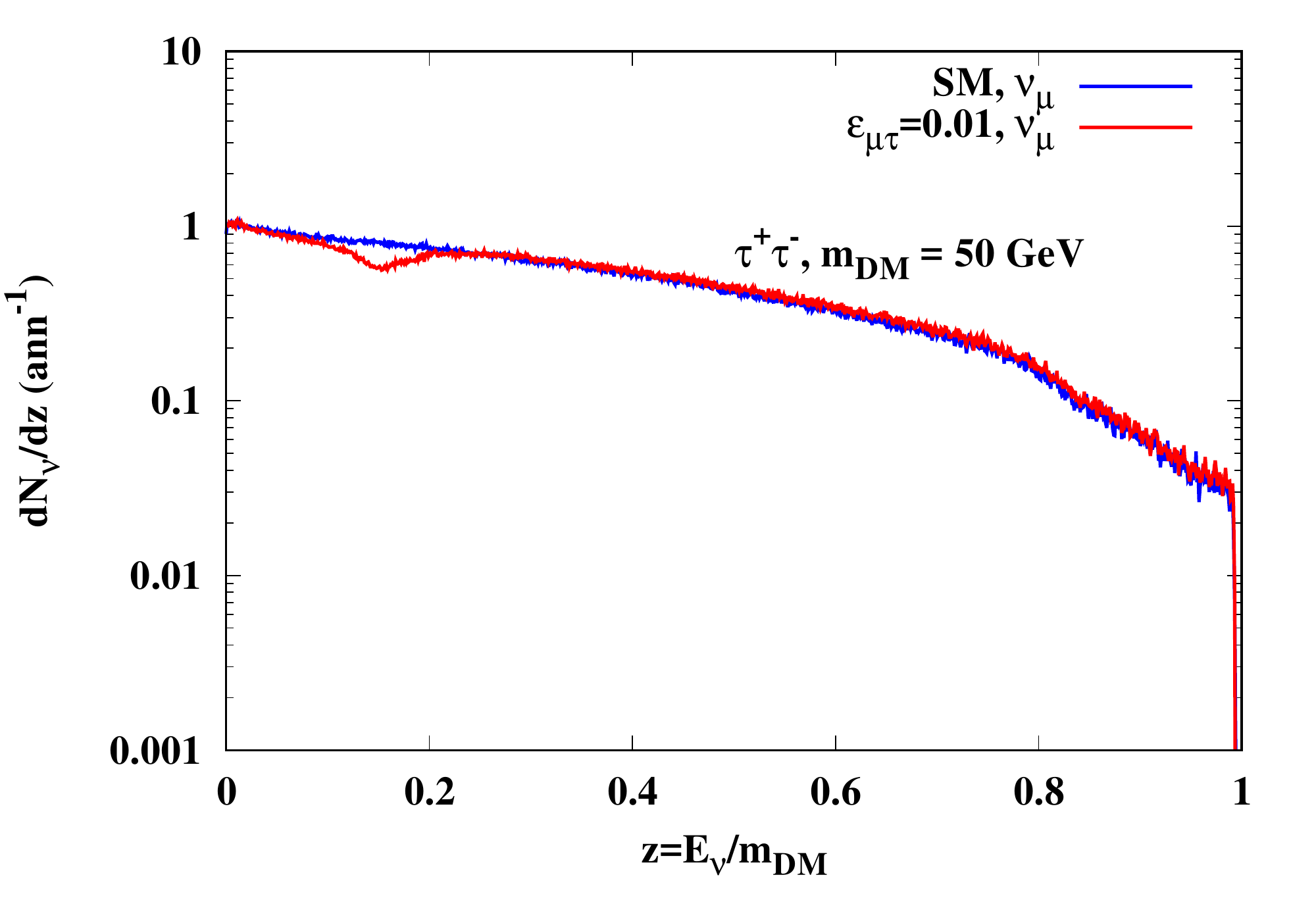}}
\end{picture}
\caption{\label{tau_mutau_inv} The same as in Fig.~\ref{tau_tautau}
  but for $\e_{\mu\tau} = 0.01$ and IH.}
\end{figure}
\begin{figure}[!htb]
\begin{picture}(300,190)(0,40)
\put(260,130){\includegraphics[angle=-90,width=0.31\textwidth]{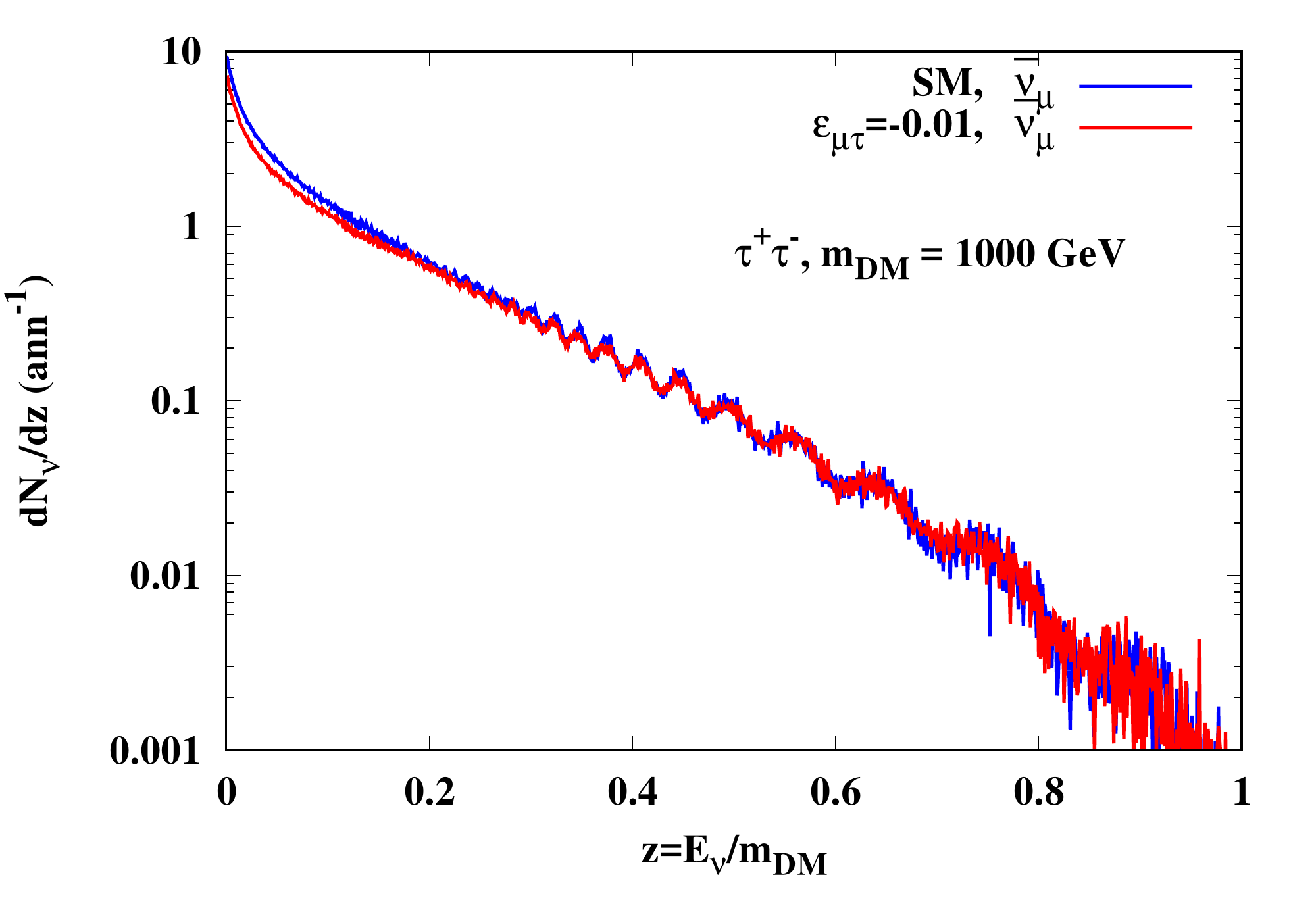}}
\put(260,240){\includegraphics[angle=-90,width=0.31\textwidth]{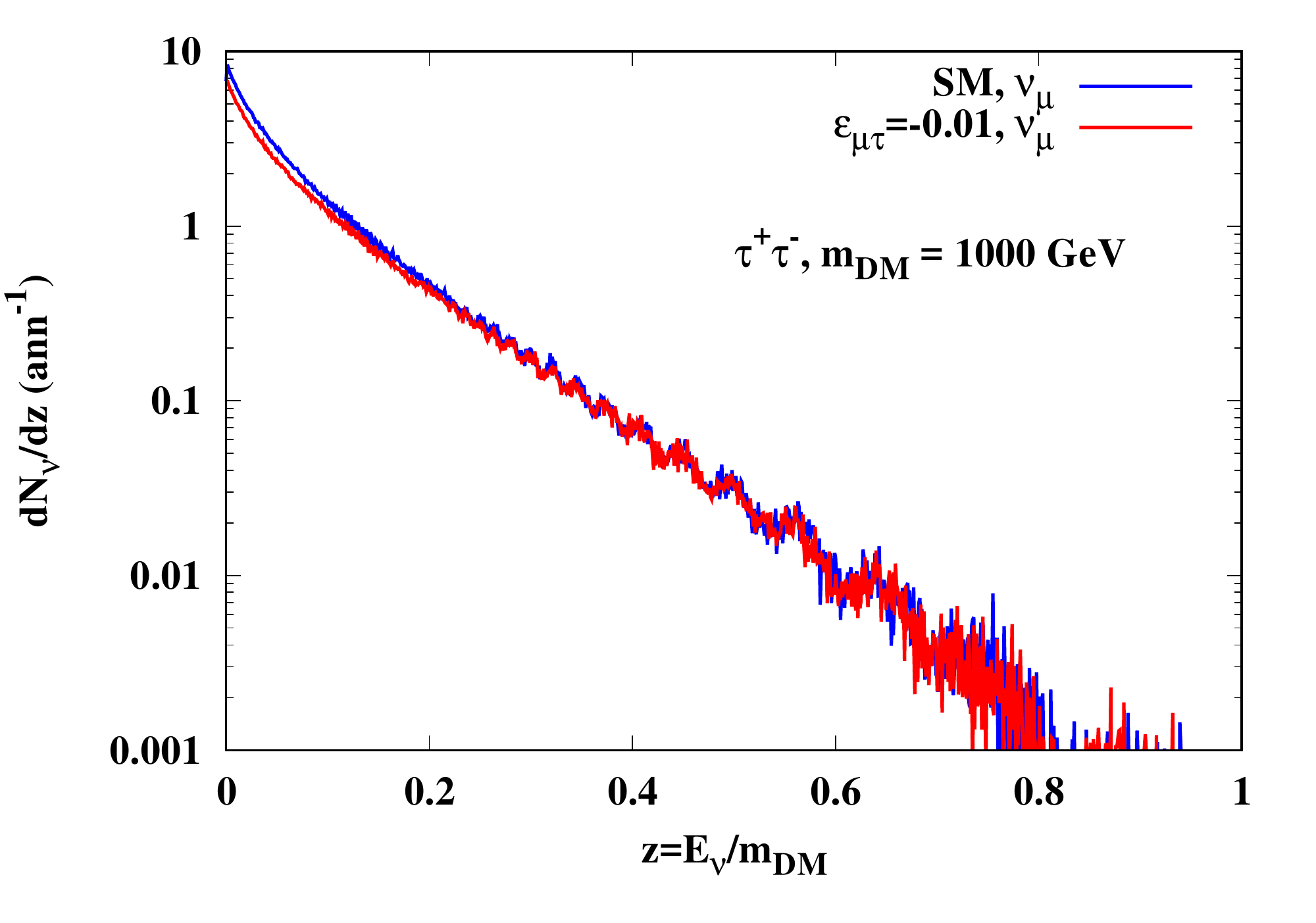}}
\put(130,130){\includegraphics[angle=-90,width=0.31\textwidth]{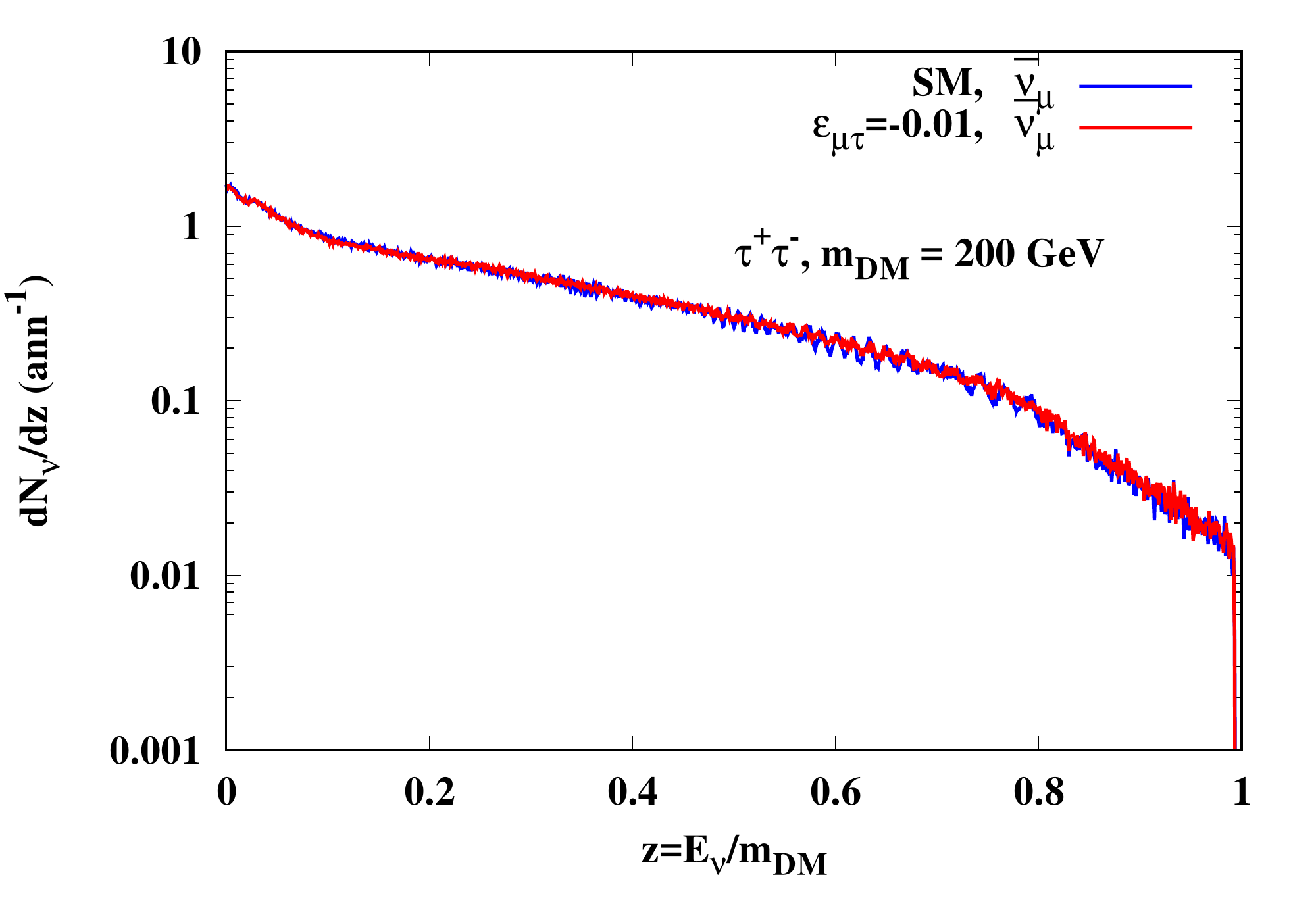}}
\put(130,240){\includegraphics[angle=-90,width=0.31\textwidth]{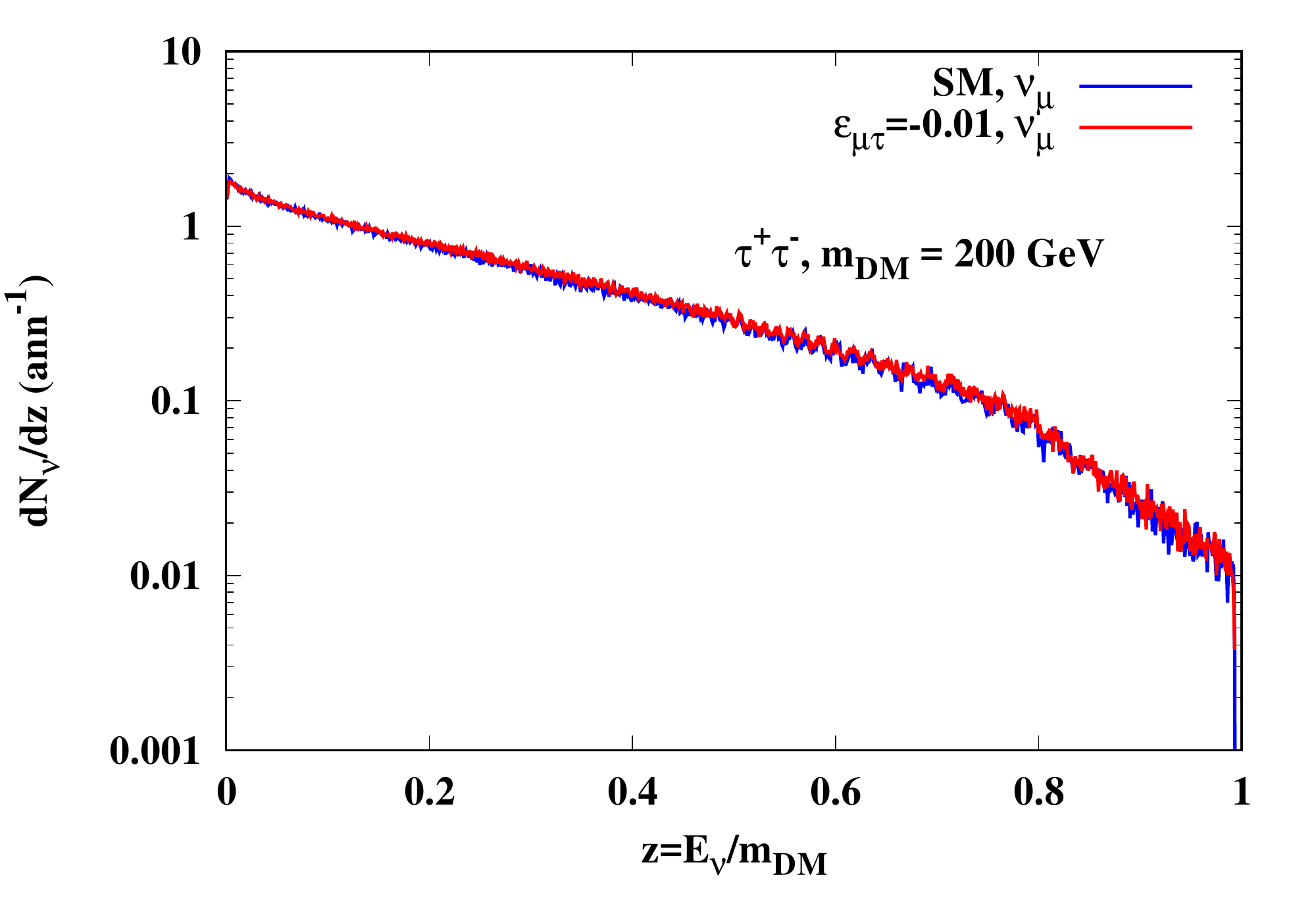}}
\put(0,130){\includegraphics[angle=-90,width=0.31\textwidth]{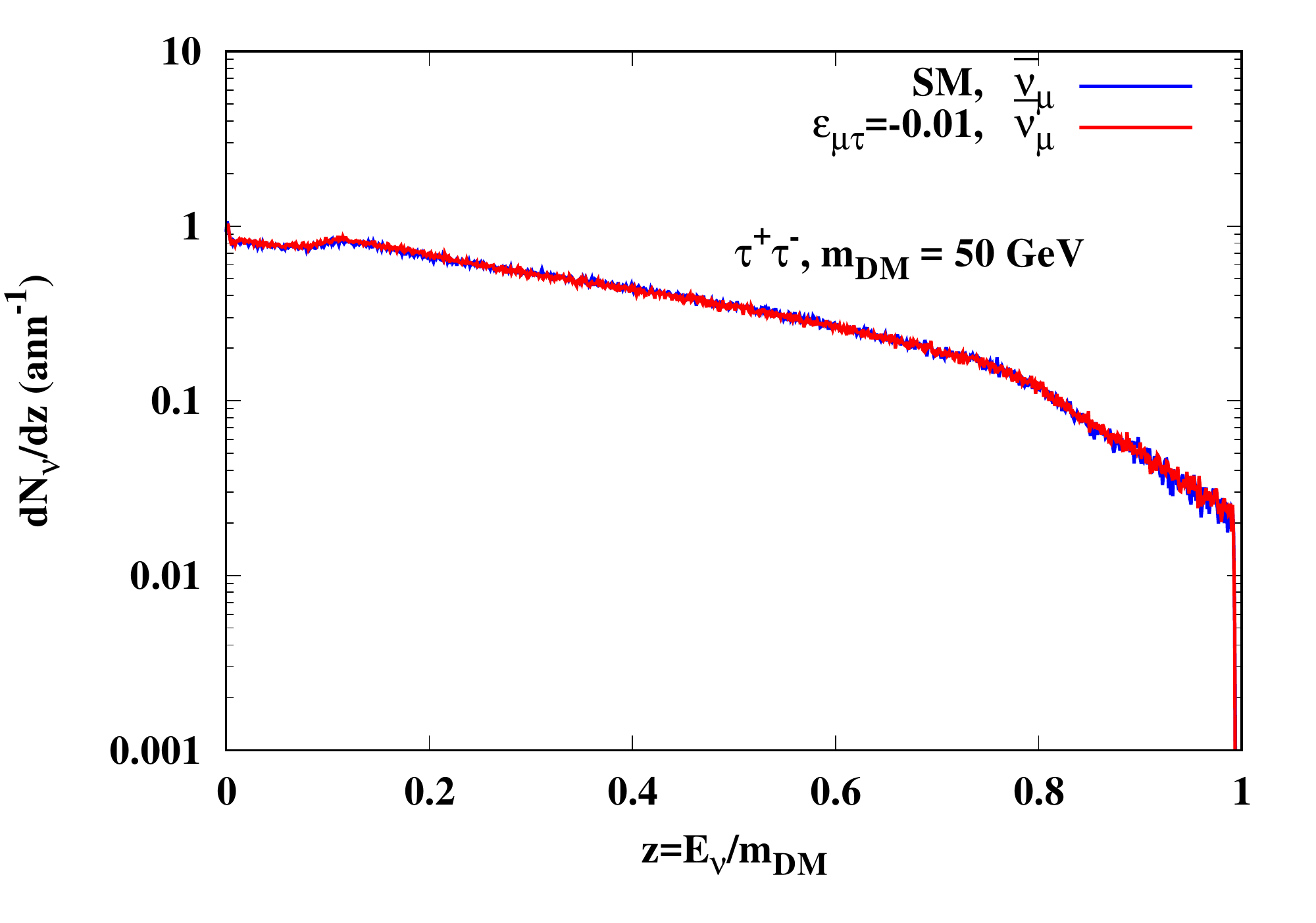}}
\put(0,240){\includegraphics[angle=-90,width=0.31\textwidth]{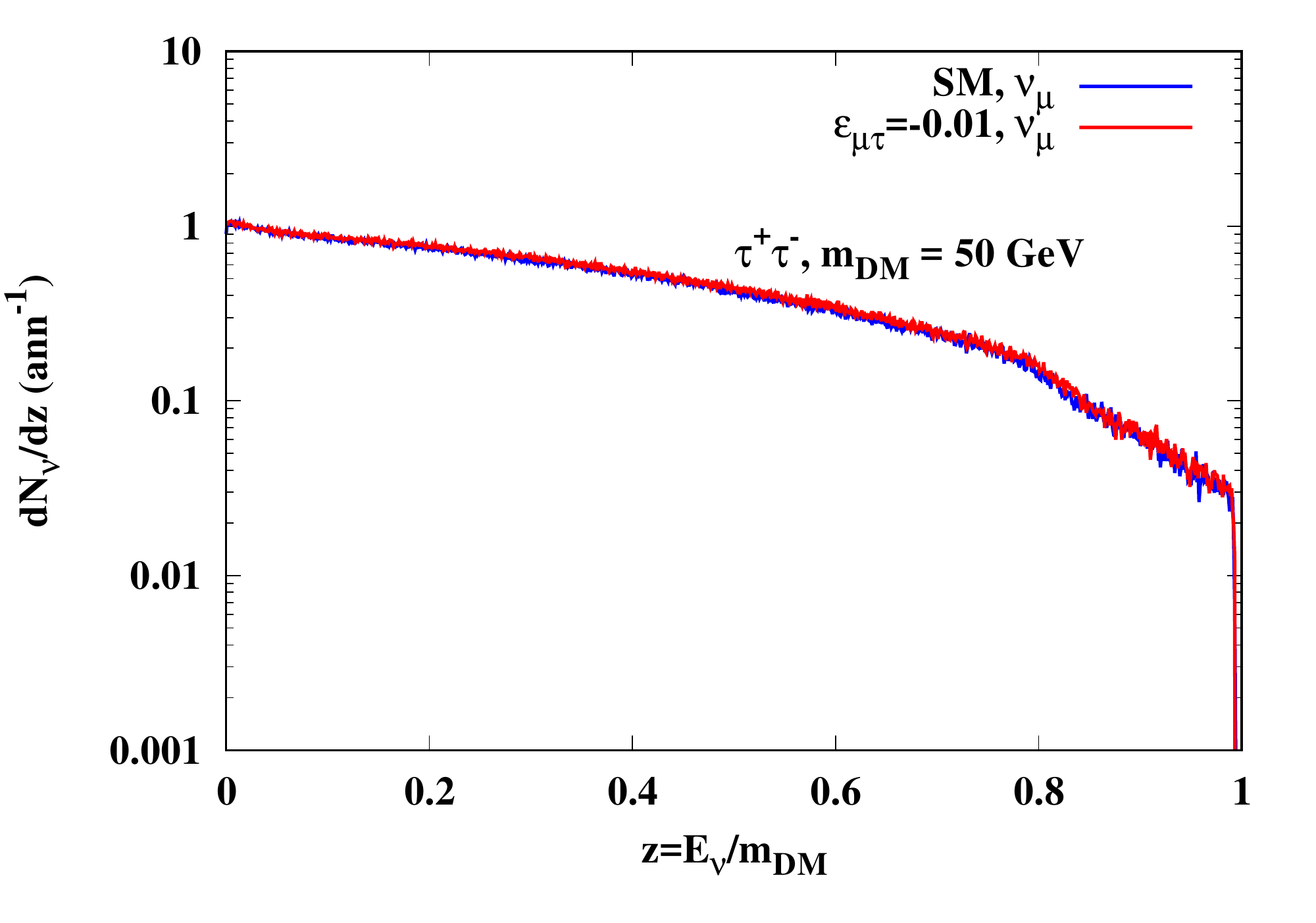}}
\end{picture}
\caption{\label{tau_mutau_inv_sign} The same as in Fig.~\ref{tau_tautau}
  but for $\e_{\mu\tau} = -0.01$ and IH.}
\end{figure}
we show muon neutrino spectra for non-zero $\e_{\mu\tau}=\pm0.01$ and
$\tau^+\tau^-$ annihilation channel. As we
have argued in the previous Section  in the absence of CC and 
NC neutrino interactions the final neutrino fluxes
should be almost the same as for no-NSI case. We see from
Figs.~\ref{tau_mutau}--\ref{tau_mutau_inv_sign} this is indeed the
case for $m_{DM}=50$ and~200~GeV. However, the energy spectra for
$m_{DM}=1$~TeV appear somewhat suppressed as compared to no-NSI
case. This is a consequence of very fast $\nu_\mu-\nu_\tau$
oscillations due to non-zero $\e_{\mu\tau}$ near the center of the Sun
and subsequent attenuation of the resulting muon flux due to CC
neutrino interactions. We remind that tau neutrinos 
regenerate at lower energies. In the no-NSI case $\nu_\mu-\nu_\tau$
oscillation length in the solar center is considerably larger and
coincides with that of for vacuum 2--3 oscillations. Thus  $\nu_\tau$
escapes interaction region almost with the same flavor (but possibly
with different energy) and no attenuation of the neutrino flux occur. 

To quantify the impact of NSI one should calculate neutrino
signal expected in a particular neutrino experiment. Here we estimate
how expected number of muon track events changes with nontrivial
NSI. Here we follow procedure of
Refs.~\cite{Gandhi:1995tf,Dutta:2000hh,Beacom:2003nh}. 
Namely, the probability
for muon neutrino to produce a muon track is given by
\be
\label{eq:4.1}
P(E_\nu,E_\mu^{th}) = \rho_NN_A\sigma_NR_\mu(E_\nu(1-y),E_\mu^{th}),
\ee
where $E_\nu$ is neutrino energy, $E_\mu^{th}$ is muon energy
threshold, $\rho_N$ is nucleon density, $\sigma_N$ is CC neutrino
nucleon cross section and $N_A$ is Avogadro number. Muon is produced
with the energy $E_\mu = E_\nu(1-y)$, where $y$ is the charged current 
inelasticity parameter and the muon range
$R_\mu(E_\nu(1-y),E_\mu^{th})$ is given by
\be
\label{eq:4.2}
R_\mu(E_\nu(1-y),E_\mu^{th}) = \frac{1}{\beta}\log{\frac{\alpha +
    \beta E_\mu}{\alpha + \beta E_\mu^{th}}}
\ee
with $\alpha = 2.0$~MeV cm$^2/$g and $\beta=4.2\cdot
10^{-6}$~cm$^2/$g. For an estimate we use $E_\mu^{th}=1$~GeV and adopt
the mean values of inelasticity $y$ which are about 0.45 for neutrino
and 0.35 for antineutrino. The rate of muon track events is
proportional to the following quantity
\be
\label{eq:4.3}
N\sim \int \frac{dN_\nu}{dE_\nu}P(E_\nu,E_\mu^{th})A^{eff}dE_\nu,
\ee
where $\frac{dN_\nu}{dE_\nu}$ is muon neutrino energy spectrum at the
detector level,
$A_{eff}$ is effective area for muon detection and sum over neutrino
and antineutrino is implied. Using obtained neutrino energy spectra
and with the help of Eq.~\eqref{eq:4.3} we calculate the
ratio of the rates of muon track events expected with and without
non-standard neutrino interactions taking a single non-zero NSI
parameter at a time. 
\begin{figure}[!htb]
\begin{picture}(300,500)(0,0)
\put(260,90){\includegraphics[angle=-90,width=0.31\textwidth]{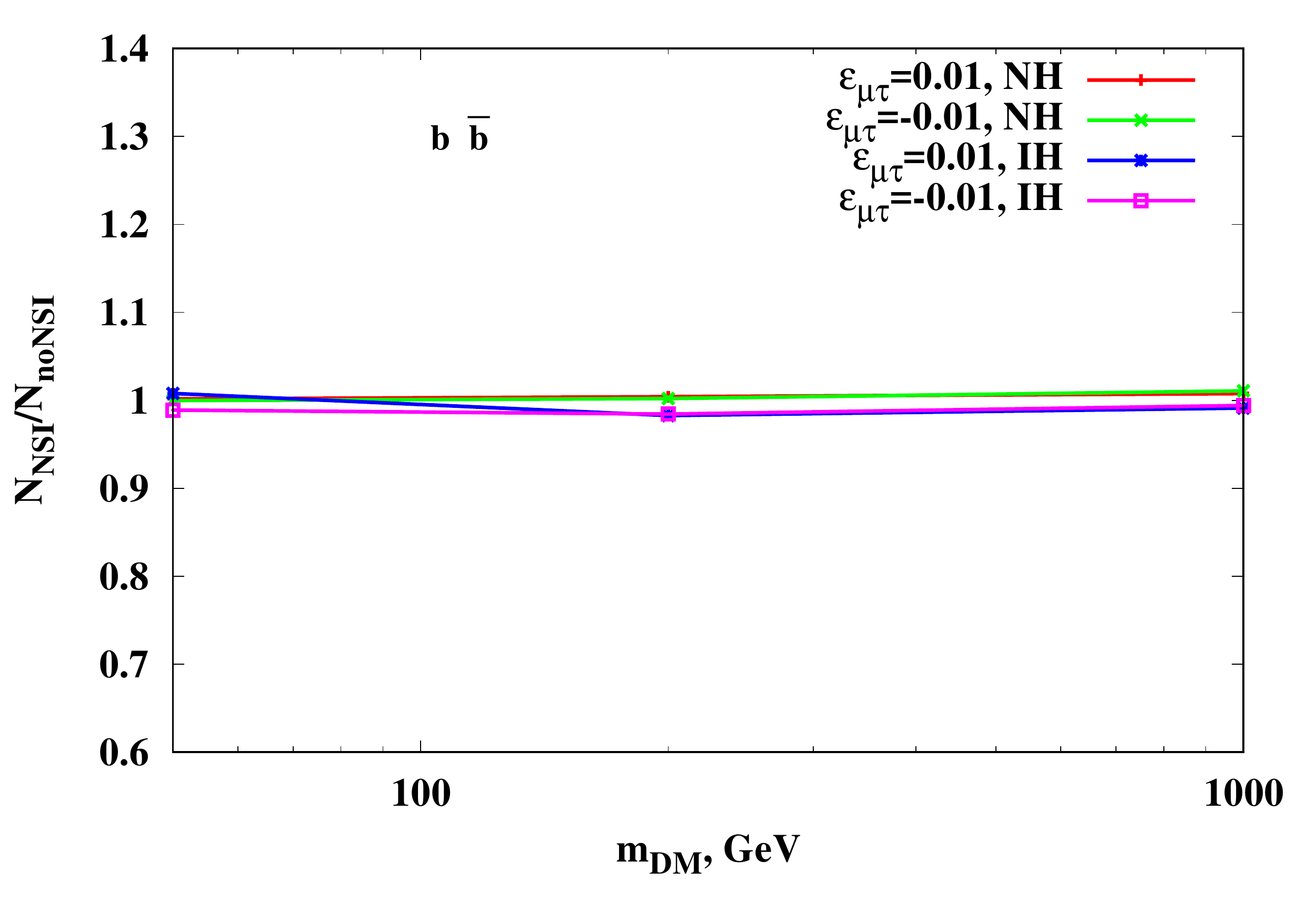}}
\put(130,90){\includegraphics[angle=-90,width=0.31\textwidth]{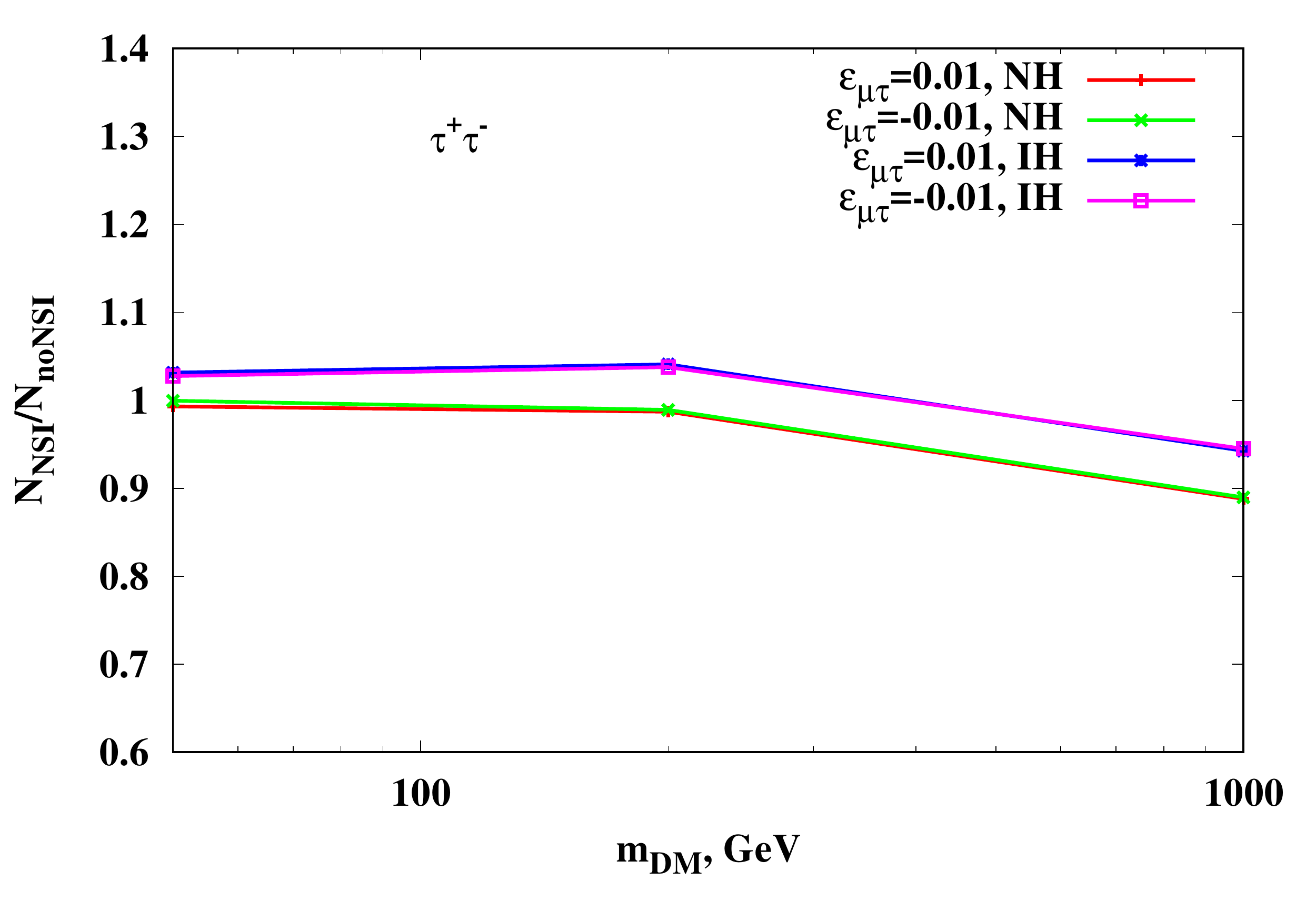}}
\put(0,90){\includegraphics[angle=-90,width=0.31\textwidth]{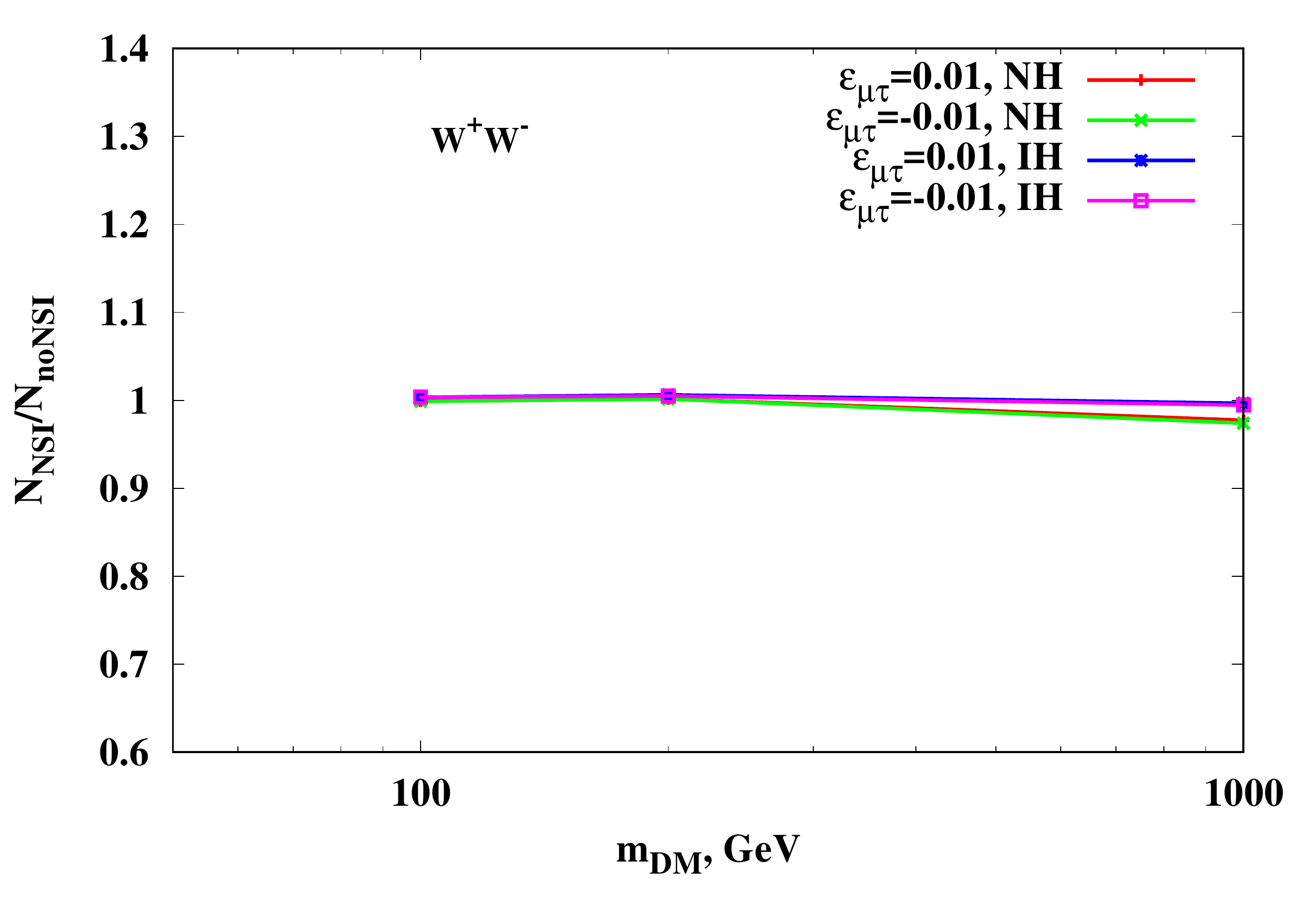}}
\put(260,190){\includegraphics[angle=-90,width=0.31\textwidth]{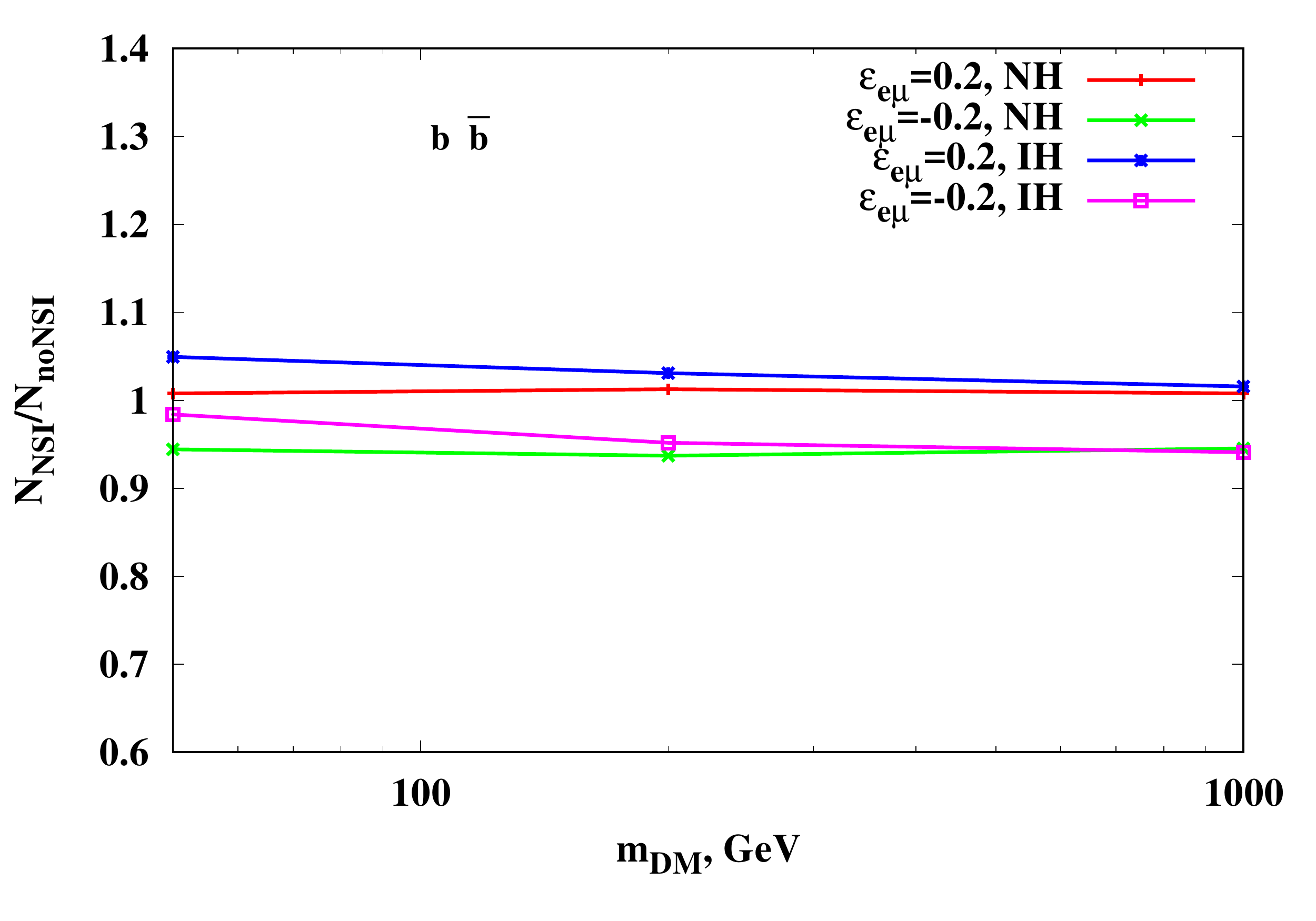}}
\put(130,190){\includegraphics[angle=-90,width=0.31\textwidth]{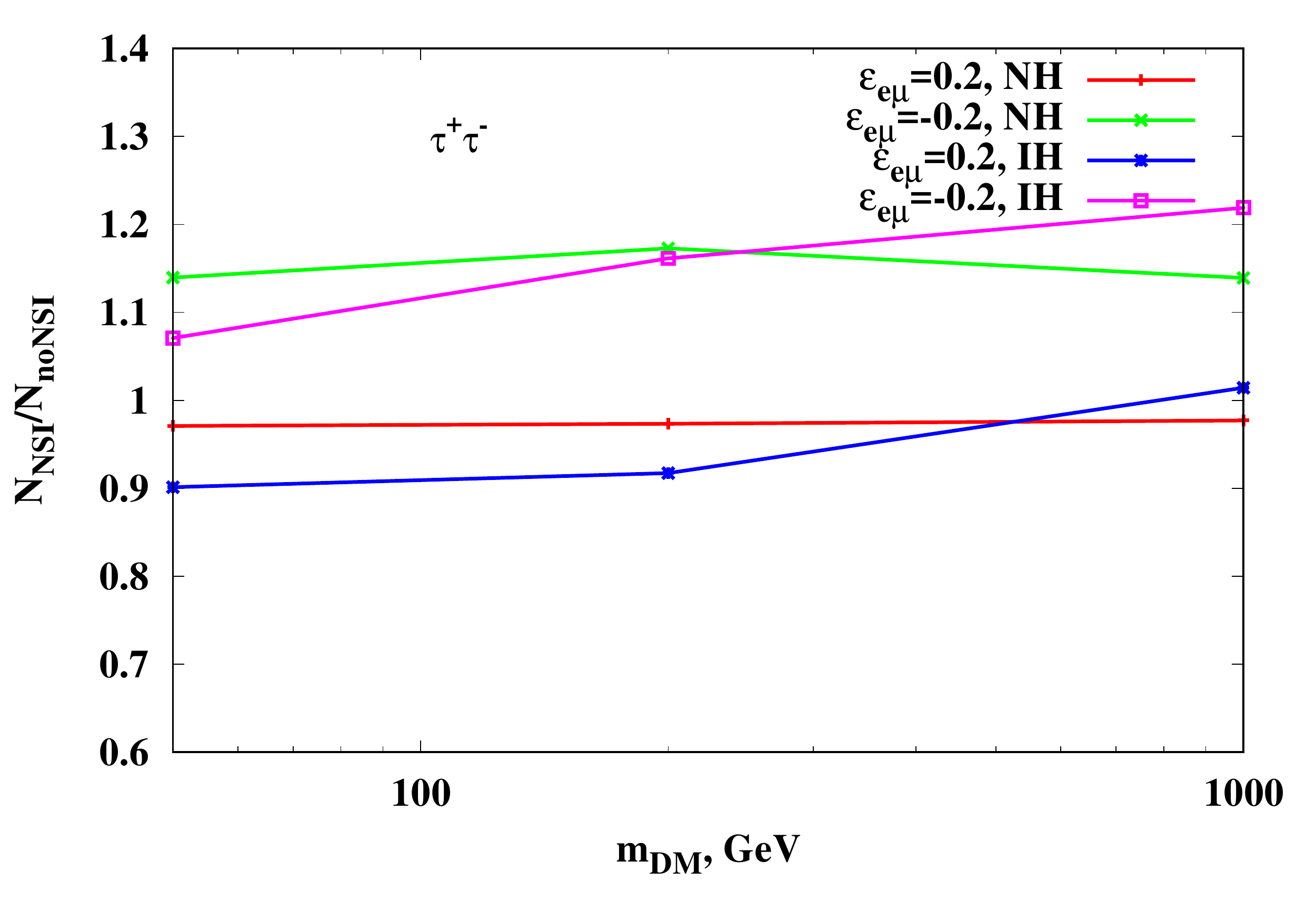}}
\put(0,190){\includegraphics[angle=-90,width=0.31\textwidth]{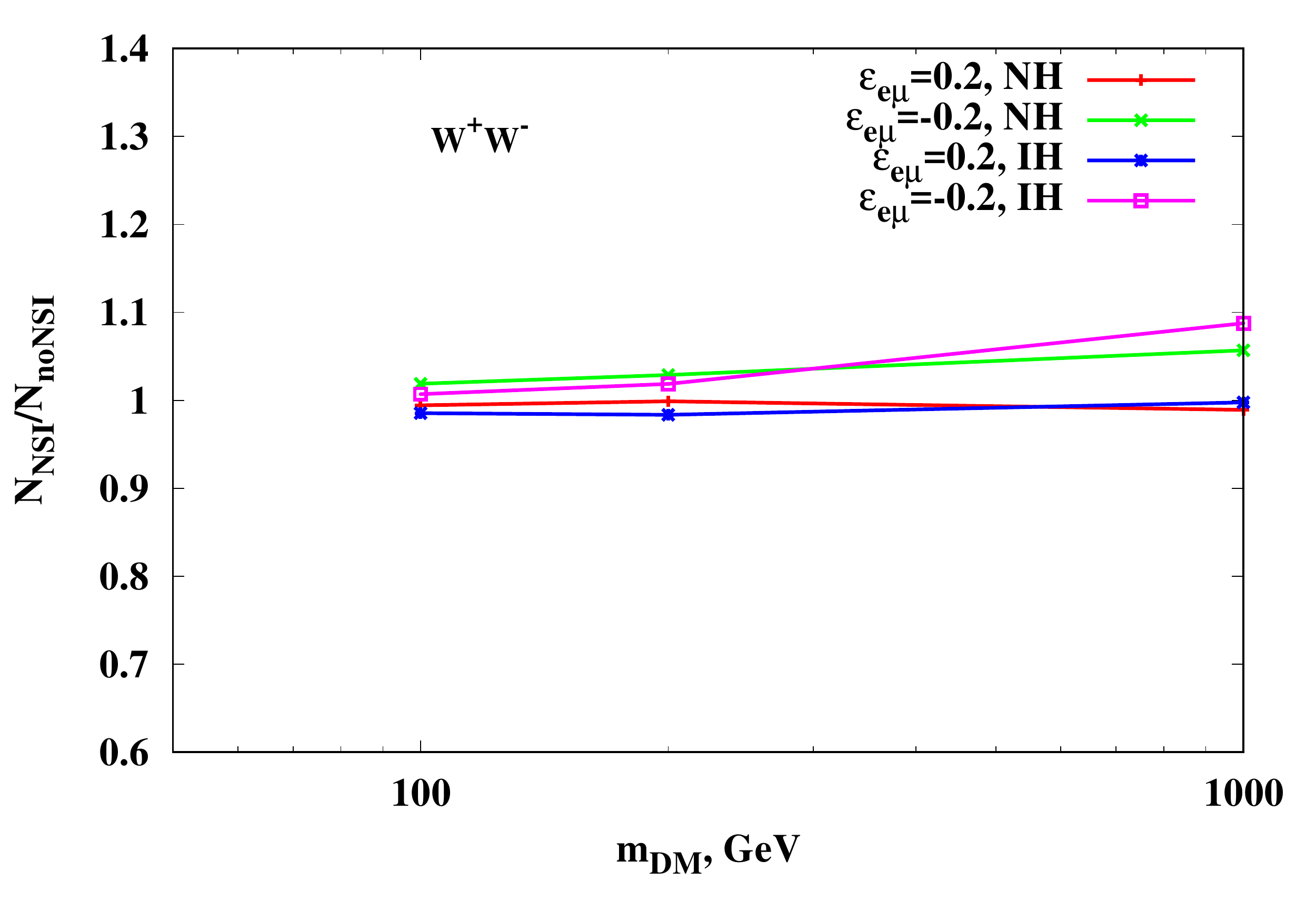}}
\put(260,290){\includegraphics[angle=-90,width=0.31\textwidth]{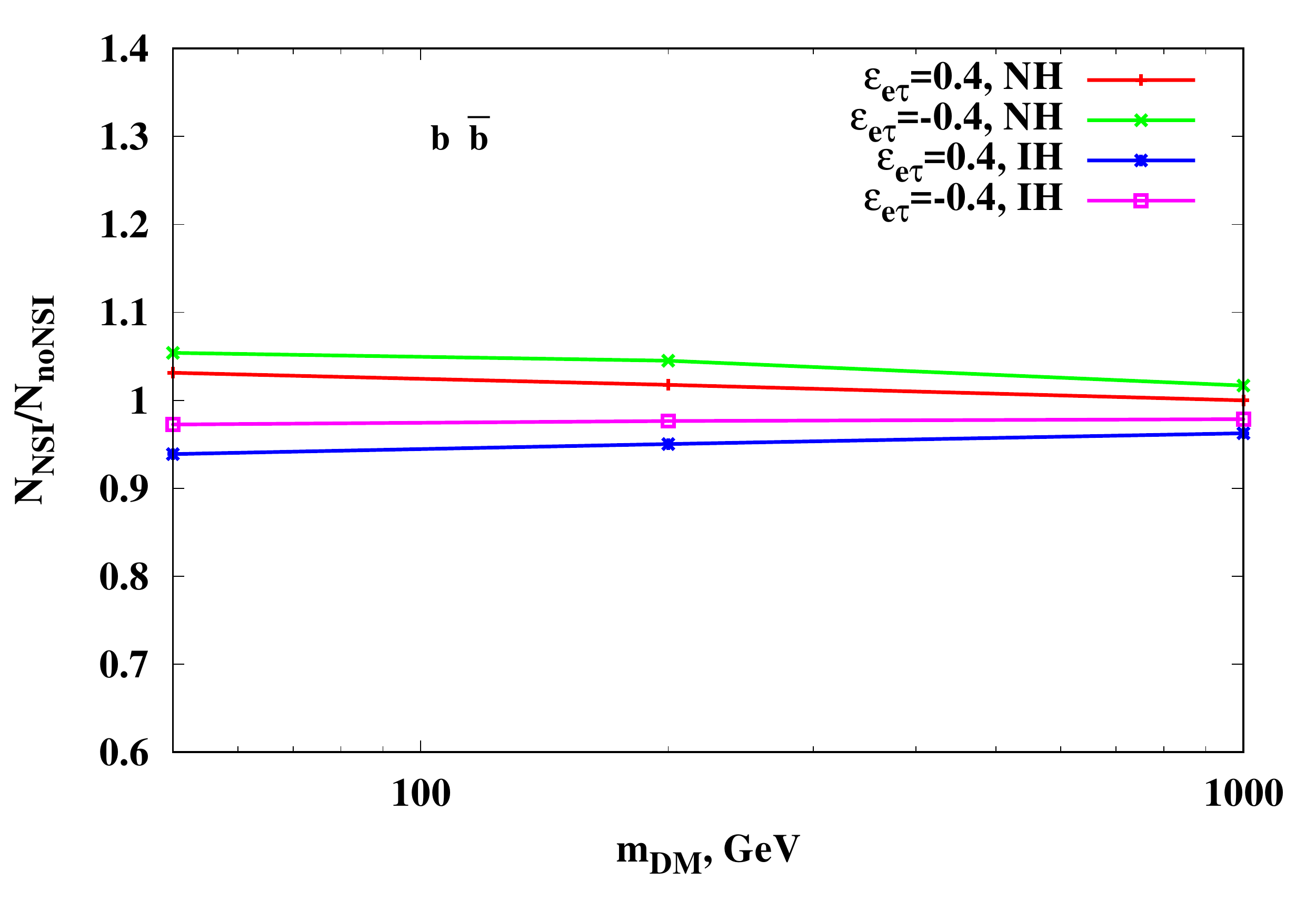}}
\put(130,290){\includegraphics[angle=-90,width=0.31\textwidth]{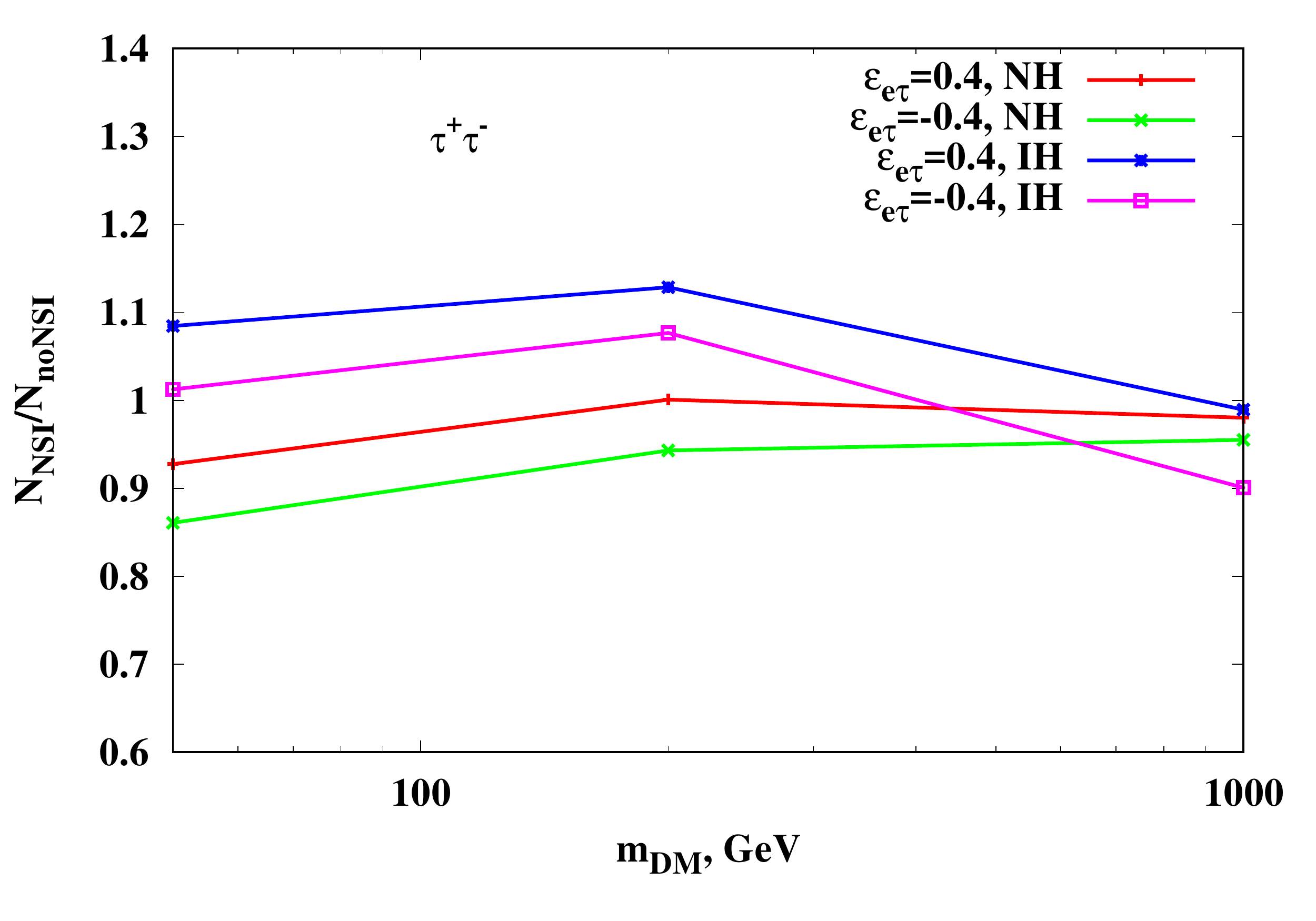}}
\put(0,290){\includegraphics[angle=-90,width=0.31\textwidth]{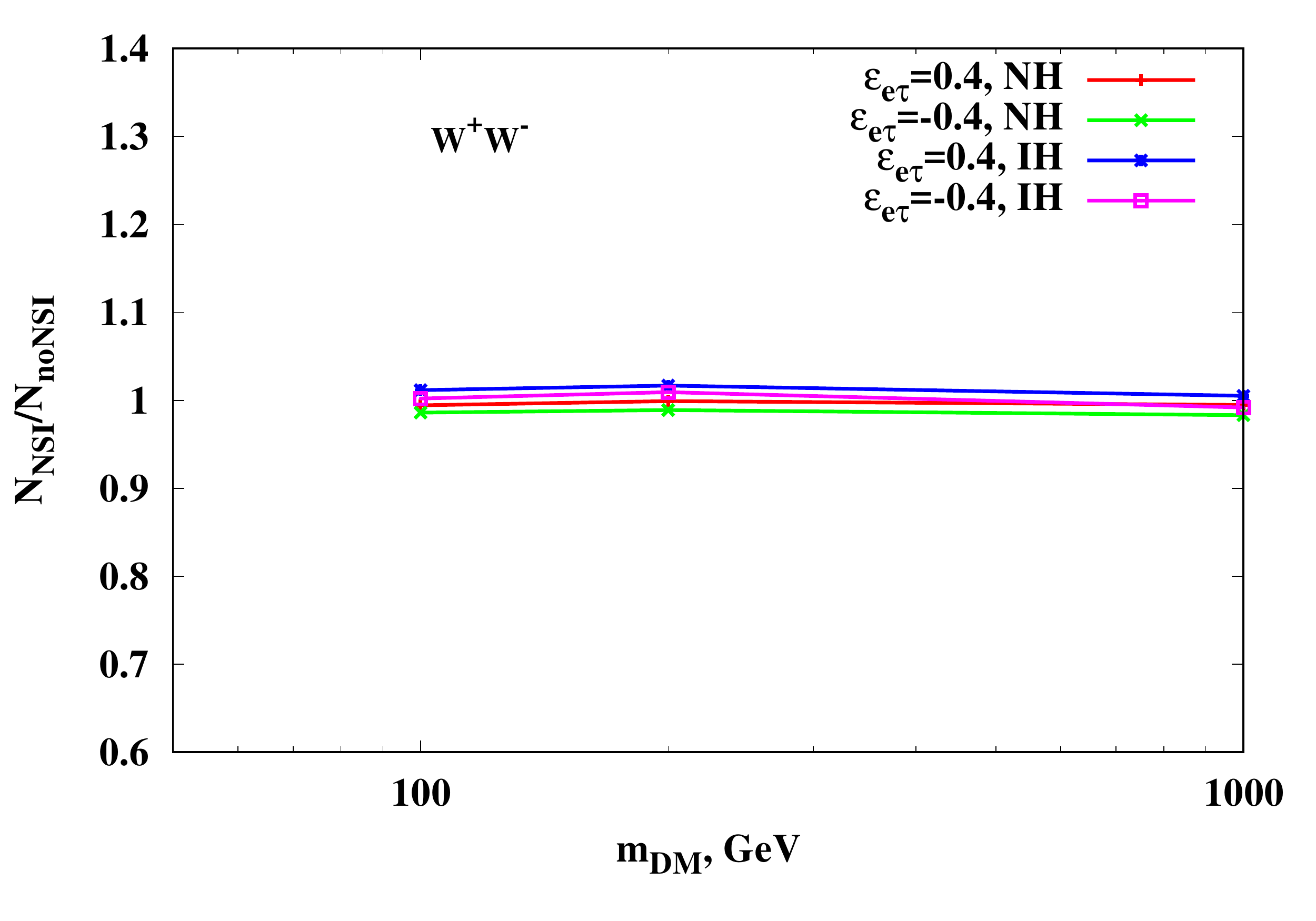}}
\put(260,390){\includegraphics[angle=-90,width=0.31\textwidth]{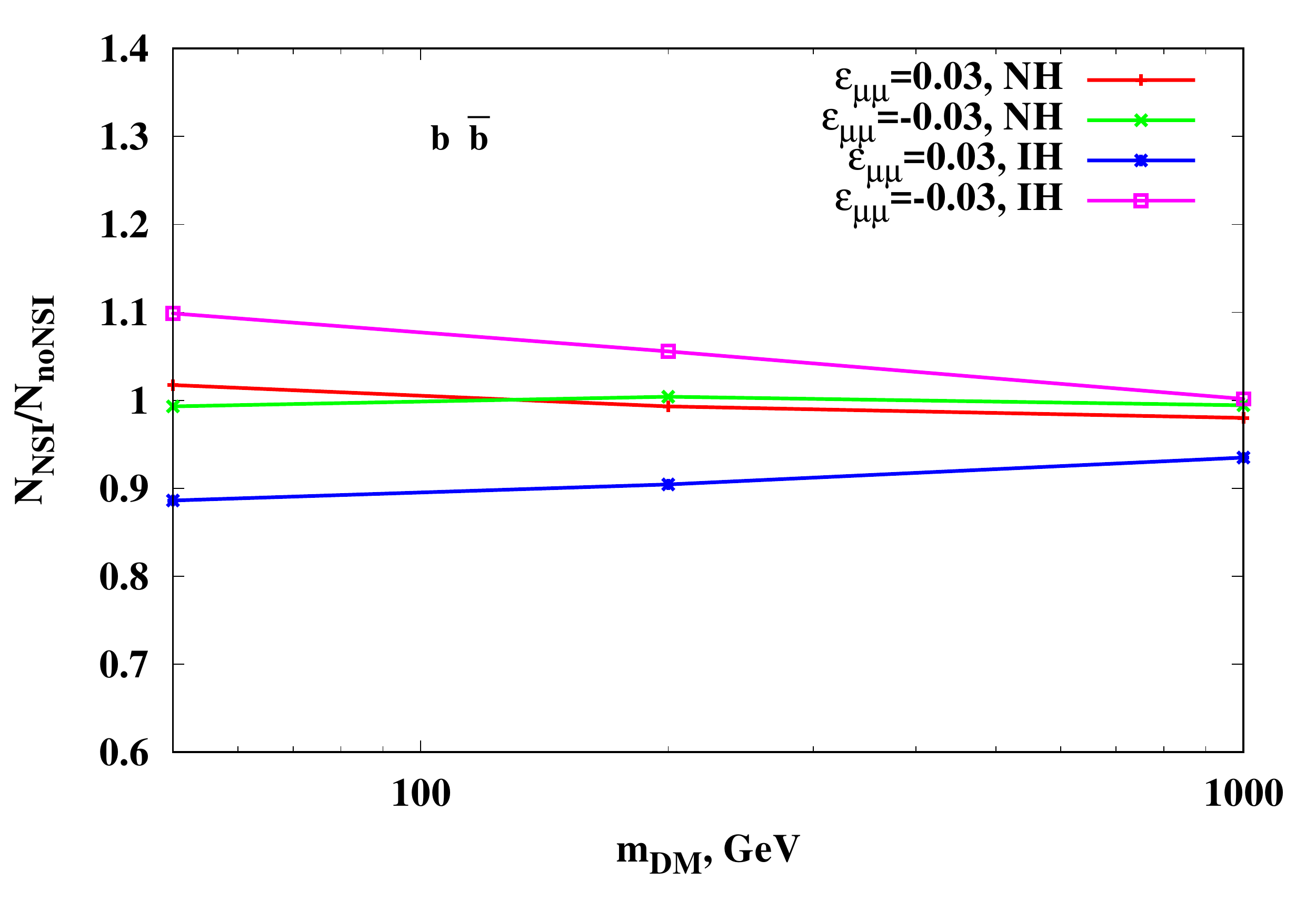}}
\put(130,390){\includegraphics[angle=-90,width=0.31\textwidth]{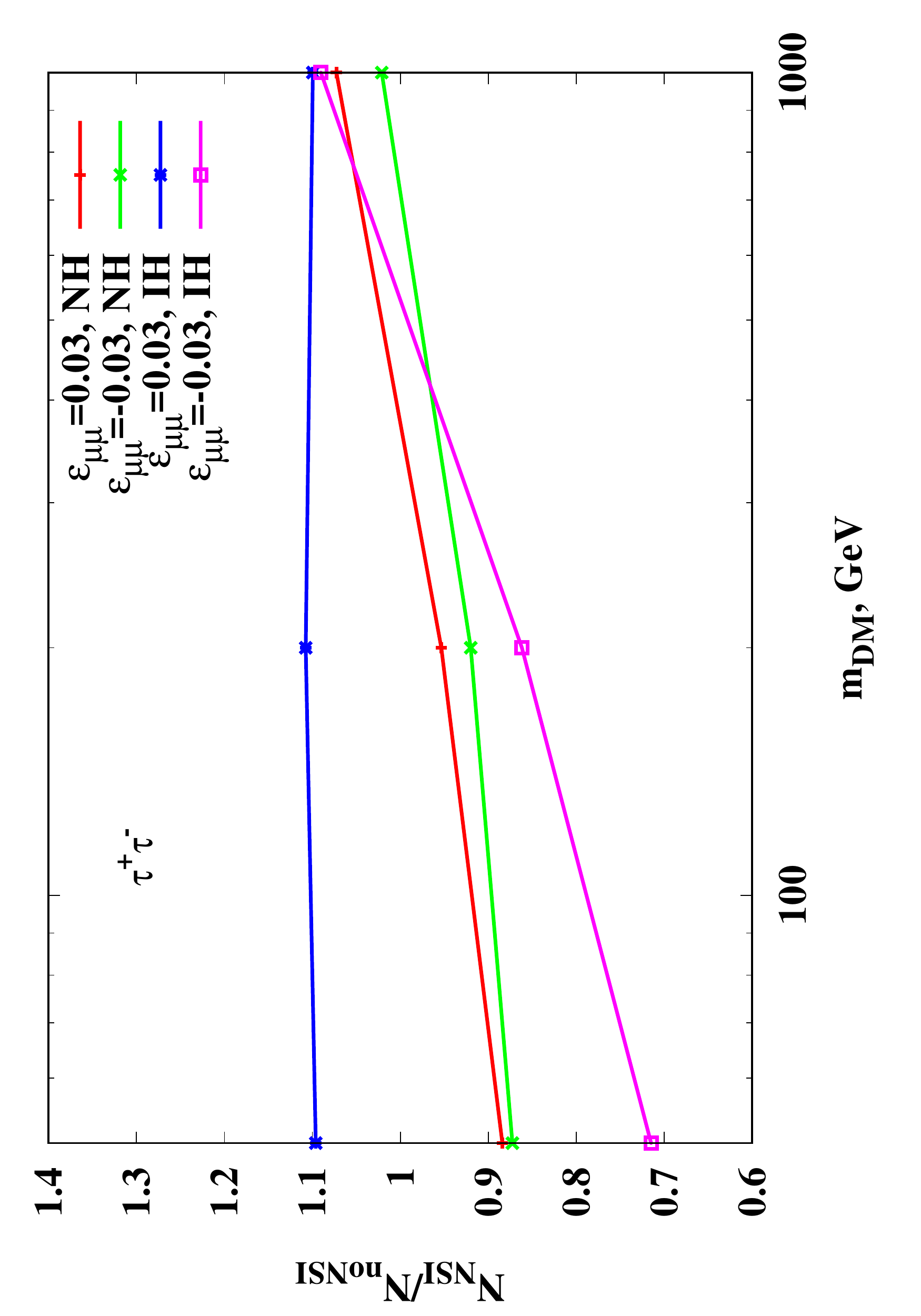}}
\put(0,390){\includegraphics[angle=-90,width=0.31\textwidth]{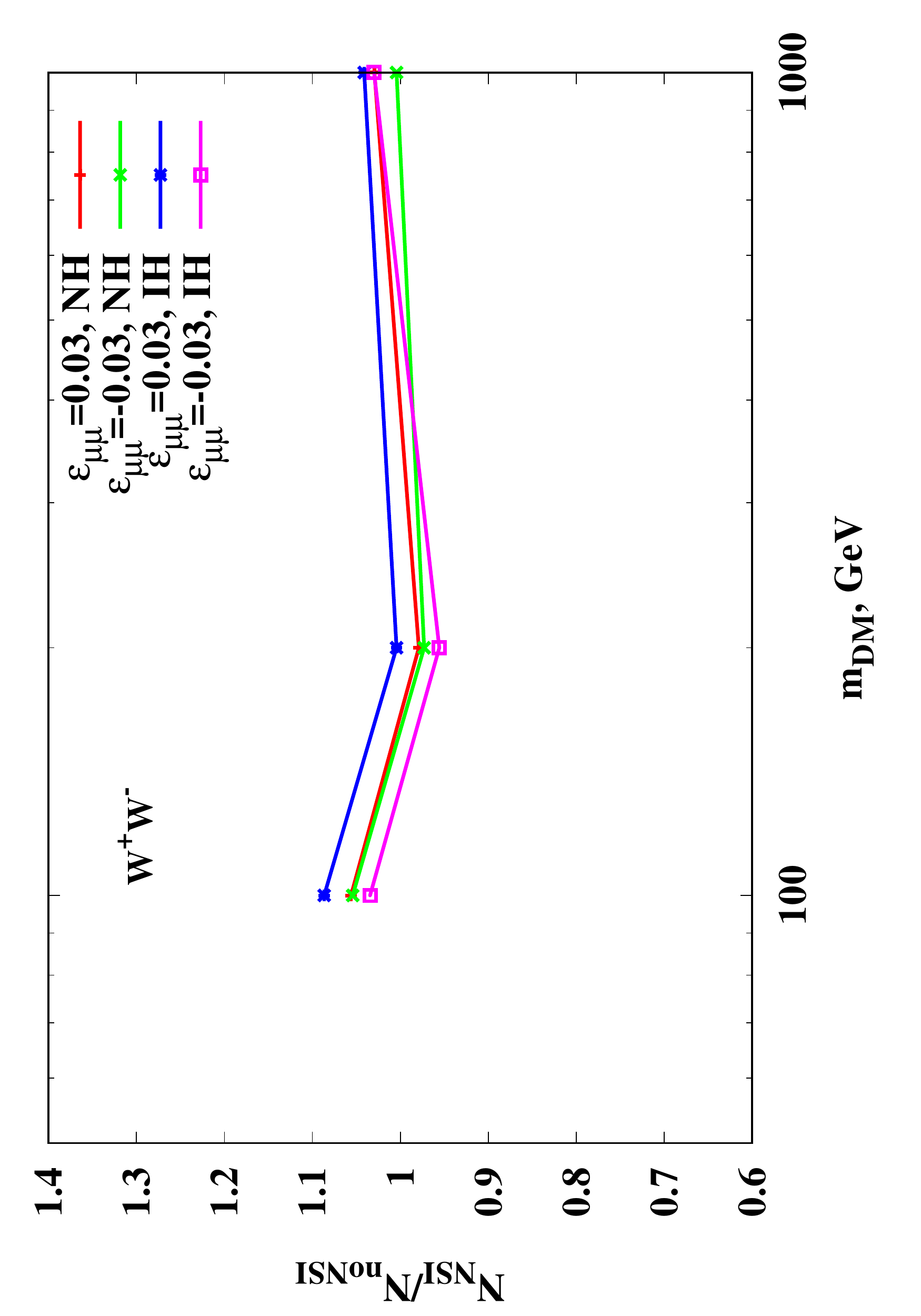}}
\put(260,490){\includegraphics[angle=-90,width=0.31\textwidth]{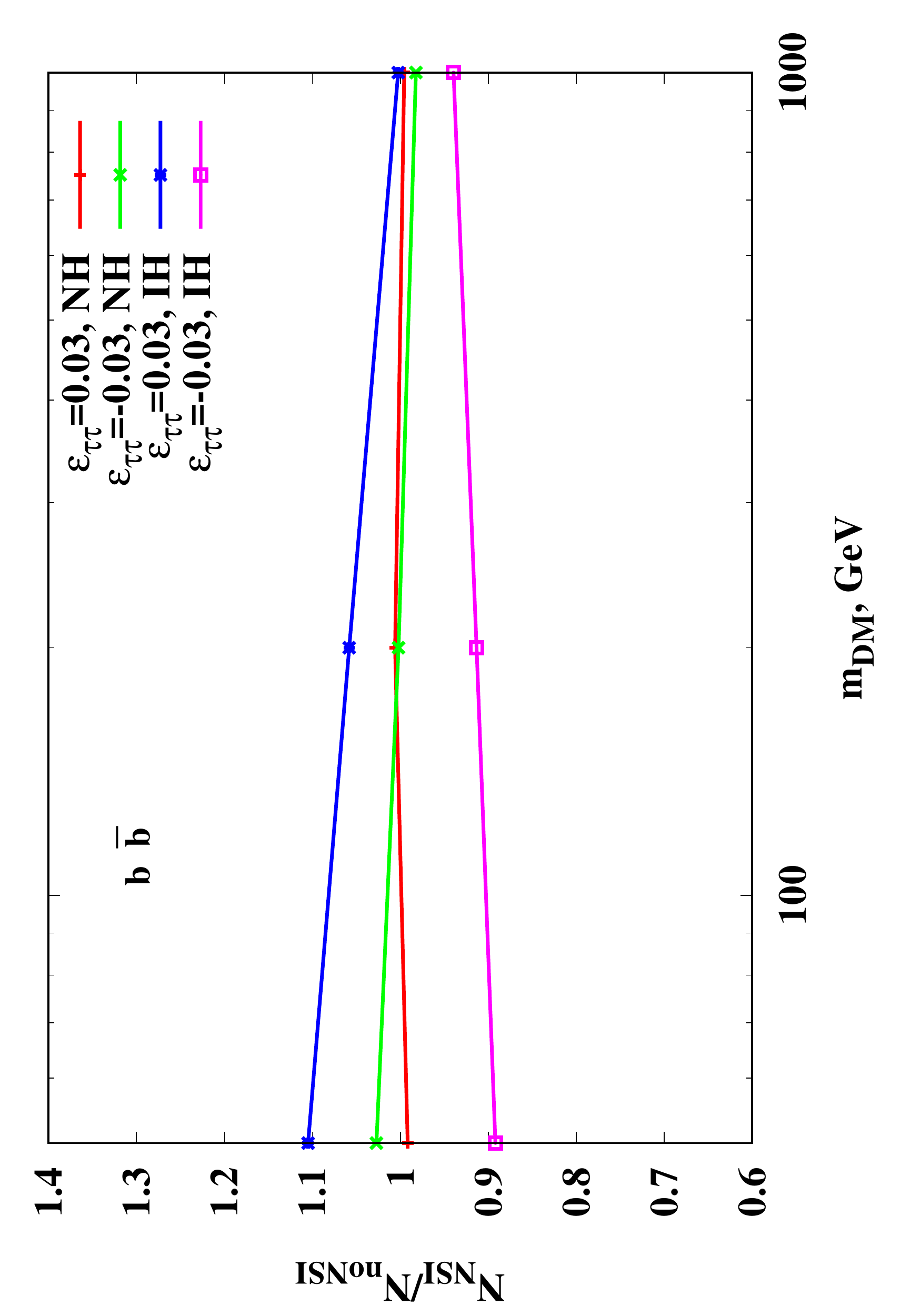}}
\put(130,490){\includegraphics[angle=-90,width=0.31\textwidth]{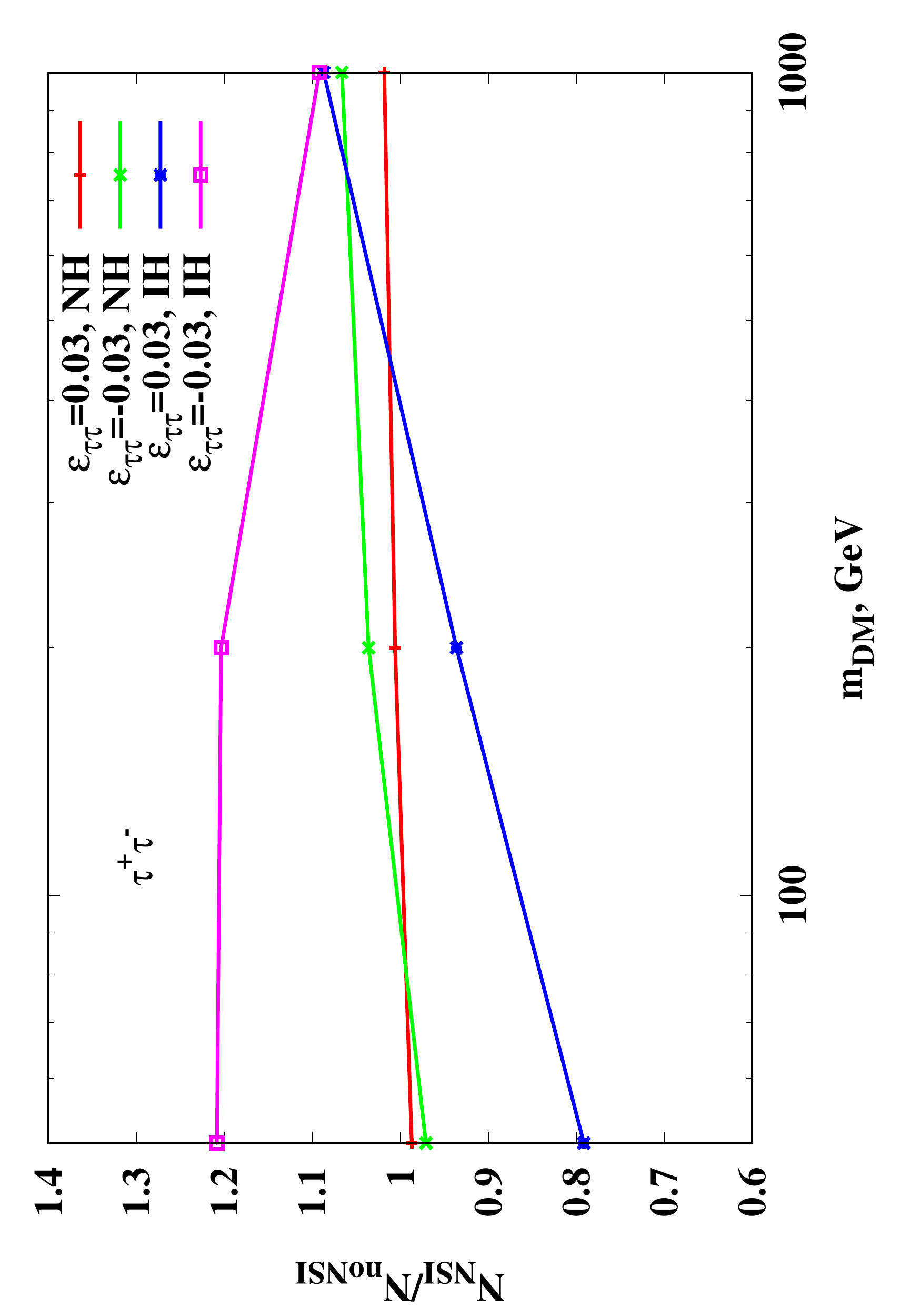}}
\put(0,490){\includegraphics[angle=-90,width=0.31\textwidth]{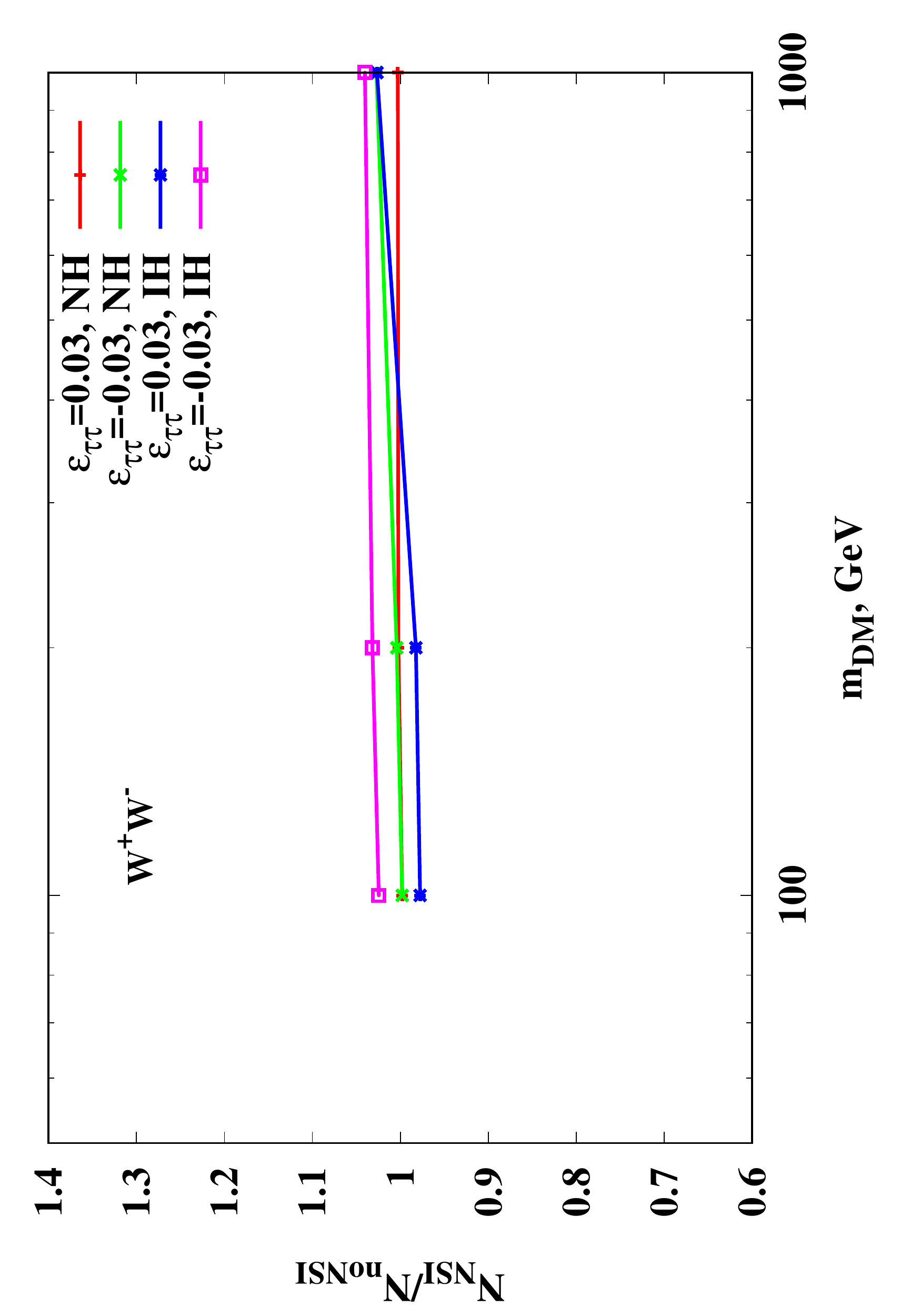}}
\end{picture}
\caption{\label{rate} The ratios of muon track events with and without
  matter NSI parameters for $\tau^+\tau^-$ (left panels), $b\bar{b}$
  (central panels) and $W^+W^-$ (right panels) annihilation channels. 
  We take the following values of matter NSI parameters (from up to
  bottom): $\e_{\tau\tau}=\pm0.03$, $\e_{\mu\mu}=\pm0.03$,
$\e_{e\tau}=\pm0.4$,  $\e_{e\mu}=\pm0.2$ and $\e_{\mu\tau}=\pm0.01$. }
\end{figure}
The results for this ratio calculated with $\e_{\tau\tau}=\pm0.03$,
$\e_{\mu\mu}=\pm0.03$, $\e_{e\tau}=\pm0.4$, $\e_{e\mu}=\pm0.2$ and
$\e_{\mu\tau}=\pm0.01$ for
both normal and inverted mass ordering are presented in
Fig.~\ref{rate} with dots. Here we show the results not only for
$\tau^+\tau^-$ dark matter annihilation channel but also for $W^+W^-$, 
and $b\bar{b}$ channels for several selected
values of $m_{DM}$. We see that the rate for $\tau^+\tau^-$
annihilation channel is most strongly affected by NSI
effects. As we already mentioned corresponding neutrino flux at
production is dominated by $\tau$ flavor. The
largest deviations appear for non-zero flavor diagonal NSI,
$\e_{\tau\tau}$ and $\e_{\mu\mu}$ and they can reach values up to
about 30\%. In this case we observe strong dependence of the result on the
mass of dark matter particle. Note that the lines between the points
in Fig.~\ref{rate} are shown for illustrative purpose only and they
are not necessarily follow real dependence on $m_{DM}$ in between.  
Nonzero $\e_{\tau\tau}$ and $\e_{\mu\mu}$ have opposite effects on the
event rate. Sizable (around 10\%) effect of non-zero $\e_{\mu\tau}$
appear for $\tau^+\tau^-$ annihilation channel and for large dark
matter masses only. Impact of the matter NSI on neutrino signal
$b\bar{b}$ annihilation channels is in general smaller and almost
never exceeds 15\% level. In the case
of $W^+W^-$ channel the neutrinos at production are approximately
equally distributed between different flavors. As it was argued in 
Ref.~\cite{Lehnert:2007fv} a flavor democratic flux of neutrinos in
the center of the Sun arrives at the Earth as flavor democratic almost
independently of complicated matter effects. Relatively small
deviations from no-NSI case for $W^+W^-$ annihilation channel in
Fig.~\ref{rate} is a manifestation of this statement.   

\begin{figure}[!htb]
\begin{picture}(300,190)(0,40)
\put(260,130){\includegraphics[angle=-90,width=0.31\textwidth]{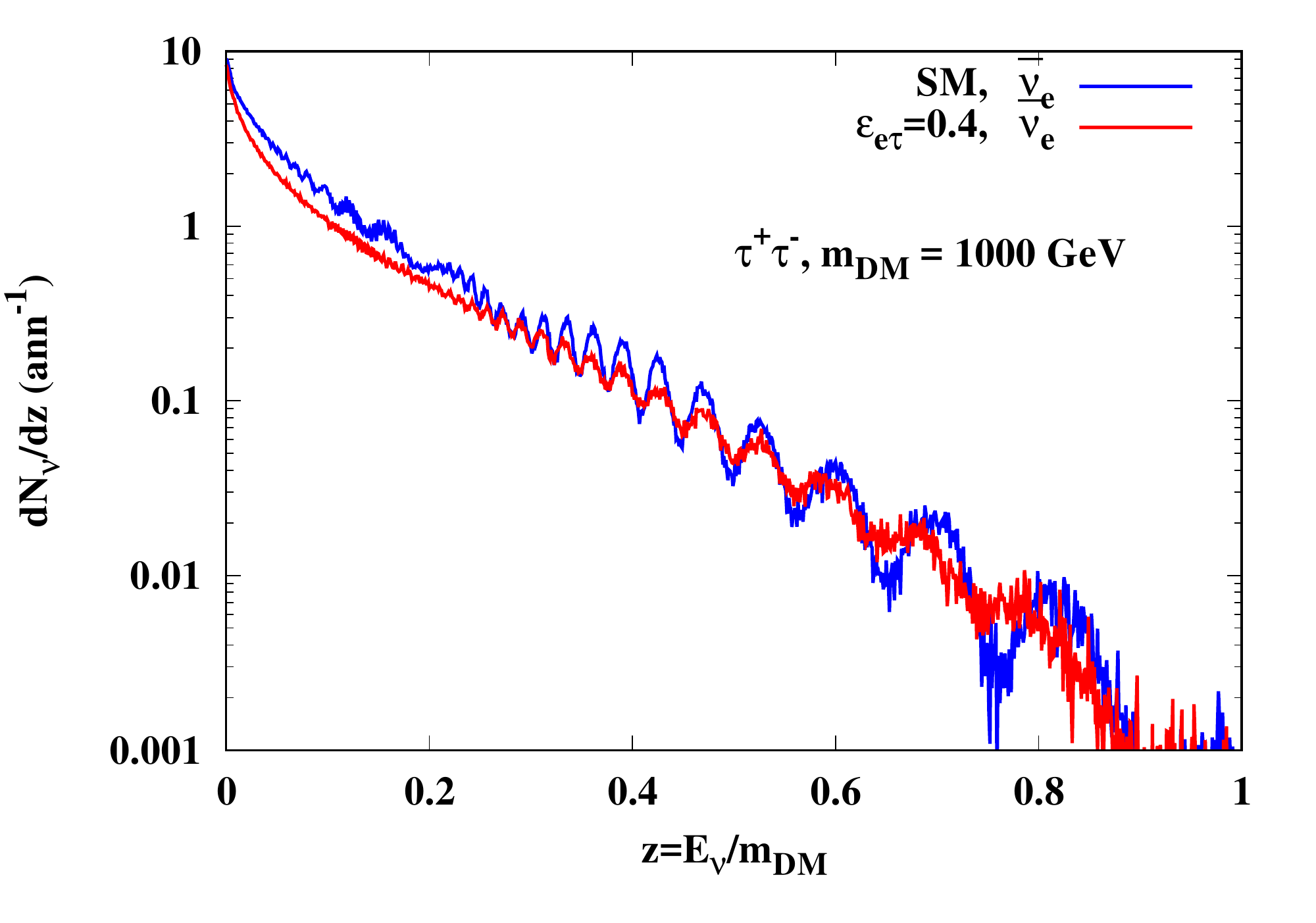}}
\put(260,240){\includegraphics[angle=-90,width=0.31\textwidth]{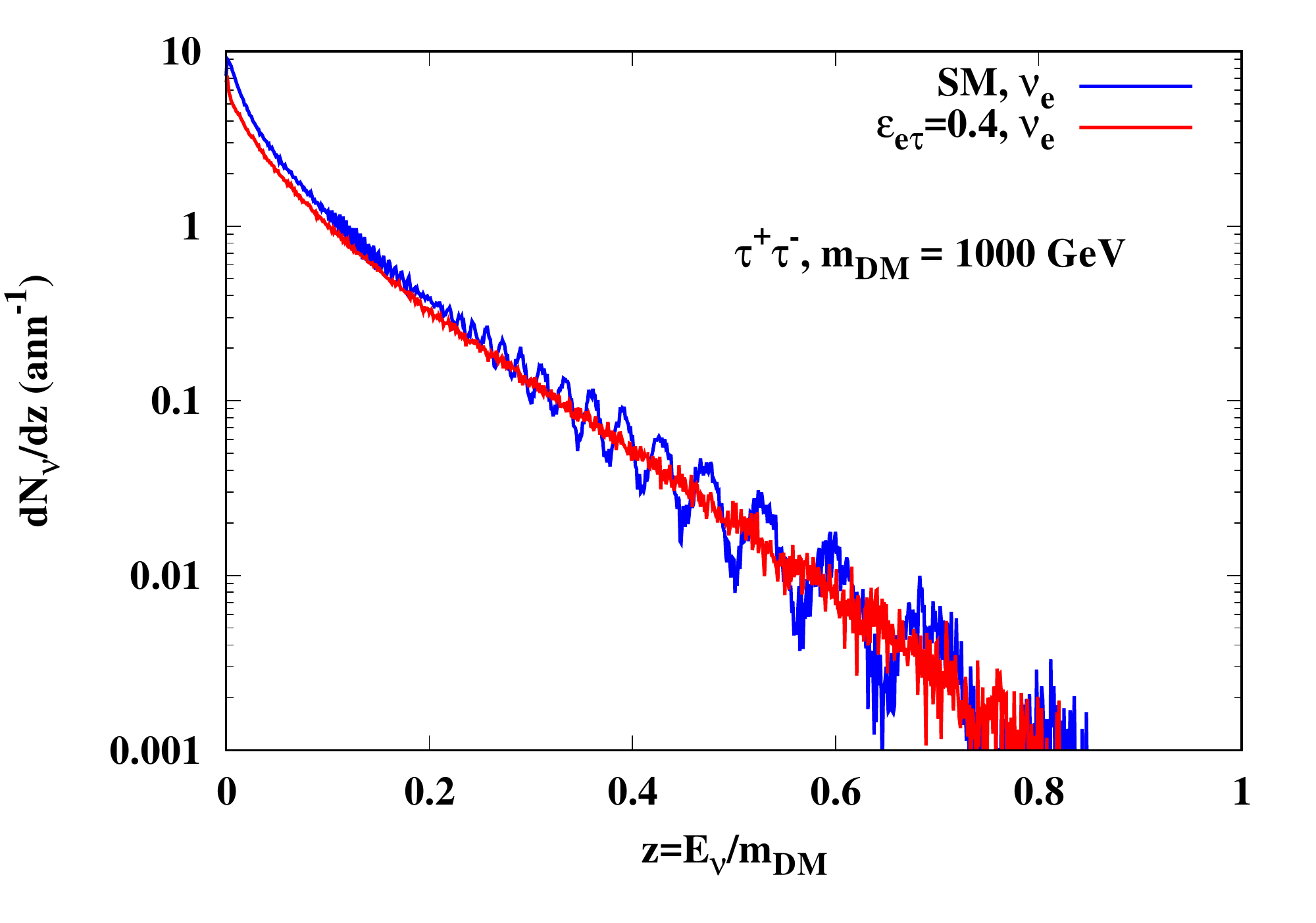}}
\put(130,130){\includegraphics[angle=-90,width=0.31\textwidth]{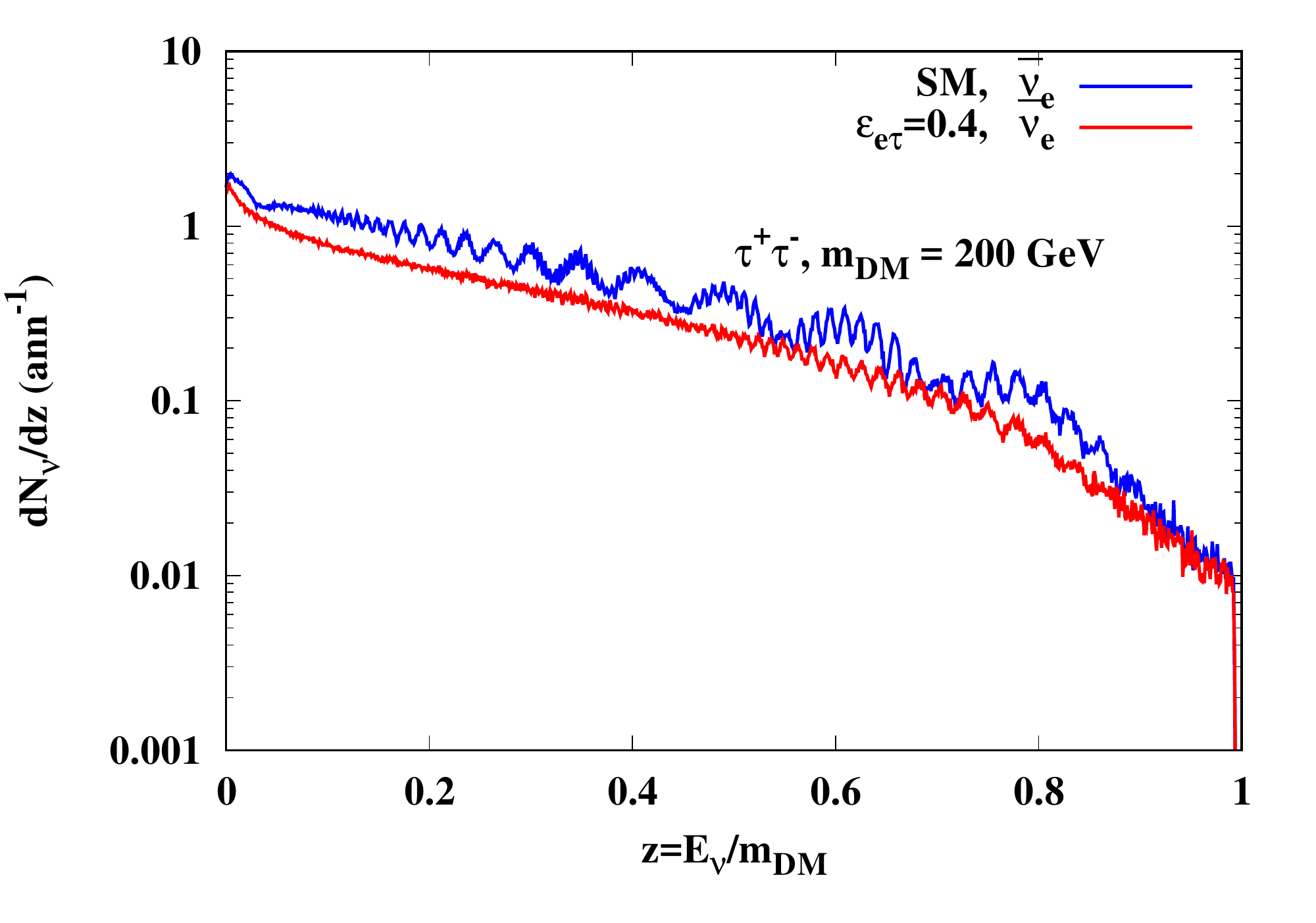}}
\put(130,240){\includegraphics[angle=-90,width=0.31\textwidth]{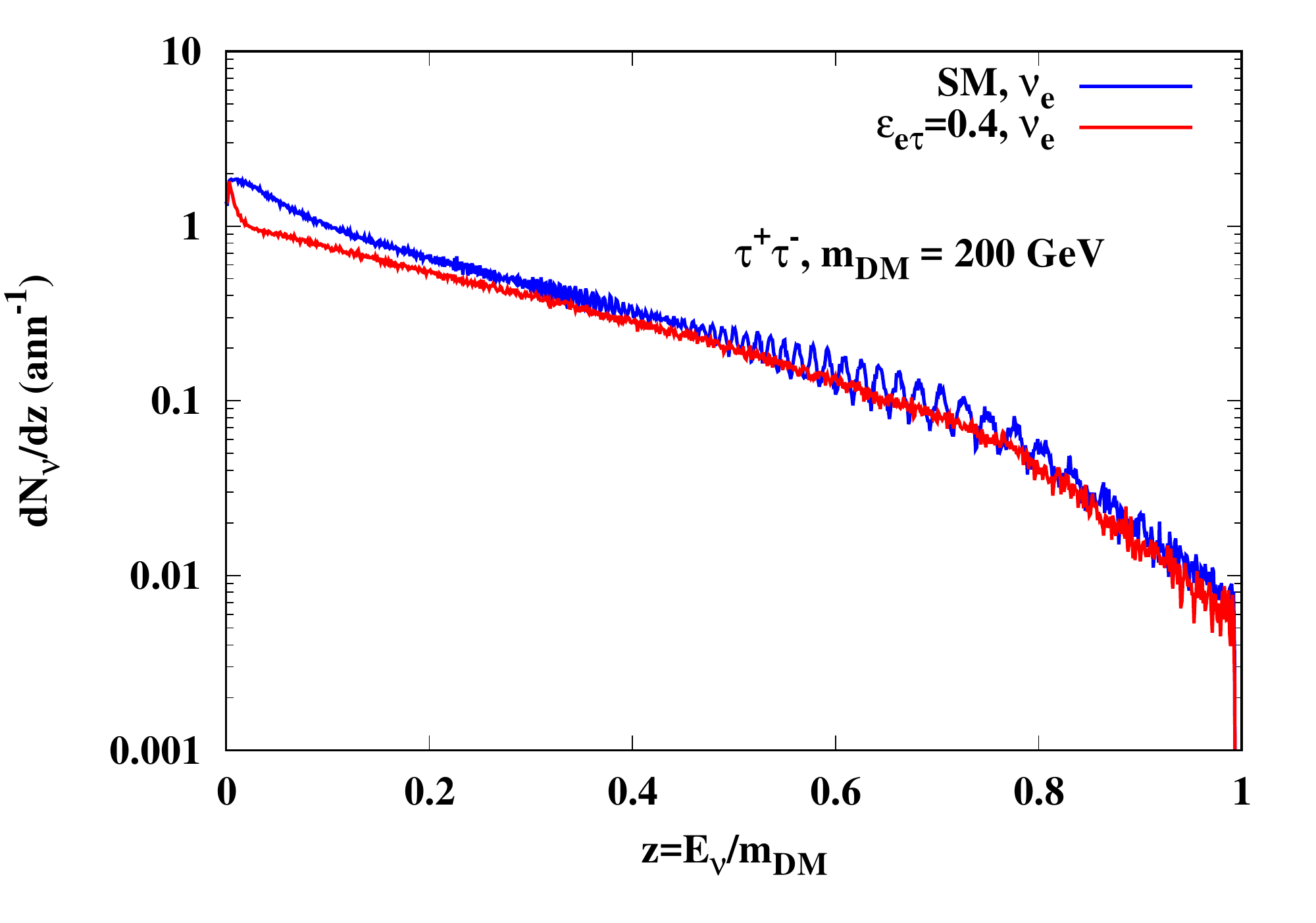}}
\put(0,130){\includegraphics[angle=-90,width=0.31\textwidth]{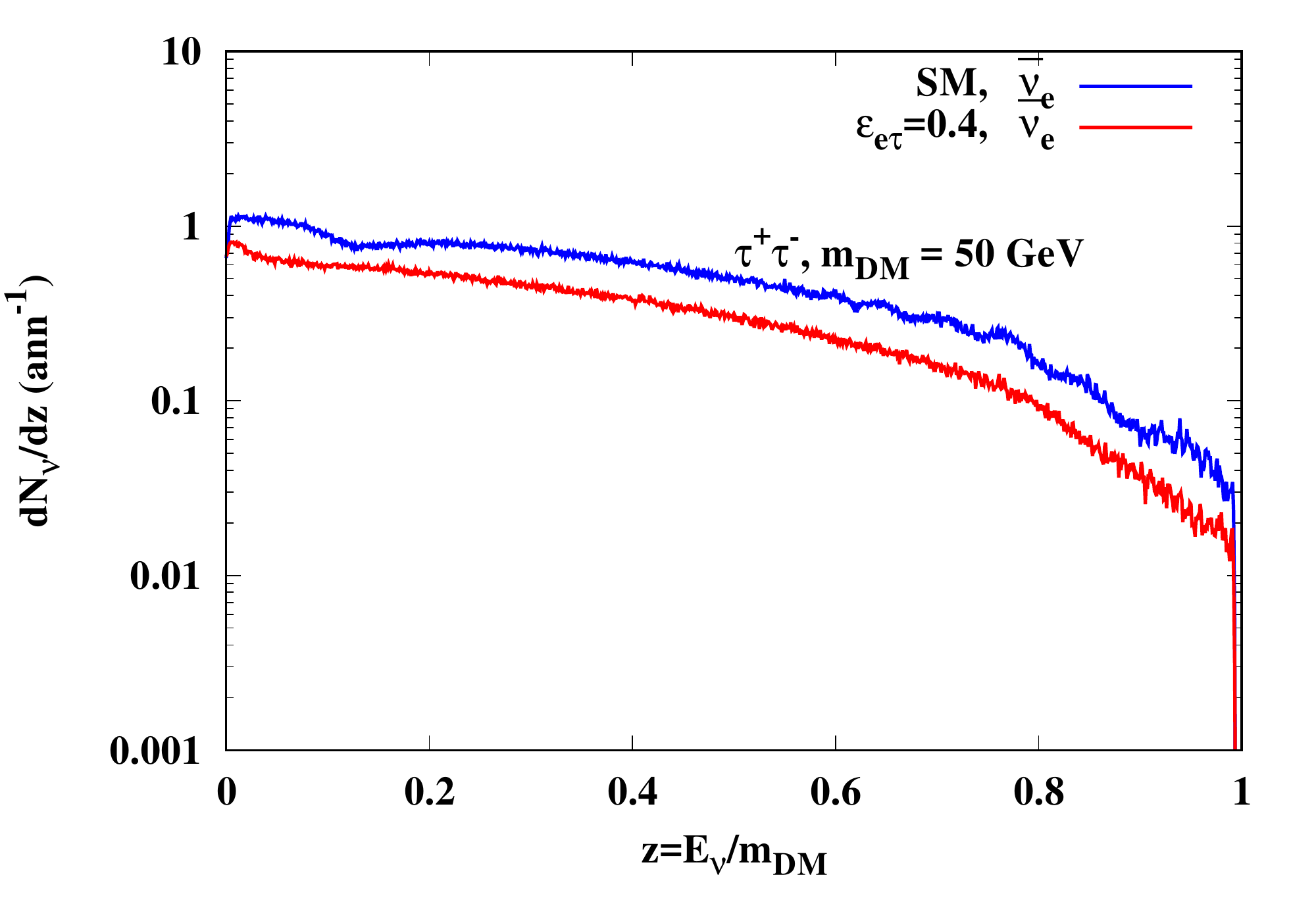}}
\put(0,240){\includegraphics[angle=-90,width=0.31\textwidth]{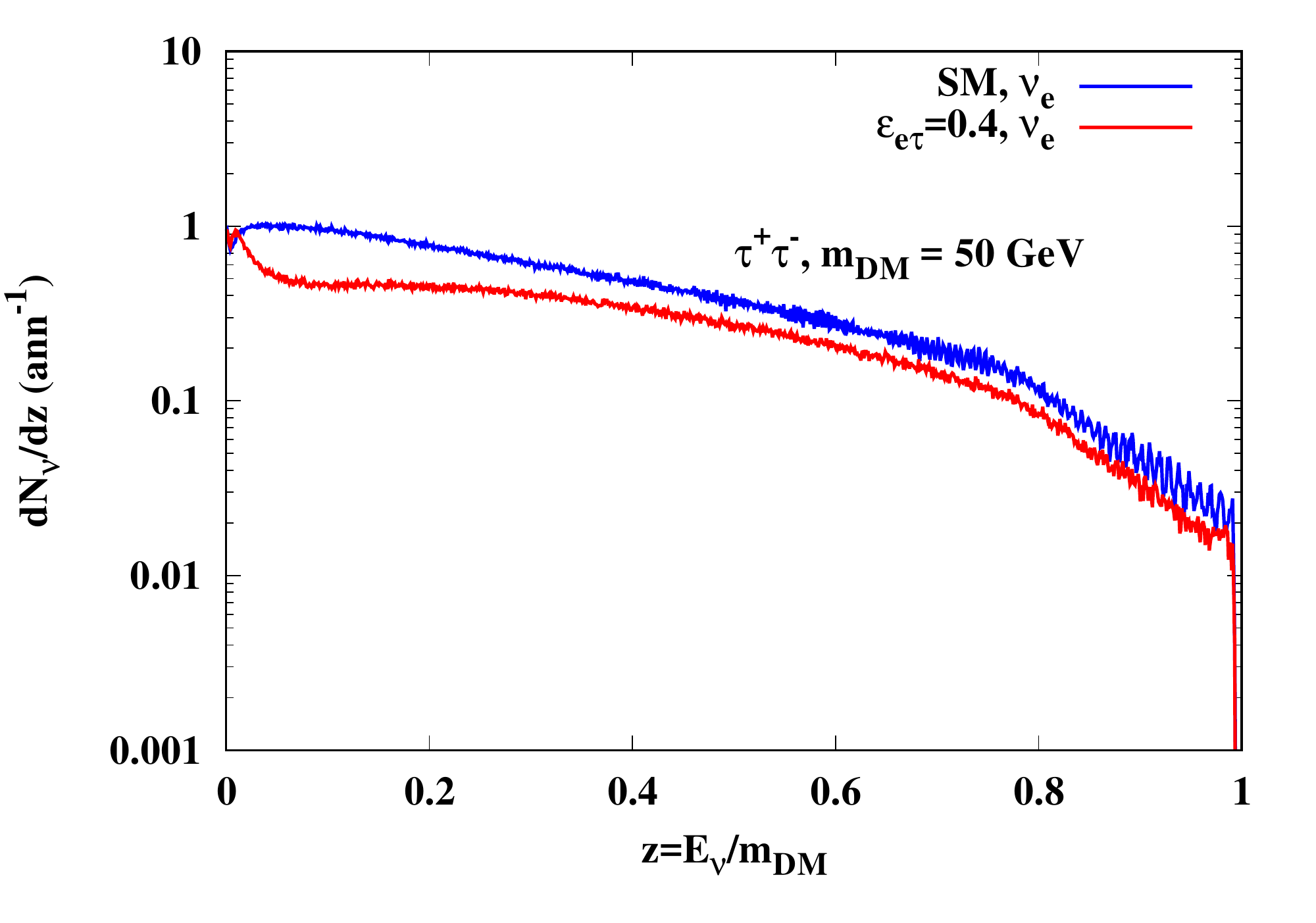}}
\end{picture}
\caption{\label{tau_etau_inv_nue} The same as in Fig.~\ref{tau_tautau}
  but for electron neutrinos,  $\e_{e\tau} = 0.4$ and IH.}
\end{figure}

\begin{figure}[!htb]
\begin{picture}(300,190)(0,40)
\put(260,130){\includegraphics[angle=-90,width=0.31\textwidth]{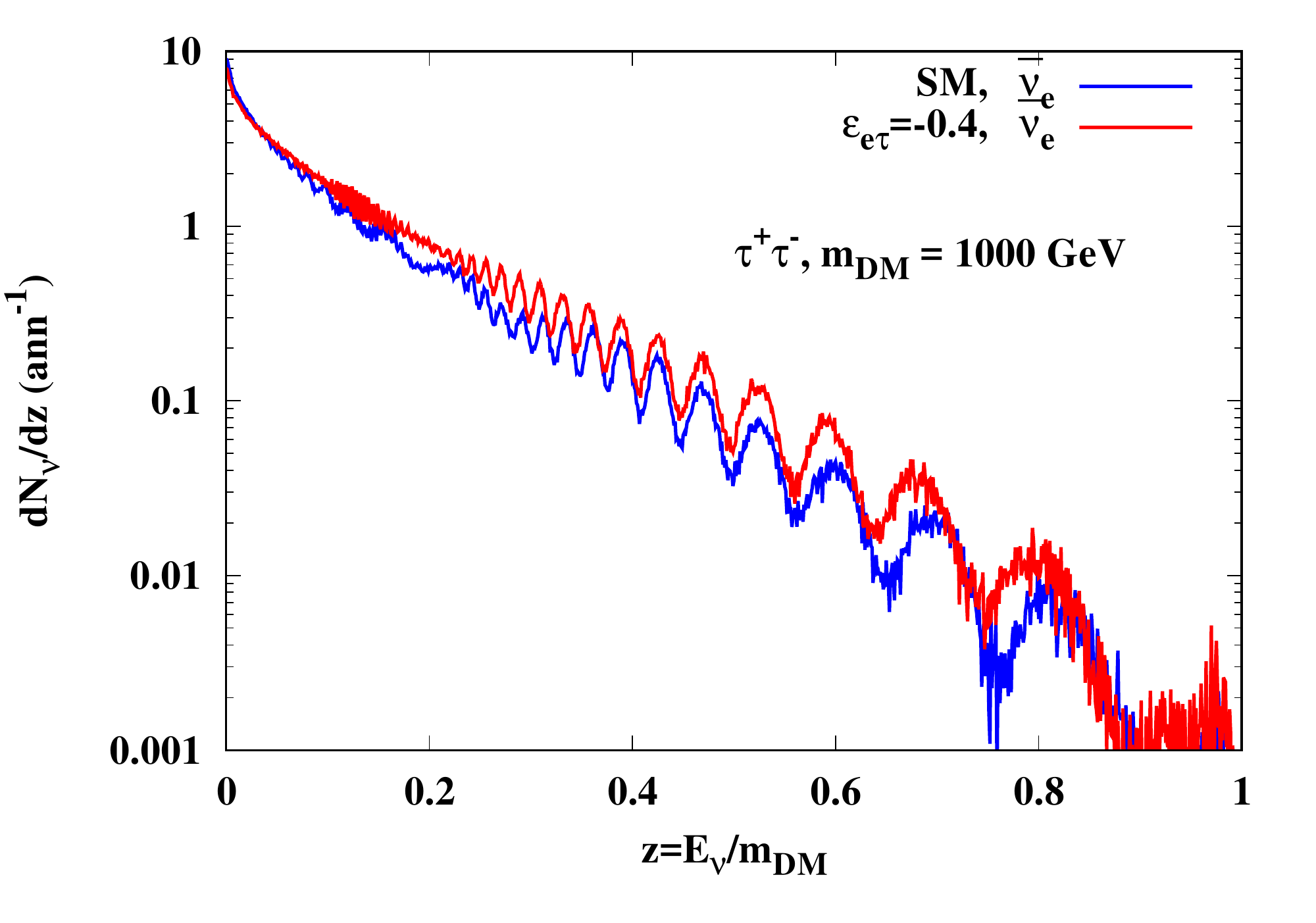}}
\put(260,240){\includegraphics[angle=-90,width=0.31\textwidth]{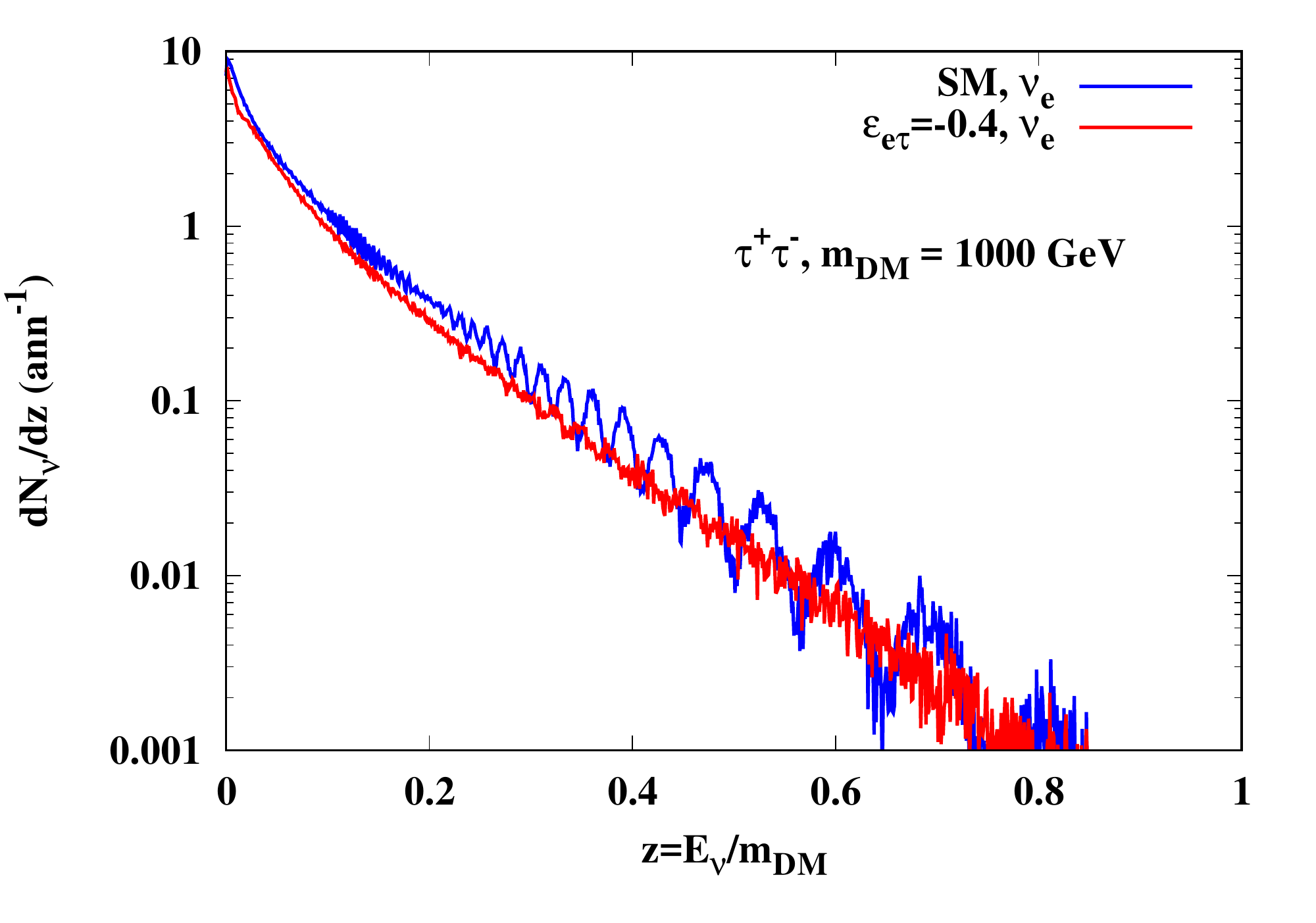}}
\put(130,130){\includegraphics[angle=-90,width=0.31\textwidth]{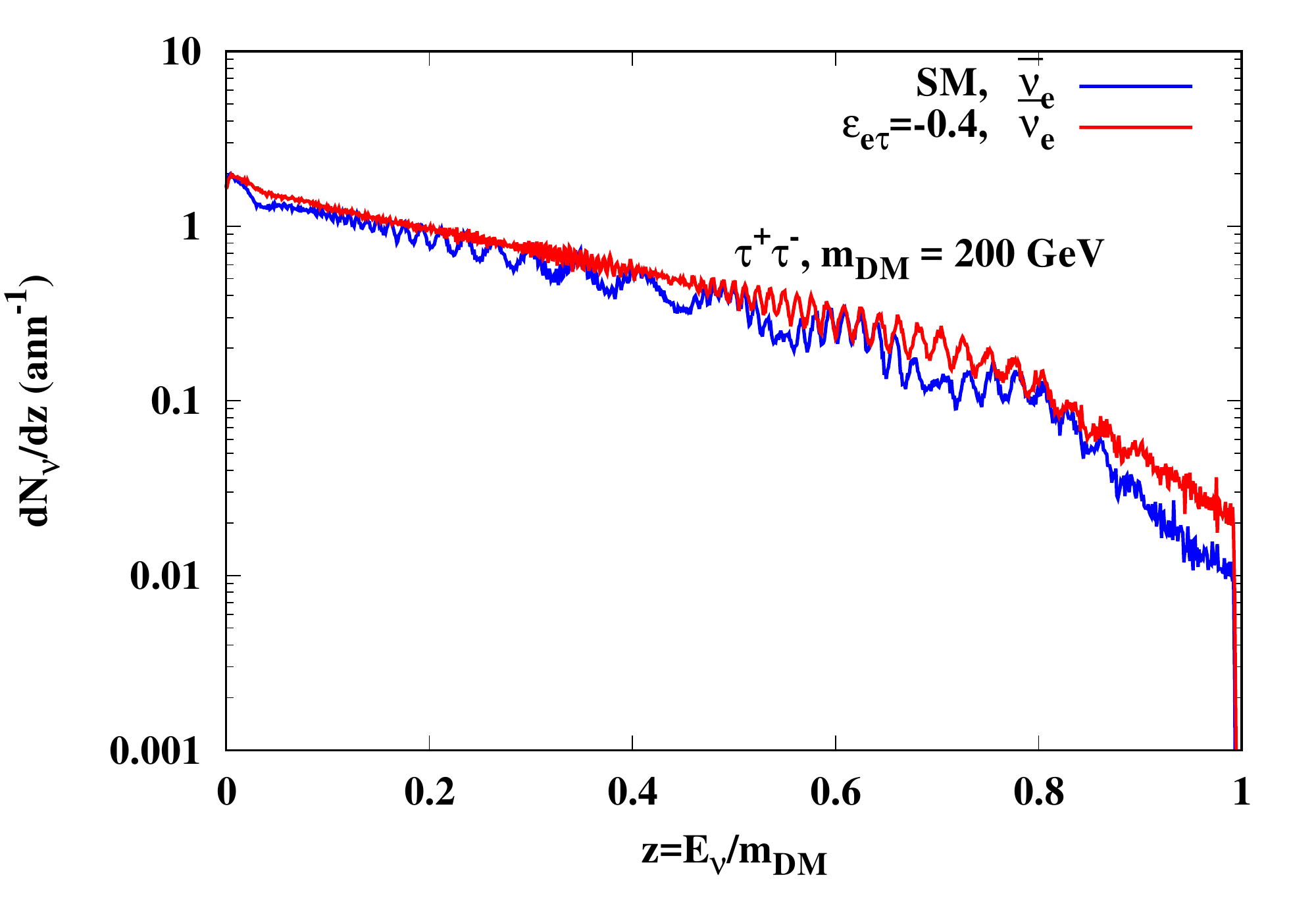}}
\put(130,240){\includegraphics[angle=-90,width=0.31\textwidth]{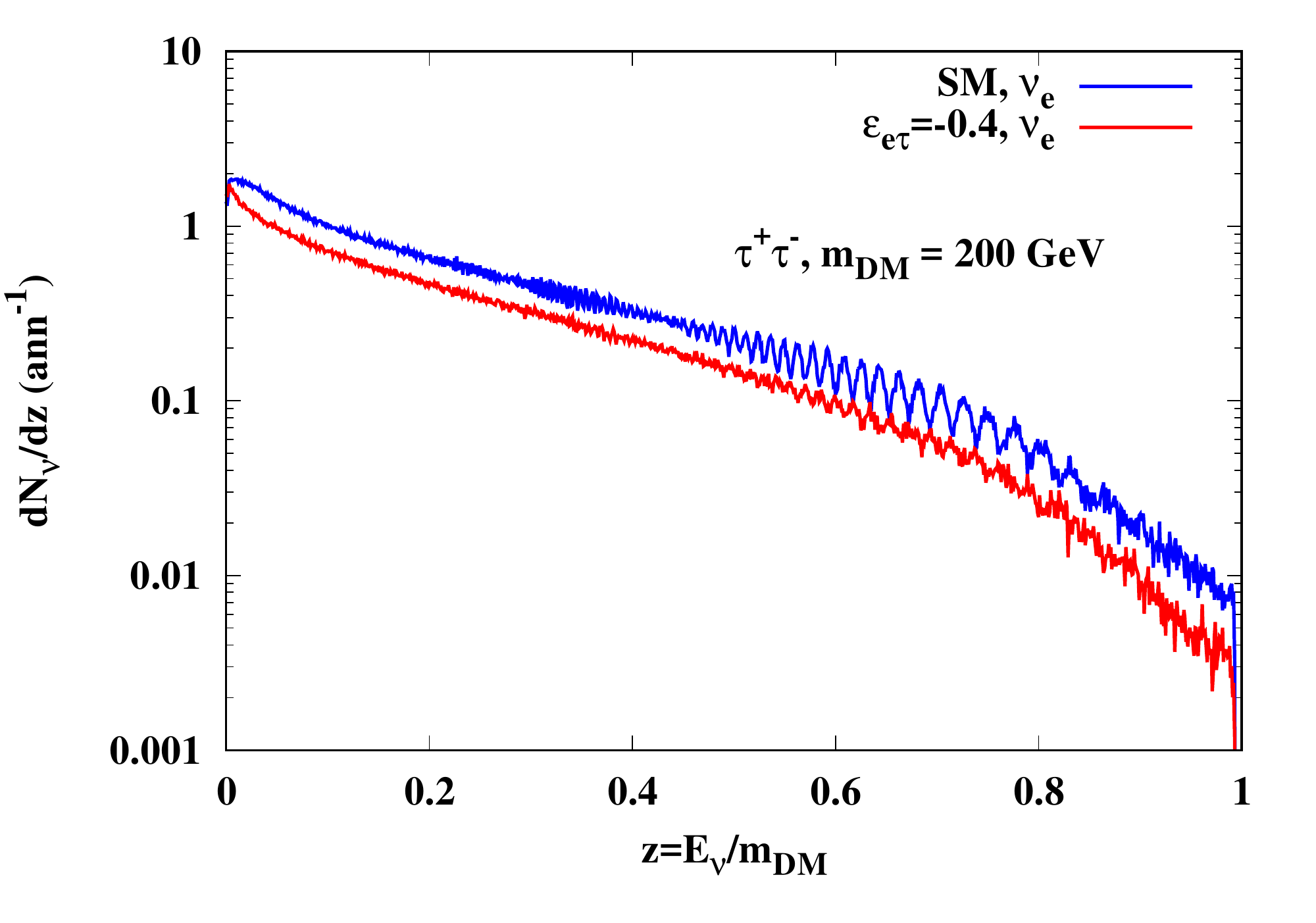}}
\put(0,130){\includegraphics[angle=-90,width=0.31\textwidth]{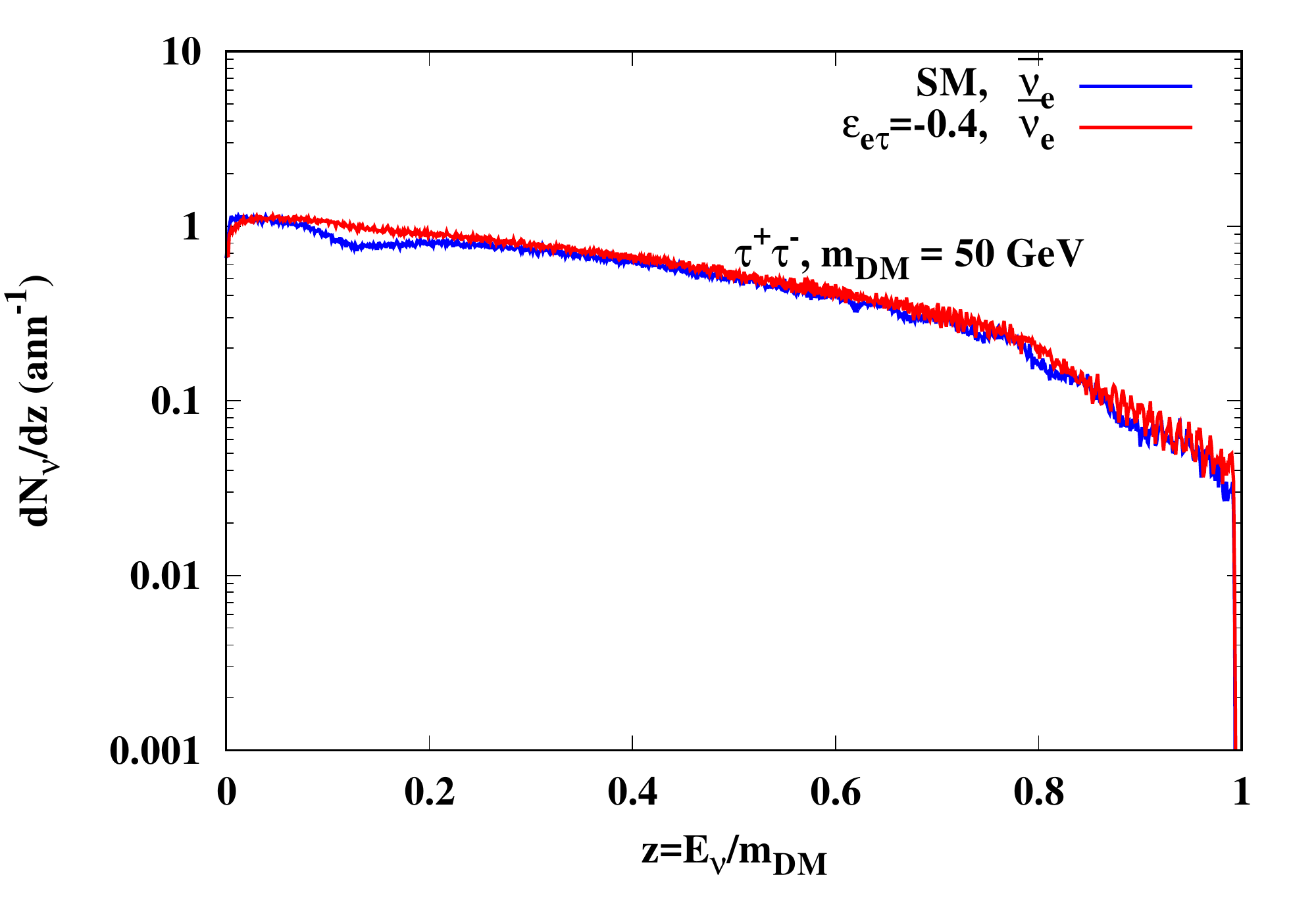}}
\put(0,240){\includegraphics[angle=-90,width=0.31\textwidth]{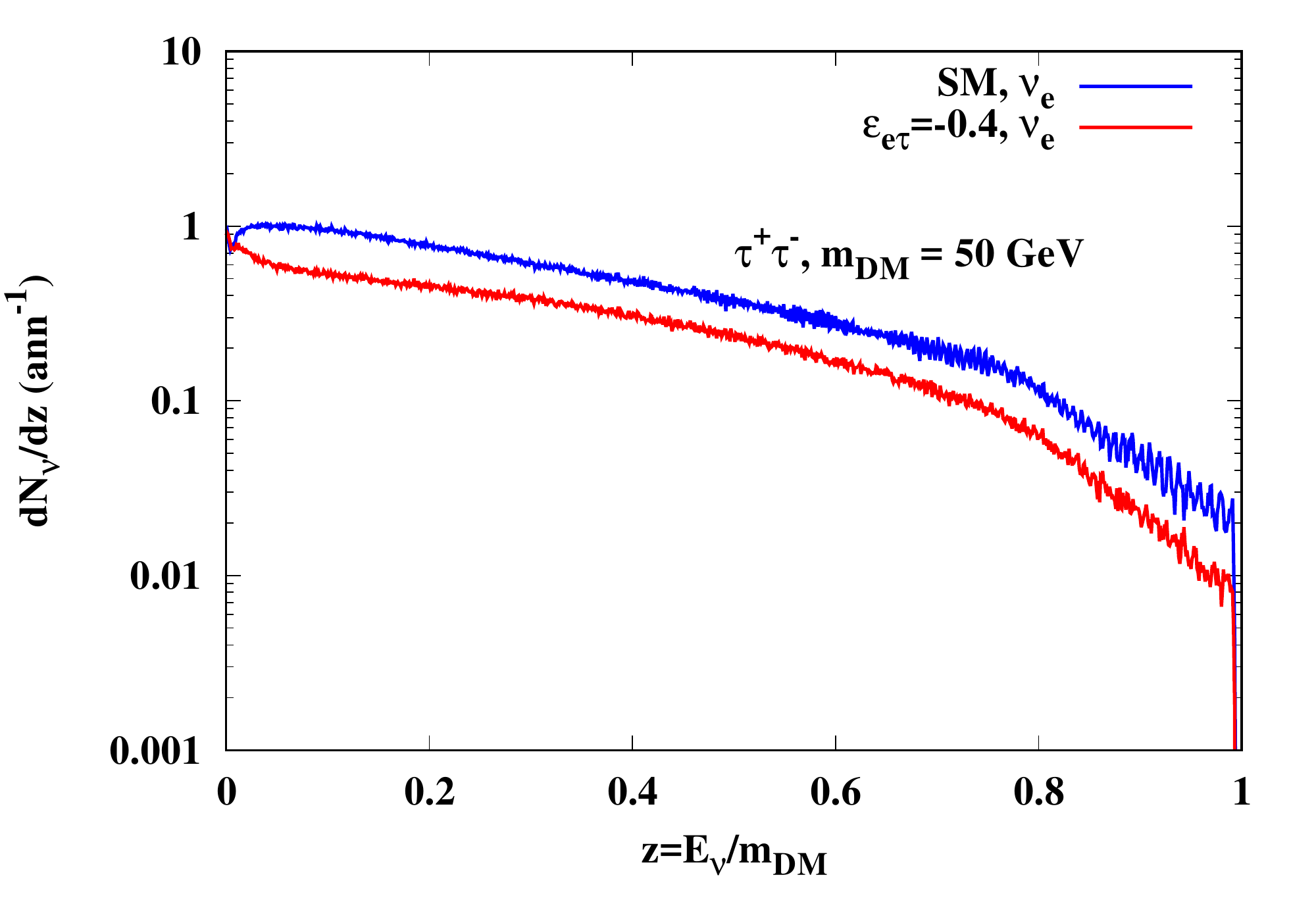}}
\end{picture}
\caption{\label{tau_etau_inv_sign_nue} The same as in Fig.~\ref{tau_tautau}
  but for electron neutrinos, $\e_{e\tau} = -0.4$ and IH.}
\end{figure}

\begin{figure}[!htb]
\begin{picture}(300,190)(0,40)
\put(260,130){\includegraphics[angle=-90,width=0.31\textwidth]{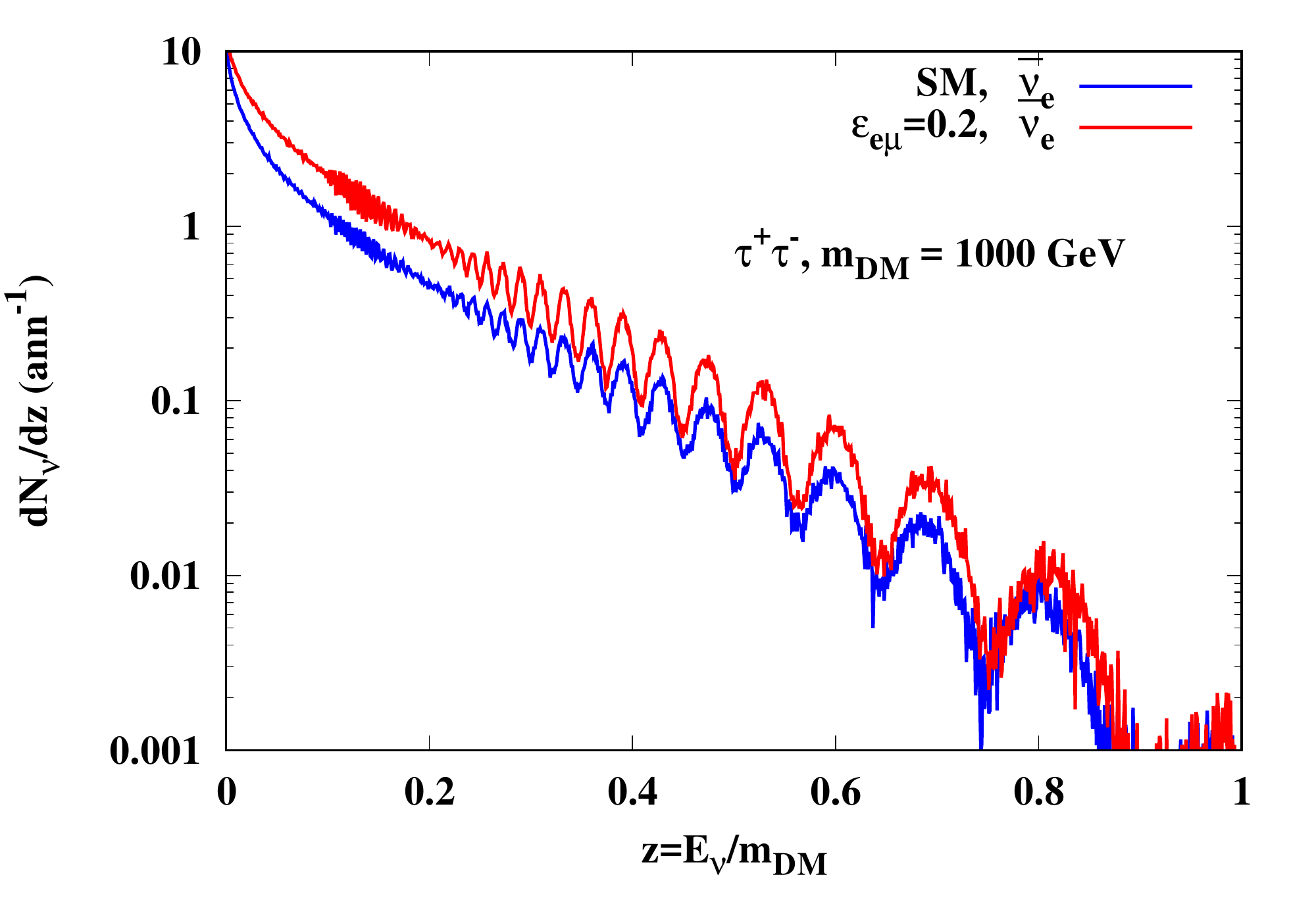}}
\put(260,240){\includegraphics[angle=-90,width=0.31\textwidth]{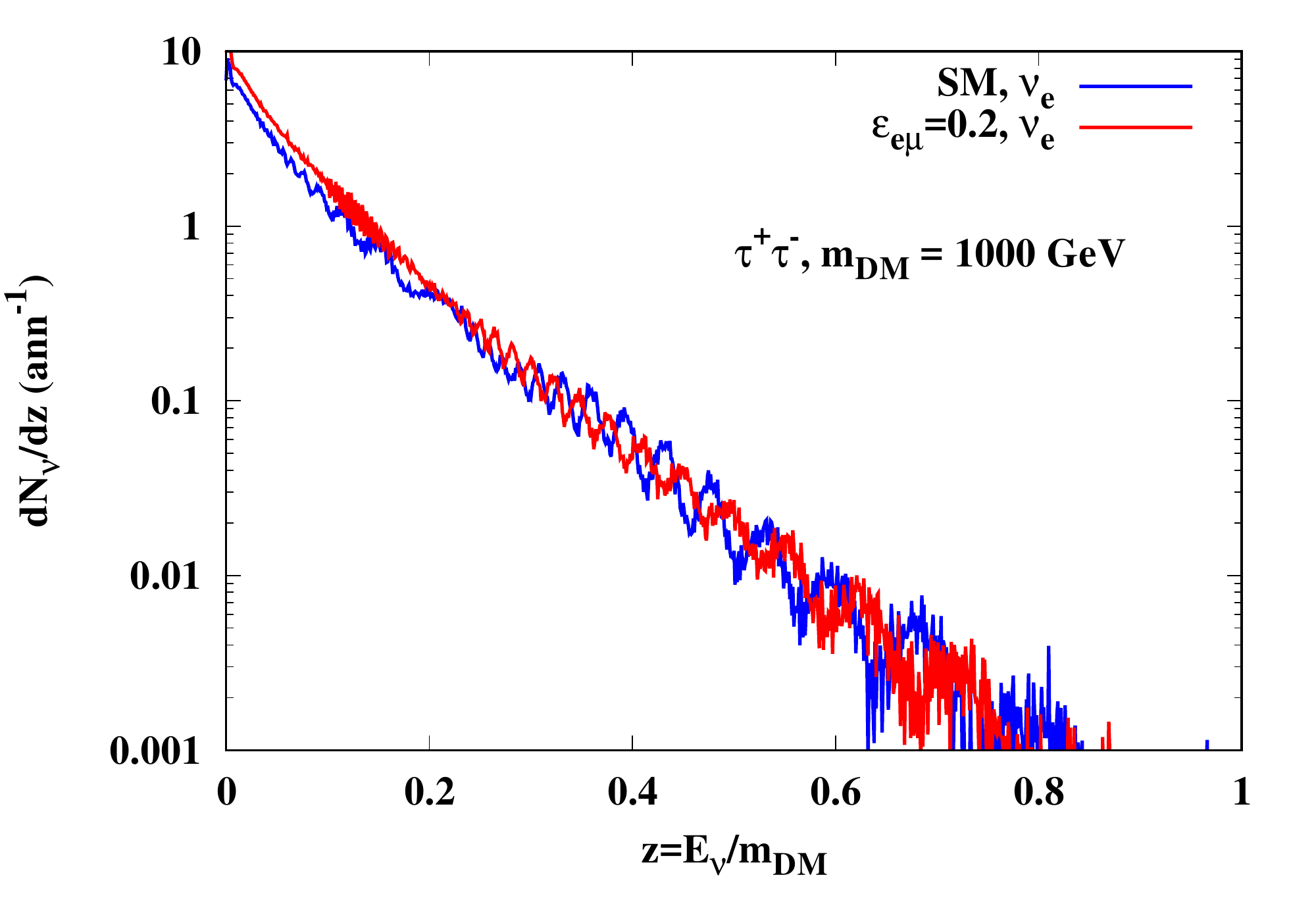}}
\put(130,130){\includegraphics[angle=-90,width=0.31\textwidth]{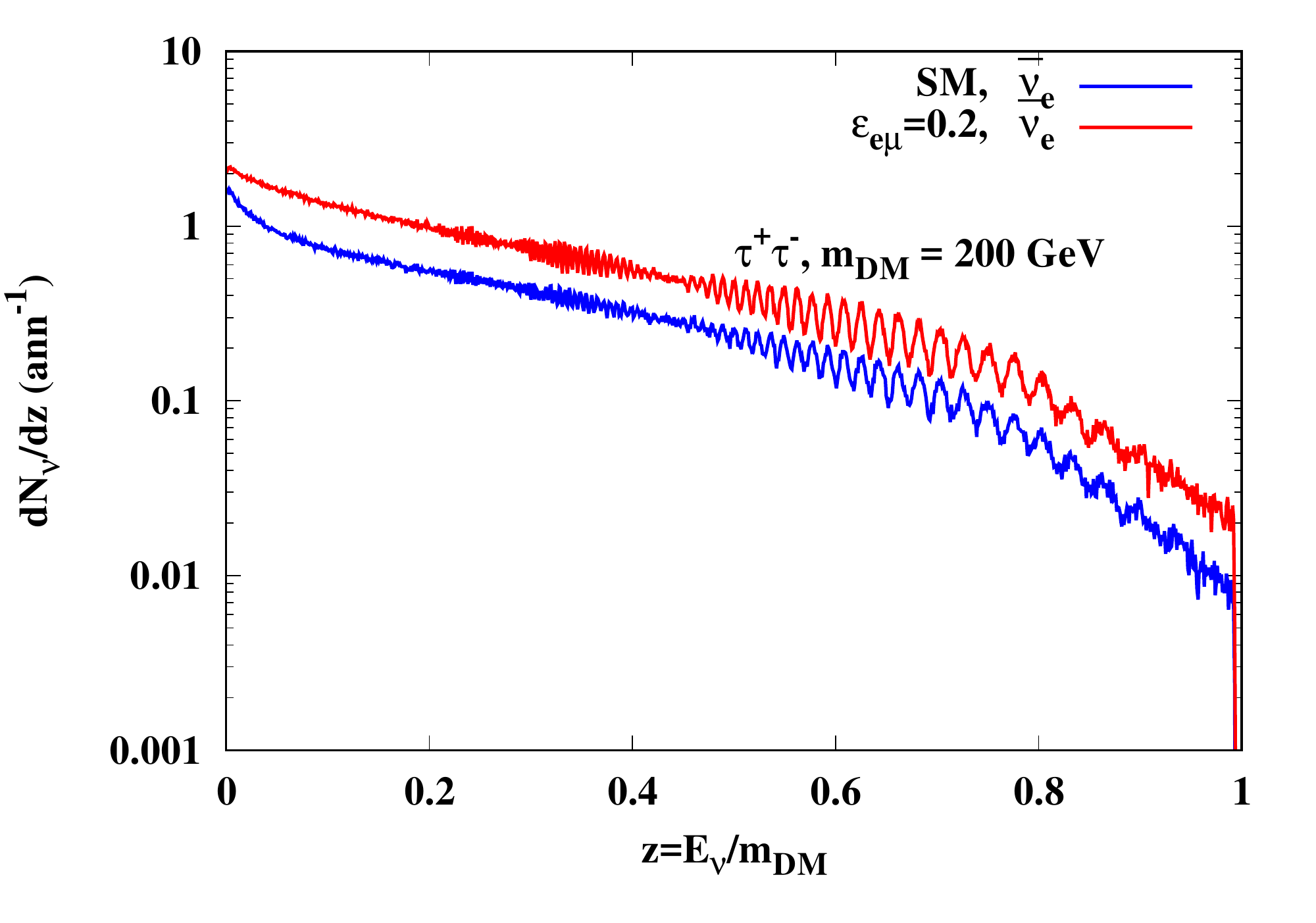}}
\put(130,240){\includegraphics[angle=-90,width=0.31\textwidth]{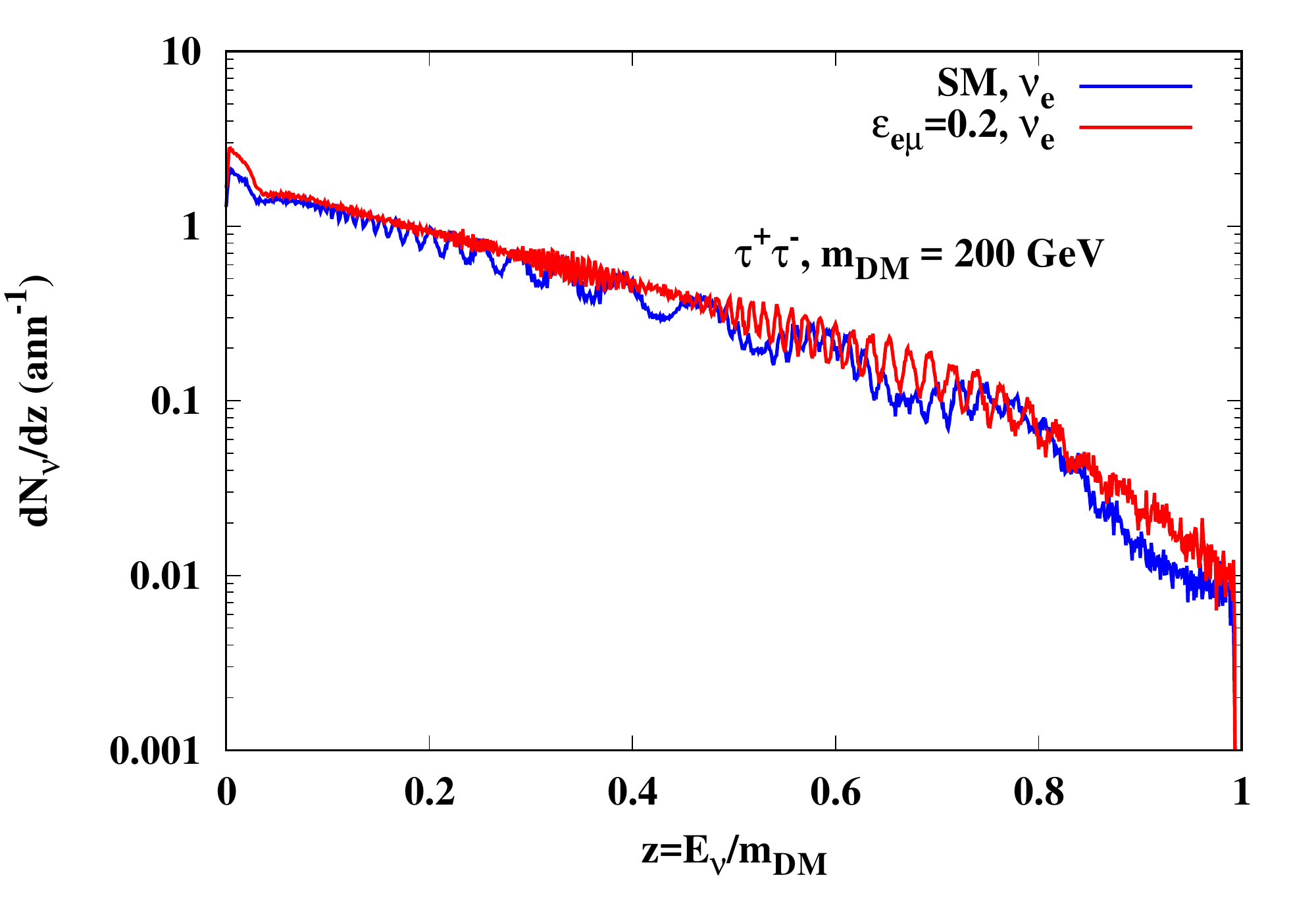}}
\put(0,130){\includegraphics[angle=-90,width=0.31\textwidth]{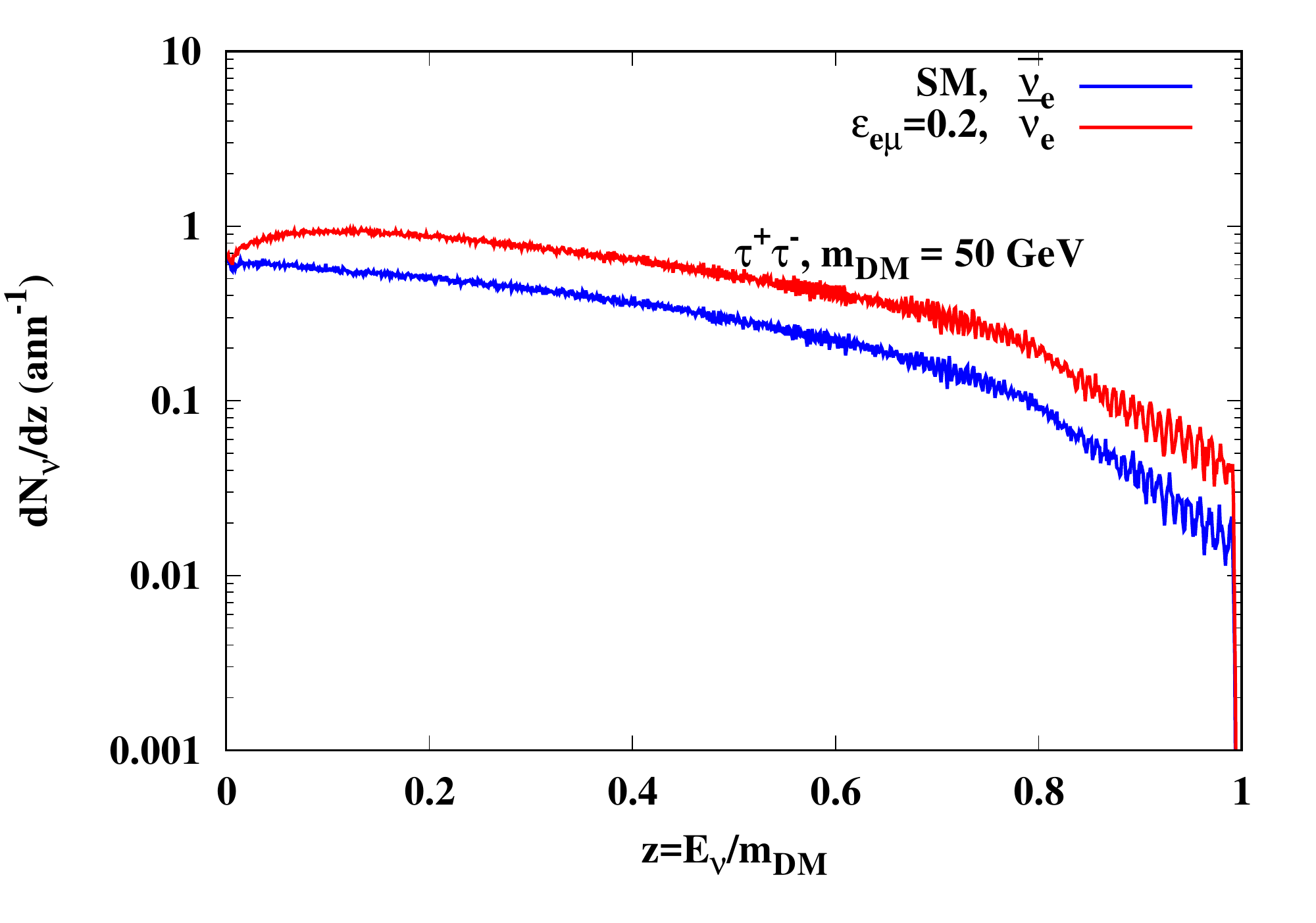}}
\put(0,240){\includegraphics[angle=-90,width=0.31\textwidth]{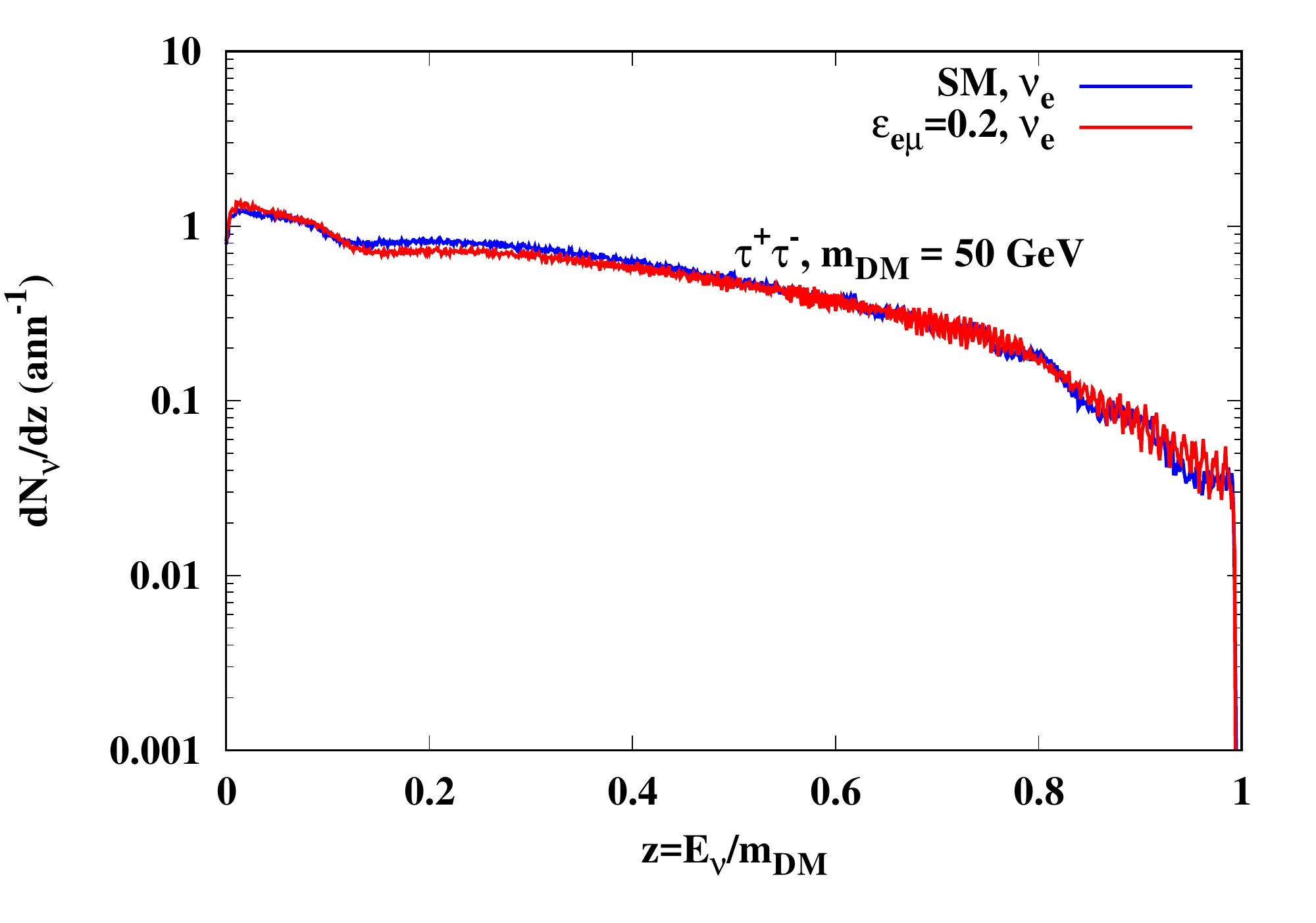}}
\end{picture}
\caption{\label{tau_emu_nue} The same as in Fig.~\ref{tau_tautau}
  but for electron neutrinos, $\e_{e\mu} = 0.2$ and NH.}
\end{figure}

\begin{figure}[!htb]
\begin{picture}(300,190)(0,40)
\put(260,130){\includegraphics[angle=-90,width=0.31\textwidth]{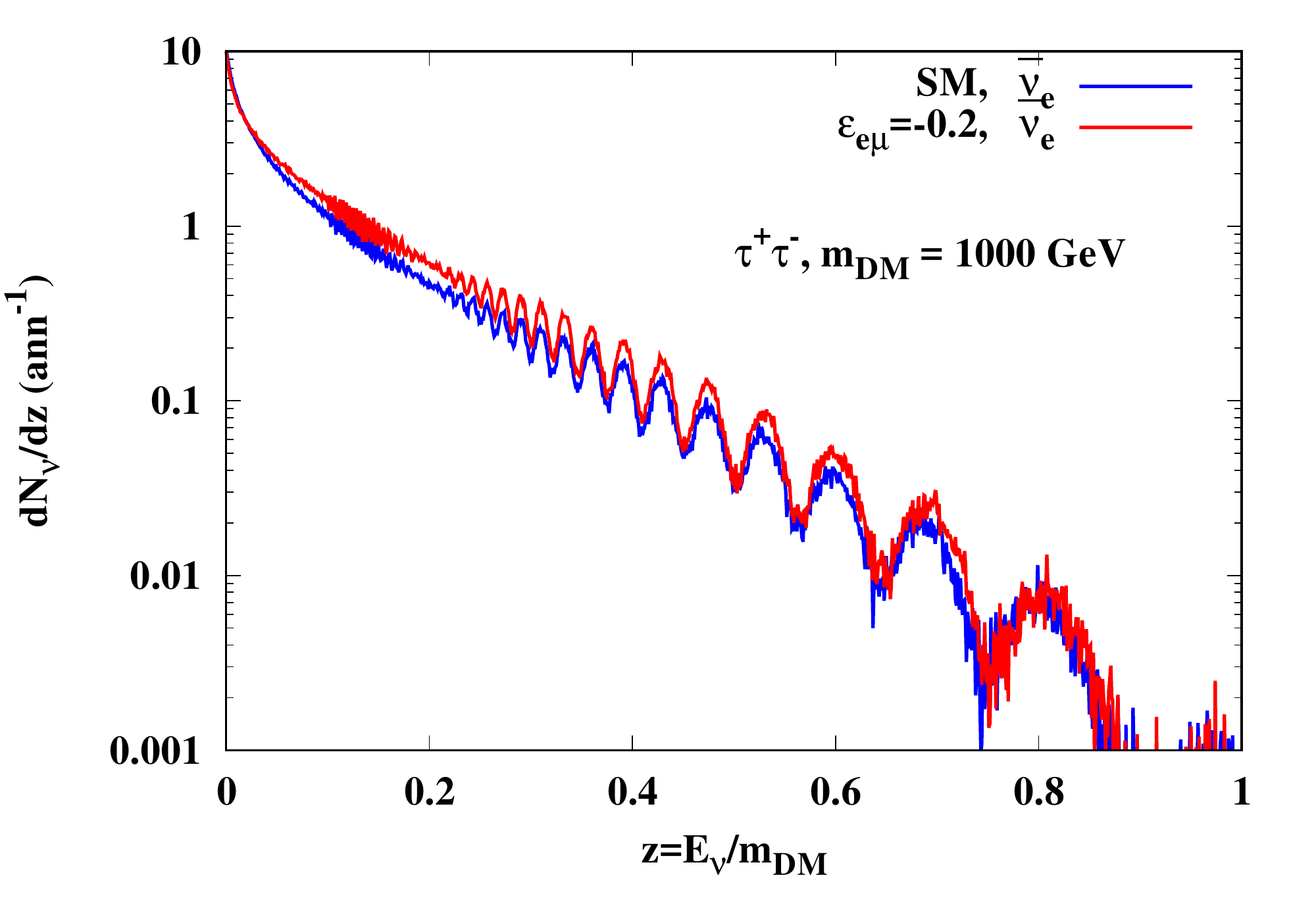}}
\put(260,240){\includegraphics[angle=-90,width=0.31\textwidth]{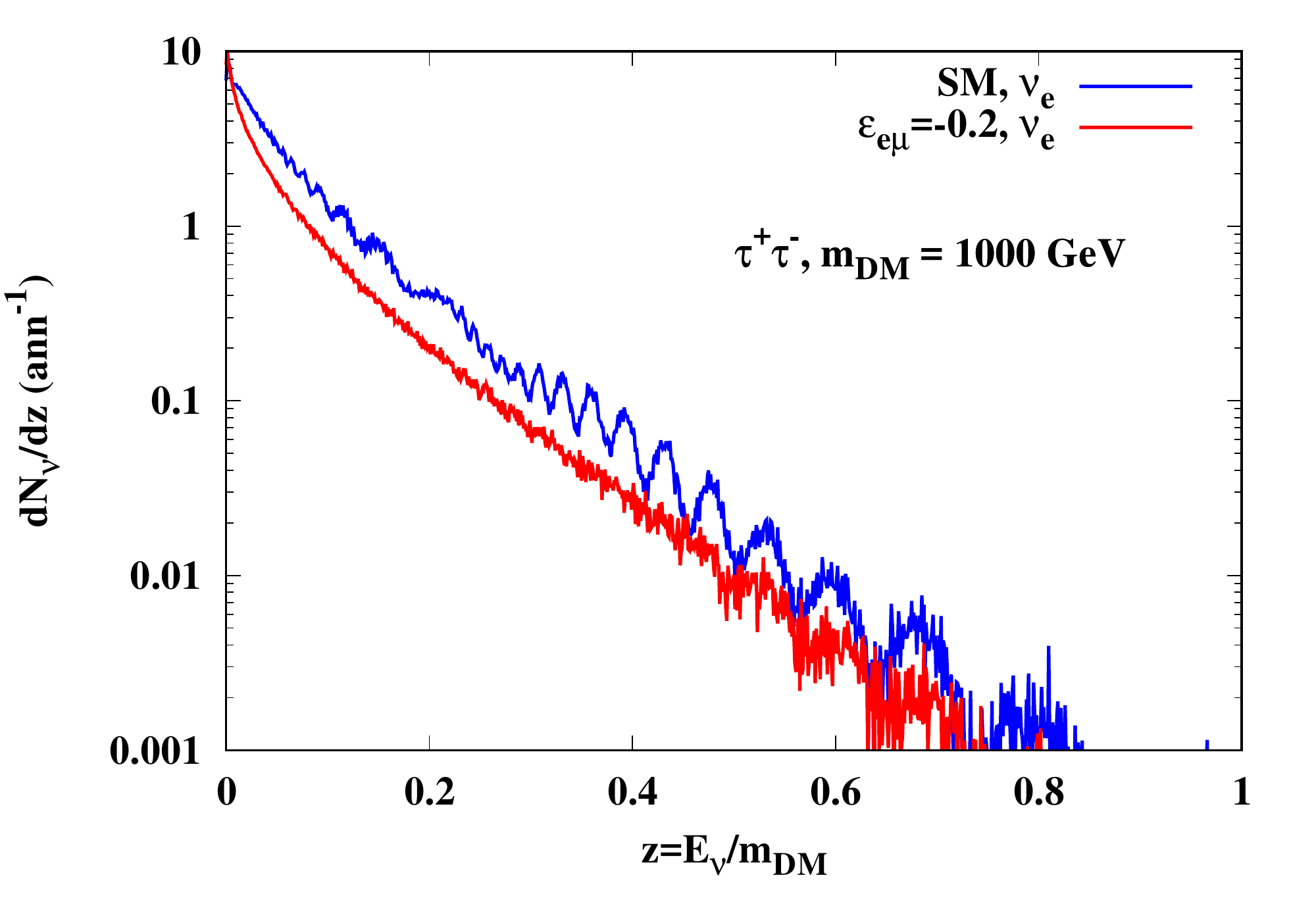}}
\put(130,130){\includegraphics[angle=-90,width=0.31\textwidth]{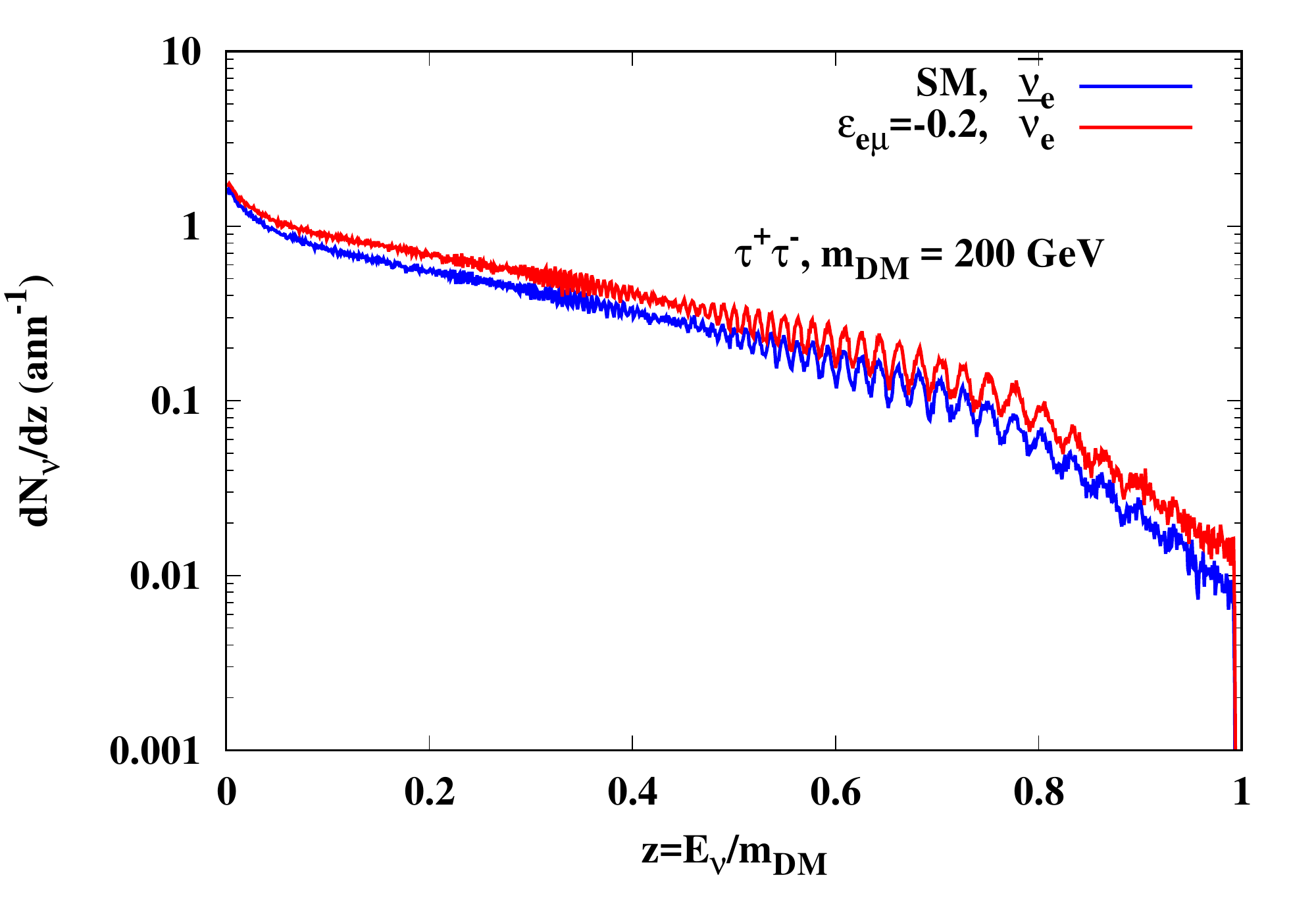}}
\put(130,240){\includegraphics[angle=-90,width=0.31\textwidth]{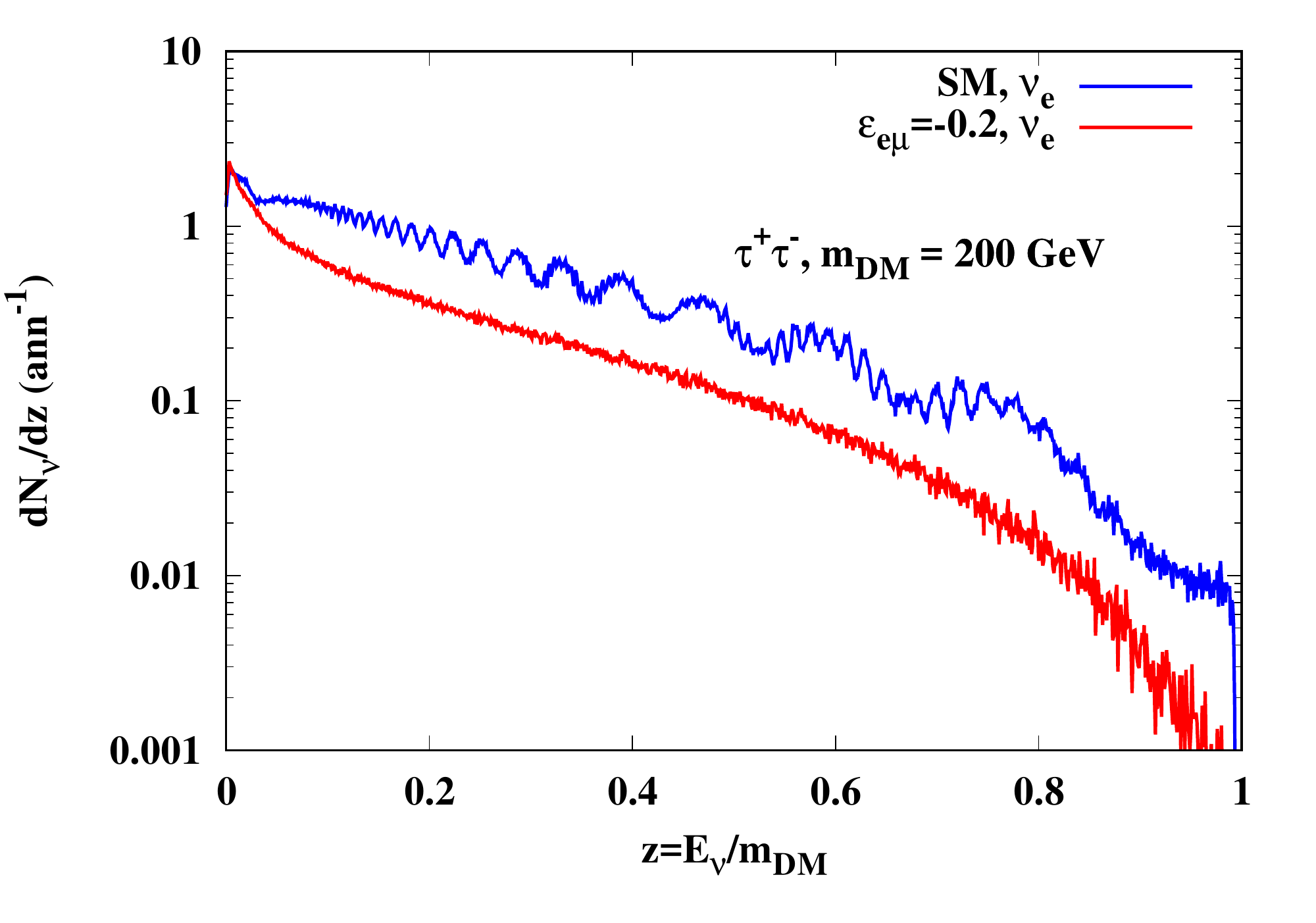}}
\put(0,130){\includegraphics[angle=-90,width=0.31\textwidth]{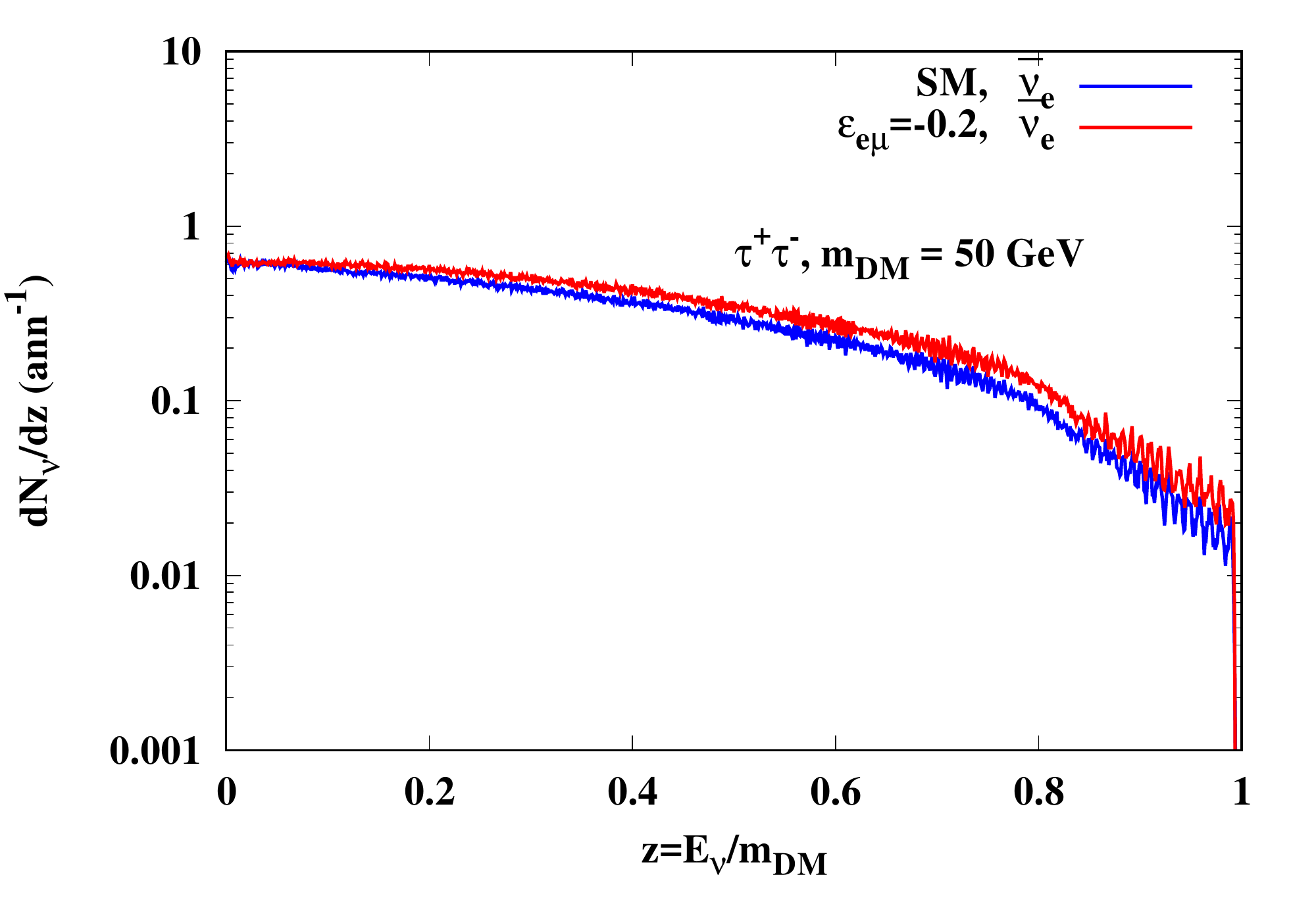}}
\put(0,240){\includegraphics[angle=-90,width=0.31\textwidth]{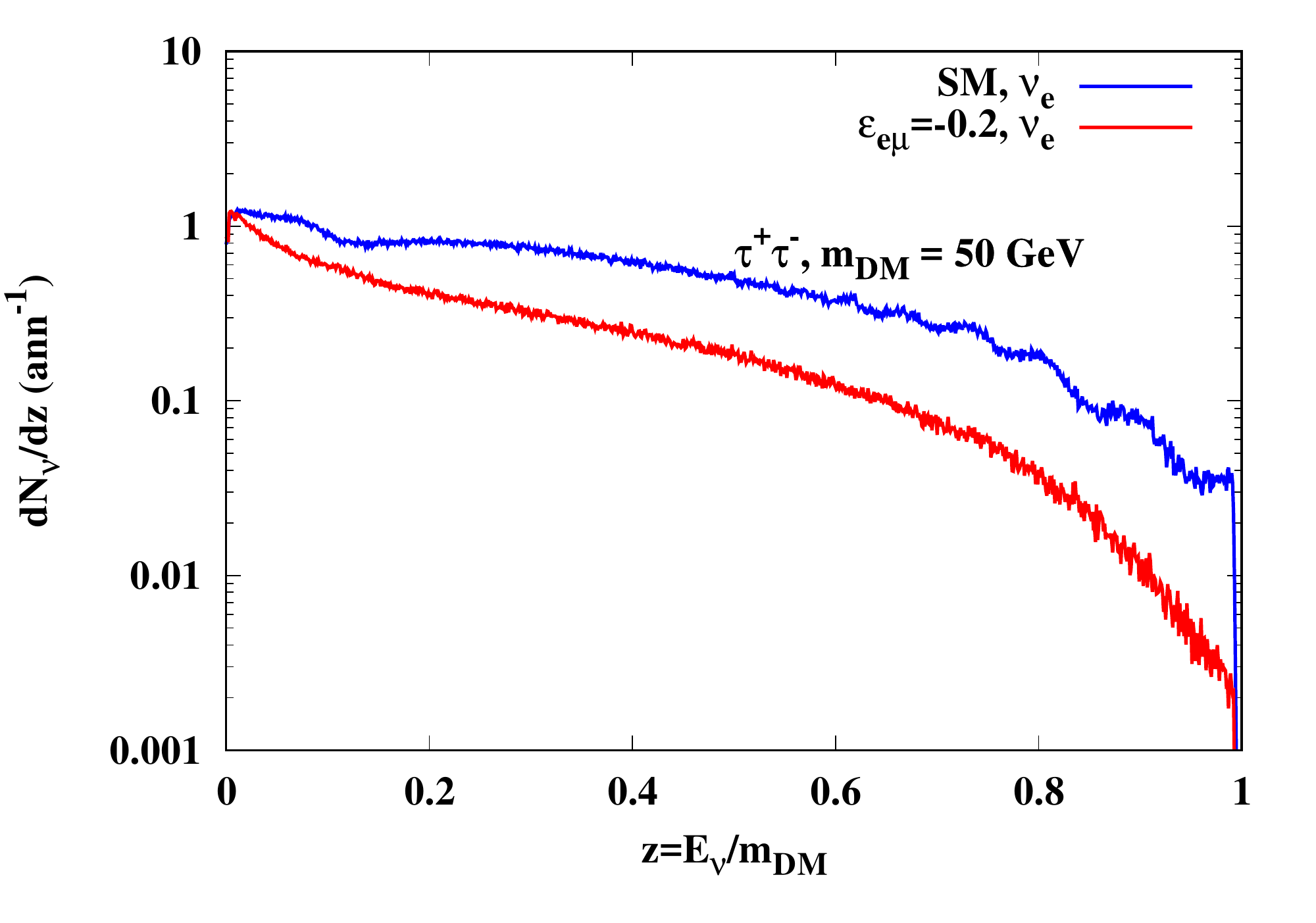}}
\end{picture}
\caption{\label{tau_emu_sign_nue} The same as in Fig.~\ref{tau_tautau}
  but for electron neutrinos, $\e_{e\mu} = -0.2$ and NH.}
\end{figure}

Let us note that directions in which NSI parameters drive the spectra
for neutrino and antineutrino are not necessarily opposite, as can be
seen in the above figures. Still the impact of the matter NSI on the
observable~\eqref{eq:4.3} is in general smaller than that of on the
neutrino and antineutrino fluxes separately. For instance, the ratio
of neutrino flux for $\e_{\tau\tau}=0.03$ and that of for no-NSI case
is about 0.5 for $\tau^+\tau^-$ channel and inverted neutrino mass
hierarchy. This can be important for experiments capable of
distinguishing between muon neutrino and antineutrino, like the Iron
Calorimeter (ICAL) detector~\cite{Kumar:2017sdq}.
Even larger deviations are found for the ratios of
electron neutrinos. In Figs.~\ref{tau_etau_inv_nue}--\ref{tau_emu_sign_nue}
we show selected results for electron neutrino and antineutrino energy
spectra for non-zero flavor changing matter NSI parameters. The cases
of $\e_{e\tau}=0.4$ and $\e_{e\tau}=-0.4$ are presented in
Figs.~\ref{tau_etau_inv_nue} and~\ref{tau_etau_inv_sign_nue} with
inverted neutrino mass hierarchy. We see that the deviation of
electron neutrino flux can be considerable and amount to a factor of
2. In Figs.~\ref{tau_emu_nue} and~\ref{tau_emu_sign_nue} we plot
electron neutrino energy spectra for $\e_{e\mu}=0.2$ and
$\e_{e\mu}=-0.2$, respectively, for  normal 
mass hierarchy. In this case ratio of the electron neutrino fluxes can
reach values about 4--5, see upper left plot on
Fig.~\ref{tau_emu_sign_nue}. Thus, NSI may considerably affect also
all-flavor searches for neutrino signal from dark matter annihilation
in the Sun performed by IceCube~\cite{Aartsen:2017mnf}.

\section{Conclusions}
In this paper we perform an analysis of possible influence of the
non-standard neutrino interactions on neutrino signal from dark matter
annihilations in the Sun. Namely, we study the influence of nonzero
$\e_{\alpha\beta}$ NSI
parameters on oscillations of GeV scale neutrinos in the Sun and
in the Earth. As an example we take experimentally allowed benchmark
values for these parameters and performed numerical analysis of
oscillations of monochromatic neutrinos in the Sun and the Earth. In
this simplified study we neglect interactions of neutrinos in the
matter of the Sun and the Earth and supply it with a simplified
analytical analysis. Next we perform full Monte-Carlo simulation of
neutrino signal from dark matter annihilations in the Sun with NSI
taking into account neutrino interactions for realistic dark matter annihilation
channels. We estimate the ratio of the muon track event rates with
and without NSI effect and find that the deviations can reach at
maximum 30\%
level for $\tau^+\tau^-$ annihilation channel and  15\% for for
$W^+W^-$  and $b\bar{b}$ channels. Besides we find that
electron neutrino flux from dark matter annihilations in the Sun can
be changed by a factor of few for non-zero flavor changing NSI
parameters $\e_{e\tau}$ and $\e_{e\mu}$. 
In a sense the results presented
in Fig.~\ref{rate} can be considered as a theoretical uncertainty to
predictions of the neutrino signal from dark matter annihilation in
the Sun related to the lack of knowledge about neutrino interactions
with the matter. As a consequence, presence of NSI affect upper limits
on dark matter annihilation rate and on elastic cross section of dark
matter particle scattering with nucleons. Still our analysis
  reveals that present experimental bounds are robust in the present
  of NSI in considerable part of their allowed parameter space. For
  instance, this is true with non-zero $\e_{\mu\tau}$ for all studied
  annihilation channels and with non-zero $\e_{e\tau}$ for $W^+W^-$
  channel. 

Finally, let us note again that the analysis performed in the paper is 
simplified in several aspects. Firstly, we consider a single non-zero
matter NSI parameter in Eq.~\eqref{eq:1:6} at a time and set all
CP-violating phases to zero. At the same time
interference between effects of several non-zero 
$\e_{\alpha\beta}$ may play an important role. Second, we assume that
the effective NSI parameters in~\eqref{eq:1:6} are the same for the
Sun and the Earth. This situation is realized if the parameters
$\e_{\alpha\beta}^{fP}$ in Eq.~\eqref{eq:1:2} are non-zero for the
case of 
electron and vanish for quarks. In more general case when neutrino can
interact with quarks in Eq.~\eqref{eq:1:2} the matter NSI parameters
$\e_{\alpha\beta}$ become not only different for the Sun and the Earth
but also become position-dependent in the case of the Sun. This may
produce new evolution patterns which have not been captured by the
present analysis.

\paragraph*{Acknowledgements}
The works is supported by the RSF grant 17-12-01547. The numerical
part of the work was performed on Calculational Cluster of the Theory
Division of INR RAS.

\bibliographystyle{JCAP-hyper}
\bibliography{nsi_sun_jcap}

\end{document}